\documentclass[letter]{ar-1col_WFIRST-HLS}

\usepackage{pdfpages}
\usepackage{amssymb}
\usepackage{natbib}
\usepackage{array}

\definecolor{pink_loc}{cmyk}{0, 0.7808, 0.4429, 0.1412}
\definecolor{coralpink}{rgb}{0.97, 0.51, 0.47}
\definecolor{carminepink}{rgb}{0.92, 0.3, 0.26}
\definecolor{burntsienna}{rgb}{0.91, 0.45, 0.32}
\definecolor{bittersweet}{rgb}{1.0, 0.44, 0.37}
\definecolor{alizarin}{cmyk}{0,1.0,1.0,0.30}

\usepackage[colorlinks=true,
        urlcolor=alizarin,
        citecolor=alizarin,
        linkcolor=alizarin,
        linkcolor=alizarin,
        filecolor=alizarin,
        setpagesize=false]{hyperref}     

\newcommand  \beq    {\begin{equation}}
\newcommand  \cm     {{\rm \,cm}}
\newcommand  \eeq    {\end{equation}}
\newcommand  \gtsim  {\lower.5ex\hbox{$\; \buildrel > \over \sim \;$}}
\newcommand  \ltsim  {\lower.5ex\hbox{$\; \buildrel < \over \sim \;$}}

\def\ben{\begin{enumerate}}
\def\een{\end{enumerate}}
\def\bi{\begin{itemize}}
\def\ei{\end{itemize}}
\def\be{\begin{equation}}
\def\ee{\end{equation}}
\def\bea{\begin{eqnarray}}
\def\eea{\end{eqnarray}}

\def\deg{\,{\rm deg}}
\def\erg{\,{\rm erg}}
\def\cm{\,{\rm cm}}

\def\CoLi{\texttt{CosmoLike }}

\begin{document}

\thispagestyle{empty}
\includepdf[offset = 70 -80,height=11.3in]{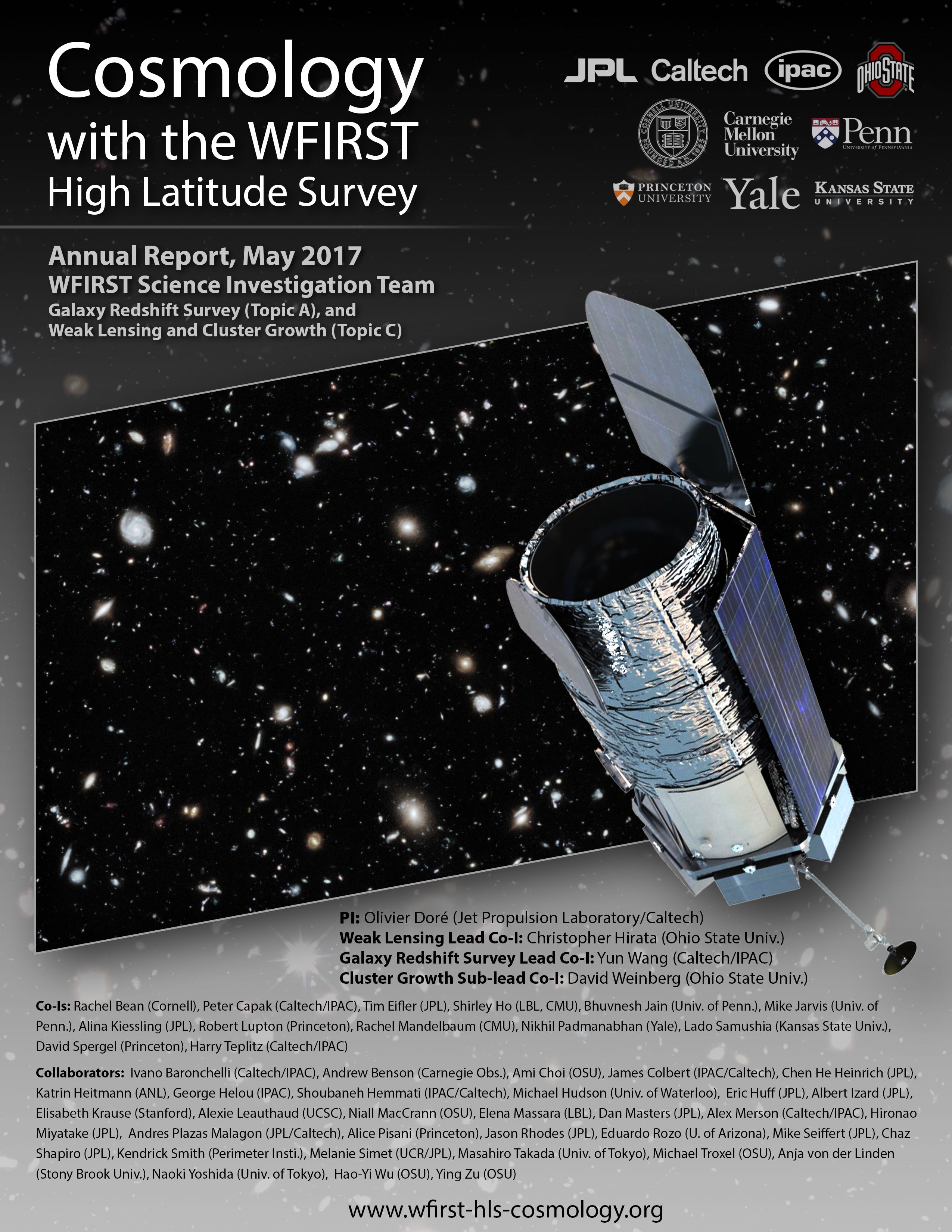}

\cleardoublepage

\markboth{Cosmology with the WFIRST HLS Survey Science Investigation Team}{Annual Report 2017}

\title{Cosmology with the WFIRST High Latitude Survey \emph{Science Investigation Team}\\
Annual Report 2017}

\pagenumbering{roman}
\def\reff@jnl#1{{\rm#1\/}}

\newcommand{\jcap}{{\rm JCAP}}
\def\aj{\reff@jnl{AJ}}                  
\def\araa{\reff@jnl{ARA\&A}}            
\def\apj{\reff@jnl{ApJ}}                        
\def\apjl{\reff@jnl{ApJ}}               
\def\apjs{\reff@jnl{ApJS}}              
\def\apss{\reff@jnl{Ap\&SS}}            
\def\aap{\reff@jnl{A\&A}}               
\def\aapr{\reff@jnl{A\&A~Rev.}}         
\def\aaps{\reff@jnl{A\&AS}}             
\def\baas{\reff@jnl{BAAS}}              
\def\jrasc{\reff@jnl{JRASC}}            
\def\memras{\reff@jnl{MmRAS}}           
\def\mnras{\reff@jnl{MNRAS}}            
\def\na{\reff@jnl{New A}}                
\def\physrep{\reff@jnl{Phys.~Rep.}}
\def\pra{\reff@jnl{Phys.~Rev.~A}}         
\def\prb{\reff@jnl{Phys.~Rev.~B}}         
\def\prc{\reff@jnl{Phys.~Rev.~C}}         
\def\prd{\reff@jnl{Phys.~Rev.~D}}         
\def\prl{\reff@jnl{Phys.~Rev.~Lett}}      
\def\pasp{\reff@jnl{PASP}}              
\def\pasj{\reff@jnl{PASJ}}              
\def\rmxaa{\ref@jnl{Rev. Mexicana Astron. Astrofis.}} 
\def\skytel{\reff@jnl{S\&T}}            
\def\solphys{\reff@jnl{Solar~Phys.}}    
\def\sovast{\reff@jnl{Soviet~Ast.}}     
\def\ssr{\reff@jnl{Space~Sci.~Rev.}}     
\def\nat{\reff@jnl{Nature}}             


\maketitle

\newpage

\topskip0pt
\vspace*{\fill}
\begin{center}
\small{NOTE: The original version of this report was submitted to the WFIRST Project Office on July 14, 2017. Some minor updates have been made in this version.}
\end{center}
\vspace*{\fill}

\newpage

\tableofcontents
\thispagestyle{empty}

\newpage
\pagenumbering{arabic}

\section{Executive Summary}
\label{sec:executive_summary}
%
%

\begin{summary}

Cosmic acceleration is the most surprising cosmological discovery in
many decades.  Even the least exotic explanation of this phenomenon
requires an energetically dominant component of the universe with
properties never previously seen in nature, pervading otherwise
empty space, with an energy density that is many orders of magnitude
lower than naive expectations. More broadly, the origin could derive from a novel, dynamically-evolving type of matter or, instead, signal deviations from General Relativity on the large scales and low densities probed by cosmological tracers. Testing and distinguishing among possible  explanations requires cosmological
measurements of extremely high precision that probe the full history of
cosmic expansion and structure growth and, ideally, compare and contrast matter and relativistic tracers of the gravity potential.
This program is one of the defining objectives of the Wide-Field
Infrared Survey Telescope (WFIRST), as set forth in the {\it New Worlds, New Horizons}
report (NWNH) \cite{NWNH2010}.  The WFIRST mission, as described in the Science
Definition Team (SDT) reports \citep[hereafter SDT13 and SDT15 respectively]{Spergel2013, Spergel2015}, has the ability to improve these
measurements by $1-2$ orders of magnitude compared to the current
state of the art, while simultaneously extending their redshift grasp,
greatly improving control of systematic effects, and taking a unified
approach to multiple probes that provide complementary physical information
and cross-checks of cosmological results.

We described in this document the activities of the Science Investigation Team (SIT) \emph{Cosmology with the High Latitude Survey}.
This team was selected by NASA in December 2015 in order to address
the stringent challenges of the WFIRST dark energy (DE) program through the
Project's formulation phase.  This SIT has elected to address Galaxy Redshift Survey (GRS), Weak Lensing (WL) and Cluster Growth (CL) of the WFIRST Science Investigation Team (SIT) NASA Research Announcement (NRA) with a unified team, because the two investigations are tightly linked at both
the technical level and the theoretical modeling level. Our team thus fully embrace
the fact that the imaging and spectroscopic elements of the High Latitude Survey
(HLS) will be realized as an integrated observing program, and they jointly
impose requirements on instrument and telescope performance, operations, and
data transfer. We also naturally acknowledge that the methods for simulating and interpreting weak lensing and
galaxy clustering observations largely overlap. Many members of our team
have expertise in both areas.

WFIRST is designed to be able to deliver a definitive result on the origin of
cosmic acceleration.  If the growth rate of structure is inconsistent with the
evolution of the Hubble constant, this would be the signature of  the breakdown
of General Relativity on cosmological scales.  If the evolution of the Hubble
constant is consistent with the growth rate of structure but inconsistent with
vacuum energy, then this would imply that dark energy is dynamical.  Either
result would have a profound impact on our understanding of physics.  WFIRST is
not optimized for “Figure of Merit” sensitivity but for control of systematic
uncertainties in the astronomical measurements and for having multiple
techniques each with multiple cross-checks.  Our SIT work focuses on
understanding the potential systematics in the WFIRST dark energy measurements.

In our proposal, we structured our planning around the series of deliverables
described in \S \ref{sec:deliverables}. We will present in this detailed report our progress on
these deliverables and illustrate that we either reached or exceeded our proposed
expected milestones.

Because the development of the science requirements is at the core of our proposed
investigation, we present some broad aspects of our strategy in \S
\ref{sec:reqt_philosophy} before giving a summary of the High Latitude Imaging
Survey (HLIS) and of the HLS Spectroscopic Survey (HLSS) science requirements as we formulated them to support the WFIRST Project Office in \S \ref{sec:wl_gal-clusters} and \S
\ref{sec:gc}. We present our revised cosmological forecasts and associated trade studies in
\S \ref{sec:forecast}. We also address questions of survey operations and
optimization in \S \ref{sec:operation}, our actions towards broad community engagement
in \S \ref{sec:engagement} and discuss in \S \ref{sec:other_contributions} the
other ways in which our SIT supported the WFIRST mission.
\end{summary}

\section{Proposed and Actual Deliverables}
\label{sec:deliverables}

\begin{figure}
\includegraphics[width=0.99\textwidth]{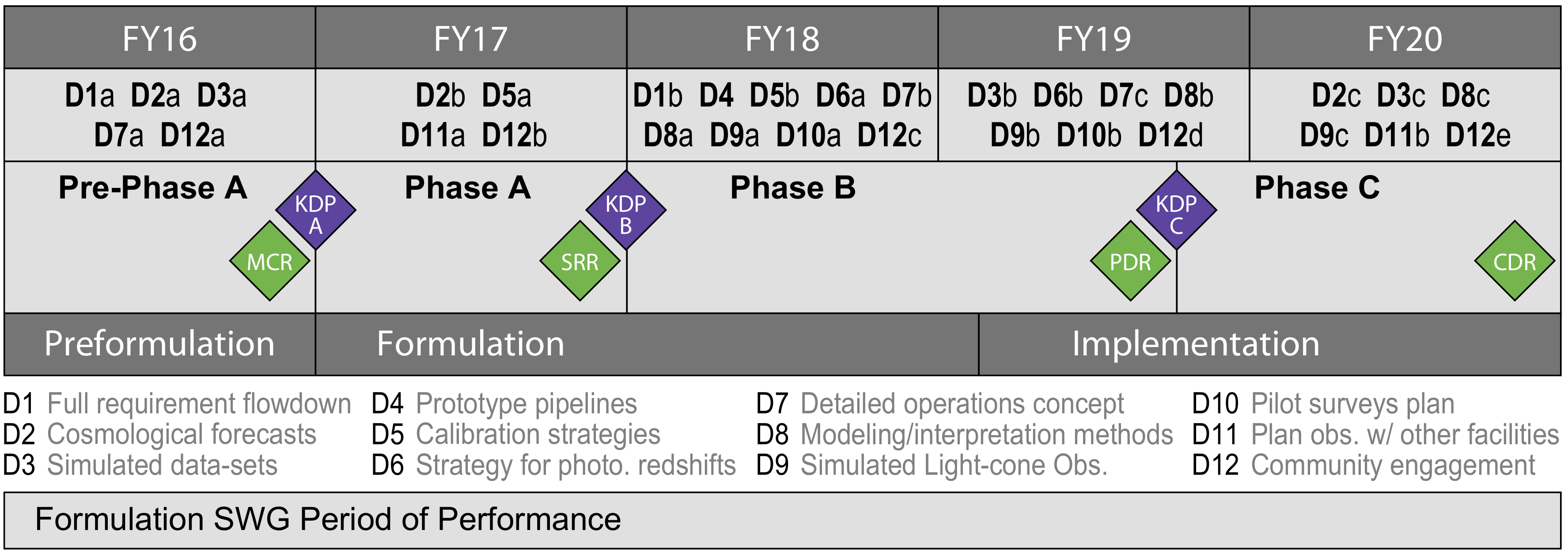}
\caption{Our proposed deliverable schedule in concordance with the WFIRST
project timeline as displayed in the WFIRST SIT call \S 3.2. Deliverables that
are made in multiple stages are labeled (a, b and c) and appear in multiple
years. This figure is taken from our original proposal and the schedule has in
fact shifted since our investigation started on January 2016, thus well into
FY16.} \label{tab:milestones_mgt}
\end{figure}

\begin{summary}
In our proposal, we structured our planning around a series of deliverables
numbered D1-12. We will use throughout this report the same nomenclature and
report on our progress on each of these deliverables when compared to the
proposed calendar visible in Figure~\ref{tab:milestones_mgt}. We will
illustrate that we compare favorably on all deliverables. We give in this section a quick summary but will give more details in the relevant sections. We will also
explicitly reference  the deliverables (D1-12) in the relevant section titles of our report. In the text below, the definition of each deliverables is quoted directly from our proposal and we summarize the progress briefly in italic.
\end{summary}





\paragraph*{(D1) Full requirements flow-down} from the high-level science goals
of the HLS galaxy clustering and weak lensing survey to detailed performance of
the telescope, wide field instrument, software, operations, and data transfer.
\emph{Throughout the year, we delivered three versions of the level 2 science
requirements for both the HLSS (\S \ref{sec:gc}) and the HLIS (\S
\ref{sec:wl_gal-clusters})}.

\paragraph*{(D2) Forecasts of the cosmological performance of the HLS Imaging
and Spectroscopy data sets,} including expected constraints on dark energy,
modified gravity, neutrino masses, and inflation, from analyses that include the
measurement of the location of the Baryon Acoustic Oscillations (BAO),
Redshift-Space Distortions (RSD), galaxy power spectrum and higher order
statistics, cosmic shear, galaxy-galaxy lensing, and cluster demographics. These
forecasts incorporate realistic assessments of observational systematics and
theoretical modeling systematics, and they examine the expected constraints from
different probes individually, in concert with each other, and in concert with
expected constraints from the WFIRST supernova program, CMB experiments, and
other cosmological surveys such as DESI, LSST, and Euclid. We use our
forecasting tools to investigate trades, e.g., the impact of survey or
instrument design choices (area, depth, pixel size, spectral resolution, etc.)
on cosmological performance. \emph{We developed a unique software package (\CoLi) that
enables us to jointly forecast all the WFIRST cosmological probes, including
their covariance (\S \ref{sec:forecast}). We used this framework to conduct trade studies \S\ref{sec:forecast}. We released to the community the
associated WFIRST chains (\S \ref{sec:other_contributions})}.

\paragraph*{(D3) Simulated imaging and spectroscopic data sets} for testing pipeline
performance and evaluating systematic biases --- e.g., from confusion,
noise, and incompleteness in images and spectra, or errors in Point
Spread Function (PSF) determination or shape measurement.
These data sets will be created with varying levels of complexity
in the source catalogs and instrumental effects, to allow isolation
of individual contributions to statistical and systematic uncertainties.
Some of these artificial data sets will be made publicly available,
and some will take the form of data challenges, where the underlying
parameters are initially known only to the creators of the data set,
in the spirit of the Shear Testing Program (STEP) and Gravitational
Lensing Accuracy Test (GREAT) weak lensing data
challenges \cite{Heymans2006, Massey2007, Bridle2010, Kitching2012, Mandelbaum2015}. \emph{We implemented a WFIRST dedicated module in the state-of-the-art simulation image simulation pipeline GalSim and will release it to the community (\S \ref{sec:hlis_image_sim}). We will contributed mock observations to the SOC based image simulation effort for the HLSS}.

\paragraph*{(D4) Proto-type imaging and spectroscopic pipelines}, including weak lensing shape measurement and galaxy redshift measurement, tested against the
above artificial data sets. These proto-type pipelines will provide
building blocks for development of full pipelines during the implementation
phase, and they will allow us to sharpen definitions of software
requirements and to identify challenges to and strategies for meeting
these requirements. \emph{We started to develop dedicated quick tools that will allow us to built and evaluate a GRS pipeline (\S \ref{sec:grs_algo})}.

\paragraph*{(D5) Calibration strategies} for photometry, shape measurement, spectroscopy, and redshift completeness. Evaluation of the expected performance of these strategies against the science requirements. The requirement on knowledge of the dark current and the calibration approaches are fully defined, based on analysis done during the dark filter trade (October 2016 -- February 2017). \emph{We contributed extensively to the WFIRST WFI Calibration Plan. This includes extensive quantitative analysis of proposed calibration techniques (\S \ref{sec:calibration_plan})}.

\paragraph*{(D6) A strategy for the determination and calibration of photometric redshifts} using WFIRST data and anticipated external data (e.g., LSST optical
photometry), and defining ground-based data that are needed to
implement this strategy (e.g., spectroscopic training sets, large
redshift surveys for calibration via cross-correlation).
Evaluation of the impact of remaining photometric redshift uncertainties
on statistical and systematic errors in weak lensing and clustering analyses.
Definition of requirements for WFIRST photometric redshifts informed
by this strategy and evaluation. \emph{We made substantial progress by co-leading a large dedicated spectroscopic observation program (C3R2), generating mock WFIRST and LSST observations based on HST CANDELS data (\S \ref{sec:wl_photoz}), by devising calibration strategies based on Self-Organized Maps (\S \ref{sec:wl_photoz}), and by studying the importance of the Integral Field Channel (IFC) to calibrate photometric redshifts (\S \ref{sec:forecast})}.

\paragraph*{(D7) A detailed operations concept for the HLS Imaging and Spectroscopy
program,} extending the work presented in SDT13 and SDT15. \emph{In collaboration with the relevant WFIRST WG, which we are co-leading, we delivered to the project multiple detailed updates to the operation concept (\S \ref{sec:operation}) and propagated it into image simulations (\S \ref{sec:hlis_image_sim}) and forecasts (\S \ref{sec:forecast})}.

\paragraph*{(D8) Development of methods for modeling and interpreting the cosmological
measurements anticipated from WFIRST}. Determination of the effects of non-linear gravitational clustering, realistically complex relations between the
galaxy and dark matter distributions, and the influence of the baryon
component on matter clustering. The study of techniques to
remove systematic biases, e.g., by marginalization over nuisance
parameters. Utilization of cosmic shear, galaxy-galaxy lensing, cluster mass functions and cluster weak lensing, BAO, RSD, the galaxy
power spectrum, and higher order statistics for galaxy clustering, weak
lensing, and various combinations. Identification of areas where
further improvements of theoretical modeling would significantly
enhance the cosmological return from WFIRST. \emph{We have not started to work on this deliverable yet besides generating realistics mock observations (\S \ref{sec:light-cone})}.

\paragraph*{(D9) Simulated light-cone observations} based on cosmological simulations for guiding this methodology development and testing its performance.
Most of these data sets will be at the level of galaxy redshift and shape
catalogs rather than the pixel-level imaging and spectroscopy simulations
described above.  They will incorporate varying degrees of complexity
regarding galaxy bias, redshift evolution, survey geometry, and
observational systematics such as incompleteness, shape measurement errors,
and photometric redshift biases.  Many of these artificial data sets
will be made publicly available, and some will take the form of data
challenges, where the underlying parameters are initially known only
to the creators of the data set. \emph{We started assembling multiple light-cone observations dedicated to GRS, but also WL+GRS (\S \ref{sec:light-cone}) and expect to release the catalogs in the coming months. We published one dedicated paper (\S \ref{sec:other_contributions})}.

\paragraph*{(D10) Pilot survey proposals with associated figures of merits,} to
be executed during the first months of WFIRST operations. These would become
part of the final dark energy data set but also pin down remaining astrophysical
or instrument performance uncertainties at the level needed to optimize the HLS.
We will develop the figures of merit required to quickly assess the data-quality
and make operational decisions regarding the cosmological surveys. \emph{This
activity has not started yet beyond discussions of the deep fields, in
conjunctions with the other SITs and other major observational efforts during
our community workshop (\S \ref{sec:engagement})}.

\paragraph*{(D11) A prioritized program of observations from other facilities,}
ground and space-based, needed to calibrate or finalize strategy decisions on
the WFIRST dark energy program. \emph{Members of our SIT are leading an
ambitious spectroscopic observations campaign (C3R2) aiming at calibrating photometric redshifts
for WFIRST and other surveys. Members of our team are leading a major
observational program on Spitzer (the Spitzer Legacy Program (SLS)) to prepare for WFIRST and others. We expect this type of activity to be the focus of our second community workshop (\S \ref{sec:engagement}).}

\paragraph*{(D12) Broad engagement with the cosmological community,} through
workshops, talks, publications, and public release of codes and artificial data
sets, with the goals of (a) building awareness of and broad support for the
WFIRST dark energy program and (b) inspiring the community to develop methods
and carry out investigations that will maximize the cosmological return from
WFIRST. \emph{We organized in September 2016 our first community workshop in
Pasadena. It was dedicated to enabling the scientific synergies between WFIRST HLS and
LSST DESC (\S \ref{sec:engagement}). Our second community workshop dedicated to
synergies between WFIRST HLS and other surveys is scheduled for the fall 2017. We also
released new software packages, enhanced data products and forecasts (\S
\ref{sec:other_contributions}). We published 11 papers inspired by WFIRST (\S
\ref{sec:other_contributions})}.

\section{Requirement Philosophy}
\label{sec:reqt_philosophy}
%
%

\begin{summary}
The WFIRST science requirements process connect HLS
hardware and software requirements to statistical and systematic error budgets
and in turn to  cosmological constraints. While nominally a ``flow-down",  in
practice it is an iterative process as we optimize the science return within
engineering  constraints. We use different tools for each part of this process.
\end{summary}

At the highest level, we  use the \CoLi\ forecasting package to relate cosmological
constraints to data set parameters (sky coverage, galaxy density) and
parameterized descriptions of the systematic error budget. \CoLi\ is a
multi-probe analysis and forecasting pipeline that is unique in its integrated
ansatz of jointly modeling LSS probes and their correlated statistical and
systematic errors. \CoLi\ incorporates a full exploration of parameter space in
place of the Fisher formalism, and it incorporates a range of astrophysical
(e.g., intrinsic alignments, nonlinear galaxy bias, baryonic effects) and
observational (e.g., shear calibration, photo-$z$ uncertainties) systematics. It
is actively maintained and updated as part of our support of the FSWG.




WFIRST hardware capabilities (e.g., throughput, slew times) and observing
strategy/time allocation determine the HLS's statistical power, whereas the
ability to robustly constrain the instrument response model and astrophysical
nuisance parameters determine the systematic errors. Statistical errors
generally vary continuously as hardware parameters are changed, so the hardware
requirements will reflect a joint assessment of science performance and
engineering capabilities (including cost and risk). For the science assessment,
we built on our previous work on the Exposure Time Calculator (ETC) and
operations simulations codes (both written by Co-I Hirata).  Both sets of tools
are fully automated and can treat the WL and GRS surveys with a common set of
scripts. We built an interface from these tools to \CoLi\ so that
we can evaluate the science impacts of changes in WFIRST requirements (e.g., the
static wavefront error budget). Our team  work in close coordination with
project engineers to carry out a cost/benefit analysis of each such trade.

\section{Weak Lensing and Cluster Growth Investigation (D1, D3, D5, D6, D7, D11)}
\label{sec:wl_gal-clusters}
%
%

The HLS Imaging survey will (in its current design) measure the shapes of
nearly 400 million galaxies in 3 near-infrared (NIR) bands, plus fluxes in a 4th band
to improve photometric redshifts (photo-$z$).  With a data set two orders-of-magnitude
larger than the current state of the art \cite{Heymans2012,Becker2015},
the WFIRST weak lensing program will
measure the cosmic expansion history and the growth of structure with
exquisite statistical precision, demanding corresponding advances in the
control of WL systematics.  The cosmic shear power spectrum, which is the
basic WL observable, depends on both the distance-redshift relation $D(z)$
and the power spectrum of matter clustering $(\Omega_m h^2)^2 P_m(k,z)$.
The WL survey will also enable high-precision cosmological constraints
from galaxy-galaxy lensing (GGL) and from galaxy clusters, which can be
identified in either the HLS or external data sets and characterized with
the help of WFIRST WL.  The \CoLi\ forecasting tool can predict the constraints
from these methods individually and in combination with complementary
probes such as BAO, RSD, supernovae, and the CMB.

\begin{summary}
To mature the WFIRST WL investigation, our work has been organized along five main
directions:
\begin{enumerate}
  \item We developed, delivered to the project and
updated the HLII requirements;
\item We provided key contributions to the
photometric calibration plan;
\item We studied new potential detector imperfections,
developed requirements on known ones and implemented them in an accurate
WFIRST image simulation pipeline to study their effect on shape measurements;
\item We developed accurate data-driven simulations of the WFIRST lensing galaxy population and determined the
requirement on the spectroscopic samples needed to calibrate these photometric redshifts;
\item We built machinery for comprehensive cosmological forecasts for the WFIRST cluster program that
will include representations of the most significant anticipated systematic effects.
\end{enumerate}
\end{summary}

 \subsection{Developing the High Latitude Imaging Survey Requirements (D1)}

 \begin{summaryii}
   Over the last year, our main priority have been to support and guide the
   development of the WFIRST HLS imaging and in particular to identify,
   articulate and validate the scientific requirements of the instrument, the
   data reduction software, the survey and outline their flow. Responding to a calendar set by the
   Project Office, our SIT delivered three major updates to the WFIRST HLIS
   requirements to the Project Office on July 1, 2016, December 1, 2016, and
   March 2, 2017. Each of these provide progressively sharper definitions of the
   HLS requirements. We describe the main requirements and their science drivers
   below as they are included in the current Science Requirement Document.
   \emph{Disclaimer: The requirements below reflect a snapshot of the requirements formulation. The official Science Requirements Document (SRD) will always supersede the requirements written here.}
 \end{summaryii}

\subsubsection{Reference Survey and Figures of Merit}

The HLIS described in the SDT15 report covers 2200 deg2  to an imaging depth of
approximately 26.6 in Y, J, H, and 25.8 in F184 (5$\sigma$ point source, AB
magnitudes).  The predicted effective source densities are $N_{eff} \sim$ 33, 35, and 19
arcmin$^{-2}$ in J, H, F184, respectively, and $N_{eff} \sim 45$ arcmin$^{-2}$ galaxies measured
in at least one of the filters.

For the reference survey used to define baseline requirements, we back off
slightly in area to 2000 deg2 and in effective source density to $N_{eff} = 30$
arcmin$^{-2}$ to allow margin for observing inefficiencies and the possibility that
some fraction of sources cannot be used because of unreliable shape measurements
or photo-$z$ estimates.  We define the reference figure of merit in terms of the
aggregate fractional uncertainty on the amplitude of clustering $\sigma_m(z)$ for a
fixed distance-redshift relation.  Specifically, we define $FWL$ to be a constant
factor that multiplies $\sigma_m(z)$ at all redshifts, relative to the predictions of
our fiducial ΛCDM cosmological model.  The reference figure of merit is
$FoM_{WL,ref} = [\sigma(FWL)]^{-2}$ where $\sigma(FWL)$ is the forecast rms error in this quantity for the reference
survey.  With this inverse-variance definition, the FoM scales linearly with
survey area in the absence of systematic errors.  For this forecast we include
statistical errors and marginalization over a description of baryonic effects,
but we do not incorporate other systematics.  Our forecasting tools yield an
uncertainty of 0.125\% in FWL for the reference survey, or $FoM_{WL,ref} = 6400$.

In practice there is substantial degeneracy between the expansion history and
structure growth constraints from WL.  However, for characterizing the
statistical power of the survey and the impact of systematics, it is simplest,
and sufficient, to focus on a single-parameter constraint with other quantities
held fixed.  The degeneracy between growth and expansion history will be broken
largely by combining the WL measurements with SN and BAO constraints, which
depend only on expansion history.

\subsubsection{Baseline  Dark Energy Science Requirements for the HLIS}

The baseline HLIS requirements are to have sufficient observing time and
Observatory performance so that the WL constraints from the completed HLIS will
be sufficient to yield
\bea
FoM_{WL} \geq {FoM_{WL,ref}\over 2}
\eea
including statistical and systematic errors, with $FoM_{WL}$ and $FoM_{WL,ref}$ computed as described above.

While $FoM_{WL,ref}$/2 is computed based on cosmic shear alone, the $FoM_{WL}$
for the HLIS will include the constraints from galaxy-galaxy lensing and
cluster-galaxy lensing, which provide margin from additional statistical power
and their leverage for constraining systematics.  Additional margin comes from
the use of three shape measurement bands, which provides greater statistical
power than the $N_{eff}= 30$ arcmin$^{-2}$ reference case, as well as providing a method
to diagnose and mitigate systematic effects through the comparison of auto- and
cross-correlations.

Since the reference $FoM_{WL}$ is computed without contributions from shape
measurement, photo-$z$, or intrinsic alignment systematics, meeting this Level 1
requirement implies keeping the contribution of these systematics sub-dominant
relative to the statistical errors of the reference survey.

\subsubsection{Threshold Dark Energy Science Requirements for the HLIS}

The threshold HLIS requirements are to have sufficient observing time and
Observatory performance so that the WL constraints from the completed HLIS will
be sufficient to yield
\begin{eqnarray}
FoM_{WL}\geq {FoM_{WL,ref}\over 4}
\end{eqnarray}
including statistical and systematic errors, with $FoM_{WL}$ and $FoM_{WL,ref}$ computed as described above.

A factor of 4 degradation in the FoM would correspond to a factor of two
degradation in the errors on $\sigma_m(z)$.  This would still represent a factor
of $\simeq$ 20 improvement on current knowledge.

\subsubsection{Overview of Requirements Flowdown}

We define baseline Level 2 requirements for the HLIS such that, if these
requirements are satisfied, we expect the baseline requirement, $FoM_{WL}\geq
0.5 FoM_{WL,ref}$, to be satisfied. We allocate the margin relative to the
reference survey in broad categories as follows:
\begin{enumerate}
\item	A factor 0.8 in survey area (1600 deg$^2$ vs. 2000 deg$^2$)
\item	A factor 0.9 in effective source density (27 arcmin$^{-2}$ vs. 30 arcmin$^{-2}$)
\item	A factor 0.95 in shape measurement systematics
\item	A factor 0.77 in photo-$z$ systematics
\item	A factor 0.95 in intrinsic alignment systematics
\item (0.8 $\times$ 0.9 $\times$ 0.95 $\times$ 0.77 $\times$ 0.95 = 0.50).
\end{enumerate}

The margin in survey area allows for observational inefficiencies (e.g., slew
and settle times) and for time devoted to calibration observations specific to
the HLIS. In general, there is room to trade margin among these categories while
satisfying the baseline requirement.  For example, if the effective source
density exceeds 30 arcmin$^{-2}$, then there is room to accommodate larger photo-$z$
systematics or a smaller survey area.  If four filters are not required over the
full survey area to control systematics, then the number of shape measurements
can be increased (considerably) by observing a larger area in one or two bands.
Other requirements are those needed to allow the construction of galaxy catalogs
with the information needed to enable accurate WL and galaxy clustering
measurements, including accurate maps of survey depth. We do not define
individual thresholds (as opposed to baselines) for the Level 2 requirements.
The HLIS threshold is a factor of 4 lower in $FoM_{WL}$, and the best way of meeting
this threshold would likely depend on which of the baseline requirements cannot
be met.

The science requirements for the High Latitude Imaging Survey discussed above
may be summarized as:

\paragraph{HLIS 1} WFIRST shall be capable of providing HLIS science data
records over an area of at least 1600 sq. deg. (2000 sq. deg. goal) after correcting for
edge effects. [Note: losses due to bright stars or image defects at scales $<$ 1
arcmin are counted as loss of galaxy number density (see HLIS2) rather than
survey area.]

\paragraph{HLIS 2} WFIRST shall implement a High Latitude Imaging Survey to
measure galaxy shapes with a total effective z $<$ 3 galaxy density of at least
$n_{eff}$ = 27 per arcmin$^2$ in at least two bands, and three bands for at least half
of these galaxies, and photometry sufficient to provide photometric redshifts.

\subsubsection{High-Latitude Imaging Survey – Science Data Records}

High-level science products needed for weak lensing analysis include catalogs of
each source, which contains positions, classifications, photometry (aperture
photometry, adaptive moment photometry), photometric redshifts, shape
measurements for each object, links to ground-based photometric data at visible
wavelengths, etc. These catalogs should also provide error estimates for each
quantity, including covariance of output parameters. (One approach, under
investigation by LSST, is to provide posteriors on galaxy ellipticities,
effective radii, Sersic indices, etc, from an MCMC run on each object.)

\paragraph{HLIS 29a} WFIRST shall produce a mosaic image of the HLIS field using data in
each filter, and using coordinates tied to the astrometric frame defined by the
ICRF.

\paragraph{HLIS 29b} The WFIRST HLIS mosaics shall include information on the effective
exposure time for each pixel, effective PSF as a function of position, effective
depth as a function of position, data quality flags, and additional data
generated in producing the mosaics that characterize or support the mosaic
generation process.

\paragraph{HLIS 30a} WFIRST shall produce a catalog of each source in the HLIS field
containing positions, fluxes, image moments, in each filter at each epoch,
object classification information, and object-appropriate derived data. Examples
of object-appropriate derived data include photometric redshifts and
morphological parameters for galaxies, parallaxes and proper motions for stars,
limited time domain information for variable sources.

\paragraph{HLIS 30b} The WFIRST HLIS catalog shall include statistical and systematic
uncertainties for each quantity in the catalog as well as data quality flags
where numeric uncertainties are not applicable.

\paragraph{HLIS 34} WFIRST shall provide HLIS science data records that characterize the
non-Gaussian tails of the error distribution of sources, including both random
and systematic errors, to $\simeq 0.1$\% (TBD) as a function of time, location on the
sky, magnitude, and object shape.

In addition to object catalogs, the following information should be provided:
angular masks, including maps of ancillary quantities that may correlate with
the detection efficiency of galaxies and/or their photometric properties (e.g.:
effective noise per square arcsec in each filter; the effective central
wavelength of stacked images, which varies due to filter bandpass effects).

\paragraph{HLIS 35} WFIRST shall provide HLIS science data products with the angular mask
and noise map of the lensing sample.

\paragraph{HLIS 3} WFIRST shall provide HLIS science data records with additive shear
errors A limited in RMS per component over the range of angular multipoles 1.5 $<
\log_{10}\ell <$ 3.5 as specified below:

\bea
\sqrt{\sum A^2S} = 7.5 \times 10^{-5} & \rm{for} & 1.5 < \log_{10}\ell < 2.0 \\
\sqrt{\sum A^2S} = 9.9 \times 10^{-5} & \rm{for} & 2.0 < \log_{10}\ell < 2.5 \\
\sqrt{\sum A^2S} = 1.4 \times 10^{-4} & \rm{for} & 2.5 < \log_{10}\ell < 3.0 \\
\sqrt{\sum A^2S} = 1.9 \times 10^{-4} & \rm{for} & 3.0 < \log_{10}\ell < 3.5
\eea
where the sum is over independent terms in the additive systematic budget. The
total additive systematic budget, obtained via RSS of the scale bins, is
2.7$\times$ 10$^{-4}$. This requirement includes sources of additive shear with both hardware
(detector, optics) and software (biases due to data reduction pipeline) origin,
after all post-processing.

\paragraph{HLIS 4} WFIRST shall provide HLIS science data records with multiplicative shear
errors $M$ shall be known to
\bea
\sqrt{\sum M^2S} = 3.2 \times 10^{-4}
\eea
where the sum is over independent terms in the multiplicative systematic budget.
This requirement includes sources of additive shear with both hardware
(detector, optics) and software (biases due to data reduction pipeline) origin,
after all post-processing.

\paragraph{HLIS 5a} WFIRST shall provide photometric redshift codes that provide a
redshift probability distribution for an arbitrary sample of objects that
reflects a true N(z) with an error on that estimate.

\paragraph{HLIS 5b} WFIRST shall provide HLIS science data records with the
averaged redshift probability distributions $p(z)$ for objects in each tomographic
bin per the table below on the fraction of probability within
$|z_{phot}-z_{spec}|/(1+z)$ of the true redshift. .

\begin{table}[h]
\tabcolsep7.5pt
\begin{center}
\begin{tabular}{@{}|l|l|l|@{}}
\hline
Fraction of Sample & 68\% of probability within & 90\% of the probability within\\
\hline
$\sim$ 75\% (TBD) & 0.04 & 0.12\\
$\sim$ 15\% (TBD) & 0.08 & 0.24\\
$\sim$ 10\% (TBD) & 0.15 & 0.45\\
\hline
\end{tabular}
\end{center}
\end{table}
This way of phrasing the requirement takes an arbitrary $p(z)$ into account and
is more closely related to the ultimate $N(z)$ requirement than the typically used
$\sigma(z)$ and outlier fraction measurement.  It also reflects the fact that the galaxy
population is diverse, and so different populations will have different photo-$z$
properties given the photometry.

\paragraph{HLIS 6} WFIRST shall provide HLIS science data records with the $N(z)$ of each
tomographic bin of $\Delta z_{phot}=0.05$ measurable to $\Delta z/(1+z)<$0.002
(TBC).

[The 0.002 is based on the requirement that the photo-$z$ errors degrade the
aggregate precision by a factor of 1.2$^{1/2}$ (i.e., 20\% in RSS) for the Reference
survey, and assuming that the errors in the photo-z calibration are correlated
over a range of $\Delta z$ =0.2 in redshift.]

\paragraph{HLIS 8} WFIRST shall be capable of providing HLIS science data record
with S/N $\geq$ 18 (matched filter detection significance) per shape/color filter for
a galaxy with an exponential disk profile and $r_{eff}$ = 180 mas and mag AB =
24.4/24.3/23.7 (J/H/F184).

\paragraph{HLIS 9} WFIRST shall provide HLIS science data records with the PSF
ellipticity, defined by the moment ratios $e_1=(I_{xx}-I_{yy})/(I_{xx}+I_{yy})$ and
$e_2=2I_{xy}/(I_{xx}+I_{yy})$, determined to an error of $\leq$5.7$\times 10^{−4}$ RMS per component on
angular multipole scales 32 $< \ell <$ 3200.

The top-level requirement is dependent on the angular distribution (as per
HLIS3). The angular distribution from HLIS3 corresponds to placing 1.7$\times
10^{−4}$ of the shear systematic at 32 $< \ell <$ 100; 2.3$\times 10^{−4}$ at
100 $< \ell <$ 320; 3.1$\times 10^{−4}$ at 320 $< \ell <$ 1000; and 3.8$\times 10^{−4}$
at 1000 $< \ell <$ 3200. If these budgets are exceeded in one bin, the top-level
systematic error budget will have to be re-allocated.

\paragraph{HLIS 10} WFIRST shall provide HLIS science data records with the PSF
size, defined by the second moment $I_{xx}+I_{yy}$, determined to a relative
error of $\leq 7.2 \times 10^{−4}$ RMS on angular multipole scales $\ell <$
3200.

The “PSF” as defined in HLIS 9 and HLIS 10 includes the pixel response as well
as the optical PSF and image motion.

\paragraph{HLIS 14a} WFIRST shall provide a sample of spectroscopic redshifts
over the entire HLIS footprint, covering 0 $< z \leq 2.5$ and with a known selection
function.  [Note on low z: The approved 4MOST survey will do this in the
Southern Hemisphere and DESI will in the North. It might also be possible
internally to WFIRST data using rest-frame NIR lines.]

In addition to data products themselves, certain tools should be provided to
understand underlying systematic effects in the data. These include:

\paragraph{HLIS 31} The data processing system shall provide simulation packages that can
“observe” simulated fields (e.g., from a catalog of galaxies, with an $x-y-\lambda$ data
cube of each) and feed the results into the data reduction pipeline, all the way
through to simulated catalogs.

\paragraph{HLIS 32} The data processing system shall provide a simulation package that can
inject simulated galaxies (e.g., from $x-y-\lambda$ data cubes) or stars into the real
images and re-run (portions of) the data processing. This is needed to assess
completeness/selection effects, the impact of blending on objects of known
properties, and the impact of nearby stars on the measurement of galaxy
photometric properties. In principle, much of this could be done from observing
simulated skies, but these hybrid simulations are useful because they have the
correct instrument noise properties and level of crowding by construction.

\subsubsection{HLIS Calibrated Data Record Requirements}

\paragraph{HLIS 28} WFIRST shall archive HLIS calibrated images with the PSF for each
exposure specified as a function of position on the focal plane and incorporate
any World Coordinate System information needed for subsequent stages of SOC
processing.

\paragraph{HLIS 25} WFIRST shall provide HLIS calibrated images with the relative
photometric calibration in each HLIS filter on angular scales $\ell < 3200$ better than
10 (TBR) millimag RMS.

This requirement flows from the photo-$z$ requirement: variations in the
photometric calibration lead to systematic variations of the $P(z_{phot}|z_{spec})$
function across the survey, which both distorts the lensing and galaxy
clustering power spectra, and may make the photo-z calibration fields not
representative.

\paragraph{HLIS 26a} WFIRST shall provide HLIS calibrated images with the astrometric
solution in the WFI images having a relative error (offset between two images of
the same galaxy or star, possibly taken at different times during the survey) of $<$
1.3mas (TBR).

\paragraph{HLIS 26b} WFIRST shall provide HLIS calibrated images with the astrometric
solution in the WFI images having an absolute error, relative to the astrometric
frame defined by the ICRF of:

\begin{eqnarray}
<26  & \rm{mas\ RMS\ per\ component\ for} & \log_{10}\ell = 1.5 - 2.0\\
<11  & \rm{mas\ RMS\ per\ component\ for} & \log_{10}\ell = 2.0 - 2.5\\
<5.1 & \rm{mas\ RMS\ per\ component\ for} & \log_{10}\ell = 2.5 - 3.0\\
<2.2 & \rm{mas\ RMS\ per\ component\ for} & \log_{10}\ell = 3.0 - 3.5
\end{eqnarray}

\paragraph{HLIS 33} WFIRST shall provide HLIS calibrated images with the
variation of the total system response known as a function of position such that
the total flux of a source with a known Spectral Energy Distribution (SED) can
be corrected to a common filter system at the $\sim$ 0.5\% (TBC) level.

\subsubsection{HLIS Calibrated Raw Data Record Requirements}

\paragraph{HLIS 27} WFIRST shall provide HLIS raw images in the archive for each
detector exposure with each raw data record including a unique dataset
identifier for each exposure, the exposure time, the time of exposure, all
individual downlinked detector readouts used to make the exposure, the
observatory pointing orientation and any additional engineering data the Science
Center uses for subsequent processing.

\paragraph{HLIS 7} WFIRST shall have the capability of providing HLIS raw images with
photometry, position, and shape measurements of galaxies in 3 filters (J, H, and
F184), and photometry and position measurements in one additional color filter
(Y).

\paragraph{HLIS 11} WFIRST shall provide HLIS raw images with a system PSF EE50
radius $\leq$ 0.12 (Y band), 0.12 (J), 0.14 (H), or 0.14 (F184) arcsec, excluding
diffraction spikes and non-first order light and including the effects of
pointing jitter, for at least 95\% of the exposures (TBR) and over 95\% (TBR) of
the FOV.

\paragraph{HLIS 12} WFIRST shall provide HLIS raw images dithered so that
Nyquist sampling of the PSF is provided over $>$90\% (TBC) of the survey area in
the shape measurement bands.

\paragraph{HLIS 14b} WFIRST shall be capable of observing a deep field of at least 6 $deg^2$
within the HLIS with a S/N of $>$25 per filter (TBC) for the faintest galaxies in
the lensing sample. This field should be situated to overlap with other deep
multi-wavelength data.

\paragraph{HLIS 14c} WFIRST shall acquire at least 15,000 (TBC) spectroscopic redshifts in
the HLIS deep field (see HLIS 14b), representing the full extent of color space
for detected galaxies. (Spectra could be from WFIRST observations or from
ground-based observatories).

\paragraph{HLIS 19} WFIRST shall periodically acquire HLIS deep field imaging observations.
These observations will enable testing the effects of noise on the shape
measurement and photo-$z$ algorithms. The deep fields are also needed to measure
empirical noise properties in the photometric sample to ensure selection and
photo-$z$ bias can be properly characterized.

\paragraph{HLIS 20 (HLIS 15)} WFIRST shall downlink at least 3 ground-configurable linear
combinations of samples per HLIS exposure, with a goal of 6 samples. This
requirement assumes on-board subtraction of the reference frame (first read).

One means of obtaining the photometric redshift training and calibration dataset
is with an on-board spectrograph similar to that used for supernova
spectroscopy, but with a somewhat larger field of view and coarser spatial
resolution. (slitless spectroscopy may be sufficient but has not yet been
studied) Requirements for this spectrograph are as follows:

\paragraph{HLIS 15} WFIRST shall be capable of providing HLIS raw spectroscopic data with spectral
resolution $\lambda /\Delta \lambda \geq$ 100(TBR) per 2-pixel resolution element at all wavelengths.

\paragraph{HLIS 16} WFIRST shall be capable of providing HLIS raw spectral images with a
bandpass spanning at least 0.45-2.0 $\mu$m.

\paragraph{HLIS 17} WFIRST shall be capable of providing HLIS raw spectral
images with a spatial resolution of 0.3 arcsecond (TBR), with 2 or more pixels
per spatial resolution element.

\paragraph{HLIS 18} WFIRST shall be capable of providing HLIS raw spectral
images with a field of view of at least 6" $\times$ 6".

\subsection{Defining the Photometric Calibration Plan (D5)}
\label{sec:wl_calibration}

\begin{summaryii}
  In addition to articulating the science requirements, our SIT has been
  instrumental in the Calibration Working Group effort. Since precision
  cosmology measurements depend sensitively on calibration; subtle effects that
  might not be noticeable in other areas of astrophysics can become important
  when trying to measure galaxy shapes to $<0.1$\%. Activities over the past
  year have included:
  \begin{enumerate}
  \item {\em Dark filter:} Co-I's Wang, Capak, and Hirata participated
  extensively in the analyses and discussions that led the FSWG to recommend a
  dark position in the element wheel on WFIRST.
 \item {\em Calibration plan}: Our SIT has contributed extensively to the WFIRST WFI
Calibration Plan, including detailed quantitative assessments of calibration
approaches and their ability to meet requirements. In some areas, such as dark
current and the point spread function, our contributions to the calibration plan
are now traceable all the way from science measurements (WL shear) down to the
specific calibration approaches and the hardware stability requirements needed
for them to work. A major area of work leading up to SRR/MDR is to complete this
flow-down for the other areas of calibration.
\item {\em Detector characterization}: We have made use of the H4RG data provided by the Detector Characterization Laboratory and
H2RG data from the JPL Projector Laboratory to measure some of the detector-induced systematic effects relevant to weak lensing with
NIR detectors. We built toy models to study other effects.
\item {\em Simulating detector imperfections}: We started to simulated known detector imperfection in a publicly available image simulation pipeline in order to assess their effect on measured galaxy shapes. This is an important practice step
toward building calibration pipelines that will support WL science.
\end{enumerate}
In what follows, we provide some highlights from our calibration activities. The
list is not exhaustive.
\end{summaryii}

\subsubsection{Dark filter}

In the summer and fall of 2016, the FSWG was tasked with determining whether a
dark filter was needed for WFIRST calibration. This required the FSWG to
enumerate the list of calibration tests that might use the dark filter, and
establish whether alternative options were possible. We led the effort to
assemble this list of tests based on input from the SITs (both ours and others),
the SOC, and  Project personnel. The list\footnote{\tt
DarkAlternativesMatrix\_161030.docx} included 14 items: (i) the dark current
(including internal instrument backgrounds); (ii) unstable pixels; (iii)
post-reset transients; (iv) read noise correlations; (v) inter-pixel
capacitance; (vi) gain measurement; (vii) the high spatial frequency flat;
(viii) the low spatial frequency flat; (ix) persistence from previous
observations; (x) persistence from slews; (xi) classical linearity; (xii) count
rate dependent non-linearity; (xiii) the brighter-fatter effect; and (xiv)
persistence re-activation.

The problem of persistence from slews (i.e.\ streaks across the detector
following a slew from one observation to another) is of particular importance to
weak lensing, because it leads to a coherent, highly directional pattern on the
detector that has the correct symmetry to induce a coherent systematic error in
the galaxy ellipticities. This is a concern without a dark capability, or even
with a dark capability if it is not (or cannot be) used during every slew. Our
group identified two budgets in WL that flow down into slew persistence
requirements. First is the total systematic shear error budget of $2.7\times
10^{-4}$. Second is the masked pixel budget.

\begin{figure}
\includegraphics[width=5in]{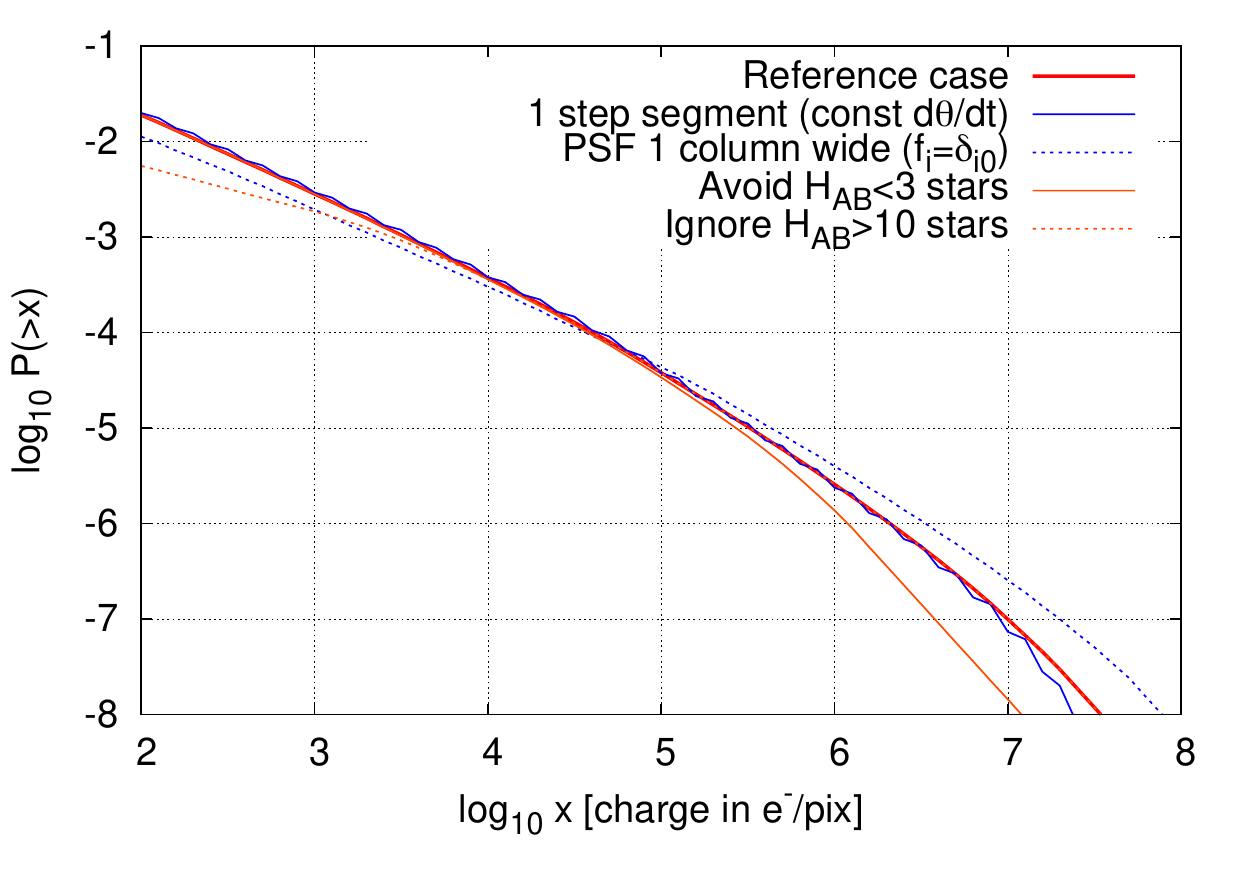}
\caption{\label{fig:slewcompare}Comparison of stimulus levels predicted
under different assumptions and approximations. The vertical axis
shows the log probability to exceed a given stimulus level during a
slew of 0.4 degrees (a step along the short axis of the field,
executed frequently during the HLS). The thick red line indicates
reference assumptions. The solid blue line treats the slews as being
at constant $\dot\theta$. The dashed blue line approximates the PSF as
1 column wide (all the flux from the star is concentrated in the
central column). The orange lines show what happens if bright ($H_{\rm
AB}<3$) or faint ($H_{\rm AB}>10$) stars are excluded from the model.}
\end{figure}

\begin{figure}
\includegraphics[width=5in]{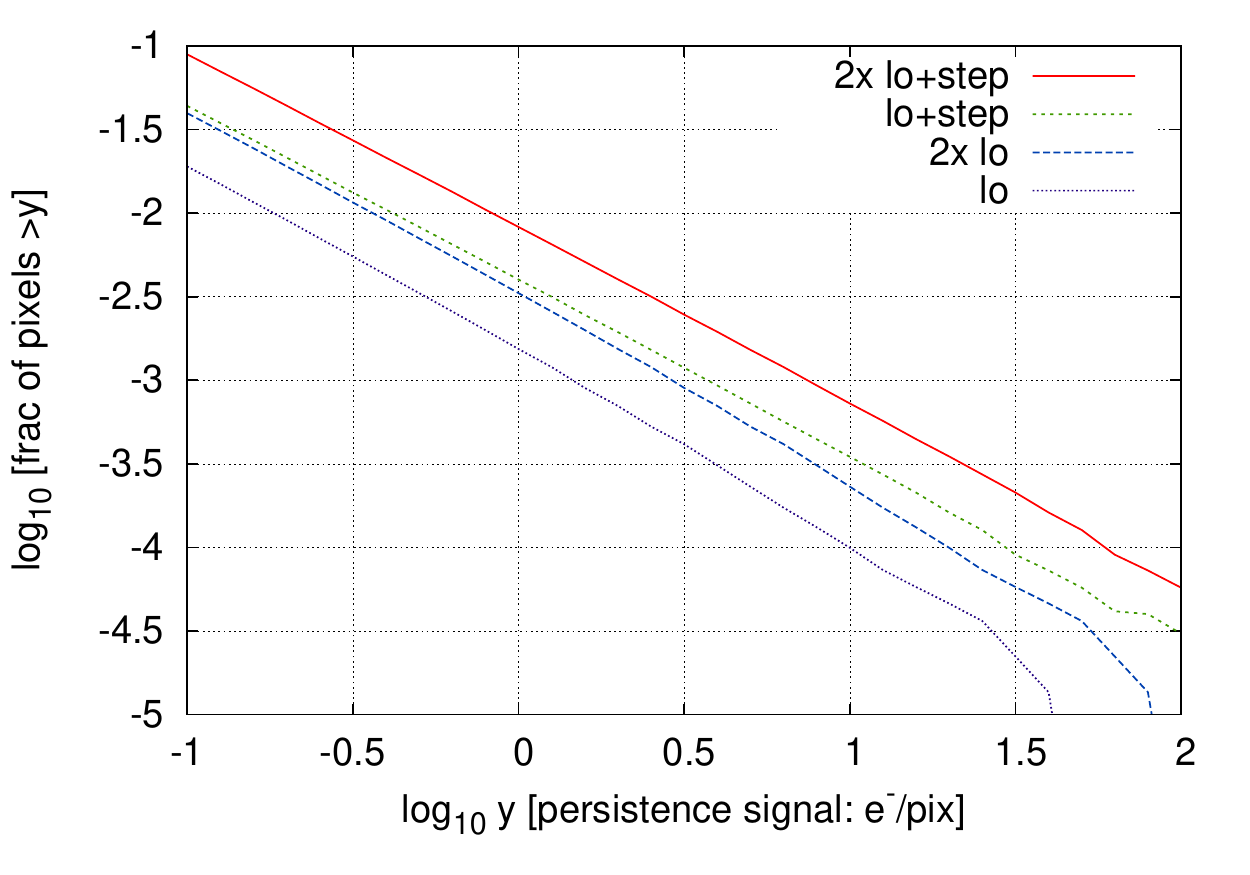}
\caption{\label{fig:sp_cdf}The cumulative distribution function of slew-induced persistence in the HLS imaging survey, $P(>y)$.
The persistence signal $y$ is estimated in electrons per pixel; the decaying persistence curve is integrated over a 160 s. Several persistence models are shown, including the ``lo'' case (typical of the development detectors), and a ``lo+step'' case (including an order of magnitude step at saturation, as seen in portions of some detectors). This figure has not been updated yet to go from the 6-year to the 5-year observing plan, although we expect only minor differences.}
\end{figure}

The details of the slew persistence study are provided in the Calibration Plan.
It consisted of several stages: first, assessing the magnitude distribution of
the stars that would be encountered in the High Latitude Survey; then assessing
the probability of stimulus levels in a slew, given the distribution of slews
from our operations model (\S\ref{sec:operation}); and then folding this through
a persistence model (based on DCL data for the development H4RG detectors) to
predict the probability distribution of persistent pixels in the HLS imaging
survey. The stimulus distribution ($x$ in e: the well depth to which a pixel is
filled during a slew) from the Calibration Plan is shown in
Figure~\ref{fig:slewcompare}, and the persistence signal distribution ($y$ in e:
the persistence signal in a pixel over the course of an exposure) is shown in
Figure~\ref{fig:sp_cdf}.

After negotiating with the Project, we settled on a mitigation strategy for slew persistence that involved saving the spacecraft orientation information from the Attitude Control System (ACS), using this to predict the locations of persistence from bright star streaks, and masking $\pm2\sigma$ on either side of these streaks. Unmasked streaks are simply accepted as part of the systematic error budget. Their impact on shape measurement is based on an analytic result derived by our SIT and tested against Monte Carlo simulations:
\begin{equation}
\Delta \gamma_1 + i\Delta \gamma_2 = \frac{M\Omega_{\rm
max}\sigma_{\rm n}^2 R^4}{2F^2N_{\rm ind}\,{\rm Res}} f_{\rm scale}
f_{\rm aniso},
\label{eq:dg}
\end{equation}
where $\Delta\gamma_{1,2}$ are the two components of spurious shear; $M$ is a margin factor; $\Omega_{\rm max} = 421.3$; $\sigma_{\rm n}^2$ is the variance of the persistence image; $R$ is the radius of the galaxy in pixels; $F$ is the signal from the galaxy in electrons per exposure; $N_{\rm ind}$ is the number of {\em independent} exposures of the galaxy\footnote{This may be less than the total number of exposures of the galaxy, since slew persistence from successive exposures will be correlated.}; Res is the galaxy resolution factor \cite{bej02}; $f_{\rm scale}$ and $f_{\rm aniso}$ are factors $\le 1$ describing the scale dependence and anisotropy of the persistence power spectrum (defined to be 1 in the worst case).

\begin{figure}
\includegraphics[width=5in]{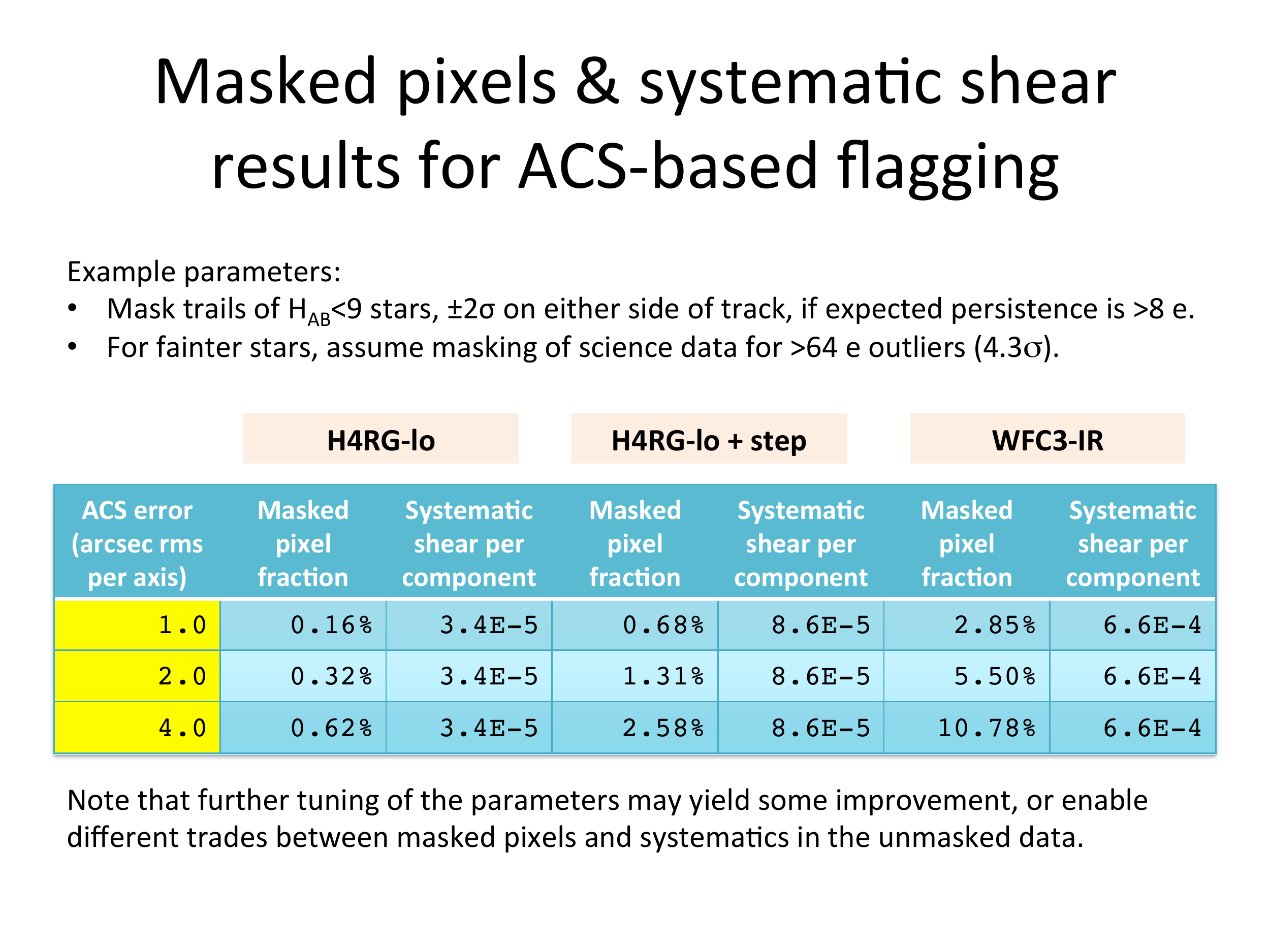}
\caption{\label{fig:slew_results_oct16}The outcome of the October 2016 slew persistence study. This shows the masked pixel fraction and the predicted systematic shear due to unmasked streaks as a function of both the persistence model and the accuracy of pointing information.}
\end{figure}

The results of this study -- shown in Figure~\ref{fig:slew_results_oct16} -- are promising, given the top-level systematic shear budget of $2.7\times 10^{-4}$ and that the modern detectors typically show ``lo'' or (in some regions) ``lo+step''-like behavior, rather than the much larger persistence characteristic of the WFC3-IR model (third column). The masking algorithm will continue to be revisited as part of the mission optimization. However, the small number of masked pixels led the FSWG to conclude that a dark shutter that operated during every slew was not required for the WFIRST HLS.

We carried out a related study, also using Eq.~(\ref{eq:dg}) and related machinery, to assess how well we need to know the dark current for WFIRST. Dark current measurements without a dark filter are possible, e.g.\ via median algorithms that combine many exposures from a survey, but are subject to: (i) a degeneracy in which the ``true'' sky brightness is unknown and hence the zero level of the dark current cannot be established, and (ii) possible correlated errors from imprinted celestial sources. The requirements, as derived in the appendix to the calibration plan, are:
\begin{list}{$\bullet$}{}
\item The error in the dark current + bias determination in a 140 s
HLS imaging exposure shall be no more than $0.0096f_{\rm corr}^{-1/2}$
e/p/s (uncorrelated part) or $0.0017 f_{\rm corr}^{-1/2}$ e/p/s
(imprinted celestial sources).
\item The error in the dark current + bias determination in a 297 s
HLS spectroscopy exposure shall be no more than $0.0059 f_{\rm
corr}^{-1/2}$ e/p/s (uncorrelated part) or $0.00072 f_{\rm
corr}^{-1/2}$ e/p/s (imprinted celestial sources).
\end{list}
Here ``$f_{\rm corr}$'' denotes the factor by which we plan to correct biases induced by errors in the dark current map (we normally choose $f_{\rm corr}=1$ to be conservative). The requirements are traceable to additive shear biases from non-circular imprinted celestial sources; multiplicative shear biases as the noise in the dark current map results in e.g.\ galaxy centroids getting ``pulled'' toward pixels whose measured dark current fluctuates below the true dark current of that pixel; and Eddington-like biases for sources detected in the GRS. While the semi-analytic estimates in the calibration plan based on source counts suggest that the HLS imaging requirement can be met without a dark filter, our SIT and the Calibration Working Group had concerns about possible degeneracies in the self-calibration procedure that can only be addressed by a detailed simulation. Moreover, the approach requires empty space in the images, which we will not have in the case of grism spectroscopy. As the imaging exposures are shorter than the spectroscopy exposures, this would require dedicated long imaging exposures (of HLS spectroscopy exposure length) just for the purpose of self-calibrating the dark. Due to sky Poisson noise, we would need many of these images -- our Feburary 2017 estimate was for $N=73$ exposures, which, if done every week, would consume 4\%\ of the wall clock time. In light of these and other issues, the Calibration Working Group recommended that WFIRST maintain the dark filter.

\subsubsection{Calibration Plan}
\label{sec:calibration_plan}
Our SIT has contributed extensively to the WFIRST WFI Calibration Plan. This includes extensive quantitative analysis of proposed calibration techniques, as detailed in the appendix to the plan. Some highlights follow.

The requirement on knowledge of the dark current and the calibration approaches are fully defined, based on analysis done during the dark filter trade (October 2016 -- February 2017).

Weak lensing was found to place demanding requirements on measurement of the count rate-dependent non-linearity (CRNL). The weak lensing program is sensitive to CRNL because it enhances the bright center of a PSF star relative to its wings, thereby making the star appear slightly smaller, but does not have a similar effect on the faint galaxies used for shape measurement. The PSF second moment is biased by a factor of $1-\alpha$ (where $\alpha$ is the CRNL exponent), and has a top-line systematic error budget of $7.2\times 10^{-4}$. This means that if $\alpha$ is measured to $\pm 3\times 10^{-4}$ (the requirement from the supernova SITs), then CRNL consumes 17\%\ of the PSF size error budget, in an RSS sense. Given that CRNL is a pernicious bias for two of the dark energy probes, we recommended a multi-faceted approach to CRNL calibration, including a lamp-on/lamp-off capability for WFIRST (this was not available on WFC3-IR).

Our team has revisited the wavefront stability requirements for weak lensing, using a set of codes and scripts on the team's GitHub site. This begins with a Fisher matrix analysis of the uncertainties in the shear power spectrum, and our top-line requirement that the systematic errors be equivalent to the statistical errors even if the survey is extended to 10,000 deg$^2$ (i.e.\ in an RSS sense, the systematic errors should be 20\%\ of the statistical errors in the nominal 2,000 deg$^2$ survey). Requirements are assessed using the significance, defined by
\begin{equation}
Z = \sqrt{\Delta{\bf C}\cdot{\bf\Sigma}^{-1}\Delta{\bf C}},
\label{eq:alpha}
\end{equation}
which is the number of sigmas at which one could distinguish the correct power spectrum from the power spectrum containing a systematic error. We built sub-allocations for multiplicative (shear calibration) errors, and for additive (spurious shear) errors in each angular bin. An early discovery was that this process depends on the redshift dependence of the shear error: some redshift dependences are ``worse'' than others by the $Z$-metric. The worst possibility is {\em not} for the error to be redshift-independent, but rather for it to change sign, as this can mimic a change in redshift evolution of the growth of structure.

In our current formalism, for each angular template, we introduce a limiting amplitude $A_0^{\rm flat}(\alpha)$, defined to be the RMS spurious shear per component $A_0$ at which we would saturate the requirement on $Z(\alpha)$ for angular bin $\alpha$ in the case of a redshift-independent systematic $w_i=1\,\forall i$ (here $\alpha$ denotes an angular bin and $i$ a redshift bin). That is, if the additive systematics did not depend on redshift, we could tolerate a total additive systematic shear of $A_0^{\rm flat}$ (RMS per component) in band $\alpha$. We also introduce a scaling factor $S[{\bf w},\alpha]$ for a systematic error
\begin{equation}
S[{\bf w},\alpha] = \frac{Z(\alpha) \,{\rm for\,this\,}w_i}{Z(\alpha)\,{\rm for\,all\,}w_i=1}
\end{equation}
that depends on the redshift dependence $w_i$. An additive systematic error that is independent of redshift will have $S=1$. A systematic that is ``made worse'' by its redshift dependence will have $S>1$, and a systematic that is ``made less serious'' by its redshift dependence will have $S<1$. The requirement that the (linear) sum of $Z$s not exceed $Z(\alpha)$ thus translates into
\begin{equation}
\sum_{\rm systematics} [A(\alpha)]^2\times S[{\bf w},\alpha] \le [A_0^{\rm flat}(\alpha)]^2,
\label{eq:A-S-sum}
\end{equation}
where $A(\alpha)$ is the RMS additive shear per component due to that systematic. We take the ``reference'' additive shear to be the additive shear in the most contaminated redshift slice; in this case, $w_i=1$ for that slice, and $|w_i|\le 1$ for the others. Under such
circumstances, we can determine a {\em worst-case scaling factor} $S_{\rm max,\pm}(\alpha)$, which is the largest value of $S[{\bf
w},\alpha]$ for any weights satisfying the above inequality. We may also determine a worst-case scaling factor $S_{\rm max,+}(\alpha)$
conditioned on $0\le w_i\le 1$, i.e.\ for sources of additive shear that have the same sign in all redshift bins. In most cases, however, something is known about the redshift dependence of the systematic error (e.g.\ for PSF errors the error scales with the size of the galaxy, and hence has a redshift dependence tied to the measured redshift evolution of galaxy sizes). In these cases, we use the correct redshift weighting factor $S$. This approach has been critical in order to set stability requirements that are consistent with the Project's integrated modeling results.

We have begun incorporating the HLS observing strategy (\S\ref{sec:operation}) in studies of self-calibration of time-dependent drifts in the response of the system (i.e.\ time dependence of the conversion from $\mu$Jy on the sky to DN/s in the digitized detector system outputs). This model is in a state of flux as we add parameters to it, but here we show a current snapshot allowing for time-dependent drifts of the response of each of the 18 SCAs making up the focal plane, with time dependence parameterized in calibration periods of $\Delta t$ (assessed down to a period of 3 hours) each. Both individual-SCA drifts and common-mode drifts are allowed, with an assumed intrinsic variation (calibration prior) of 1\%\ RMS drift in each $e$-fold of timescales. A network of randomly distributed stars with a density of 500 stars/deg$^2$ and $S/N=50$ was assumed; in self-calibration, the magnitudes of these stars are {\em not} known a priori, but are assumed to be stable across multiple repeated observations of the same field. These are preliminary parameters being used to test our tools and are not currently held as requirements. The stellar density model is very conservative since the Trilegal model predicts star counts of 572, 803, 990, and 1137 stars/deg$^2$ at $H_{\rm AB}=18-19$, $19-20$, $20-21$, and $21-22$ at the SGP, and even an $H_{\rm AB}=22$ star will have $S/N>50$. The temporal stability of the system needs further study and will be varied as an input parameter in future versions of this model. The current model uses the April 19, 2017 update to the HLS observing strategy. The number of calibration parameters varies depending on the filter, since there are no parameters for periods of time when the instrument is not observing in that filter; the current version has 17262 parameters for the H band.

\begin{figure}
\includegraphics[width=4.5in]{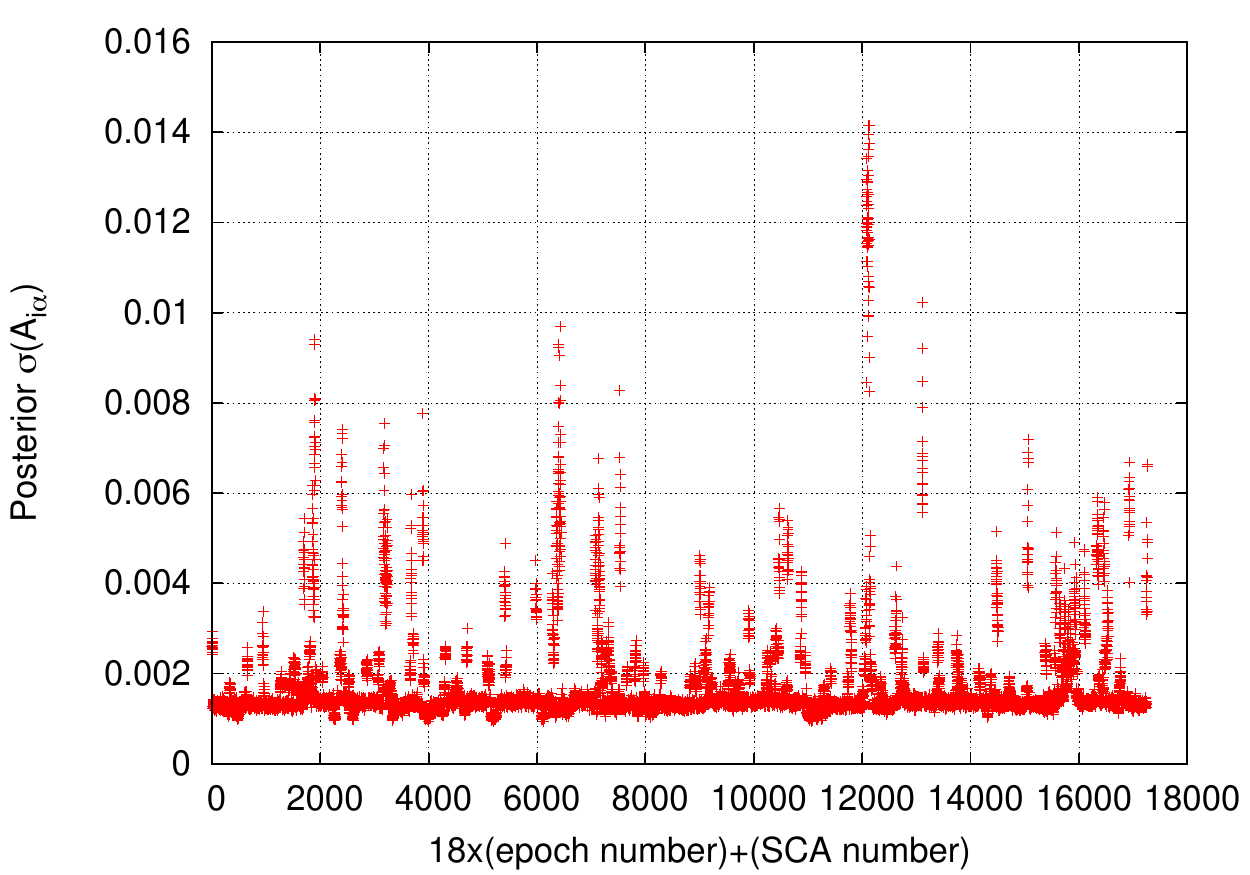}
\caption{\label{fig:hcalfig}An example of the posterior calibration error from an HLS self-calibration calculation. The horizontal scale displays the time intervals (left-to-right in time order), with 18 points per time interval indicating the various SCAs. The vertical axis shows the standard deviation of the calibration solution for that SCA at that time, $\sigma(A_{i\alpha})$, relative to the survey mean. The figure shows the case of H band with $\Delta t = 0.125$ days. This example had 17262 calibration parameters. A few epochs, mostly containing only a few observations, are poorly constrained due to minimal overlap observations. It is subject to refinement and the input parameters of the model will be varied as we work toward requirements on the stability of the detector system.}
\end{figure}

Despite the intrinsic stability assumed, in which each SCA can have its response fluctuate by 1.67\%\ RMS from one time interval to the next, the repeated observations do an excellent job of tracking these changes and reducing the posterior uncertainty. Even for $\Delta t = 0.125\,$days$=3\,$hours, the posterior calibration errors are at the level of 0.14--0.17\%\ RMS (here ``RMS'' is weighted by number of observations), depending on the filter.  An example of the model output (predicted uncertainties in the calibration parameters for each SCA at each epoch) is shown in Figure~\ref{fig:hcalfig}. It must be remembered that this analysis is overly simplistic in some ways -- particularly that we have not yet allowed for shorter-timescale variations (i.e.\ on timescales $<\Delta t$), nor have we allowed for separate gain drifts among the different readout channels. These will have to be included in a future version of the model. On the other hand, the stellar density and $S/N$ assumptions were extremely conservative (e.g.\ the full range of stellar magnitudes 18--22 should have 7 times more stars than were assumed, even at the Galactic pole), so there is margin to absorb these additional degrees of freedom. The next iteration of the model for time-dependent calibration drifts will include additional parameters, as well as updated priors reflecting expected detector system stability rather than the place-holder requirements shown here.

\subsection{Identifying and Studying the Effect of Detector Imperfection (D3, D7)}
\label{sec:wl_detectors}
\begin{summaryii}
  Since precision cosmology measurements depend sensitively on exquisite photometry; subtle effects that
  might not be noticeable in other areas of astrophysics can become important
  when trying to measure galaxy shapes to $<0.1$\%. Over the past year, we have
  studied novel possible systematic effects, implemented in an image
  simulation pipeline new and known effects and released it to the community. Our goal is to derive requirements for all these effects. Highlights include
  \begin{enumerate}
  \item The study of the effect of polarization-dependent quantum efficiency;
  \item The requirements on the interpixel capacitance;
  \item Detector characterization;
  \item Image simulation including detector imperfection and WFIRST scanning strategy to study their effect on shape measurement.
\end{enumerate}
In what follows, we provide some highlights from our detector characterization and simulation activities.
\end{summaryii}

\subsubsection{Polarization Effects}
During early 2017, work was carried out to assess the approximate level of an effect that could
cause weak lensing systematics, but that had never been previously considered by the weak lensing
community.  This effect is polarization-dependent quantum efficiency (due to e.g.\ different
reflectivity of various coatings for different polarizations of light).  Since the light from
edge-on disk galaxies typically has some low level polarization perpendicular to the disk, any
polarization-dependence of the QE could result in a preferential selection of such galaxies based on
their orientation in the focal plane.  This would violate the baseline assumption in a weak lensing
analysis, which is that all coherent galaxy alignments are due to gravitational lensing.

\begin{figure}[t]
\includegraphics[width=0.66\textwidth]{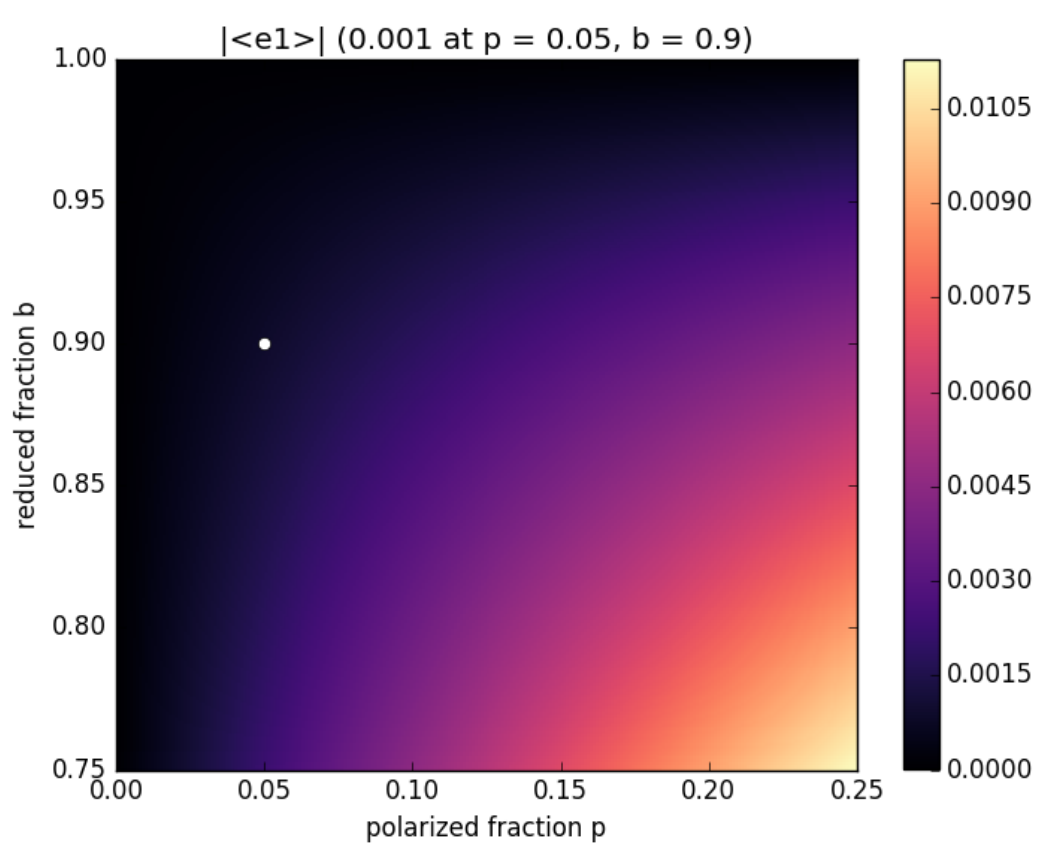}
\caption{\label{fig:polarization}
The average shear due to weak lensing selection biases due to polarization-dependent quantum
efficiency, as a function of the fraction of polarized light from edge-on disks (horizontal axis)
and the fractional attenuation of the perpendicular polarization (vertical axis).  The dot at
$(0.05,0.9)$ is our fiducial point in parameter space, and has $\langle e\rangle \approx 0.001$.
}
\end{figure}

A student at CMU, Brent Tan, worked with Rachel Mandelbaum and Chris Hirata on a simple toy model
for this effect.  The toy model had two parameters: the fraction of the disk galaxy light that is
polarized, and the relative attenuation of that perpendicular polarization component (both numbers
in the range $[0,1]$).  For each point in that parameter space, the coherent shear due to selection
bias was calculated; see results in Figure~\ref{fig:polarization}.  Finally, the results were
modified to account for the fact that not all disk galaxies are viewed edge-on and that not all
galaxies are disks, giving a net coherent shear due to this selection bias of $\sim 3\times
10^{-4}$.  The results are still quite uncertain because our fiducial values for the disk
polarization fraction were based on observations of nearby galaxies, not $z\sim 1$ disks.  However,
this is large enough to be relevant for WFIRST, so this systematic needs to be evaluated more
carefully and requirements placed in future.  A publication on this topic will be prepared during
summer 2017.

Another possible polarization-related systematic is a polarization-dependent PSF.  That will be the
subject of future work.

\subsubsection{Interpixel Capacitance Requirements}

The WFIRST detectors will suffer from electrical crosstalk between the pixels, unlike the optical
 detectors that are based on CCDs. This effect, known as the \emph{interpixel capacitance} (IPC),
 appears as a systematic effect in the weak lensing shear measurements and causes a
 bias in the measurements if not properly taken into account. The effect of IPC on the point-spread
 function (PSF) was already studied by members of our SIT in \cite{Kannawadi2016}, and requirements were placed
 on the level of uncertainty in the IPC based on how that uncertainty affects the PSF.

More recently, in late 2016, members of our SIT (Mandelbaum and student Kannawadi) carried out and
analyzed simulations to determine whether additional requirements on IPC are needed to ensure that
weak lensing shear estimation is not biased beyond our tolerances.  To calibrate the shear
multiplicative bias to an accuracy of $2\times 10^{-3}$, we find that the requirements on the IPC
placed by the PSF requirements are sufficient, so no new requirement is needed.  A paper on this
result is in preparation.

\begin{figure}[!t]
  \includegraphics[width=0.55\textwidth]{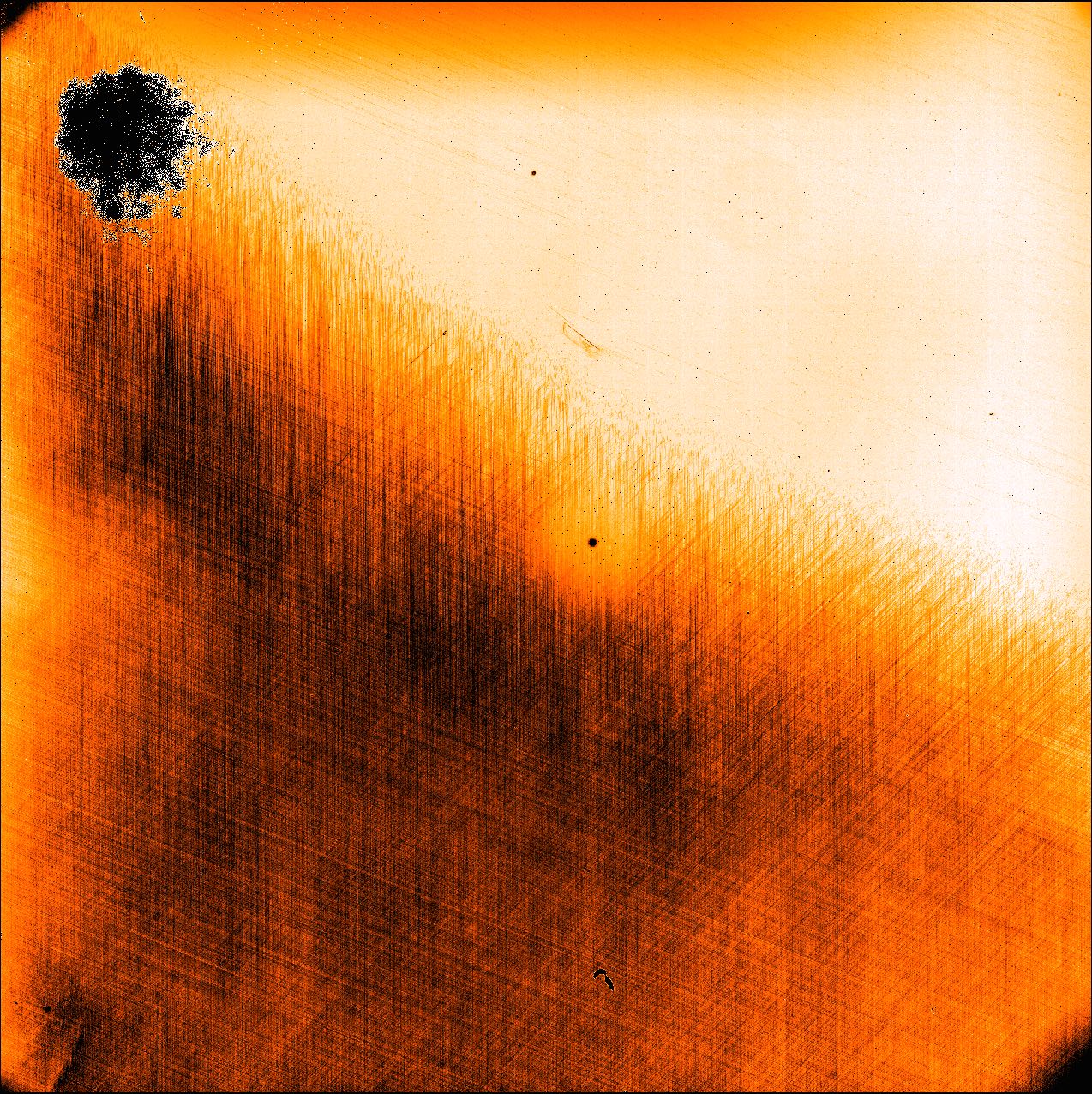}
\caption{\label{fig:crosshatch}2048 x 2048 flat-field calibration image taken with a Euclid engineering grade H2RG detector (2.4 $\mu$m cutoff) under 1 $\mu$m illumination (scaled to enhance contrast).  Flight detectors resemble the upper right region (mean QE $\sim$ 1) but still contain traces of the ``crosshatch'' pattern visible in the lower left (mean QE $\sim$ 0.8).  Scanning a grid of undersampled point sources (f/11 aperture setting) in sub-pixel increments generated 1\% RMS photometric variations in the lower left region which were not removed by the flat-field calibration.  This result provides evidence that the pattern has intra-pixel structure that potentially biases the PSF if left uncalibrated.}
\end{figure}

\subsubsection{Laboratory Detector Characterization}

The WFIRST dark energy analyses will place enormous demands on our understanding
of the detectors. Some aspects of this problem can be anticipated in advance --
for example, we know that effects such as inter-pixel capacitance,
count-rate-dependent non-linearity  will need to be carefully
characterized, and we are working as part of the Calibration Working Group to
build these measurements into the mission (collaborator Shapiro is co-leading this particular Working Group). However, with systematic error
budgets at the level of a few$\times 10^{-4}$, it is likely that WFIRST analyses
will turn up new effects that were not apparent in past missions. Therefore a
key task for our SIT is to analyze the data from development detectors and
identify these new effects early enough to inform the calibration plan.

\begin{figure}[!t]
  \includegraphics[width=6.2in]{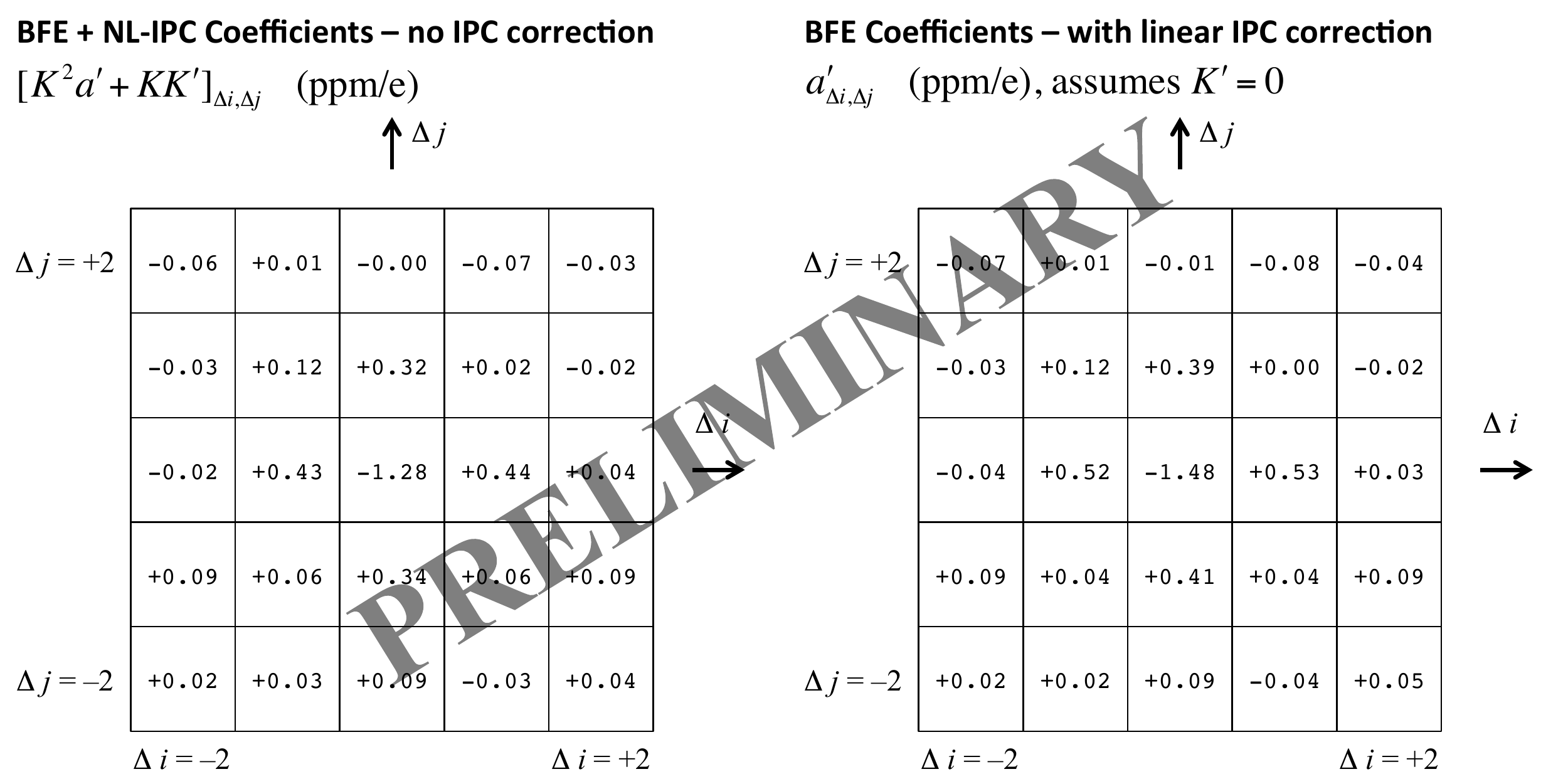}
\caption{\label{fig:kernel}The BFE + NL-IPC coefficients $[K^2a'+KK']_{\Delta
i,\Delta j}$ (left panel) and IPC-corrected coefficients $a'_{\Delta i,\Delta
j}$ (right panel), for H4RG-17940. Note that the IPC-corrected coefficents
assume that the IPC is linear, i.e.\ the non-overlapping correlations are
ascribable entirely to the BFE and not NL-IPC. The $1\sigma$ uncertainty in each
pixel is 0.07 ppm/e.}
\end{figure}

In ground-based weak lensing projects using thick CCDs (e.g.\ DES), one of the
key detector issues has been the {\em brighter-fatter effect} (BFE). This is an
electrostatic effect in which as a pixel fills up with collected charge, it
changes the electric field geometry and new charges generated are more likely to
be deflected into neighboring pixels. This has the effect of making bright stars
appear larger than faint stars, as the repulsion effect is non-linear and
increases with signal level. The field geometry is very different in a NIR
detector, but a brighter-fatter effect is still possible. \cite{2017JInst..12C4009P} and \cite{2017arXiv171206642P} report a first detection of the BFE in NIR detectors (H1RG and H2RG) using on-sky and laboratory data, respectively.

We have searched for the brighter-fatter effect in the H4RG detector arrays
using the flat fields for two devices H4RG-17940 and H4RG-18237, provided to us
by the DCL. The BFE imprints a signature in the auto-correlation function of a
flat field; using the correlations in multiple non-destructive reads in a flat
field, one can separate linear IPC from the BFE. Preliminary brighter-fatter
effect results for H4RG-17940 are shown in Figure~\ref{fig:kernel}. The BFE
coefficients are $a'_{\Delta i,\Delta j}$, which is the fractional change in
effective area of pixel $(i,j)$ when an electron is placed in pixel $(i+\Delta
i, j+\Delta j)$; they have units of parts per million per electron (ppm/e). The
flat auto-correlations are sensitive to both the brighter-fatter effect and
non-linear inter-pixel capacitance (NL-IPC); we are currently working on
distinguishing the two effects.

We are also using data from studies of more mature H2RG detectors to inform our
calibration plan.  Although these will have important differences from the WFI
flight detectors, we expect that problematic effects discovered in H2RGs will
need to be characterized in H4RGs at some level.  For instance, collaborators
Shapiro and Huff (via the Precision Projector Laboratory at JPL directed by collaborator Shapiro) have
investigated a high-frequency pattern (dubbed the ``crosshatch'') apparent in
flat-field calibrations and believed to be related to the crystal structure of
HgCdTe (see Figure~\ref{fig:crosshatch}).  Using an engineering grade H2RG
provided by the Euclid mission, tests have shown that the crosshatch pattern
affects photometry even after flat-field calibrations are applied, implying that
it has sub-pixel structure that can bias PSF measurements.  Data was shared with
this SIT to investigate the dependence of the pattern on polarization and angle
of incidence.  The same H2RG detector is also being used to conduct a PSF-based
test of BFE to compare with our flat-field analysis.

\subsubsection{Simulating Detector Imperfection and Their Effect on WFIRST Shape Measurements}
\label{sec:hlis_image_sim}
Building on the existing GalSim framework and WFIRST module, Michael Troxel and Ami Choi (in collaboration with Hirata, Jarvis, and Mandelbaum) are developing an image
simulation pipeline to assess the impact of various physical effects on the fidelity of the measured galaxy shapes.  The end product will provide realistic simulations containing all pertinent effects and conditions from the observational process specific to the WFIRST mission that may affect the quality of the lensing shear extracted
from the real WFIRST images.  These simulations will provide a foundation to characterize the relative impact of undesired effects and to validate the shear measurements themselves.  As the distribution and density of galaxies are realistically incorporated, the resulting multi-epoch images can also be used to test different dither strategies.  An intermediate level goal is to update the module with the most recent hardware and survey parameters describing the mission, to do some GalSim development to make the pipeline more efficient, and to estimate the relative impacts from detector effects such as the IPC described in earlier sections.

\begin{figure*}[!t]
  \includegraphics[width=0.9\textwidth]{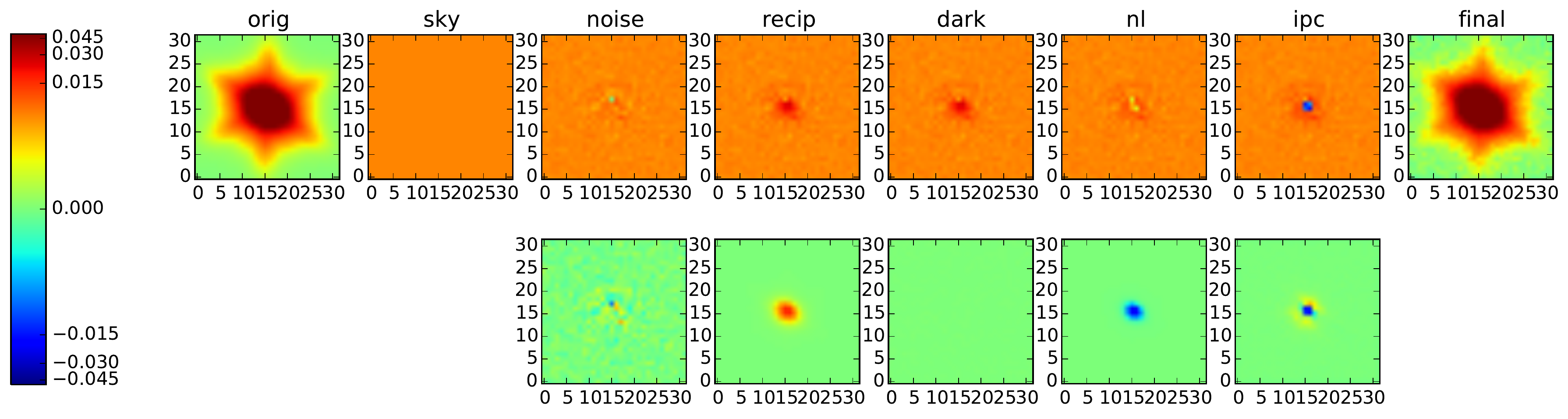}
\caption{\label{fig:wl_imsim} An example of a simulated bright, elliptical galaxy produced by the image simulation pipeline.  The top row of subpanels begins with the original simulated bright elliptical galaxy (far left).  Each subpanel to the right of the original galaxy is a ``difference'' image that has the original galaxy subtracted off after the given effect (top label) has been applied.  These effects are: sky background, noise, reciprocity, dark current, nonlinearity, inter-pixel capacitance, respectively.  The bottom row of subpanels corresponds to the individual, isolated effects (top label).  The top, far right subpanel shows the final galaxy image after all effects have been applied.  The sub panels have been normalized by the maximum flux value of the original galaxy image, and the colors have been mapped to $\log_{10}$ values with a small range right around zero mapped to linear values as shown on a single colorbar corresponding to all of the subpanels.}
\end{figure*}

The pipeline is currently capable of simulating galaxies on individual postage stamps with a size and flux distribution drawn from the CANDELS catalog from Capak and Hemmati, a dither pattern from Hirata, realistic noise (see below), correct layout of SCAs, and options to dial a range of detector-level effects such as IPC, non-linearity, and reciprocity, among others.  The output simulated galaxies and truth tables are saved in Multi Epoch Data Structures (MEDS), which is a format commonly used in DES.  A publicly available shape measurement software, ngmix (\href{https://github.com/esheldon/ngmix/}), has been interfaced to measure shapes of the simulated galaxies.  Figure~\ref{fig:wl_imsim} illustrates a few of the effects in the context of a simulated bright, elliptical galaxy on a 32x32 pixel postage stamp.  The software is maintained in a repository available at \href{https://github.com/matroxel/wfirst_imsim} on GitHub.

The H4RG read noise model developed by \cite{Rauscher2015} can be straightforwardly incorporated into the simulations in order to study the effects of correlated noise on shape measurement.  Hirata has used simulations and a semi-analytic formalism to show that anisotropic noise induces additive ellipticity measurement errors, with the most damaging contributions coming from noise power at spatial wavelengths $\sim\pi R$ where $R$ is the observed scale radius of the galaxy or star.  With these tools, we will be able to derive requirements on detector noise and calibrate shape measurement pipelines to correct for remaining correlated noise.

\begin{figure}
\centering
\begin{minipage}{.4\textwidth}
  \centering
  \includegraphics[width=.7\linewidth]{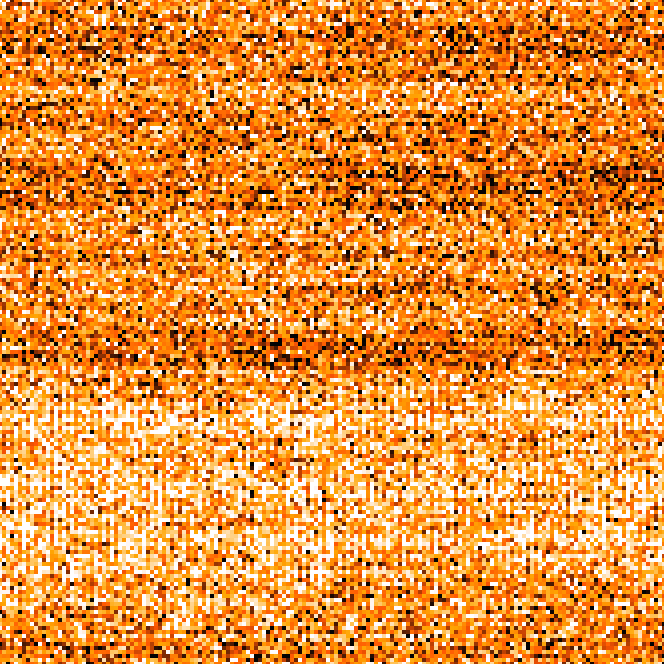}
\end{minipage}%
\begin{minipage}{.4\textwidth}
  \centering
  \includegraphics[width=1\linewidth]{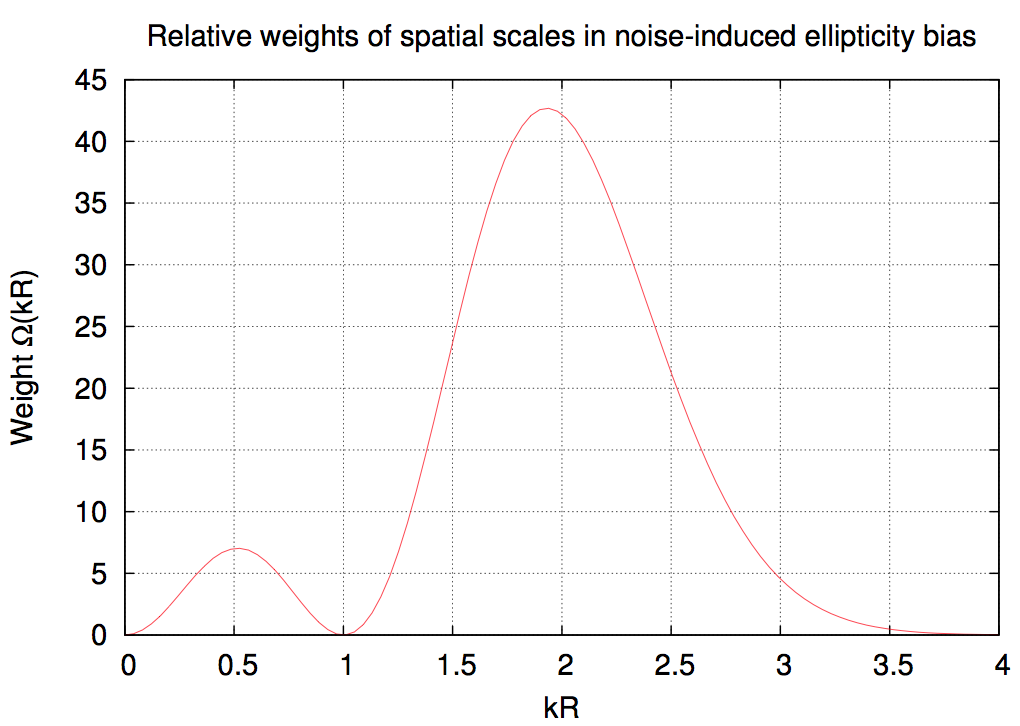}
\end{minipage}
\caption{LEFT: Simulated noise in an H4RG subregion using the model of \cite{Rauscher2015}.  This image demonstrates the spatial effect of 1/f noise (horizonal bands) and alternating column noise (high frequency vertical striping).  RIGHT: Weight function showing the relative contribution of noise with spatial wavenumber $k$ to the ellipticity bias of a source with scale radius $R$.  For a given signal to noise level, the read noise shown would bias $e_1$ by different amounts for an undersampled PSF versus a larger galaxy.  This weight function assumes an adaptive moments shape measurement algorithm.}
\label{fig:correlated_noise}
\end{figure}

\subsection{Enabling Photometric Redshifts with WFIRST (D6, D11)}
\label{sec:wl_photoz}

\begin{summaryii}
Accurate photo-$z$s are crucial to all WFIRST probes of dark energy. In the
first year of SIT activity we have focused on developing accurate data-driven
simulations of the WFIRST lensing galaxy population and determining the
requirement on the spectroscopic samples needed to calibrate these photometric
redshifts. We proposed a plan to calibrate this sample and study the importance of the IFC. In the process, we generate new
data products that we released to other SITs and to the community.
\end{summaryii}

\begin{figure}
 \includegraphics[width=0.45\textwidth] {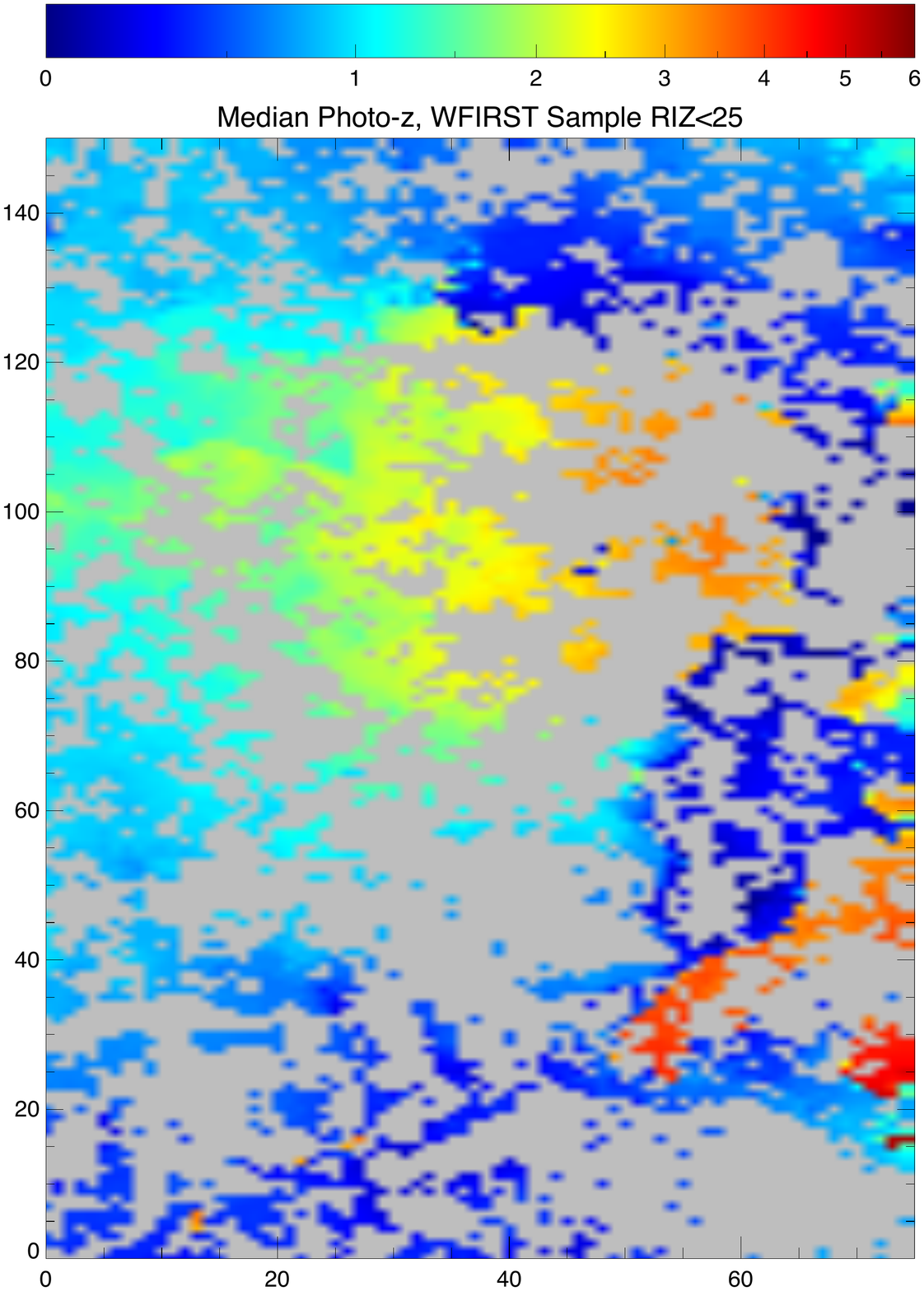}
\caption{A Self Organizing Map (SOM) \citep{Masters2015} of color space generated from $\sim6$ square degrees of data in the VVDS 2h, SXDS/UDS, COSMOS, and EGS survey areas colored by photometric redshift is shown.  Only regions occupied by the CANDELS survey are colored, with the remaining area grey.  CANDELS only covers 42\% of the color space, representing 49\% of the overall galaxy population.}
\label{fig:EuclidCandelsRep}
\end{figure}

\subsubsection{Generating Data-driven Simulations of WFIRST Galaxy Population}
\label{sec:candels}

The closest analogs to WFIRST data are the COSMOS and CANDELS HST surveys, however neither is fully analogous to
 WFIRST HLS data.  The COSMOS data cover 1.7 square degrees with HST-ACS (F814W)
 with ground based data analogous to LSST.  However, WFIRST analogous infrared
 data are not available over the majority of the field and extrapolations to the
 WFIRST lensing cuts from the F814W data over-estimate the number-density of
 sources usable for lensing.  In contrast, CANDELS has WFIRST analogous infrared
 data, but covers only 0.2 square degrees,which means it does not sample the
 full WFIRST galaxy population, and has very heterogeneous optical coverage.
 Specifically, a comparison between the CANDELS and COSMOS data to R,I,Z$<$25
 found that only 42\% of COSMOS galaxy colors (representing 49\% of the galaxy
 population) are present in CANDELS.  Figure~\ref{fig:EuclidCandelsRep} shows a
 Self Organizing Map (SOM) \citep{Masters2015} of the galaxy color space with
 regions where CANDELS galaxies fall marked.  The empty regions are shown in
 grey and correspond to cells with low galaxy density in COSMOS.  So these
 galaxies are simply less likely to be found in the relatively small area of
 CANDELS.

\begin{figure}
\centering
  \includegraphics[trim=0cm 0cm 0cm 0cm, clip,width=0.50\textwidth] {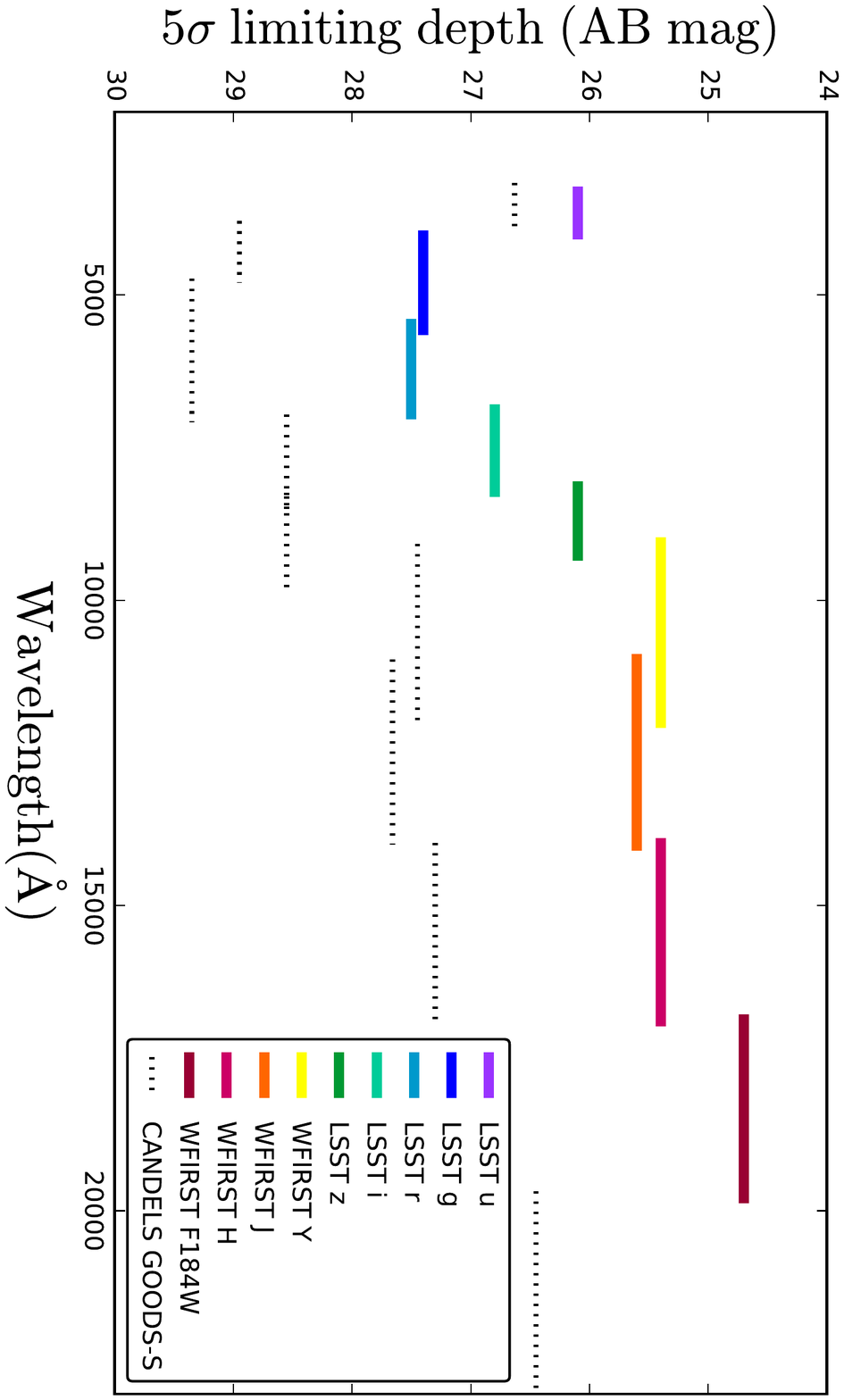}
\caption{The 5$\sigma$ limiting AB magnitude of LSST and WFIRST filters plotted
as solid color lines and the 5$\sigma$ limiting AB magnitude of CANDELS GOODS-S
filters are plotted with dotted black lines (see Table \ref{tbl:filters} for
filter names).  Note the significant differences in the filter system which
necessitates conversion from one to the other.}
\label{fig:filters}
\end{figure}

To overcome these limitations we have taken several approaches.  First, we have
collected a homogeneous 0.3-2.5$\mu$m data set over $\sim6$ square degrees in
the VVDS 2h, UDS/SXDS, COSMOS, and EGS fields.  These data are not as deep as
WFIRST, but are analogous to the LSST and Euclid data and allow us to estimate
the cosmic variance in galaxy population and estimate requirements on
spectroscopy.  These have been combined with the CANDELS catalogs which probe
WFIRST depth but are sample size and variance dominated.  We then adapted the
simulations described in \citet{stickley2016} developed for the SPHEREx mission
to assign a R$\sim$600 spectra to each object.  The CANDELS data are very
heterogeneous (see Table \ref{tbl:filters}), so we converted the various
photometric systems to a LSST+WFIRST system.  Figure~\ref{fig:filters} shows an
example of the LSST+WFIRST system along with the CANDELS filters on GOODS-S as
an example.  For the conversion we compared the converted photometry from other
bands to actual CFHT-LS $u^*$, $g^*$, $r^*$, $i^*$, $z^*$ and VISTA $Y$, $J$,
$H$, $K_s$ photometry in fields where they are available.   We found a simple
linear interpolation between filters in flux produced the best agreement.  Using
the \citet{stickley2016} or other template fits produced discretized value  in
the output fluxes which biased further analysis.

The combination of these catalogs produces a reasonable estimate of the WFIRST
galaxy population for the purposes of assessing variance and the effects of
cuts.  However for some analysis a fully simulated catalog is required so that
the inputs are known perfectly.  To provide this we further adapted the methods
described in \citet{stickley2016} to produce simulated WFIRST+LSST photometry.
These three sets of simulated samples are being provided to other WFIRST SIT
teams for their analysis. Specifically we have been working with the Foley SNe focused SIT
team to simulate photo-$z$ performance for supernova cosmology.


\begin{table}
\tabcolsep 2.8pt
\footnotesize
\caption{CANDELS filters in each field used to create the LSST+WFIRST catalog}
\begin{center}
\begin{tabular}{@{}*{15}{c}@{}}
\hline
\hline
\\
Field&&&&&&& Filters\footnote{\footnotesize Refer to CANDELS catalog papers for detailed description of observations in each filter, GOODs-S: \citealt{Guo2013}; GOODS-N: Barro et al. in prep; EGS: \citealt{Stefanon2017}; UDS:\citealt{Galametz2013}; COSMOS: \citealt{Nayyeri2017}}&&\\
\\
\hline
\\
GOODS-S& $\rm U_{VIMOS}$& $\rm F435W$& $\rm F606W$&$\rm F775W$ & $\rm F814W$ & $\rm F850lp$ & $\rm F098W$& $\rm F105W$ & $\rm F125W$ &$\rm F160W$&$\rm Ks_{HAWK-I}$\\
\\
GOODS-N&$\rm U_{KPNO}$& $\rm F435W$& $\rm F606W$&$\rm F775W$ & $\rm F814W$ & $\rm F850lp$ &$\rm F105W$ & $\rm F125W$ &$\rm F160W$& $\rm Ks_{CFHT}$\\
\\
EGS& $\rm U_{CFHT}$& $\rm g_{CFHT}$& $\rm F606W$ &$\rm r_{CFHT}$& $\rm i_{CFHT}$& $\rm F814W$ &$\rm z_{CFHT}$ & $\rm F125W$&$\rm F160W$&$\rm Ks_{CFHT}$\\
\\
UDS& $\rm U_{CFHT}$& $\rm B_{subaru}$&$\rm F606W$ &$\rm Rc_{subaru}$& $\rm i_{subaru}$&$\rm F814W$ &$\rm z_{subaru}$ & $\rm Y_{HAWK-I}$&$\rm F125W$&$\rm F160W$&$\rm Ks_{HAWK-I}$\\
\\
COSMOS&$\rm U_{CFHT}$& $\rm B_{subaru}$& $\rm F606W$&$\rm r_{subaru}$&$\rm i_{CFHT}$&$\rm F814W$ &$\rm z_{CFHT}$ & $\rm Y_{UVISTA}$&$\rm F125W$&$\rm F160W$& $\rm Ks_{UVISTA}$\\
\\
\hline
\end{tabular}
\end{center}
\label{tbl:filters}
\end{table}

\begin{figure}
\centering
\includegraphics[trim=0cm 0cm 0cm 0cm, clip,width=0.95\textwidth] {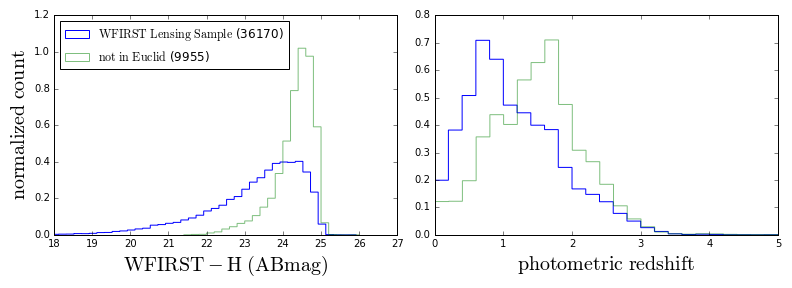}
\caption{{\bf Left:} A magnitude histogram of the WFIRST lensing sample split
into a C3R2 Euclid like lensing sample (blue, $RIZ<25$) and those only in WFIRST
(green).  Roughly 20\% of the WFIRST sample consists of galaxies fainter than
what C3R2 is calibrating for Euclid.  Note the Euclid lensing sample is cut at
$RIZ<24.5$, shallower than the proposed calibration \citep{Masters2015}.  {\bf
Right:} The redshift distribution of the C3R2-Euclid sample (blue) and the
WFIRST only sample (green) are shown normalized to an integral of 1. Even though
the WFIRST galaxies are fainter than the calibration limit they cover a redshift
range similar, just with with more galaxies at high-redshift.   }
\label{fig:EuclidVsWFIRST}
\end{figure}

\begin{figure}
\centering
\includegraphics[trim=0cm 0cm 0cm 0cm, clip,width=0.75\textwidth] {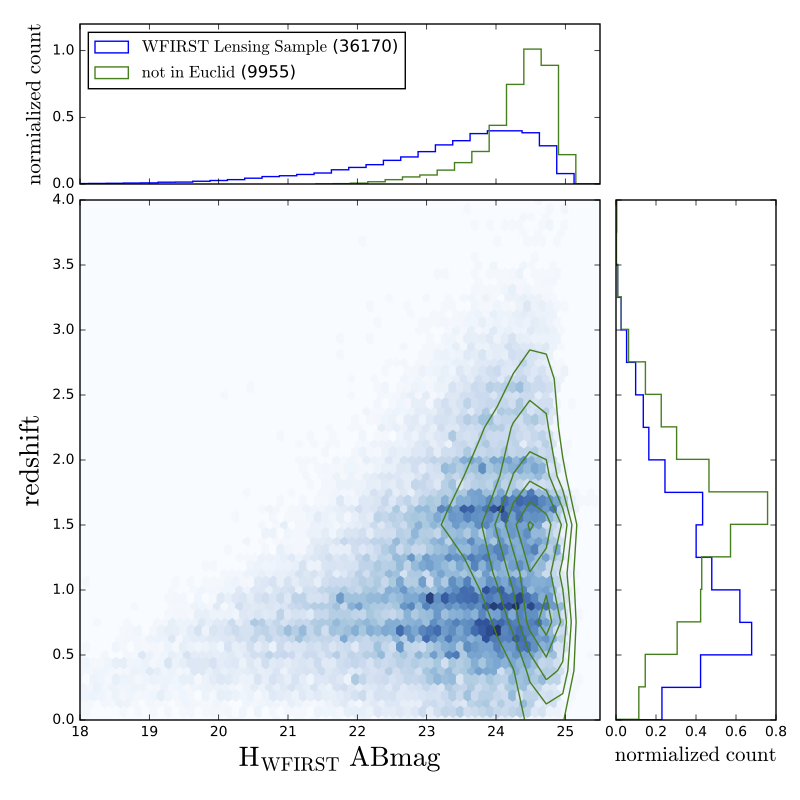}
\caption{$H$ band magnitude vs redshift is plotted for the WFIRST lensing sample
shown in Figure \ref{fig:EuclidVsWFIRST} with shading indicating relative
density.  The WFIRST only sample shown as green contours and the histograms from
from Figure \ref{fig:EuclidVsWFIRST} are plotted on the axis.   }
\label{fig:EuclidVsWFIRST2}
\end{figure}

\subsubsection{Calibrating the Photometric Redshifts of WFIRST Weak Lensing Galaxy Population}

\begin{figure}
\centering
 \includegraphics[trim=0cm 0cm 0cm 0cm, clip,width=0.98\textwidth] {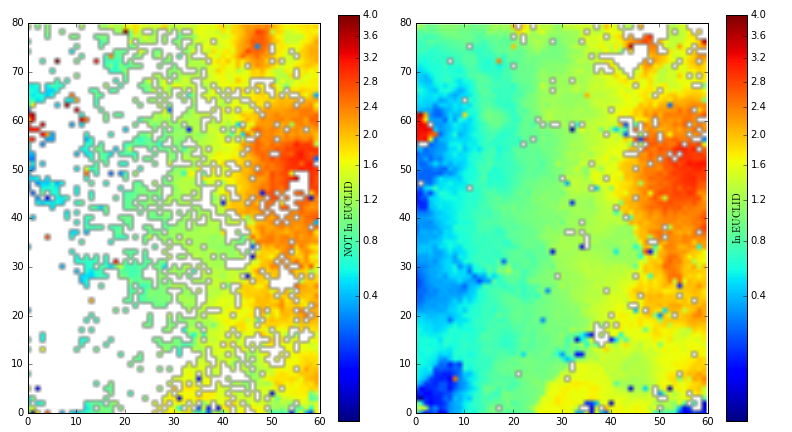}
\caption{A Self Organizing Map (SOM) \citep{Masters2015} of the WFIRST color
space based on the CANDELS data cut to the WFIRST lensing sample color coded by
their median redshift.  The left panel shows SOM cells that contain galaxies too
faint to be in the C3R2 RIZ$<25$ Euclid like sample
\citep{Masters2015,Masters2017}.  The right panel shows SOM cells with galaxies
that are in the C3R2 sample.  $\sim96\%$ of the WFIRST color space is occupied
by galaxies in the C3R2 sample, however $\sim20\%$ of the WFIRST sample is
fainter than the C3R2 limit.  This means their calibration would have to be
verified to ensure no magnitude dependence on redshift. }
\label{fig:WFIRSTSOM}
\end{figure}

These simulations have been used for several analyses within our SIT.  Figure~\ref{fig:EuclidVsWFIRST} shows the relative differences in the magnitude and
redshift distribution of the total Euclid and WFIRST faint lensing samples.
WFIRST clearly adds fainter and higher-redshift systems to the weak lensing
sample.  However, Figure \ref{fig:WFIRSTSOM} shows a Self-Organizing-Map
analysis \citep{Masters2015} of the WFIRST lensing sample compared with the
Euclid sample. Even though WFIRST is significantly fainter than Euclid, 96\% of
galaxies fainter than the Euclid sample have color analogs at brighter
magnitudes.  The implication is that while WFIRST is seeing fainter galaxies
than Euclid, these galaxies are very similar to less numerous but brighter
systems seen by Euclid.

\begin{figure}
\centering
 \includegraphics[trim=0cm 0cm 0cm 0cm, clip,width=0.45\textwidth] {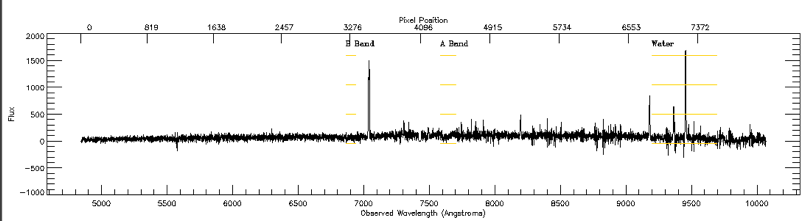}\\
   \includegraphics[trim=0cm 0cm 0cm 0cm, clip,width=0.45\textwidth] {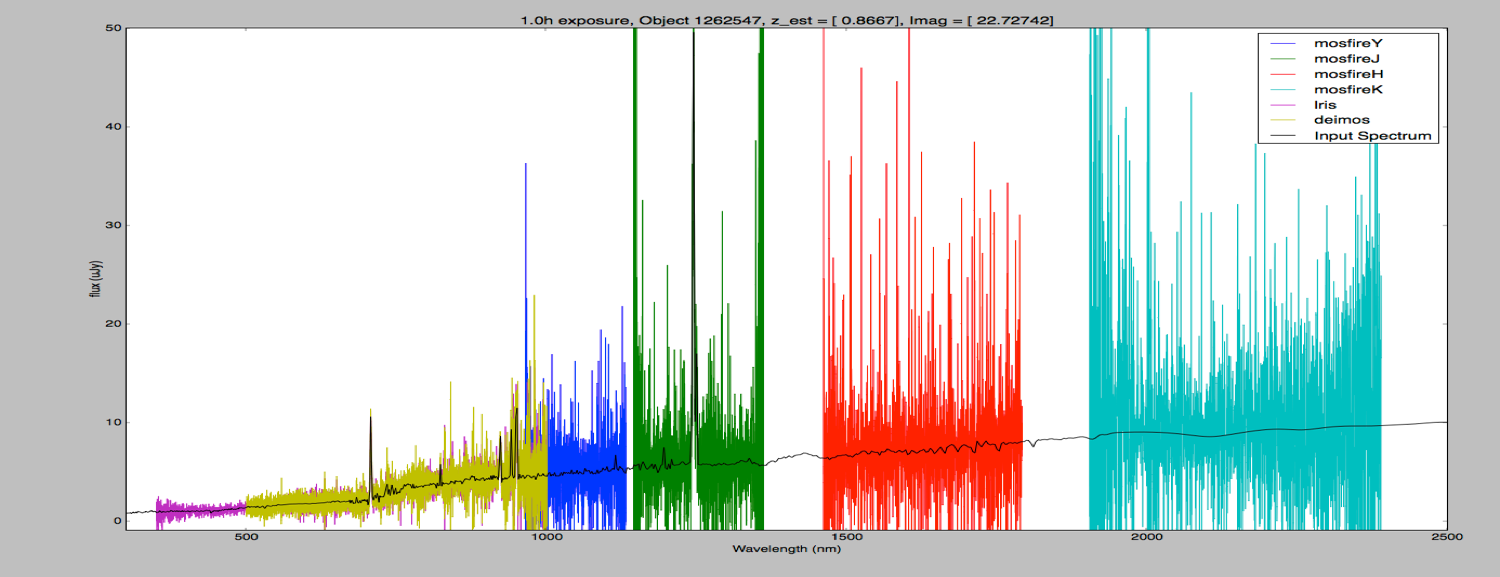}

\caption{ A real (top) and simulated (bottom) spectra of a $I=22.7$ galaxy in the C3R2 sample.}
\label{fig:SpecSim}
\end{figure}

To determine how difficult it would be to obtain spectra for these faint systems
we conducted an analysis of the R$\sim$600 SEDs fit to the photometry.  Based on
the C3R2 survey spectra \citep{Masters2017} we developed a spectral simulator
which accurately re-produces ground based spectra for Keck DEIMOS, LRIS, and
MOSFIRE.  In addition to these instruments the simulated response of the
WFIRST-IFC was simulated.  Example simulated spectra based on the model fits
along with actual Keck spectra obtained for those sources are shown in Figure
\ref{fig:SpecSim}.

We found that indeed most of the faint WFIRST lensing galaxies were analogs of
brighter systems. This alleviates the need to obtain spectroscopic redshifts to
this population since the color-redshift relation will be known.  However, steps
must be taken to validate that the redshift distribution does not change at
fainter magnitudes in ways not apparent in the WIFRST+LSST colors.

The simplest method would be to extend a survey such as C3R2 \citep{Masters2017}
to fainter magnitudes. However, these faint galaxies are difficult to obtain
high-quality spectra for from the ground.  For the purposes of this analysis we
define high quality as an SNR$>$7 on two emission features or an SNR$\sim5$ on
an underlying continuum.  Based on this criteria, 20\% of the WFIRST color space
requires $>5$h spectra from Keck.  It is important to note this is in terms of
color space, and the exact number of spectra required to calibrate this color
space will require further analysis.   Of these, 1\% are sources that require
long ground exposures due to strong emission lines falling between ground based
observing windows and spectra could be obtained with the WFIRST grism.  A
further 15\% would have high-quality redshifts with WFIRST-IFC parallel
observations based on simulating the spectra and assessing the number of
features with SNR$>$7.  However, due to the low-resolution of the WFIRST-IFC
further analysis may be merited.   The remaining 4\% of galaxies could not be
calibrated by WFIRST or from the ground with 10m telescopes and would require
either ELTs or JWST.

\subsection{Cluster Cosmology with WFIRST}

\begin{summaryii}
Our work during the past year has focused on building machinery
for comprehensive cosmological forecasts for the WFIRST cluster
program that will include representations of the most significant
anticipated systematic effects.
\end{summaryii}

Cluster cosmology is generally considered to be less demanding in terms of
hardware requirements than cosmic shear, since the large galaxy over-densities
and shear signals are not as easily masked by subtle optical aberrations or
detector behaviors. Nevertheless, it may place new requirements on survey
footprint/operations (to ensure overlap with other data sets); pipeline behavior
in crowded fields (e.g., \cite{2015MNRAS.449.1259S}); and ancillary data
products and simulations to describe, e.g., changes in selection effects and
source redshift distributions in the presence of blending and magnification.

 Clusters can be identified
 using galaxy counts in WFIRST and external data, or from X-ray or
 Sunyaev-Zeldovich surveys.  WFIRST yields high-precision measurements of
 cluster weak lensing shear profiles, which can be combined with cluster
 abundances and cluster-galaxy cross-correlations to derive cosmological
 parameter constraints, most notably on the amplitude of matter clustering.

We have verified our ability to reproduce previous forecasts quantitatively with
new and independent code -- a non-trivial exercise that required resolving
ambiguities about halo mass definitions, source redshift distributions, and so
forth. We have extended the new code so that it can simultaneously model weak
lensing signals from small scales (the ``one-halo'' regime) out to large scales
described by linear theory. To achieve accurate results in the transition
between these regimes, we have developed a numerically calibrated prescription
for mass profiles in the ``splashback'' zone beyond the cluster virial radius.
These numerical calibrations are based on a suite of cosmological N-body
simulations that we are using to create full numerical ``emulators'' for
cluster-mass and cluster-galaxy cross-correlation functions, using parameterized
halo occupation distributions to relate galaxy populations to the underlying
dark halo population. We are exploring the degree to which cluster-galaxy
cross-correlations can sharpen cosmological constraints when combined with
cluster weak lensing; we anticipate including these cross-correlations in our
cosmological forecasts at a later time.

Our current focus is on incorporating a realistic description of photometric
redshift distributions and nuisance parameters that describe systematic
uncertainties in those distributions. Our hope is that the combination of cosmic
shear and cluster weak lensing measurements will prove much more robust to
photometric redshift uncertainties than either technique individually, because
the two methods have somewhat different dependence on source redshifts, and
because the redshifts of the clusters themselves are accurately known. We will
then turn to nuisance parameters that describe uncertainties in the clusters
themselves, e.g., contamination, incompleteness, mis-centering, and projection
biases in cluster selection.

Collaborator Anja von der Linden presented a technical overview of the WFIRST
cluster program at the January 2017 WFIRST Science meeting at the Center for
Computational Astrophysics, with an emphasis on the opportunities and
requirements for joint analyses with LSST and Sunyaev-Zeldovich surveys.
Collaborator Eduardo Rozo, together with Eli Rykoff, has been leading efforts in
cluster identification in the Dark Energy Survey (DES), building on their
earlier work with the Sloan Digital Sky Survey.  Rozo is also leading the
cluster cosmology analyses in the DES, with Year 1 DES results expected in a few
months. WFIRST cluster identification and analysis methods will build on the DES
techniques, and the lessons from applying these techniques to state-of-the-art
wide-field survey data will be invaluable for WFIRST planning.

\section{Galaxy Redshift Survey Investigation (D1, D4, D8, D9)}
\label{sec:gc}





The defining goal of HLS spectroscopy is to derive constraints on dark energy
from a slitless spectroscopic (grism) redshift survey of approximately 20
million emission line galaxies (ELG) in the redshift range $z=1-3$. The galaxy
redshift survey will enable high-precision measurements of the cosmic expansion
history via BAO and structure growth via RSD. Acoustic oscillations in the
pre-recombination universe imprint a characteristic scale on matter clustering,
which can be measured in the transverse and line-of-sight directions to
determine the angular-diameter distance $D_A(z)$ and Hubble parameter $H(z)$,
respectively \citep{Blake03,Seo03,CW12}.  Anisotropy of clustering caused by
galaxy peculiar velocities constrains (in linear perturbation theory) the
combination $\sigma_m(z) f_g(z)$, where $\sigma_m$ describes the rms amplitude
of matter fluctuations and $f_g(z) \equiv d\ln\sigma_m(z)/d\ln a$ is the
fluctuation growth rate. Thus the GRS on its own can address the key questions
identified by NWNH\@: whether cosmic acceleration is caused by modified gravity
or by dark energy, and whether (in the latter case) the dark energy density
evolves in time \citep{Guzzo08,Wang08}.  These tests become more powerful in
combination with weak lensing and cluster measurements from HLS Imaging and
high-precision relative distance measurements from the Supernova Survey
\citep{dePutter:2013xda,dePutter:2013nha}. The broadband shape of the galaxy
power spectrum and higher order measures of galaxy clustering provide additional
diagnostics of dark energy, neutrino masses, and inflation, and insights on the
physics of galaxy formation. There are two largely distinct sources of
systematics in the galaxy clustering program, associated with the uniformity of
the GRS and with astrophysical modeling uncertainties. While all aspects of our
GRS investigation are interconnected, we worked on science requirements, image simulations,
and prototype pipelines, cosmological forecasting, modeling, and cosmological simulations.

\begin{summary}
To mature the WFIRST GRS, our work has been organized along four main directions.
\begin{enumerate}
\item We developed, delivered to the project and updated the GRS requirements;
\item We generated new WFIRST specific light-cone simulations;
\item Using HST measurement, we started a new data analysis effort to improve our knowledge of the H$\alpha$ luminosity function, a critical element to plan the GRS;
\item We developed quick and agile analysis tools that will help us develop a pseudo-pipeline in the coming years.
\end{enumerate}
\end{summary}

 \subsection{Developing the GRS Requirements (D1)}

 \begin{summaryii}
   Over the last year, our main priority have been to support and guide the development of the WFIRST HLS spectroscopy and in particular to identify, articulate and validate the scientific requirements of the instrument, the data reduction software, and the survey. Responding to a calendar set by the Project Office, our SIT delivered three major updates to the WFIRST GRS requirements to the Project Office on July 1, 2016, December 1, 2016, and March 2, 2017. Each of these provide progressively sharper definitions of the  GRS requirements. We describe the main requirements and their science drivers below. \emph{Disclaimer: The requirements below reflect a snapshot of the requirements formulation. The official Science Requirements Document (SRD) will always supersede the requirements written here.}
 \end{summaryii}


 \subsubsection{Science Requirements (Level 2a)} In this section we present the current level 2 science requirements as delivered to the Project Office. This section should be considered a snapshot as we will refine this requirements further in the coming years.
\label{sec:sr2a_grs}

 \paragraph{HLSS 1} The area to be surveyed shall be $\sim$1500 deg$^2$ (2000 deg$^2$ goal) after correcting for edge effects.  This area will be contiguous to the extent practical, and at least 90\% of the survey area must also be covered by the high latitude imaging survey.

 The survey area should be contiguous and large enough to reduce edge effects in
 the BAO/RSD measurements.  The $>90\%$ overlap with the HLIS enables joint
 analysis of 90\% of WL and GRS data, which maximizes the dark energy science
 from WFIRST.  Imaging also provides undispersed galaxy positions, improving
 redshift determination.  The statistical precision of the dark energy
 constraints is sensitive to the survey area as well as the survey depth; a trade
 study of depth versus area will need to be carried out to optimize both, in the
 context of Euclid and LSST. We also need to investigate the impact of dividing
 the area into two equal patches near the NEP and SEP respectively, to take
 advantage of potential ground-based telescope resources.  A survey of one or
 two large, contiguous areas has smaller edge effects and better window functions
 than a survey comprised of many smaller areas.

 We have carried out trade studies of the HLSS survey design. We note that it
 will be important to conducting these trade studies in the context of the joint
 science return of HLSS and HLIS that properly accounts for correlations among
 spectroscopic and imaging observables and accounts for their correlated
 systematics.  We are in the progress of implementing a corresponding forecasting
 effort.  Here, we have carried out a trade study of area versus depth for the
 HLSS only, starting from a baseline survey of 2227 deg$^2$ and a wavelength
 range of 1.05-1.85 microns. We consider two alternative scenarios, i.e. a survey
 twice as wide and shallower and a survey half as wide but correspondingly
 deeper.  The galaxy redshift distributions were computed using the WFIRST
 Exposure Time Calculator ETC v14. The H$\alpha$ forecasts are based on the average
 of the 3 models in \citet{Pozzetti:2016}, and the [O III] forecasts are based
 on the \citet{Mehta:2015} luminosity function.

 We extend the \CoLi framework \citep{Eifler:2014,Krause2016} to compute the constraining power of all scenarios on cosmic acceleration, closely following \citet{Wang2013}. We run 500,000 step
 MCMC simulated likelihood analysis in a 23 dimensional parameter space. We
 simultaneously vary 7 cosmological parameters and 16 ``nuisance'' parameters
 describing uncertainties due to the linear galaxy bias model, the non-linear
 smearing of the BAO feature, peculiar velocity dispersion, power spectrum shot
 noise, and redshift errors. We assume priors on cosmological parameters from the
 current state of the art experiments, i.e. the Planck mission, the Baryon
 Oscillation Spectroscopic Survey (BOSS), the Joint Lightcurve Analysis (JLA)
 supernovae, as described in \citet{Aubourg:2015}.

 The information gain is quantified using the standard Dark Energy Task Force FOM
 and an extended cosmology FOM, which measures the enclosed volume in the full
 7-dimensional cosmological parameter space, not just in the 2 dark energy
 parameters. We will refer to these FOMs as DE-FOM and Cosmo-FOM.  Compared to
 the baseline survey, we find a decreased DE-FOM of 32\% and a decreased
 Cosmo-FOM of 45\% for the shallow/large area survey. For the deep/small area
 survey we find an increased DE-FOM of 5\% and an increased Cosmo-FOM of 2\%.
 While our trade study validates the design of the baseline survey, we note that
 these findings are model and prior dependent and will carry out further studies
 varying the input parameters. In particular, the [OIII] galaxy number density
 will be updated pending inclusion of the results from the latest observational
 data from HST grism observations.

 We also investigated whether the survey area needs to be contiguous. We
 constructed two identical sets of Gaussian simulations, one set covering
 contiguous 2000 deg$^2$ and a second set consisting of two $1000$ deg$^2$
 disjoint fields. The BAO signal was then measured in the 2D power spectrum using the
 most recent techniques applied to the BOSS DR12 data. The BAO positions measured
 in disjoint fields were biased by 1\% on average compared to the contiguous
 field in both line-of-sight and transverse directions. This bias persists even
 after properly correcting for the window effects and is unlikely to be coming
 from the sample variance since our sets consisted of close to one thousand
 independent simulations. This bias could be a result of either bigger than the
 box-size modes or various edge effects. The window of the real data will be more
 involved than we considered in our test case and the biases may be larger. This
 investigation is ongoing but our preliminary results seem to support the
 conclusion that a contiguous area is preferable for the standard BAO analysis.

\paragraph{HLSS 2} The comoving density of galaxies with measured redshifts shall satisfy $n > 3\times10^{4}\ (h/\textrm{Mpc})^3$ at z=1.6.

 This is set by $nP_{0.2} \sim1$ at $z=1.6$, with 20\% margin. Requiring $nP_{0.2}
 \sim1$ implies $n> 3\times10^{-4}\ (h/\mathrm{Mpc})^{-3}$ at $z=1.3$, and $n >
 6.5\times 10^{-4} (h/\mathrm{Mpc})^{-3}$ at $z=1.8$. Given the Hirata
 forecast of H$\alpha$ ELG counts (Model 3 in \citet{Pozzetti:2016}, $nP_{0.2}\sim0.6$ at $z=1.8$,
 and $nP_{0.2}>2$ at $z=1.3$.  We cannot require a higher galaxy number density than
 what nature provides, given fixed observing time and area coverage. Here we have
 chosen a characteristic high redshift, $z=1.6$, at which it is impossible for a
 ground-based survey to obtain spectra for a large number of galaxies. There
 remain large uncertainties in the H$\alpha$ LF due to the limited availability of
 uniform data. It is likely that the actual number of H$\alpha$ ELGs is higher than
 assumed here; thus we have additional margins for this requirement. We have
 assumed a bias for H$\alpha$ ELGs of $b(z) = 1+0.5z$. The bias relation has been rescaled
 to agree with \citet{Geach2012} measurement of $b=2.4$ at $z=2.23$ for $f >
 5\times 10^{-17} \, \mathrm{ erg\,s^{-1}cm^{-2}}$.

 This is significantly deeper than the Euclid GRS survey, which ensures that the
 WFIRST GRS is deep enough for carrying out robust modeling of systematic effects
 for BAO/RSD, higher order statistics, and the combination of weak lensing and
 RSD as tests of GR. This number density requirement could in principle be met
 using either H$\alpha$ or [OIII] ELGs, depending on the survey strategy. There is no
 need to set a separate requirement for [OIII] ELGs; this depth ensures high
 number densities for both [OIII] and H$\alpha$ ELGs.

 Galaxy number density is a key input in the dark energy Figure-of-Merit. It
 is very sensitive to the H$\alpha$ LF, which still has large uncertainties but will
 become better determined as more data become available and more comprehensive
 analyses are done. The flow down of the galaxy number density requirement here
 to the minimum survey depth depends on the LF of ELGs.
 Co-I Teplitz is a key member of the WISP team. He is supervising a postdoc, Ivano Baronchelli, in deriving more precise LFs for H$\alpha$ and [OIII] ELGs using WISP data.

\paragraph{HLSS 3} The wavelength range of the HLSS will allow measurement of
H$\alpha$ emission line redshifts over the redshift range 1.1$<$z$<$1.9.

 The corresponding wavelength range is 1.38 $\mu$m to 1.9 $\mu$m.  This wavelength
 coverage also allows measurements of [OIII] emission line redshifts over the
 range $1.8 < z < 2.8$.  A wider wavelength range that allows H$\alpha$ emission line
 detection over a wider redshift range is desirable, as it increases the survey
 volume and therefore adds margin for meeting other baseline requirements.  It is
 also critical that the WFIRST GRS redshift range is complementary to that of
 Euclid, with its red cutoff at 1.85 $\mu$m, or $z < 1.8$.

 The key consideration is that a space mission should focus on what cannot be
 accomplished from the ground, and be complementary to other space missions in
 wavelength coverage. Ground-based GRS can reach $z\sim1$ without great difficulty,
 thus we should focus on $z>1$. Euclid GRS can only reach $z \sim2$; its shallow depth
 does not enable a high enough number density of observed [OIII] ELGs. WFIRST GRS
 is deep enough to observe both H$\alpha$ (656.3nm) and [OIII] (500nm) ELGs, with the
 number density of the latter sensitive to the survey depth (the deeper the
 survey the higher their number density).

 For the nominal wavelength range of 1-2 microns, WFIRST GRS covers $0.52 < z <
 2$ using H$\alpha$ ELGs, and $1 < z < 3$ using [OIII] ELGs. Thus the redshift range
 requirement is met including both types of ELGs.

 In addition to the trade studies in HLSS 1 we examine the impact of an extended
 wavelength range on the DE-FOM and the Cosmo-FOM. We follow the same methodology
 as detailed in the HLSS 1 description in extending the wavelength range from
 1.05-1.85 microns for the baseline model to 1.00-1.89 for the extended model.
 We find a decreased DE-FOM of 2\% and a decreased
 Cosmo-FOM of 11\% for the extended wavelength survey with respect to our
 baseline scenario. While the FoM trade study seems to favor a narrower redshift
 range, we emphasize that these findings are model and prior dependent and will
 conduct further studies varying the input parameters. The reduction in the
 telescope temperature to 260K will have a major impact on this trade study. In
 addition, the FoM comparison is quantitative but simplistic; it does not reflect
 how the various future surveys will complement each other. Euclid GRS covers the
 wavelength range of 0.92-1.85 microns using the same BAO/RSD tracers as WFIRST,
 thus there is unique scientific value in WFIRST having a wavelength cutoff
 longer than 1.85 microns. We note however that the Euclid Wide Survey of 15,000 sq. deg, which will measure BAO/RSD, will only use the red grisms,
covering 1.25-1.85 microns. The blue grism (0.92-1.3 microns) will likely be used in the deep fields (~40 sq. deg.).

\paragraph{HLSS 4} Redshift measurement errors $\sigma_z$ shall satisfy $\sigma_z < 0.001(1+z)$, excluding outliers, for galaxies smaller than $0.54''$ in radius. The fraction of outliers with $|z_\mathrm{obs}-
 z_\mathrm{true}|/(1+z_\mathrm{true})>0.003$ shall be less than 10\%. The incidence of outliers shall be known to a fraction of $2\times 10^{-3}$ of the full sample at each redshift.

 This is a requirement on the rms error of the redshift measurements, and not on
 every redshift measurement.

 We justify the requirement on the knowledge of the outlier fraction as follows.
 If you add a contaminant to the galaxy power spectrum with contamination
 fraction $\alpha$, then you leak in an amount of power from the ``wrong'' line
 with amplitude $\alpha^2$, but you dilute the power spectrum of the ``real''
 signal by a factor of $(1-\alpha)^2$. For BAO the dilution is a minor issue (it
 reduces S/N), but for RSD it is a problem because it reduces the galaxy bias by
 $(1-\alpha)$ without changing the linear redshift-space distortion parameter
 $f\sigma_8$. So your inferred rate of growth of structure $f\sigma_8(z)$ is
 reduced by a factor of $1-\alpha$.  This leads to a stringent requirement on
 knowledge of $\alpha$ -- if it is 9.8\% but you think it is 10\% you have a
 0.2\% systematic error, which is a reasonable budget for this contribution.

\begin{figure}
\includegraphics[width = 3.5in]{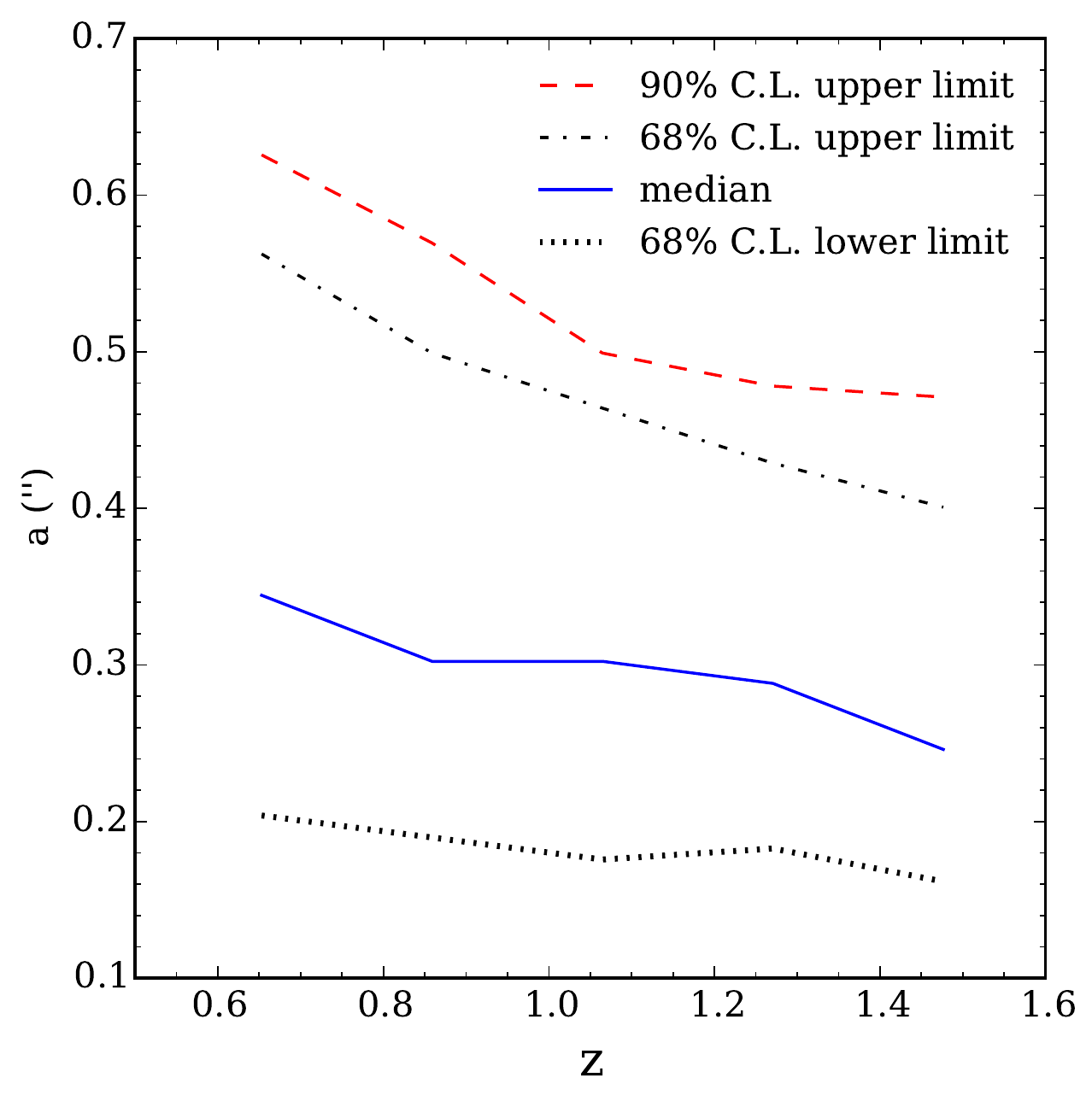}
\label{fig:size_zbin}
\caption{The 90\% upper limit of the semi-major axis
continuum size of H$\alpha$ galaxies with H$\alpha$ line flux $>10^{-16}
\mathrm{erg/s/cm}^2$, $0.55 < z < 1.85$), based on 1773 galaxies from WISP (WISP
team, private communication).} \end{figure}

 Larger size galaxies have larger redshift errors. To assess redshift accuracy,
 a realistic pixel level grism simulation covering at least 2 square degrees
 needs to be carried out and processed. The current data from the HST grism
 survey WISP finds that 90\% of galaxies that would be observed by WFIRST
 (H$\alpha$ flux $> 10^{-16} \mathrm{erg/s/cm}^2$, $0.55 < z < 1.85$) have a
 size less than $0.54''$ (semi-major axis continuum size). Since the maximum
 redshift of the WISP sample is $\sim1.6$, WFIRST H$\alpha$ ELGs will likely
 have smaller sizes on average, see Figure \ref{fig:size_zbin}.

 The [OIII] ELGs are more compact, so this requirement is also sufficient for
 the redshift precision for [OIII] ELGs.

\paragraph{HLSS 5} Relative position measurement uncertainties shall be less than $3.4''$ over the entire survey area.

 We need to measure galaxy positions to better than $\sim0.1 \mathrm{Mpc}/h$ (which corresponds to $3.4''$, assuming that 105 $\mathrm{Mpc/h}$ subtends 1 degree), in order to measure galaxy clustering accurately. This should be met easily if HLIS 26 is met, which makes systematic errors in the astrometry negligible. Given the pixel scale of $0.11''$, this requirement is automatically met within each field, and is tied to the precision of astrometry across different fields.

\paragraph{HLSS 6} The survey completeness shall be 50\% (TBC), and the redshift purity
 shall be 90\% (i.e., the outlier fraction is less than 10\%). Completeness is
 defined as the fraction of H$\alpha$ ELGs with measured redshifts flagged as reliable, and purity is defined as the fraction of measured redshifts flagged as reliable that are actually within 2$\sigma$ of the true redshifts.

 A requirement on completeness and purity is needed to translate the H$\alpha$ ELG
 number counts predicted by the H$\alpha$ LF to the galaxy number density that can be used to measure BAO/RSD by WFIRST GRS.  The completeness of 50\% and redshift
 purity of 90\% are put in as crude estimates based on extrapolations from
 Euclid.  Since WFIRST has a higher spatial and spectral resolution compared to
 Euclid, and more rolls (4 versus 3) per field, we expect a higher completeness
 and purity for WFIRST.  The actual requirements will need to be validated by
 grism simulations, since these are determined by what are feasible given the
 instrumentation and the true universe.  The requirement on the knowledge on the
 contamination fraction is set by HLSS 4.

 \subsubsection{Implementation Requirements (Level 2b)} In this section, we present the implementation requirements as delivered to the Project Office.

\paragraph{HLSS 7} The observatory shall provide a slitless spectroscopy mode, with a spectral dispersion no larger than 10.85 $\AA$/pix.

 Gratings tend to give constant dispersion in linear space, rather than R-space.
 The above dispersion would give point-source spectral resolution
 $R=\lambda/\Delta\lambda$ in the
 range $550 <R < 800$, for a 2-pixel resolution element.

 The grism resolution requirement is set by requiring the redshift precision to
 be 0.1\% (set by BAO/RSD science), thus is not sensitive to which ELGs (H$\alpha$
 vs [OIII]) we use as tracers. Going to lower spectral resolution would degrade
 the redshift precision and put BAO/RSD science goals at risk. The number density
 of [OIII] ELGs may be significantly higher than previously assumed; this gives
 some margin in the spectral resolution requirement due to the smaller sizes and
 less line blending of [OIII] ELGs.

 Given the margin from a likely higher [OIII] ELG number density than previously
 assumed, we have removed the requirement on resolving H$\alpha$ and NII for all
 galaxies of radius $0.3''$, and 90\% of galaxies of radius $0.54''$, which would drive
 the grism resolution higher. The blending of H$\alpha$ (6563$\AA$) and NII (6584$\AA$) leads to
 a metallicity-dependent shift in line centroid for larger sources;
 this would lead to a systematic bias in the measured redshifts, which can
 propagate into the BAO/RSD measurements. Having a higher grism resolution would
 alleviate this problem, at the cost of a reduction in survey depth, and more
 overlapping of spectra for galaxies. This is a trade study that we will carry out
 as the required grism simulations become available.

 \paragraph{HLSS 8} Spectra shall achieve $S/N \geq 5$ for $r_\mathrm{eff} = 300
 \mathrm{mas}$ for an emission line flux $~ 1.0\times10^{-16}\ \mathrm{erg/cm}^2\mathrm{/s}$, from a source at 1.8 $\mu$m.

 This sensitivity is sufficient to meet the comoving space density requirement
 HLSS 2 with some margin given best estimates of the H$\alpha$ luminosity function
 \citep{Pozzetti:2016} at these redshifts. The use of a $\mathrm{S/N} \geq 5$ threshold for an arbitrary spectrum pre-spectral-decontamination gives margin for detection of sources whose spectra overlap others, or for loss of some exposures to cosmic ray hits or other artifacts, as $\mathrm{S/N} \geq 5$ post-decontamination is expected to be
 sufficient for meeting the redshift accuracy requirement HLSS 4, and the
 post-decontamination S/N should be significantly higher than the
 pre-decontamination S/N for a given spectrum. Current calculations of
 observatory performance indicate that the sensitivity specified here is achieved
 in a total exposure time of  $\sim1200$ seconds per field.

 The median continuum size (semi-major axis) of H$\alpha$ ELGs is $0.3''$ (see Figure \ref{fig:size_zbin}).
 This sensitivity requirement is phrased in parallel with the sensitivity requirement
 of the WL survey. This depth is a factor of two to three deeper than the Euclid
 GRS. The depth is sufficient to give the required galaxy number density in HLSS 2.

\paragraph{HLSS 9} The uncertainty of the wavelength measurement $\lambda$ shall satisfy $\Delta\lambda/\lambda \leq 0.001$.

 Although this is redundant since it is essentially the same as HLSS 4, it is
 necessary to keep it since it flows HLSS 4 into a dataset requirement.

\paragraph{HLSS 10} The spectroscopic bandpass shall satify $\lambda_\mathrm{max} \geq 1.9\ \mathrm{\mu m}$, and $\lambda_\mathrm{max}/\lambda_\mathrm{min} > 1.82$.

 We need $\lambda_\mathrm{max} > 1.9\ \mathrm{ \mu m}$ for redshift reach, in order to be complementary to Euclid and ground-based surveys.  Furthermore, we need $\lambda_\mathrm{max}/\lambda_\mathrm{min} > 1.82$ for line identification using multiple lines: to ensure that we cannot have [OII] (373nm) falling off the blue end of our coverage while H$\alpha$ (656.28nm) falls off the red end. We have
 assumed that the actual bandpass extends 1.5\% from either end, since it is
 problematic to use emission lines that fall within 1.5\% of the bandpass edges.

\paragraph{HLSS 11} 50\% of the energy (excluding diffraction spikes and non-1st order light) shall be enclosed in a circle of radius $<0.21^{\prime\prime}$ over 95\% of the field.

 This limit of $0.21^{\prime\prime}$ is required by source separation in the input
 catalog for spectral extraction, and is enabled by the addition of the phase
 mask corrector, and leaves some margin on the wavefront error.

\paragraph{HLSS 12} The filter used to define the bandpass of the grism shall have cutoff transition widths $\sigma < 1\%$ (0.7\% goal) after including the effects of broadening
 by the range of incident ray angles at each position in the FoV, where $\sigma$ is
 defined by $\sigma= (\lambda(T=0.90)- \lambda(T=0.10))/\lambda(T=0.50)$. $T$ is the transmission of the
 grism bandpass.

 This is based on the grism guiding considerations; the assessment of grism
 guiding (and its positive outcome) assumed $\sigma < 1\%$.

 \subsubsection{Implementation (Operations Concept) Requirements} We present here the implementation requirements related to the operation concept. We note that Co-I Hirata is the co-lead for the WFIRST Operations Working Group. We did not update Requirements HLSS 13-15, but include them here for completeness.

\paragraph{HLSS 13} Exposures of each field shall be obtained at a minimum of 3 dispersion directions, with two being nearly opposed.

\paragraph{HLSS 14} The observatory shall be able to place the WFC at a commanded
 orientation with an accuracy of $0.64^{\prime\prime}$ ($3\sigma$) in pitch and yaw, and $87^{\prime\prime}$ ($3\sigma$) in
 roll (TBR these were arbitrary values that give a net $3\sigma$ position uncertainty
 of 10 pixels. For the HLSS, the primary driver is that the position uncertainty
 is small with respect to chip gaps, which gives larger uncertainties than
 specified above. The smaller values quoted here are consistent with efficient
 target acquisitions, which would flow down from an observing efficiency spec.)

\paragraph{HLSS 15} The observatory pointing jitter and drift shall not exceed 100 mas in the spectral direction on the WFC focal plane (goal of 60 mas) and 50 mas in the cross- dispersion direction (TBR).

\paragraph{HLSS 16} Imaging observations shall be obtained of the fields in the HLSS that reach JAB=24.0, HAB=23.5, and F184AB=23.1 for an $r_\mathrm{eff}=0.3^{\prime\prime}$
 source at 10$\sigma$ to achieve a reference image position, in 3 filters.

 Provided the HLSS covers area already observed in the HLIS, this
 requirement will be met automatically.  This requirement applies to any HLSS
 fields that are counted toward the minimum survey area requirement but are not
 covered by the HLIS.  Imaging in at least three filters is required to build a
 minimal spectral template for grism spectral decontamination.

\paragraph{HLSS 17} There shall be 40 observations of two deep fields, each 11 deg$^2$ in area, sufficient to characterize the completeness and purity of the overall galaxy redshift sample. The 40 observations repeat the HLSS observing sequence of 4 exposures 10 times, with each deep field observation having the same exposure time as a wide field observation of the HLSS. The dispersion directions of the 40 observations should be roughly evenly distributed between 0 and 360 degrees.

To calibrate the HLS GRS, we need a spectroscopic subsample, with the same selection criteria as that of the HLS GRS, containing more than 160,000 galaxies  that have a redshift purity $>99\%$. We need 160,000 galaxies to know the redshift purity to 1\% (which requires 10,000 objects, assuming noise of $1/\sqrt{2N}$ from Poisson statistics) in at least four categories (low z, high z, faint, luminous).

Based on the estimated galaxy number density of $>7273$ per deg$^2$ at the flux limit for the GRS, $10^{-16} \mathrm{erg} \, \mathrm{s}^{-1}\mathrm{cm}^{-2}$, we need a total area for the deep fields of 160,000/7273=22 deg$^2$.  These can be split into two subfields of 11 deg$^2$ each.  Smaller subfields prevent the testing of galaxy clustering statistics in each subfield. Each deep field should be part of the HLS footprint, so they are representative of the GRS as a whole.

 The visits to the deep field should consist of 10 sets of HLS-GRS-like visits,
 matching the integration time, dither pattern, and observational time-sequence
 of the HLS-GRS strategy, with each set of HLS-GRS-like visits covering the same
 areas of 22 deg$^2$. Assuming a completeness of 50\% and uncorrelated sets, the
 completeness after 10 sets of visits is (1-0.5)10=0.001, leading to a 99.9\%
 complete sample for calibrating the GRS. Since each set of observation consists
 of 4 roll angles, the total number of deep field observations is 40. The
 dispersion directions of the 40 visits should be roughly evenly distributed
 between 0 and 360 degrees, in order to map out possible sources of systematic
 errors due to inhomogeneity.

\paragraph{HLSS 18} The observing efficiency of the HLSS, defined as the total science exposure time divided by the total time allocated to the survey, shall be TBD\%.

 The total time includes slew, settle, target acquisition, and
 calibration observations that are specific to the HLSS, including the
 extra-depth observations of the deep fields described in HLSS 17.  This minimum
 observing efficiency, together with a 0.67 year total allocation of observing
 time, allows science exposures of 1600 deg$^2$ (TBC) with the exposure time
 indicated in the comment to HLSS 8; this provides a 7\% (TBC) margin over the
 1500 deg$^2$ requirement (HLSS 1) to allow for data that may be unusable because of instrumental artifacts, bright sky objects, etc. This is a high level
 requirement that will need to be revisited as the mission implementation details
 become more solid; it should be set such that the core science goals for the GRS
 are achieved without putting mission success at risk.

\subsubsection{Calibration Requirements} In this section, we present the calibration requirements as delivered to the Project Office.

\paragraph{HLSS 19} The relative spectrophotometric flux calibration shall be known to 2 percent relative accuracy (with the goal of 1\%), in order to understand the effective sensitivity limit for each redshift bin for each area surveyed.

 The requirement here is only on the {\it relative} spectrophotometry, which impacts the selection function of galaxies. Absolute line flux calibration will only change the overall number of objects and the dN/dz, but will not introduce
 density variations.  Large scale structure measurements require precise
 knowledge of the selection function of galaxies. Although the overall redshift
 distribution may be determined by averaging over the entire survey, fluctuations
 in the selection function can easily contaminate the underlying cosmological
 density fluctuations.

 The spectroscopic sample for the GRS is expected to be defined by a line flux
 limit of 10$^{-16}\,$erg$\,$s$^{-1}$cm$^{-2}$. Spatial errors in the spectrophotometric calibration
 will introduce artificial spatial fluctuations in the number density of
 galaxies, which could contaminate the cosmological signal.

 We start by setting a requirement on the spatial uniformity of the mean number
 density as a function of physical scale. We require that the non-cosmological
 fluctuations in the mean number density (or the selection function of the
 survey) be $< 1\%$ (sqrt variance) when averaged over spatial scales between 10
 Mpc/$h$ to 200 Mpc/$h$. At small scales, this is $\sim$ two orders of magnitude smaller
 than the cosmological signal, while at the $\sim$ BAO scale of 100 Mpc/$h$, this
 is $\sim$ one
 order of magnitude smaller than the cosmological signal.  These fluctuations
 equal the cosmological signal at $\sim400 \mathrm{Mpc}/h$.  These physical scales correspond
 to $\sim0.5$ degrees to 6 degrees at a redshift of 1.5.

 We convert the above requirement to a requirement on the spectrophotometric
 calibration accuracy, assuming the Model I luminosity function of \citet{Pozzetti:2016}. At the flux limit of WFIRST, this yields a requirement of 1\% relative spectrophotometric calibration, averaged over angular scales of 0.5 degrees to 6 degrees.

 This is a very stringent requirement. We have relaxed this requirement from 1\%
 to 2\% to add margin for mission success, assuming that we will achieve 1\%
 relative spectrophotometric flux calibration in post-processing by projecting
 out problematic modes in the analysis.

 We plan to make this requirement more precise, in the form of ``The relative
 spectrophotometric flux shall be known to 2\% relative accuracy in TBD (probably
 the spectral resolution) wavelength bins with a goal of 1\% on scales larger
 than TBD (per pointing, 0.3 deg) and TBD\% on scales smaller than 0.3 deg.''
 We are working on deriving and justifying these numbers.
 Co-I's Capak, Hirata, and Padmanabhan are members of the WFIRST Calibration Working Group, working on a detailed calibration strategy for WFIRST.

\paragraph{HLSS 20} The uncertainty in the wavelength calibration shall not introduce biases in the wavelength measurement by amounts greater than
 $\Delta\lambda/\lambda = 10^{-4}$ on any
 angular scales exceeding 0.064 degrees within a field, and
 $\Delta\lambda/\lambda = 2\times10^{-5}$ from
 field to field.

 Variations in the wavelength calibration within a field, and from field to field
 on large scales, wash out the clustering signal by de-correlating the projected
 component of the clustering signal on those angular scales.

 Within a field, the acceptable level of wavelength error is
 $\Delta\lambda/\lambda \sim 10^{-4}$, which
 is 10\% of the errors on individual redshift measurements (0.001), to avoid
 increasing the overall redshift error by a significant factor. The angular scale
 is set by the optimal smoothing scale for BAO reconstruction, $\sim 5 \,
 \mathrm{Mpc}/h$. At $z=3$, this subtends 0.064 degrees for a flat universe with
 $\Omega_\mathrm{m}=0.3$ and a cosmological constant.

 For field to field, the acceptable level of wavelength error is $2\times10^{-5}$, which comes from comparing two adjacent fields.
Since we expect $\sim 10^4$ galaxies per deg$^2$, we have $\sim 2810$ galaxies per FOV of 0.281 deg$^2$.  If the galaxies have a redshift error of $10^{-3}$ each, then one can measure systematic offsets between fields (statistically) at the $10^{-3}/\sqrt{2810}$ level, which is $1.9\times10^{-5}$. At that level the power from the systematics is sub-dominant to the power from the redshift error.

 \subsubsection{Requirements on Science Data Products:} In this section, we present the GRS requirements on science data products. We are in the process of studying HLSS 21-25. These depend on the structure and responsibilities of the SOCs and the SITs. Co-Is Teplitz and Capak have extensive experience in data processing for space missions, and have provided detailed comments on these requirements to the WFIRST Project Office.

\paragraph{HLSS 21} The raw data for each grism exposure shall be available through the
 archive, with each dataset including identifying information such as time of
 exposure, observatory pointing orientation, a unique dataset identifier, and any
 engineering information needed for subsequent processing. Each detector readout
 for a given exposure shall be included in the dataset.

\paragraph{HLSS 22} Calibrated data for each grism exposure shall be available through the archive. Each detector readout shall be calibrated at the appropriate level, and the individual calibrated readouts will be combined to produce a net spectral image. These datasets shall include information on the effective PSF as a
 function of position and incorporate any World Coordinate System information
 needed for subsequent stages of processing. As sources are not yet identified,
 association of a pixel with a source position and wavelength is not yet
 possible.

\paragraph{HLSS 23} Source catalogs of the same field derived from WFC imaging data shall be combined with observatory pointing information for each grism exposure to produce a segmentation map that associates each catalog source with a range of spectral image pixels. The spectral images of bright stars in each detector shall be used to refine the astrometric solution.  These segmentation maps shall be used to extract 1D spectra for each source, and to flag pixels that may contain flux from multiple sources. The extracted spectra shall include
 information on the effective exposure time for each pixel, effective PSF as a
 function of position, data quality flags, and any other information needed to
 interpret the data.

\paragraph{HLSS 24} Extracted spectra of each source from multiple roll angles shall be
 combined to produce a single net spectrum of each source. For sources that are
 spatially resolved, the result shall be provided as a data cube of position and
 wavelength. The spectra obtained at nearly opposing roll angles shall be used to
 account for possible offsets of the emitting region from the center of the
 broad-band image. The data from all roll angles shall be used, to the extent
 possible, to resolve ambiguities in the proper source to associate with pixels
 illuminated by overlapping spectra. These net spectra shall include information
 on the effective exposure time for each pixel, statistical and systematic
 uncertainties in the measured fluxes and wavelengths, effective PSF as a
 function of position, data quality flags, and any other information needed to
 interpret the data.

\paragraph{HLSS 25} The data processing system shall have the capability of inserting fake sources into the spectral image data and re-executing the generation of
 high-level science products. These tests are essential for verifying the proper
 operation of the tools that generate high level science products and for
 understanding the sensitivity of the survey and systematic effects that may be
 present in the survey sample.

\paragraph{HLSS 26} The data processing system shall provide sufficient knowledge of the 3D selection function so that the artificial correlations due to inaccuracies in the 3D selection function are less than 10\% of the statistical error bars on
 scales smaller than 2 degrees, and less than 20\% on larger angular scales.

 This requirement is only meaningful in terms of the contribution to the total
 error budget by the uncertainties in the 3D selection function. The BAO scale is
 less than 2 degrees in the redshift range for the HLSS.

 To convert the positions of observed galaxies in the large-scale structure
 into clustering measurements (correlation function, power spectrum, higher order
 statistics) we need to know how the ``average'' number density of objects (in the
 absence of clustering) changes in the observed volume. The mean number density
 will vary significantly both in redshift and with angular position due to
 effects of target selection, data reduction and observing conditions. Previous
 surveys were able to separate the selection function in two independent parts:
 the radial selection function and the angular selection function. It is likely
 that the WFIRST selection function will not be separable in this way, i.e.
 different parts of the sky will have different radial profiles. For now we will
 assume that this type of separation is possible. This assumption is reasonable
 for preliminary investigation since most effects are either mostly radial (e.g.
 target selection, data reduction) or angular (e.g. imaging quality, galactic
 extinction).

 The knowledge about 3D selection function is usually encoded into sets of random
 catalogues. When computing clustering statistics, the random catalogues remove
 the systematic effects of varying mean number density (due to target selection,
 data reduction or observing conditions). If the 3D selection function is not
 correct, the effects will not be completely removed and will generate spurious
 correlations that can bias the true cosmological signal. The angular mask of the
 WFIRST data will vary pixel to pixel on the infrared detector. The full
 description of the angular mask may turn out to be computationally intractable.
 For the core science goals we require the description of the mask to be correct
 with an angular resolution of approximately 3 arcmin. This corresponds to a
 spatial resolution of $3 \,h^{-1} \mathrm{Mpc}$ at $z=1.5$. This is driven by the fact that we
 need to be  able to resolve the BAO peak. In principle, our requirements on the
 knowledge of the 3D selection function are driven by the main requirement that
 the spurious correlations should be no more than TBD per cent of statistical
 errors between the scales of 10 and 150 h$^{-1}$Mpc in clustering signals (either
 in correlation function multipoles or power-­spectrum). For the galaxy sample
 expected from WFIRST, this corresponds to TBD per cent uncertainty in the
 knowledge of the radial distribution and the angular mask.

 To further quantify the effect of systematics offset in the angular mask on
 clustering measurements we have performed tests on mock catalogues representing
 BOSS CMASS sample. This is justified by the fact that the BAO and growth rate
 measurements from WFIRST GRS in redshift bins of $z\sim0.1$ are expected to be
 roughly equal to the CMASS constraints with $z\sim0.2$.

 The mock surveys are generated from N-body simulations, with a median redshift
 of 0.6, with galaxies of halo mass range about $7\times10^{13}\,M_\odot$. The mock
 surveys have proper BOSS 3D selection function (which we take as truth here).
 Now we distort the selection function in the following scenarios:

\begin{figure}
 \includegraphics[width =0.75\textwidth]{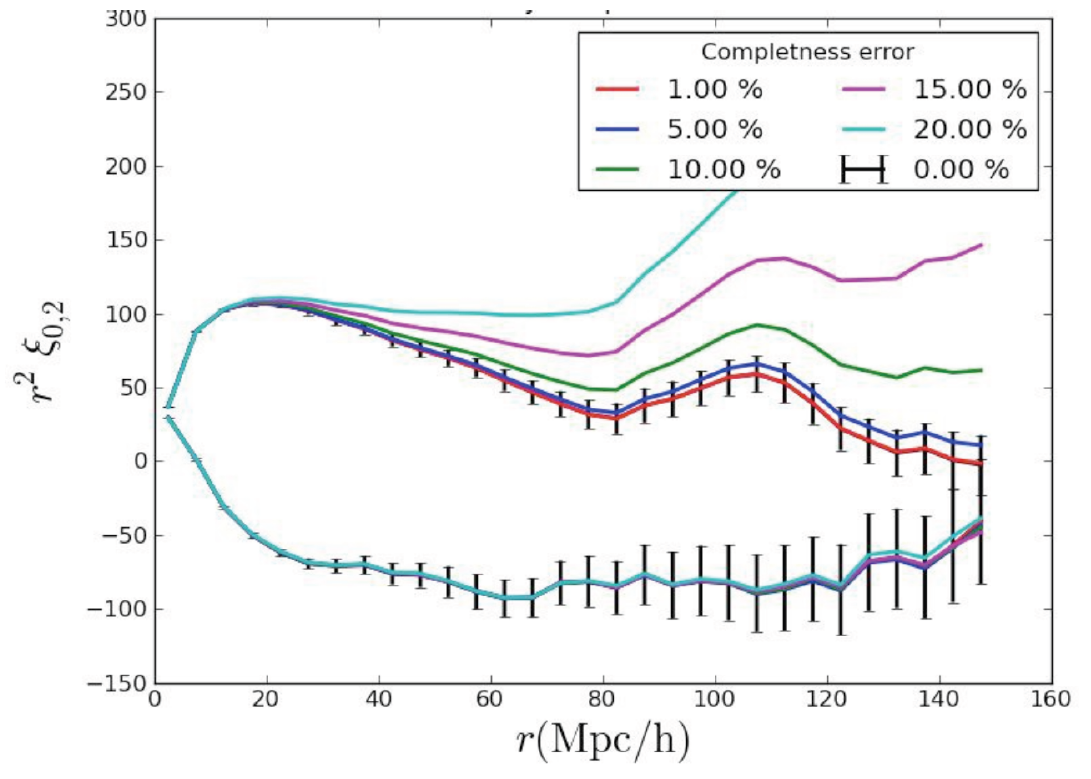}
 \label{fig:selection_function}
 \caption{The monopole (upper curves) and quadrupole (lower curves) of a single mock survey and how they respond to various completeness errors in the scenario described in scenario. We can see from the level of two point correlation function, even a 5\% error on the completeness can change the monopole   and quadrupole significantly that we expect RSD to be affected, while BAO is not significantly affected, as this mostly changes the amplitude of the correlation function.}
  \end{figure}

\begin{enumerate}
\item The survey region is divided into two equal area along RA and one area
 has true completeness whereas the other one has $1-x$ completeness where x is a
 given completeness error (as shown in the legend of Figure~\ref{fig:selection_function}).
 We show in Figure \ref{fig:selection_function} the resulting monopole and quadrupole with varying completeness error. This
 is an interesting limiting case as surveys can sometimes be affected by large
 scale systematics generated by either calibration of two parts of the sky, or
 due to large scale effects caused by the galactic foregrounds.
\item The true survey completeness is multiplied with a gaussian function. The
 gaussian function has mean 0 and variance as the denoted completeness error. We
 then fold both positive and negative side of the Gaussian to the negative side
 and hence allow the completeness to be only smaller than its true value. We vary
 the scale at which we change the 3D selection function, starting with 1 degree
 to 4 degrees (4 deg. is approximately the BAO scale at this redshift).
 \end{enumerate}

 From the preliminary analysis shown here, we expect that we will need to
 accurately model the 3D completeness function down to a few \% level.  At 5\%
 the effects can already be very detrimental to our large scale structure
 analyses using BAO and RSD. Scaling from these, we arrive at the requirement
 that artificial correlations due to inaccuracies in the 3D selection function
 are less than 10\% of the statistical error bars on scales smaller than 2
 degrees, and less than 20\% on larger angular scales.

 \subsubsection{Requirement on Cosmological Volume Simulations:} We now present a new category of requirements for a space mission. These are not needed
 for mission success (data acquisition by the spacecraft), but only for meeting the
 high level science requirements (level 1). We summarize these as follows:

 \begin{enumerate}
 \item a few accurate mocks with galaxies included using semi-analytical
 galaxy formation model, to verify and validate WFIRST GRS pipeline;
 \item $\sim 100$ mocks with high mass resolution of 109 solar masses, to inform theoretical modeling of the data;
 \item $\sim 10,000$ mocks with low resolution, to derive the covariance matrices for the WFIRST data.
 \end{enumerate}

 To quantify the requirement on cosmological volume simulations, we consider only
 the case for galaxy clustering science (which includes BAO and RSD) for now for
 simplicity. For WFIRST, simulations are required for the following three
 objectives:
\begin{enumerate}
\item establishing the basic correctness of the pipeline;
\item informing the theoretical modeling of small scale clustering as a function of
 tracer properties;
 \item calculating the covariance matrix for each GRS probe and
 across many probes.
 \end{enumerate}

 In order to establish the basic correctness of the WFIRST pipelines and
 predictions, sophisticated synthetic mock galaxy catalogs are essential.
 These catalogs, which must realistically emulate WFIRST both in sky area and
 depth, are typically constructed by running large gravity-only simulations and
 then ``painting'' realistic galaxies on top.  Populating a simulation with
 galaxies can be done in several ways: (i) empirically, using statistics such as
 the ``halo occupation distribution'', (ii) by placing the normal, baryonic matter
 in the simulation ab initio and explicitly solving the hydro-dynamical
 equations, or (iii) by using a semi-analytical galaxy formation model (SAM),
 whereby the astrophysical processes and formation histories of galaxies are
 described using physically motivated, parameterized equations. The advantage of
 SAMs over alternative methods is their ability to meet the demands from next
 generation cosmological surveys for large (suites of) galaxy mock catalogues
 that are both accurate and can be constructed rapidly. In contrast, full
 hydro-dynamical simulations are far too slow and empirical methods are limited
 by the availability of existing high redshift observations, which are necessary
 for the calibration of these methods. SAMs also require some observations for
 calibration but, once tuned to fit observations at low redshift, they are able
 to make predictions out to high redshift without the need for further
 observational input. Furthermore, empirical methods are often limited in that
 they are calibrated in one or two photometric bands, whilst SAMs are designed to
 model the star formation history of a galaxy and so have the ability to make
 predictions for a wide variety of multi-wavelength data simultaneously. This
 feature of SAMs is vital to ensure that we can examine cross-correlations
 between the spectroscopically-selected dataset for galaxy clustering analysis
 and the photometrically-selected dataset for weak lensing analysis. Besides
 testing the pipeline and making (limited) cosmological forecasts, these galaxy
 mock catalogs would also be a valuable resource for science working groups
 focusing on legacy science (e.g. galaxy evolution, active galactic nuclei). Note
 that,  compared to the large number of approximate mock catalogs necessary for
 covariance estimation, only very few accurate galaxy mocks are required to
 verify and validate the WFIRST pipeline.

 To inform the theoretical modeling of clustering especially at non-linear
 scales as a function of the tracer properties would require a significant number
 of simulations that have relatively realistic modeling of the tracer properties
 at the relevant redshift.  For WFIRST, we can take the current number density of
 emission line galaxies (for H$\alpha$ galaxies only) from our baseline
 calculation and used the \citet{Tinker2008} halo mass function, along with the
 \citet{Giocoli2008} subhalo mass functions to compute the total number of
 halos and subhalos above some mass threshold and then match that to the baseline
 GRS number densities. This maps back to approximately 10$^12$ solar masses from z=1 to z=2. Assuming that we need to have at least 100 particles to resolve halos at
 1012 solar masses, and another factor of 10 particles to resolve properties of
 the halo progenitors, we will need dark matter particle mass resolution of
 approximately 10$^9$ solar masses. The extra factor of 10 is due to the galaxy
 formation model that depends on the properties of the progenitors which is an
 approximation that may change as we understand the galaxy properties better and
 as more observations of the tracers arrive. We expect to require of order 100
 simulations to reduce the shot noise of the correlation function in order to
 compare the theoretical modeling to the simulated correlation function. These
 realistic mock surveys may also require the modeling of non-standard
 cosmological  models, such as extensions to non-zero total neutrino masses, or
 modified gravity models.

 Finally, we will need to calculate the covariance matrices of the main probes of
 clustering, namely BAO and RSD, and the cross-covariances among these probes
 (or across different methods as in recent BOSS analyses). We can approach the
 calculation of the covariance matrices through multiple avenues. One can
 generate (in principle) a large number of approximate mock surveys using
 relatively fast approximate methods (eg., PTHalos, QPM, FastPM, etc), and apply
 the relevant survey properties onto these mock surveys. The small scale modeling
 of the clustering may not be 100\% accurate, but is likely to be adequate for
 the linear RSD modeling and BAO analyses where medium to large scales are most
 important. The number of approximate simulations required can be on the order of
 O(10,000) depending on the number of parameters we will be estimating using
 these covariance matrices, but the time requirement of these approximate mocks
 is relatively modest. One can also envision using more theoretical approaches
 (such as \citet{OConnell:2015src,Padmanabhan2016,Friedrich:2015nga}, which only require a relatively modest number of realistic mock surveys which are required for (b).

\subsection{GRS Light-cone Simulations (D8, D9)}
\label{sec:light-cone}

 \begin{summaryii}
Light-cone cosmological simulations are a critical tool to design the GRS survey design, to develop and to validate analysis tools and theoretical predictions. We developed four complementary simulation approaches to tackle multiple questions relevant to the GRS:
\begin{enumerate}
  \item A lognormal simulation to generate quickly large cosmological volumes;
  \item A fast approach using simple galaxy-halo prescriptions to generate joint GRS and WL simulations;
  \item A realistic emission line galaxies modeling to study the confusion between H$\beta$ and [O III] emitters;
  \item A realistic semi-analytic galaxy evolution model to make robust H$\alpha$ mock catalogs.
 \end{enumerate}
 \end{summaryii}

 In order for WFIRST HLS to reach its  high level science requirements, we have
 proposed to ({\bf D9})  produce simulated light-cone observations to ({\bf D8})
 develop both the methods for modeling and interpretation of cosmological
 measurements from WFIRST.  Most of these data sets will be at the level of
 galaxy redshift and shape catalogs rather than the pixel-level imaging and
 spectroscopy simulations described above. They will incorporate varying degrees
 of complexity regarding galaxy bias, redshift evolution, survey geometry, and
 observational systematics such as incompleteness, shape measurement errors, and
 photometric redshift biases. Many of these artificial data sets will be made
 publicly available, and some will take the form of data challenges, where the
 underlying parameters are initially known only to the creators of the data set.
 Here we report on our first year simulation efforts  in a 4 prong approach,
 ranging from the largest volume to the highest resolution below:

\subsubsection{Lognormal Simulations}

 Samushia and his postdoc have produced a suit of few thousand fast
 ``enhanced log-normal simulations'' for the WFIRST GRS expected samples. While
 these simulations do  not correctly reproduce the small scale structure and
 higher order statistics of the field, they can be used for studying various
 large scale effects and implement light-cone effects. The simulations have so
 far been used to study the effect of splitting the WFIRST footprint into two
 non-contiguous areas. We plan to use these simulations in the future to study
 systematic effects in the measurements (e.g., window effect correction) and to
 validate the BAO/RSD proto-pipeline. These simulations are very well-suited for
 such tasks since their input two-point signal is known exactly.

 \subsubsection{Fast simulations with galaxy positions and shapes}

 Kiessling, and Huff are working with Postdoc Izard to develop a fast
 pipeline to provide galaxy mock catalogues with weak lensing. The goal of our
 initial project is to model effects of various systematics and determine their
 impact on galaxy clustering and weak lensing observables and their covariance
 matrices. The starting point for this project is  300 ICE-COLA fast simulations
 \citep{Tassev:2013pn,Izard:2015dja} generated by Izard during his PhD that
 provide the dark matter density field and halo catalogs (both in the light cone
 geometry). The former is used to compute maps of weak lensing distortions in
 the Born approximation using a technique developed by Izard (Izard et al, 2017;
 in prep). Galaxies are attached to halos using a new pipeline developed for
 this project. This pipeline takes the halo catalogues in the light cone
 geometry generated by the fast simulations and produces galaxy samples with
 broad band luminosities and weak lensing properties. The produced catalogs are
 all sky and span 0 $<$ z $<$ 1.4, with a minimum dark matter halo mass of
 $10^{12}  M_{sun}/h$.

 \subsubsection{Testing the effect of line confusions on BAO using N-body simulations}

  \begin{figure}
   \includegraphics[width = 0.4\textwidth]{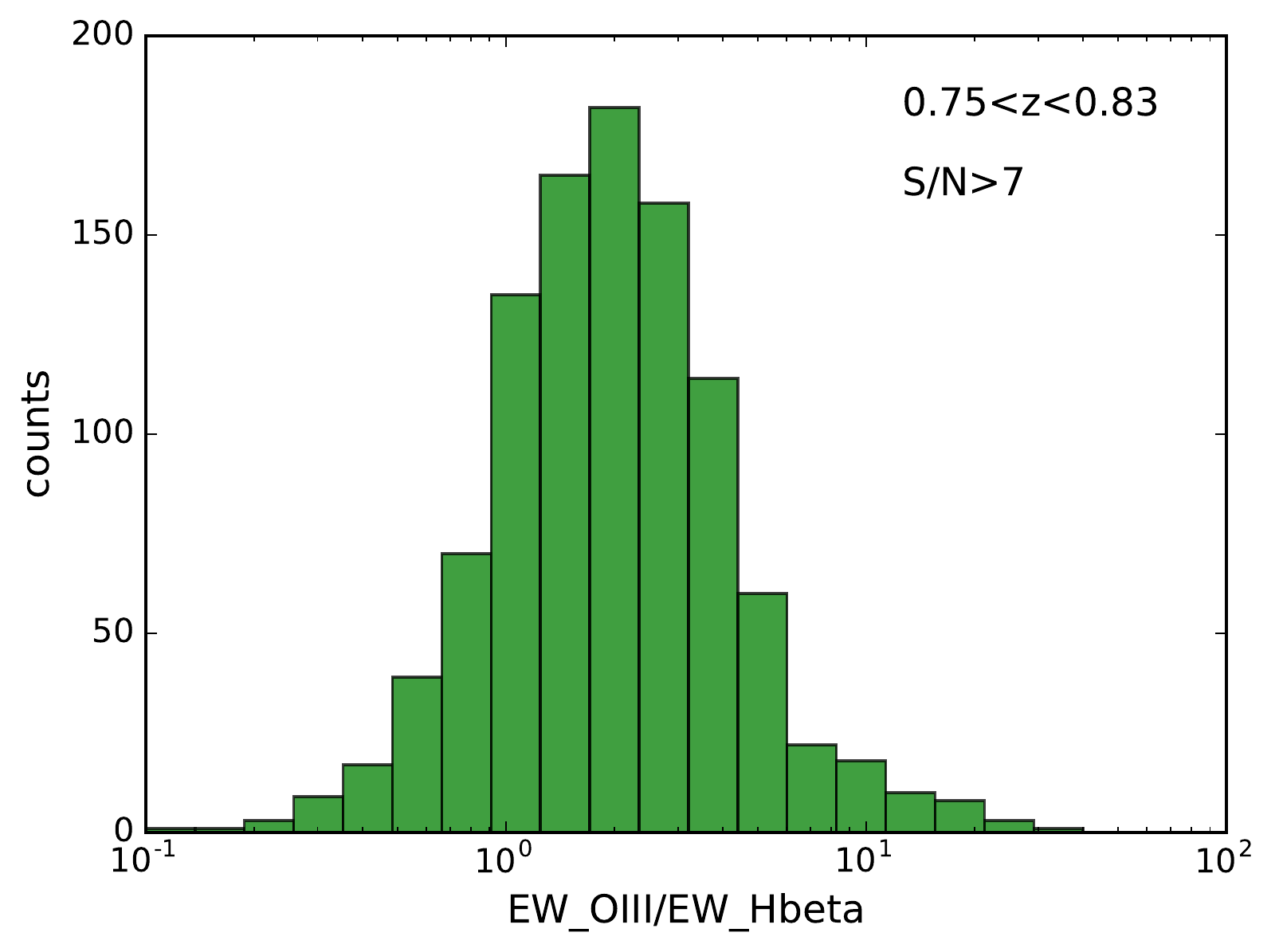}
  \label{fig:line-ratios}
   \caption{The line ratios between H$\beta$ and [O III] using the  dataset from DEEP2 survey, which has acquired over 50,000 emission line galaxies, but a significant fraction of them have H$\beta$ and/or [O III]. The plot shows $\approx$1,000 galaxies from DEEP2 sample that has both H$\beta$ and [O III], while each line is detected at least at 7-$\sigma$.}
  \end{figure}

 Ho and Massara concentrates on one particular goal in generating their simulations. Their goal is to investigate the effect of confusion between H$\beta$ and [O III] emitters, in particular in the high level science goal in BAO. In WFIRST,  the primary science targets for the redshift survey will be H$\alpha$ and [O III]. There is a special concern of H$\beta$ vs. [O III] confusion due to their proximity in wavelength (hence the inability of photo-z's to distinguish them reliably). This is particularly true given that an H$\beta$ emitter mis-identified as [O III] will have an inferred radial position different by 8900 km/s (or: 89 $h^{-1} Mpc$ * (1+z)/$\sqrt{\Omega_\Lambda + \Omega_m (1+z)^3}$, which in our range of redshifts is near the BAO scale). We will also get some [O II] emitters -- these might be useful directly for cosmology, or for disentangling other line emitters (see Pullen et al. 2014). Note that at WFIRST resolution 3726 and 2729 are a blend.

\begin{figure}
  \includegraphics[width = 0.5\textwidth]{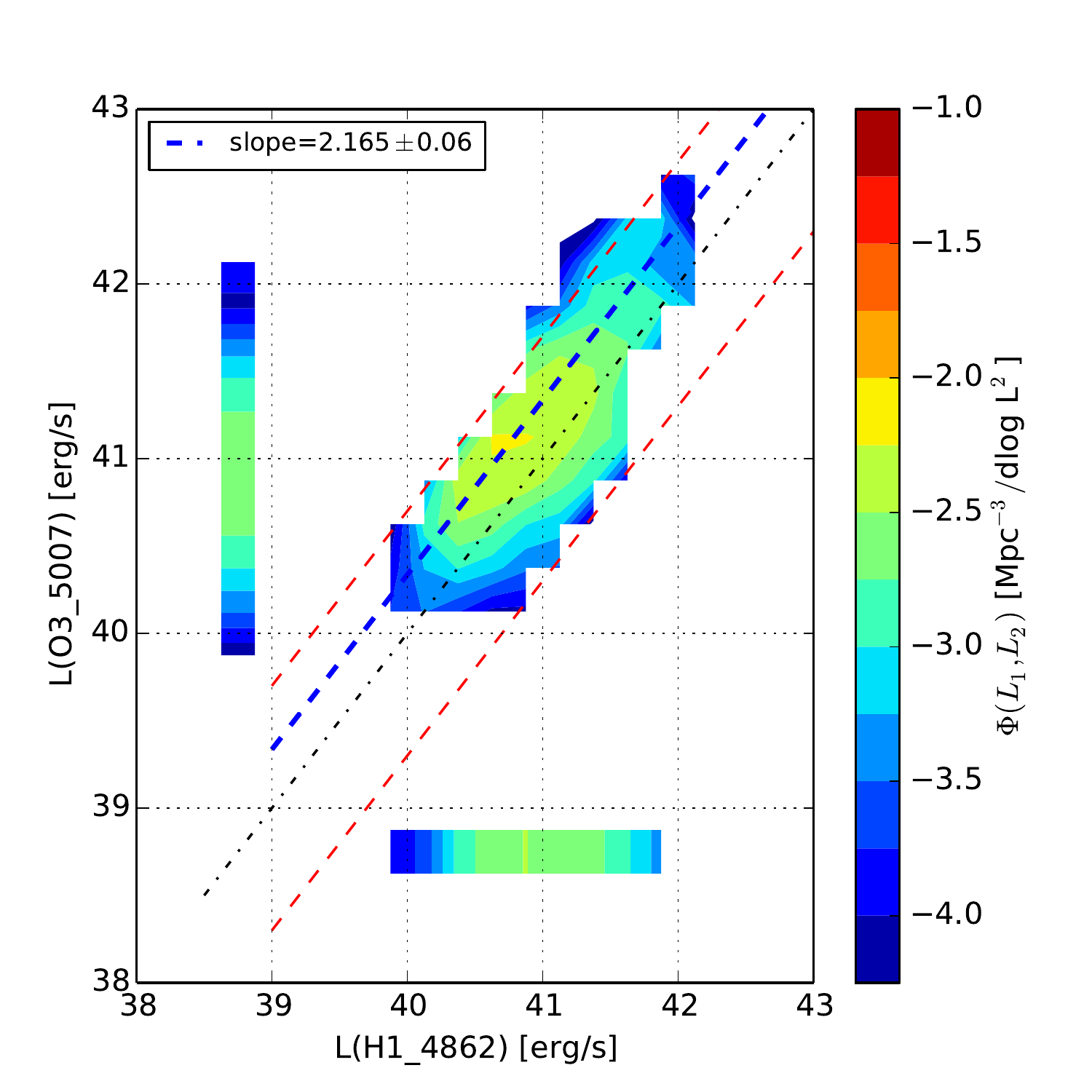}
 \label{fig:cond-lum}
  \caption{The conditional luminosity function between H$\beta$ and [O III] using the same galaxies as in Figure~\ref{fig:line-ratios}.  The black and red lines are the conditional luminosity function (and 1 sigma) from DEEP2 survey).
 }
 \end{figure}

 A key challenge in the mocks will be making sure the populations of each line and the correlations among the different line strengths and with environment are sufficiently realistic for the tests we are doing. This has historically proven to be very difficult due to the heterogeneous nature of the observational constraints. Therefore, we have investigated the equivalent width ratio between H$\beta$ and [O III] using the dataset from DEEP2, which has acquired over 50,000 emission line galaxies, but a significant fraction of them have H$\beta$ and/or [O III]. This is so far one of the few statistical samples that can be used to look at the line ratios (see Figure~\ref{fig:line-ratios}). We also look at the conditional luminosity function between the two line luminosity observed in DEEP2 (see Figure~\ref{fig:cond-lum}). We now proceed to use these line ratios to create the emission line galaxy catalog.

 We start with dark matter simulations generated by Stanford group (led by Risa
Wechsler, which is part of WFIRST Simulation WG and belongs to the EXPO SIT)
that contain galaxies up to 5-sigma detection limit of grizY = [27.5, 27.0,
26.4, 25.9, 24.0] over 10313 sq. degrees.  Each of the galaxy contain its own
spectral energy distribution which are generated to fit most updated luminosity
function and color evolution measurement. Galaxy magnitudes and shapes are
affected by shear and magnification. We then measured and validated the
correlation function and luminosity function of [O III] selected galaxies (with
WFIRST detection limits). Next, we plan to apply
the measured line ratios as described in Figure~\ref{fig:line-ratios} to all the
galaxies and reapply the WFIRST detection limits to create a realistic WFIRST
emission line galaxy catalog (with  [O III] as the main galaxy targets), and
assess the effects on BAO peak position due to the possible smearing by
mis-identifying the H$\beta$ as [O III]. In addition to the mocks themselves, it
will be important for us to identify which aspects of the emission lines we
think are close to reality, which are of the right order of magnitude, and which
could be qualitatively different from the real Universe. We envision that we
would continue improving these up to the point where we have the real WFIRST
data.

 In addition, we have also released a code that calculates the interloper
 fraction (Wong, Pullen \& Ho): a Python-based program that applies secondary
 line identification and photometric cuts to mock galaxy surveys, in order to
 simulate interloper identification. We also have a module specifically designed
 to do WFIRST and predict interloper rates for WFIRST \citep{Wong:2016eku}.

 \begin{figure}
 \label{fig:corr-func}
 \caption{The correlation function of [O III] galaxies with WFIRST detection limit within our 10,000 sq.deg simulations.
 }
 \end{figure}

\subsubsection{H$\alpha$ emitter number density forecasts} 

In work led by Merson, Wang, Benson, Masters, Kiessling and Rhodes
the open source semi-analytical galaxy formation
model, \textsc{Galacticus} \citep{Benson2012}, was used to predict the
H${\rm \alpha}$-emitter number counts and redshift distributions for
the WFIRST GRS. This work is published in \citet{Merson2018}.

A four square degree lightcone catalogue was constructed by processing
the dark matter merger trees of the Millennium
Simulation \citep{Springel05} with the \textsc{Galacticus}
model. Emission linesare modelled in \textsc{Galacticus} by
interpolating over a library of emission line luminosities obtained
from the \textsc{Cloudy} \citep{Ferland13} code and stored as a
function of hydrogen (HI), helium (HeI) and and oxygen (OII) ionising
luminosities, as well as the hydrogen gas density and metallicity of
the interstellar medium (ISM). The emission line luminosities are then
processed to incorporate attenuation due to interstellar dust, which
can be modelled using several different methods. Merson and
collaborators consider three dust methods
from \citet{Ferrara99}, \citet{Charlot00}
and \citet{Calzetti00}. However, it is worth noting that any
user-specified dust method can be used in conjunction
with \textsc{Galacticus}.

\begin{figure}
  \centering
  \includegraphics[width=3.5in]{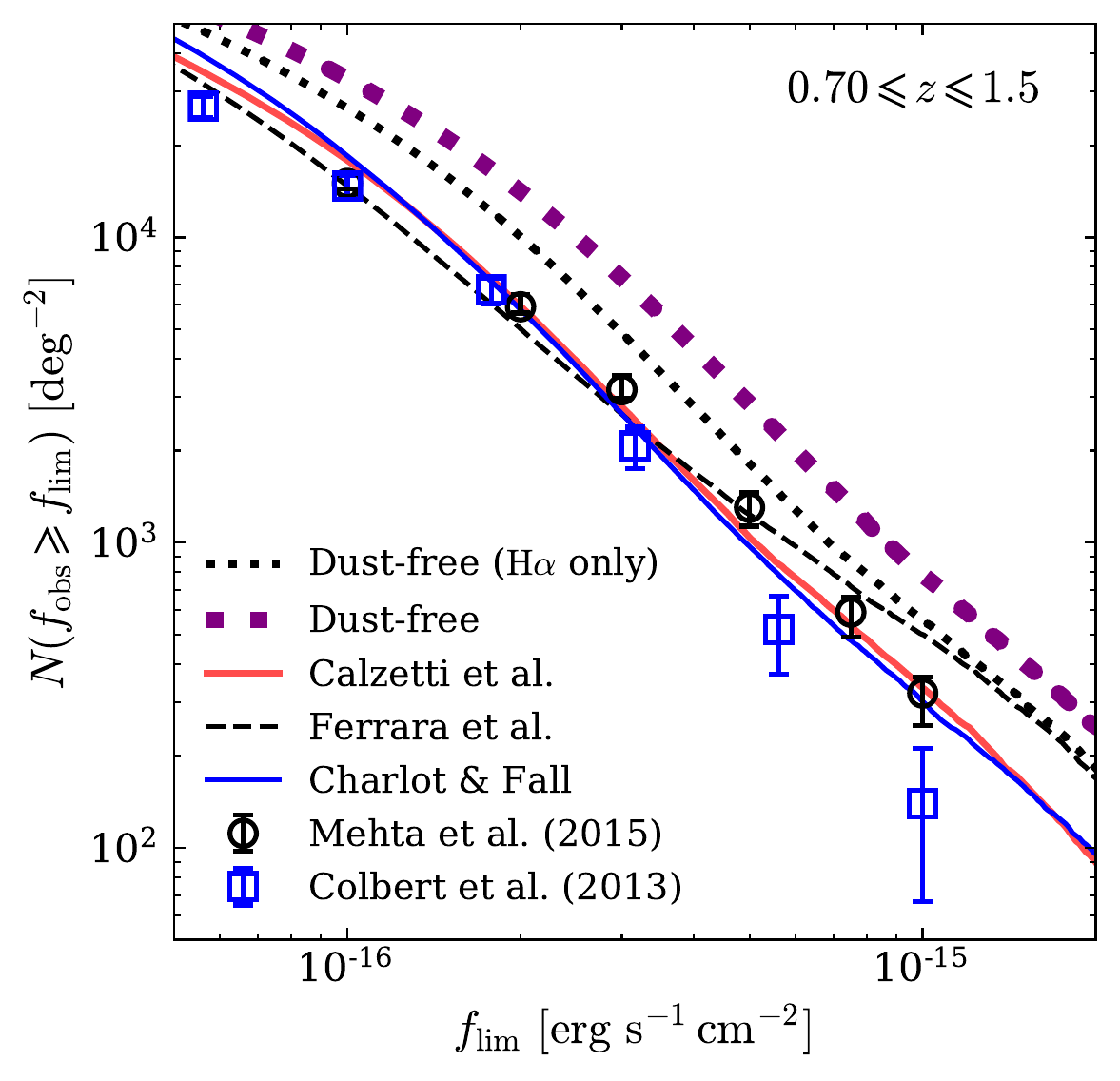}
  \caption{Predictions for the cumulative H${\rm \alpha}$ flux counts
    for the redshift range $0.7<z<1.5$ from a \textsc{Galacticus}
    lightcone mock catalogue. The dotted lines show the counts
    assuming dust-free fluxes. Apart from the thin dotted line, which
    shows the counts for the pure H${\rm \alpha}$ luminosities, all
    lines show counts corresponding to the ${\rm H\alpha}+{\rm N\left
    [II\right ]}$ blended counts. When adopting one of the dust
    attenuation
    methods; \protect\cite{Ferrara99}, \protect\cite{Charlot00}
    or \protect\cite{Calzetti00}; the dust parameters are optimized
    using a chi-squared minimization technique comparing
    the \textsc{Galacticus} to those
    of \protect\cite{Mehta:2015}. Full details can be found
    in \protect\cite{Merson2018}. The open data points show the
    observed blended flux counts from \protect\cite{Colbert13}
    and \protect\cite{Mehta:2015}.}
  \label{fig:halpha_flux_counts}
\end{figure}

First, the \textsc{Galacticus} predictions for the cumulative counts of H${\rm
\alpha}$-emitting galaxies over the redshift range $0.7\leqslant z\leqslant 1.5$
obtained with each dust method are compared with the latest WISP counts from
\citet{Mehta:2015}. The H${\rm \alpha}$ luminosities from \textsc{Galacticus} are
corrected to introduce contamination due to N[II] by cross-matching
the \textsc{Galacticus} with galaxies from the SDSS catalogue
of \citet{Masters2016}. For each \textsc{Galacticus} galaxy, the five
nearest neighbour galaxies in stellar mass versus specific
star-formation rate space are identified from the SDSS catalogue. The
H${\rm \alpha}$ luminosity of the \textsc{Galacticus} galaxy is then
corrected according to the median ${\rm N\left [II\right ]}/{\rm
H\alpha}$ ratio from these neighbors. Further modelling of the ${\rm
N\left [II\right ]}/{\rm H\alpha}$ ratio as a function of redshift and
stellar mass is presented by \citet{Faisst2018}.

The blended ${\rm N\left [II\right ]}+{\rm H\alpha}$ luminosites are
then attenuated using each of the three dust methods in turn. A
chi-squiared minization technique is used to identify the optimum dust
parameters needed so that the number counts from \textsc{Galacticus}
are consistent with the observed counts of \citet{Mehta:2015} down to
a flux limit of $1\times 10^{-16}{\rm erg}\,{\rm s}^{-1}{\rm cm}^{-2}$
(see Figure~\ref{fig:halpha_flux_counts}). The chi-squared values
indicate that the counts obtained from the \citet{Calzetti00} dust
method lead to the best agreement with the observed counts from the
WISP survey. By using the optimized dust parameters, \cite{Merson2018}
find that down the expected flux limits of the WFIRST mission,
the \textsc{Galacticus} model is able to predict number densities
that are consistent with existing observations from the WISP
survey. Although the lightcone used in this analysis has a small
area on the sky, it is expected that cosmic variance has little effect
on the predicted number counts. Since the lightcone analysis is
time-consuming and expensive in computing resources, the lightcone
used in this study was limited in size to 4 sq deg, in order to
provide timely input to WFIRST. However, Merson, Wang and Benson plan to build
significantly larger lightcones in future work.

After setting the strength of the dust attenuation for each method,
\cite{Merson2018} then examine whether the optimized attenuation strengths
yield reasonable matches to the observational estimates for the
distribution of optical depths (at the observer-frame H${\rm \alpha}$
wavelength) and the H${\rm \alpha}$ line luminosity function. For
each of the dust models, \textsc{Galacticus} predicts optical depths
that are consistent within error with the optical depth estimates
from WISP
\citep{Dominguez13}, though the \textsc{Galacticus} optical depths do
not show any increase with observed H${\rm \alpha}$ luminosity, as has
been suggested in the literature. The predicted luminosity function
from \textsc{Galacticus} is consistent with the H${\rm \alpha}$
luminosity function from WISP as measured by \citet{Colbert13} over
the redshift range $0.9\leqslant z \leqslant 1.5$. Comparison with the
luminosity functions from HiZELS \citep{Sobral13} at $z\simeq 0.84$,
$z\simeq 1.47$ and $z\simeq 2.23$
 shows that with these chosen dust
attenuations \textsc{Galacticus} is able to reproduce the $z\simeq
0.84$ luminosity function, but becomes progressively a worse fit
towards higher redshift, especially at $z>2$. This suggests that the
dust methods may be lacking some redshift evolution or dependence on
other galaxy properties, or that the \textsc{Galacticus} emission line
luminosities are the incorrect\ strength. Investigating these
possibilities requires rigorous calibration of the \textsc{Galacticus}
model, which will be carried out in future work.

\begin{table*}
\centering
\caption{Predicted cumulative number of H${\rm \alpha}$-emitting
  galaxies per square degree for the WFIRST GRS, assuming a redshift
  range of $1\leqslant z\leqslant 2$. The fluxes were attenuated by,
  for each dust method, Monte-Carlo sampling likelihoods for the dust
  parameters 1000 times for each galaxy. This table shows the mean and
  standard deviation from these 1000 realizations. Further details can
  be found in \cite{Merson2018}. The upper half of the table shows the
  counts assuming blended ${\rm H\alpha}+{\rm N\left [II\right ]}$
  fluxes, whilst the lower half of the table shows the counts from
  de-blended ${\rm H\alpha}$ fluxes. Predicted counts are reported for
  three dust methods: \citet{Calzetti00}, \citet{Charlot00}
  and \citet{Ferrara99}. Note that the flux limits correspond to
  blended ${\rm H\alpha}+{\rm N\left [II\right ]}$ fluxes. The
  efficiency of the survey is instrumentation dependent, and has not
  been included.}
\begin{tabular}{|c|c|c|c|}
\hline
Flux limit&\multicolumn{3}{|c|}{Dust method used with \textsc{Galacticus}}\\
${\rm erg}\,{\rm s}^{-1}{\rm cm}^{-2}$&\cite{Ferrara99}&\cite{Calzetti00} & \cite{Charlot00}\\
\hline\hline
\multicolumn{4}{|c|}{\textsc{Cumulative counts for blended ${\rm H\alpha}+{\rm N\left [II\right ]}$ fluxes}}\\
$1\times 10^{-16}$&$10403\pm138$&$15176\pm528$&$12195\pm987$\\
$2\times 10^{-16}$&$3797\pm60$&$4307\pm170$&$3059\pm230$\\
$3\times 10^{-16}$&$2263\pm33$&$2084\pm73$&$1414\pm75$\\
\hline
\multicolumn{4}{|c|}{\textsc{Cumulative counts for ${\rm H\alpha}$ de-blended fluxes}}\\
$1\times 10^{-16}$&$7467\pm102$&$10566\pm402$&$8365\pm765$\\
$2\times 10^{-16}$&$2827\pm42$&$2877\pm108$&$2025\pm140$\\
$3\times 10^{-16}$&$1771\pm24$&$1479\pm46$&$1032\pm45$\\
\hline
\end{tabular}
\label{tab:cumulativeFluxCounts}
\end{table*}

 Finally, the \textsc{Galacticus} lightcone is used to present
predictions for the redshift distribution and the differential and
cumulative blended ${\rm H\alpha}+{\rm N\left [II\right ]}$ flux
counts for the WFIRST GRS, as well as two surveys mimicking a
Euclid-like selection. Dust attenuation is applied by converting the
results from the chi-squared minimization into a dust parameter
likelihood. For each dust method the flux counts are estimated 1000
times by Monte-Carlo sampling from these
likelihoods. \cite{Merson2018} find that for a WFIRST GRS with
redshift range $1\leqslant z\leqslant 2$ and blended flux limit of  $1\times 10^{-16}{\rm erg}\,{\rm s}^{-1}{\rm
cm}^{-2}$, \textsc{Galacticus} predicts a number density between
10,000 and 15,000 galaxies per square degree. Number counts for the
three dust methods are shown in
Table~\ref{tab:cumulativeFluxCounts}. The predicted cumulative flux
counts are compared to forecasts from empirical models originally
presented by \citet{Pozzetti:2016}. At at flux limit of $1\times
10^{-16}{\rm erg}\,{\rm s}^{-1}{\rm cm}^{-2}$, comparable to the flux
limit of the WFIRST GRS, the \textsc{Galacticus} counts from the three
dust methods are in closer agreement with each other than the counts
from the three \citet{Pozzetti:2016} models. \cite{Merson2018} find
that these faint fluxes the \textsc{Galacticus} counts are most
consistent with \citet{Pozzetti:2016} model three, which of the three
empirical models gives the lowest counts for faint fluxes. At brighter
fluxes, approaching $1\times 10^{-15}{\rm erg}\,{\rm s}^{-1}{\rm
cm}^{-2}$, the \textsc{Galacticus} counts show closer agreement
with \citet{Pozzetti:2016} model two, which predicts the highest
counts for bright fluxes. Note that all the H${\rm \alpha}$-emitter
counts discussed in \cite{Merson2018} are expected number counts of
target galaxies for spectroscopy, and \emph{not} the counts of
galaxies with redshift measurements. The latter will depend on the
redshift purity and completeness for each survey, which in turn
depends on instrumentation and noise parameters.
\begin{summary}
In this section, we have illustrated that our multiple simulation efforts
complement each other in different ways. LogNormal and Fast simulations with
both positions and shapes,  are useful for pipeline testing and constructing
covariance matrices. The more realistic simulations will test various assumption
and requirements we make for WFIRST HLS. For example, do we need higher
resolution in the grism to disentangle the [O III] and  H$\beta$? Do any of
these line ratios or galaxy properties depend on environment?  Would it lead to an environment dependent BAO smearing if we do not take these into account?  As
we develop more realistic galaxy simulations such as those with \textsc{Galacticus}, we will be able to use the appropriate prescription for
galaxy properties we cannot measure with current observations and make our final
WFIRST simulation as realistic and physically driven as possible. We will examine a variety of other properties of emission line galaxies, including the distribution of OIII luminosities fluxes and the contamination from NII.
This will allow us to have the best tool to develop both the methods for modeling and interpretation of cosmological measurements from WFIRST, and release the most
realistic WFIRST galaxy catalog to the public before WFIRST launch.
\end{summary}

\subsection{Improving our Knowledge of the H$\alpha$ Luminosity Function}
\label{sec:LF}
\begin{summaryii}
Knowing the H$\alpha$ and [OIII] luminosity functions is critical in order to optimize the WFIRST GRS. We are addressing current limitations in our knowledge using a newly available set of HST spectroscopic data within the WISP collaboration.
\end{summaryii}

\begin{figure}[!t]
\includegraphics[width=0.5\textwidth]{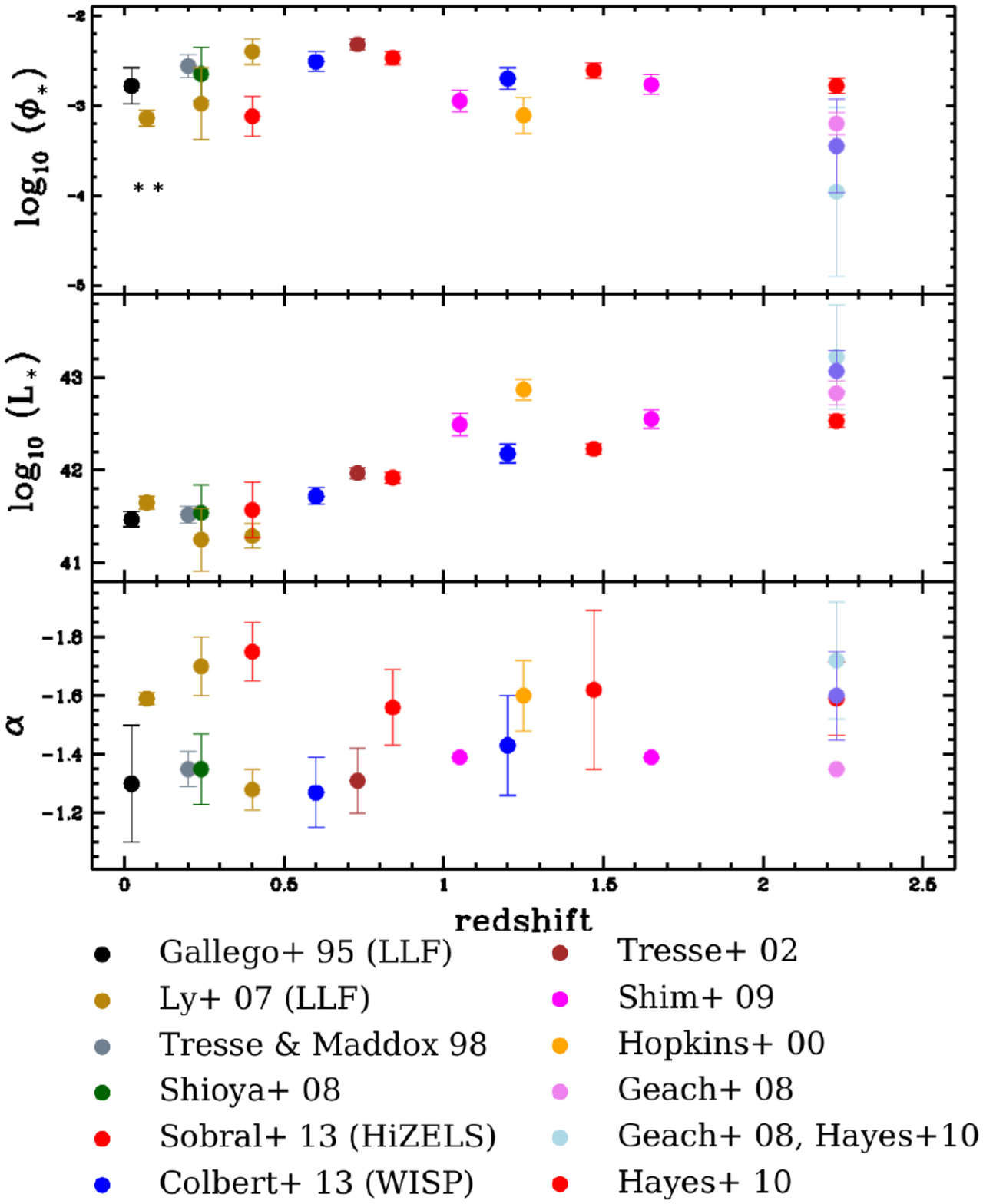}
\caption{H$\alpha$ LF as estimated in various surveys. The uncertainties on the different parameters defining the LF are considerable. The aim of this project is to measure the most precise H$\alpha$ and [OIII] luminosity functions to-date, in the redshift range 0.7-1.6 and 2-2.2 respectively.}
\label{fig:img_depths}
\end{figure}

In order to trace the galaxy distribution through z$\sim$2.7, the future WFIRST
surveys will make use of H$\alpha$ and [OIII] selected emission line galaxies.
These two lines will allow WFIRST to cover the redshift range 1$<$z$<$2
(H$\alpha$) and 2$<$z$<$2.7 ([OIII]) respectively. Therefore, knowing the
expected number of H$\alpha$ galaxies in the survey volume is required and the
study of the H$\alpha$ luminosity function (LF) become fundamental in order to
optimize the planned redshift surveys. However, despite the numerous works on
this topic, given the high uncertainties associated to the H$\alpha$ LF
determination, a sufficiently precise measure is still lacking
(Figure~\ref{fig:img_depths}). Additionally, the samples of emission-line
galaxy are expected to be affected by a complex selection function that depends
on galaxy luminosity, size, line equivalent width and also on contamination from
mis-identified single emission-line galaxies at different depths.

A newly available set of HST spectroscopic data, gathered in the context of the
WFC3 Infrared Spectroscopic Parallel (WISP) survey \citep{Atek:2010},
of which team members Teplitz and Baronchelli are members, has recently become
publicly available. Currently, this survey covers a total area of more than
$\simeq$ 2000 sq. arcmin., with thousands of emission lines from a total
of $\sim$500 randomly distributed HST pointings identified between z=0.3 and
z=1.5. Thousands of emission lines from a total of $\sim$500 randomly
distributed HST pointings have already been identified by the WISP team, between z=0.3 and z=1.5. The spectroscopic capabilities of HST+WFC3 are comparable to those that WFIRST can reach over a 2-order of magnitude wider area. This fact makes the WISP survey a particularly suitable test bench to optimize the future WFIRST surveys.

The accurate constraint of the parameters defining the line LF ($\phi^{*}*, L^{*}, \alpha$) is hindered by their high correlation. Therefore, estimating these parameters requires uniform sampling of both the bright and faint ends of the luminosity distribution. In the next months, in the context of a collaboration involving both WISP and SIT members, Teplitz, Baronchelli and Wang, are planning to obtain the most accurate estimate, to-date, of the H$\alpha$ and [OIII] LFs in the redshift range 0.7-1.6 and 2-2.2 respectively. This result will be achieved by combining deep and shallow data from various HST-WFC3 slitless spectroscopic surveys: WISP (\citet{Atek:2010}, PI: Malkan, M.), the SN followup of the CANDELS survey (SN-CANDELS, PI: Riess, A.), 3D-HST (\citet{VanDokkum:2011,Brammer:2012}, PI: Van Dokkum) and AGHAST (A Grism H-Alpha SpecTroscopic survey in the GOODS-N, PI: B. Weiner).
Since previous studies suggest a substantial brightening of $L^{*}_{H\alpha}$ (Figure~\ref{fig:img_depths}), this study will be performed by dividing the sample in three redshift bins between z=0.7 and z=1.6. Even if split, the combined sample is large enough to allow for an uniform coverage of the $H\alpha$ luminosity range, from few \% of $L^{*}_{H\alpha}$ to many times $L^{*}_{H\alpha}$.

Above z=1.6, the H$\alpha$ line is redshifted outside the detectable wavelength
range and since single line emitters are considered H$\alpha$ by default, the
ability to confirm the identification of [OIII] lines is limited to bright lines
with clear shapes. Greatly extending the small dataset used in
\citet{Colbert:2013ita}, the newly released WISP data will allow us to better
constrain the contamination of the H$\alpha$ selected sample, and the loss of
completeness of the [OIII] sample. We will use the multi-wavelength dataset on
3D-HST + AGHAST to identify a combination of colors that could possibly mitigate
this problem.

The identification of the spectral lines and the measure of their fluxes is
fundamental for all the successive part of the LF project here described, but it
can not be completely performed by an automatic algorithm. Thanks to the same
collaboration involving both the SIT and WISP team members, automatic softwares
have already been optimized to i) remove low S/N detections and some of the
contaminating false detections and ii) measure the total line fluxes after their
identification. However, the human interaction is required to visually identify
each detected line (mostly H$\alpha$ and [OIII], but also OII, H$\beta$ and
SII), to perform high quality continuum and line fits, and to remove not
automatically identified contaminants. In the past weeks, the work has been
focused on these tasks. As a result, a large dataset of thousands of objects
with measured line fluxes and corresponding redshift is now available.

At the end of the work described in this section, for which more than a semester
is still reasonably required, the results will lead to a publication in one of the
major journals, with the co-authorship of Teplitz, Baronchelli and Wang.

 \subsection{Cosmological Forecasting and Data Analysis Algorithms (D4)}
\label{sec:grs_algo}

 \begin{summaryii}
  We have developed software package to make quick forecasts on basic dark energy parameters from higher order statistics, incluging power spectrum and bispectrum multipoles. This light and agile software package will be the backbone of our GRS proto-pipeline effort.
  \end{summaryii}

 The key dark energy constraints from the WFIRST GRS will result from the BAO and
 RSD measurements from the two-point statistics of the observed galaxy field.
 Similar measurements from the higher order statistics are weaker and currently
 are considered less robust. Cosmological constraints from higher order
 statistics scale very steeply with the number density and since WFIRST GRS will
 provide very dense galaxy samples they may significantly enhance the yield from
 the standard two-point BAO/RSD analysis. We also expect the methods of analyzing
 higher order statistics to become more robust and standardized by the time of
 WFIRST launch. Because of these considerations it would be helpful to have a
 higher order statistics forecasting tool. We have developed software package to
 make forecasts on basic dark energy parameters from higher order statistics.
 The main assumptions are similar to the ones made in the standard
 power-spectrum forecasting tool used for baseline WFIRST predictions. We will
 work on integrating this software with the standard forecasting tool developed
 by our SIT. While the key design decisions will still be based on the
 two-point statistics forecasts, knowing how different choices will effect higher
 order analysis will be very informative.

 The KSU group has assembled a fast and lightweight set of tools for analyzing
 the WFIRST GRS data. Currently this toolset starts from the redshift catalogue
 and the visibility cube and produces the measurements of power spectrum
 multipoles. The multipoles are then analyzed to extract the BAO and RSD signal
 from them. The BAO extraction algorithms replicate the analysis of the final
 BOSS DR12 sample. The RSD analysis is currently simplistic and uses the linear
 model. We will update this toolset by implementing more realistic RSD models.
 The toolset will eventually be linked to the redshift catalogue and visibility
 cube producing software. This software will provide the backbone of our BAO/RSD
 proto-pipeline and will be validated with high fidelity WFIRST simulations.

 \begin{summary}[]
 \begin{enumerate}
 \item We will keep refining and solidifying the scientific requirements;
 \item We will pursue our light-cone simulations and release mock catalogs to the community when publishing our results;
 \item We will pursue our effort with image simulation, relying on the light-cone simulation to incorporate realistic galaxy distributions and properties;
 \item We will continue our pseudo-pipeline effort;
 \end{enumerate}
 \end{summary}

\section{Cosmological Forecasts (D2, D6, D7)}
\label{sec:forecast}
\newcommand{\nn}{\nonumber}
\newcommand{\vpi}{\mathbf \pi}
\newcommand{\vecd}{\mathbf d}
\newcommand{\matC}{\mathbf C}
\newcommand{\matQ}{\mathbf Q}

\newcommand{\om}{\Omega_\mr m}
\newcommand{\omb}{\Omega_\mr b}
\newcommand{\sig}{\sigma_8}
\newcommand{\ns}{n_s}
\newcommand{\w}{w_0}
\newcommand{\wa}{w_a}

\renewcommand{\d}{{\rm d}}
\newcommand{\pd}{P_{\delta}}
\newcommand{\pe}{P_\mr E}

\newcommand{\vt}{\vartheta}
\newcommand{\vp}{\varphi}
\newcommand{\eps}{\epsilon}
\newcommand{\abs}[1]{| #1 |}
\newcommand{\mr}{\mathrm}

\renewcommand{\d}{{\rm d}}

\newcommand{\like}{L}
\newcommand{\prob}{P}
\newcommand{\probr}{P_r}
\newcommand{\p}{\mathbf p}
\newcommand{\pco}{\mathbf p_\mr{c}}
\newcommand{\pnu}{\mathbf p_\mr{n}}
\newcommand{\plf}{\mathbf p_\mr{LF}}
\newcommand{\D}{\mathbf D}
\newcommand{\Del}{\mathbf \Delta}
\newcommand{\M}{\mathbf M}
\newcommand{\N}{\mathbf N}
\newcommand{\U}{\mathbf U}

\begin{summary}
We developed a unique and coherent cosmological forecast software package, \CoLi. Using this single unified framework, we revised the HLSS and HLIS cosmological forecasts, including systematics effects. We also pioneered a joint HLSS-HLIS analysis. This led us to emphasize the importance and the power of a \emph{multi-probe} cosmological approach. We also performed several trade studies to support Project Office activities, in conjunction with the SNe SITs.
\end{summary}

In this section we will describe the results a simulated likelihood analysis that forecasts science return and quantified systematics for different choices in the WFIRST HLS survey. We first introduce the \CoLi software framework that was used to generate our forecasts. We then present forecasts for HLSS and HLIS specific observables. We conclude with a summary of ongoing research to explore WFIRST multi-probe analysis strategies, i.e., to combine all WFIRST observables and their correlated systematics consistently in simulated likelihood analyses. We also present ongoing work that will assess synergies between WFIRST and other contemporary surveys, e.g., the ground-based, optical Large Synoptic Survey Telescope (LSST) and future missions studying the Cosmic Microwave Background, e.g. CMB-Stage4 (CMB-S4) experiments.

\subsection{\CoLi Introduction}
\label{sec:cosmolike}

\begin{summaryii}
  \CoLi is unique in its approach that jointly models LSS probes as well as their correlated systematics. It is the only code that computes multi-probe covariances, including the (dominant) higher-order terms of the matter density field.
\end{summaryii}

\CoLi has been used in several science efforts, e.g. the combined probes forecasts presented in \citep{Eifler:2014}, an analysis of SDSS shear data \citep{Huff2014}, and efforts to develop new mitigation strategies for baryonic physics \citep{Eifler2015} and galaxy intrinsic alignment \citep{Krause2016}. It is one of two cosmology analysis pipelines used by the ongoing Dark Energy Survey's Year 1 analysis and it has previously been used in the DES science verification data analysis \citep{DES2015}. Beyond DES,
\begin{figure}
  \includegraphics[width=16.0cm]{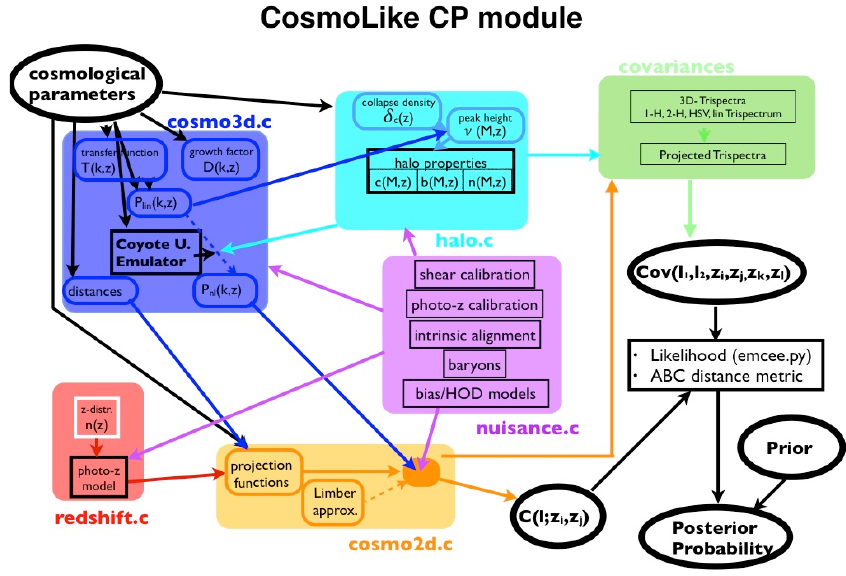}
   \caption{Illustration of a subset of CosmoLike's code structure. Starting from cosmological parameters we show how different parts of the code interface and how projected summary statistics for various observables are obtained. This particular workflow corresponds to the multi-probe analysis as described in Sect. {\ref{sec:multi-probe}}. For the HLSS survey a subset of these routines are being used.
}
  \label{fi:fcosmolike}
\end{figure}
the code has been used to simulate a realistic multi-probe likelihood analyses for LSST \citep{Krause2017} that include cosmic shear, galaxy clustering, galaxy-galaxy lensing, cluster number counts and cluster weak lensing.
The software relies on a massively parallelized computation of fine-tuned look-up tables with a targeted sub-second run-time per point in parameter space. Given the large number of parameters describing systematic effects in future LSS analyses, this efficient computation ensures an acceptable turnaround time for analyses. In particular, it allows one to generate a large number of simulated likelihood analyses which is necessary to optimize future surveys. Figure~\ref{fi:fcosmolike}
illustrates the \CoLi combined probes module we plan to use for the proposed analysis. Starting from a cosmological model, this schematic illustrates how projected power spectra for various LSS probes are computed:
\begin{enumerate}
\item In the first step \CoLi computes the 3D-density power spectrum using one of 3 different methods. The density power spectrum can be modeled using the latest suites of numerical simulations, i.e. the Coyote Universe emulator \citep{Heitmann2014}, or the latest implementation of Halofit \citep{tsn12}, or using a state-of-the-art analytic halo model \citep{Krause2013}.
\item Given a survey specific redshift distribution \CoLi computes projected (tomographic) 2D power spectra for WL, galaxy-galaxy lensing, and galaxy clustering.
\item  For galaxy-galaxy lensing and galaxy clustering, the connection between dark and luminous matter is modeled using either a linear bias model or a sophisticated (but computationally slower) HOD model \citep{Krause2013}.
\item The covariance is computed analytically using perturbation theory to include the higher-order moments of the density field, including the Halo Sample Variance contribution that dominates the statistical error budget. This covariance takes all cross-correlations between the different probes into account.
\item \CoLi has various systematics modeling capabilities, in particular it can account for baryonic effects, intrinsic alignment, shear calibration and photo-z uncertainties \citep{Eifler2015, Krause2016, Krause2017}.
\item If one chooses to use two-point correlation functions instead of power spectra CosmoLike offers an extremely fast Fourier transformation using a Hankel transform. The code employs a parallel MCMC sampling technique \citep{Goodman2010}, implemented using the emcee python package \citep{Foreman-Mackey2013}.
\end{enumerate}

\subsection{HLS Spectroscopic Survey Forecasts and Trade Studies}

\begin{summaryii}
  We implemented in the \CoLi package a state of the art spectroscopic survey forecasting formalism. We used this tool to update the HLSS forecast and to conduct two trade studies. We validated and explored the validity of science requirements HLSS1 and HLSS3. (\S~\ref{sec:sr2a_grs}).
  \begin{enumerate}
  \item We carried out a trade study of area versus depth for the HLSS only, starting from a baseline survey of 2,227 sq. deg. and a wavelength range of 1.05-1.85 microns. We considered two alternative scenarios, i.e., a survey twice as wide and a survey half as wide but correspondingly deeper.
  \item We examined the impact of an extended wavelength range on the cosmological constraining power.
\end{enumerate}
\end{summaryii}

\subsubsection{Varying the HLSS Area} 
\begin{figure*}
  \includegraphics[width=16.0cm]{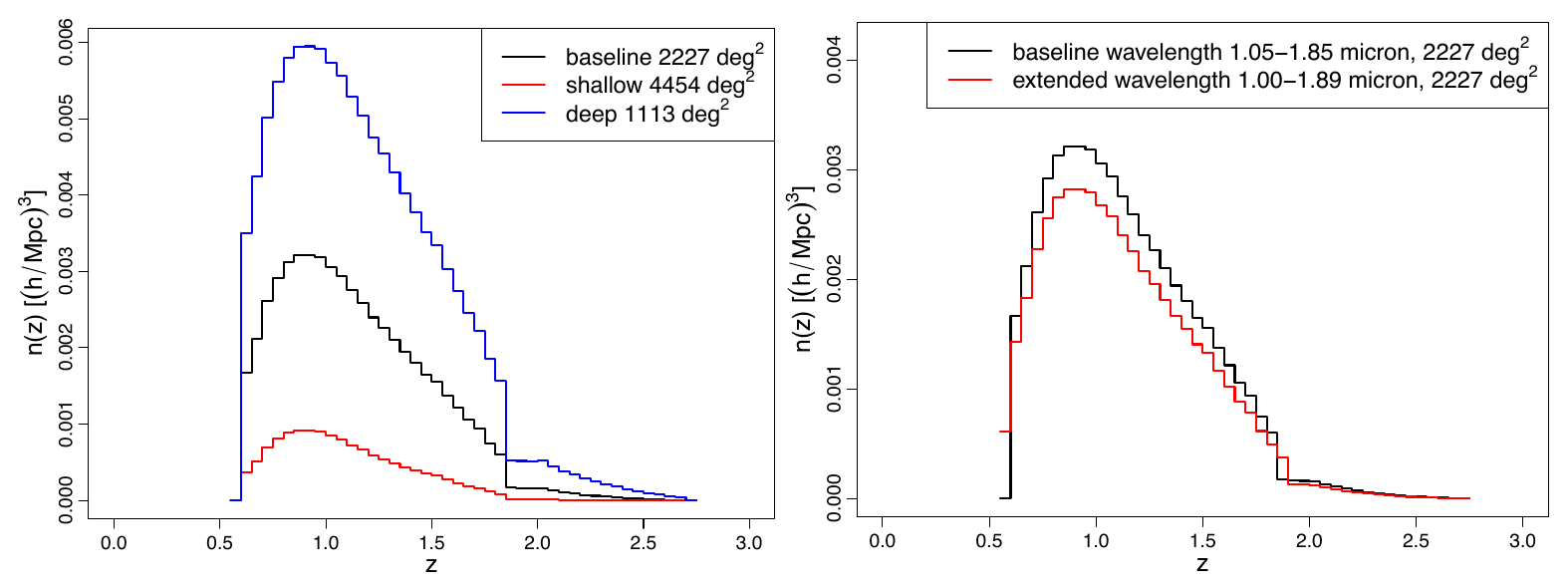}
    \caption{\textit{Left:}Redshift distribution of galaxies in the baseline, the shallow/large area, and the deep/small area survey. \textit{Right:} Redshift distribution of galaxies in the baseline and extended wavelength survey.
}
  \label{fi:forecast1}
\end{figure*}

The galaxy redshift distributions were computed using the WFIRST \texttt{Exposure Time Calculator ETC v14}. The H$\alpha$ forecasts are based on the
average of the 3 models in \citet{Pozzetti:2016}, and the [O III] forecasts are
based on the \citet{Mehta:2015} luminosity function. The resulting redshift
distributions are displayed in Figure~\ref{fi:forecast1} (left panel). We extended the \CoLi framework \citep{Eifler:2014,Krause2017} to compute the constraining power of all cosmic acceleration scenarios, closely following \citet{Wang2013}. We ran a
500,000 step MCMC simulated likelihood analysis in a 23 dimensional parameter
space. We simultaneously varied 7 cosmological parameters and 16 nuisance
parameters describing uncertainties due to the linear galaxy bias model, the
non-linear smearing of the BAO feature, peculiar velocity dispersion, power
spectrum shot noise, and redshift errors. We assumed priors on cosmological
parameters from the current state of the art experiments, i.e. the Planck
mission, the Baryon Oscillation Spectroscopic Survey, the Joint Lightcurve
Analysis, as described in \citet{Aubourg:2015}.

The information gain is quantified using the standard Dark Energy Task Force FOM and an extended cosmology FOM, which measures the enclosed volume in the full 7-dimensional cosmological parameter space, not just in the 2 dark energy parameters. We will refer to these FOMs as DE-FOM and Cosmo-FOM. Compared to our baseline scenario we find a decreased DE-FOM of 32\% and a decreased Cosmo-FOM of 45\% for the shallow/large area survey. For the deep/small area survey we find an increased DE-FOM of 5\% and an increased Cosmo-FOM of 2\%. We note that these findings are model and prior dependent and recommend further studies to confirm these forecasts. We also note that the [O III] numbers are pending a future update in part due to the reduction in the baseline telescope temperature to 260 K.


\subsubsection{Varying the HLSS Redshift Coverage} In addition to the trade studies in HLSS 1 we examined the impact of an extended wavelength range on the DE-FOM and the Cosmo-FOM. We follow the same procedure as detailed in the HLSS 1 paragraph extending the wavelength range from 1.05-1.85 microns for the baseline model to 1.00-1.89 for the extended model. The corresponding redshift distributions of the galaxy samples computed from the ETC v1.14 are depicted in  Figure~\ref{fi:forecast1} (right panel). We find a decreased DE-FOM of 2\% and a decreased Cosmo-FOM of 11\% for the extended wavelength survey with respect to our baseline scenario. We iterate that these findings are model and prior dependent and recommend further studies varying the input parameters.

\subsection{HLS Imaging Survey Forecasts}
\label{sec:HLISforecasts}

\begin{summaryii}
We extended the \CoLi package to accurately forecast the HLIS weak gravitational lensing signal, including multiple additional systematics. This extension was then used to study multiple modifications to the survey and its requirements with a focus on the effect of systematics. In particular, we studied
\begin{enumerate}
  \item The science gain when extending the survey to 10,000 deg$^2$, instead if 2,200 deg$^2$;
  \item The combined impact of uncertainties in shape and photo-$z$ measurements;
  \item The impact of uncertainties in shape measurements only;
  \item The impact of uncertainties in photo-$z$ estimation only;
  \item The impact of uncertainties in baryonic physics modeling;
  \item The impact of uncertainties in intrinsic alignment modeling.
\end{enumerate}
\end{summaryii}

\begin{figure}
\includegraphics[width=14cm]{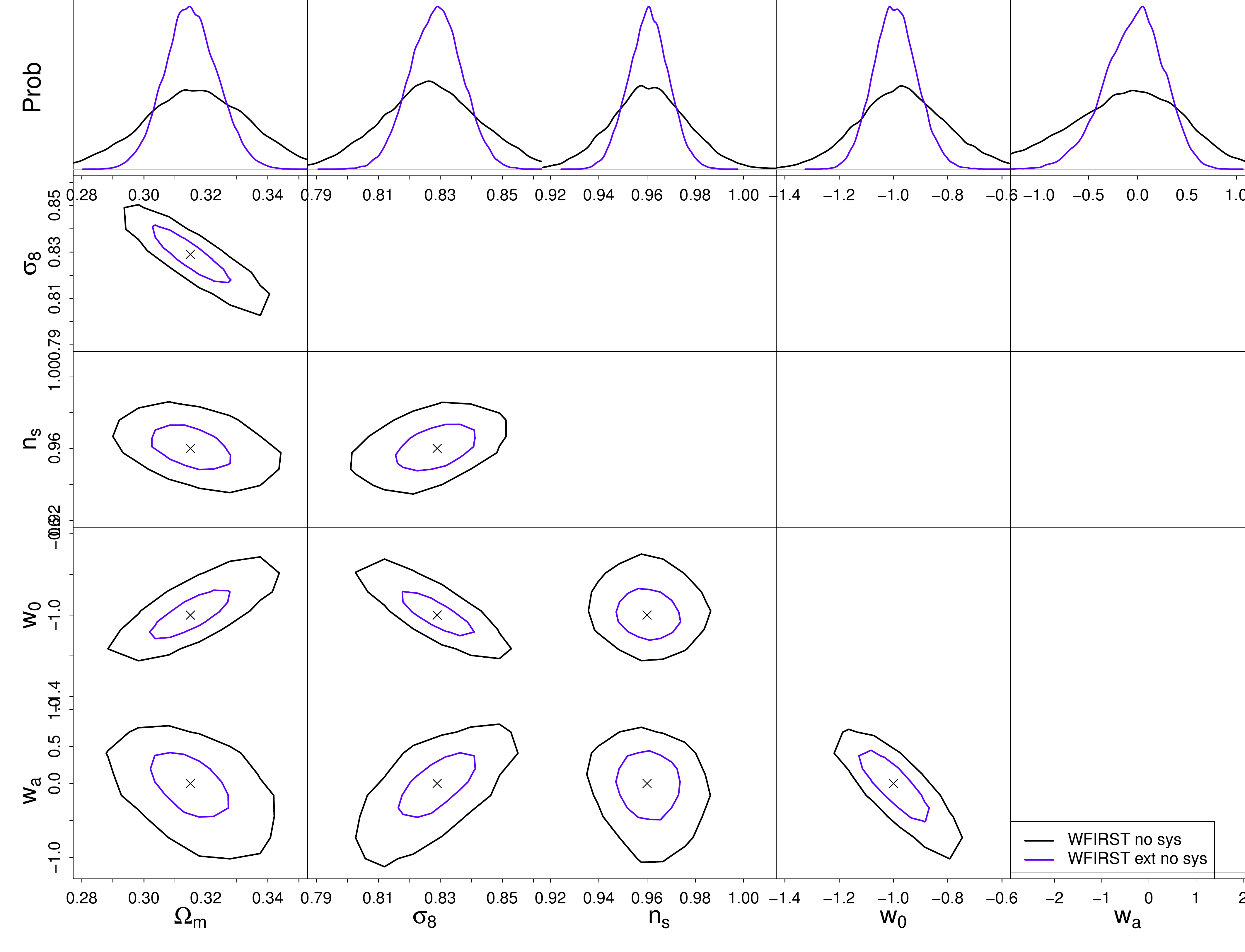}
\includegraphics[width=14cm]{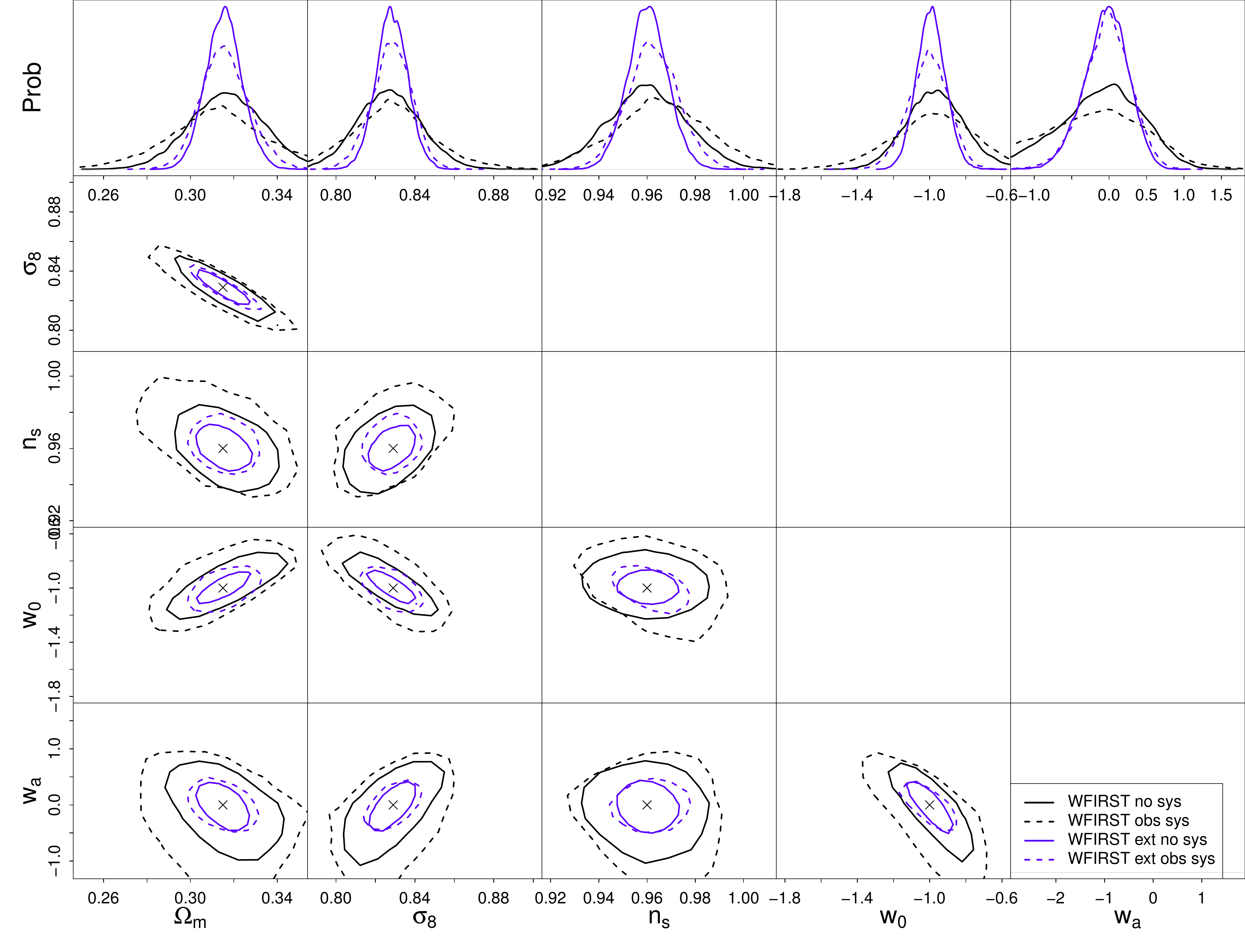}
\caption{\textit{Top:} WFIRST forecasts: statistical errors. Extended 10,000 $\mr{deg^2}$ mission in blue, regular 2,200 $\mr{deg^2}$ mission in black. \textit{Bottom:} Broadening of WFIRST error bars accounting for shear calibration and photo-z errors.}
         \label{fi:extended}
\end{figure}

\begin{figure}
\includegraphics[width=14cm]{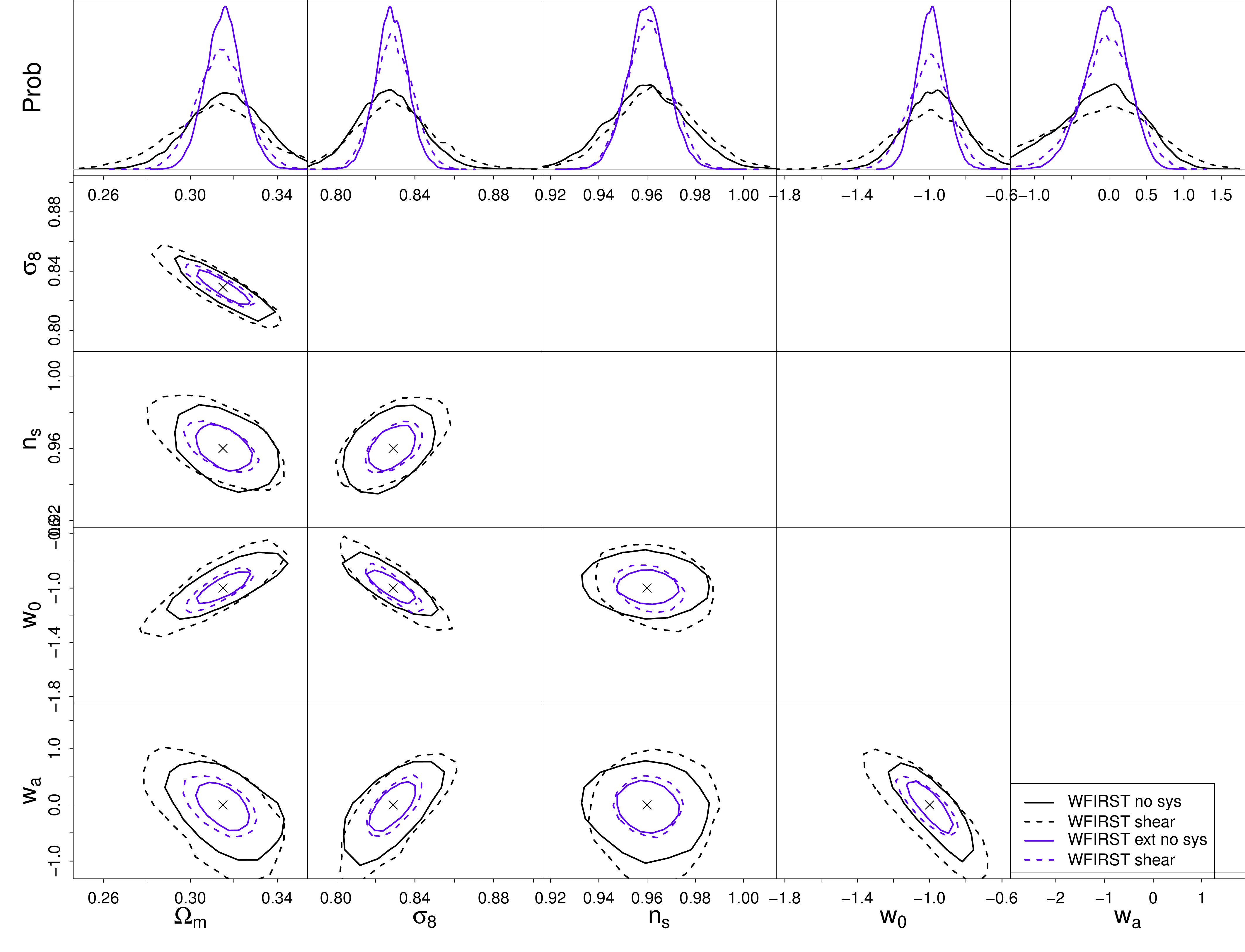}
\includegraphics[width=14cm]{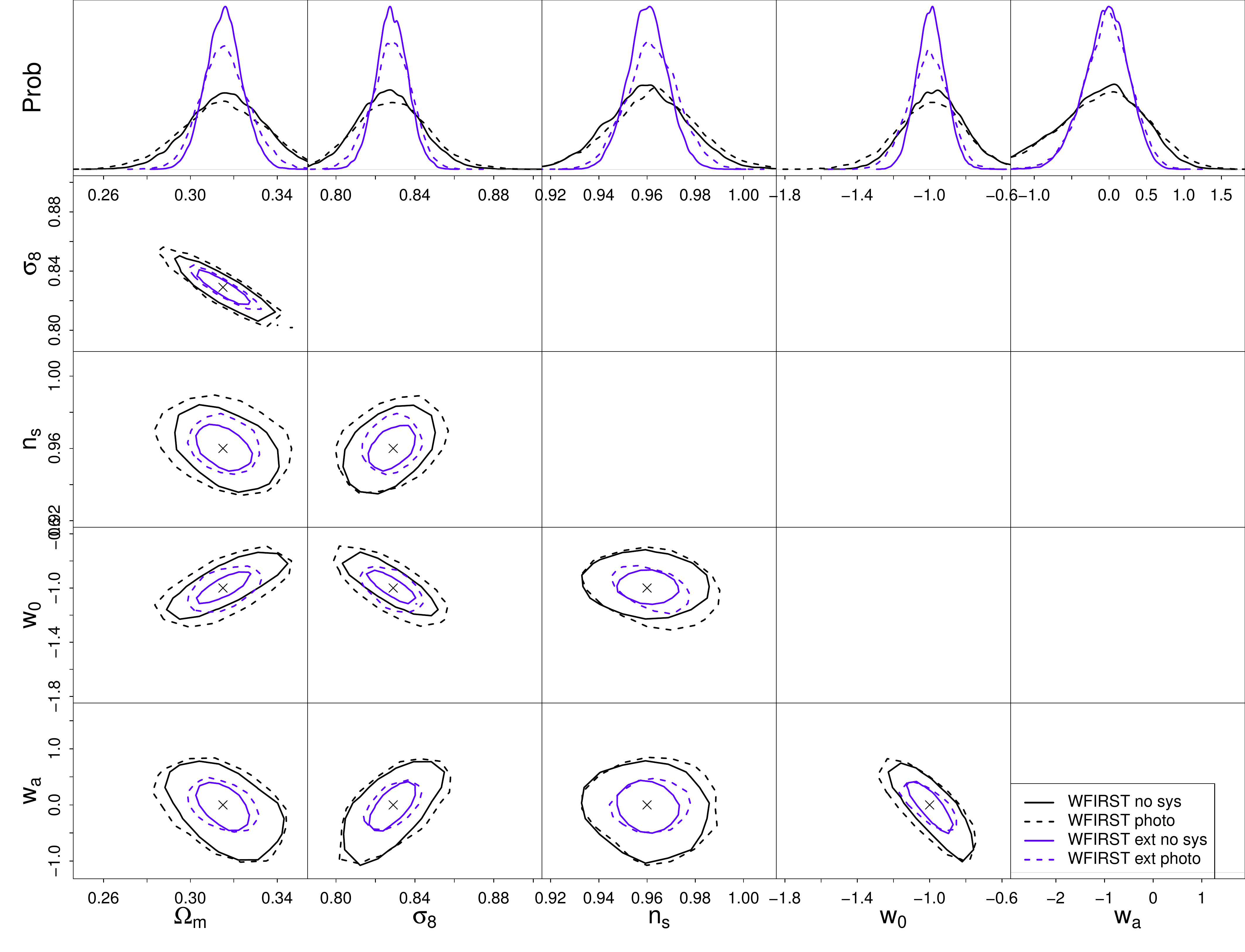}
\caption{\textit{Top:} WFIRST statistical errors (solid) compared to errors when including uncertainties from multiplicative shear calibration errors (dashed).  Extended 10,000 $\mr{deg^2}$ mission in blue, regular 2,200 $\mr{deg^2}$ mission in black
\textit{Bottom:} WFIRST contours when accounting for photo-$z$ uncertainties; see text for details.}
         \label{fi:sys_obs}
\end{figure}

\begin{figure}
\includegraphics[width=14cm]{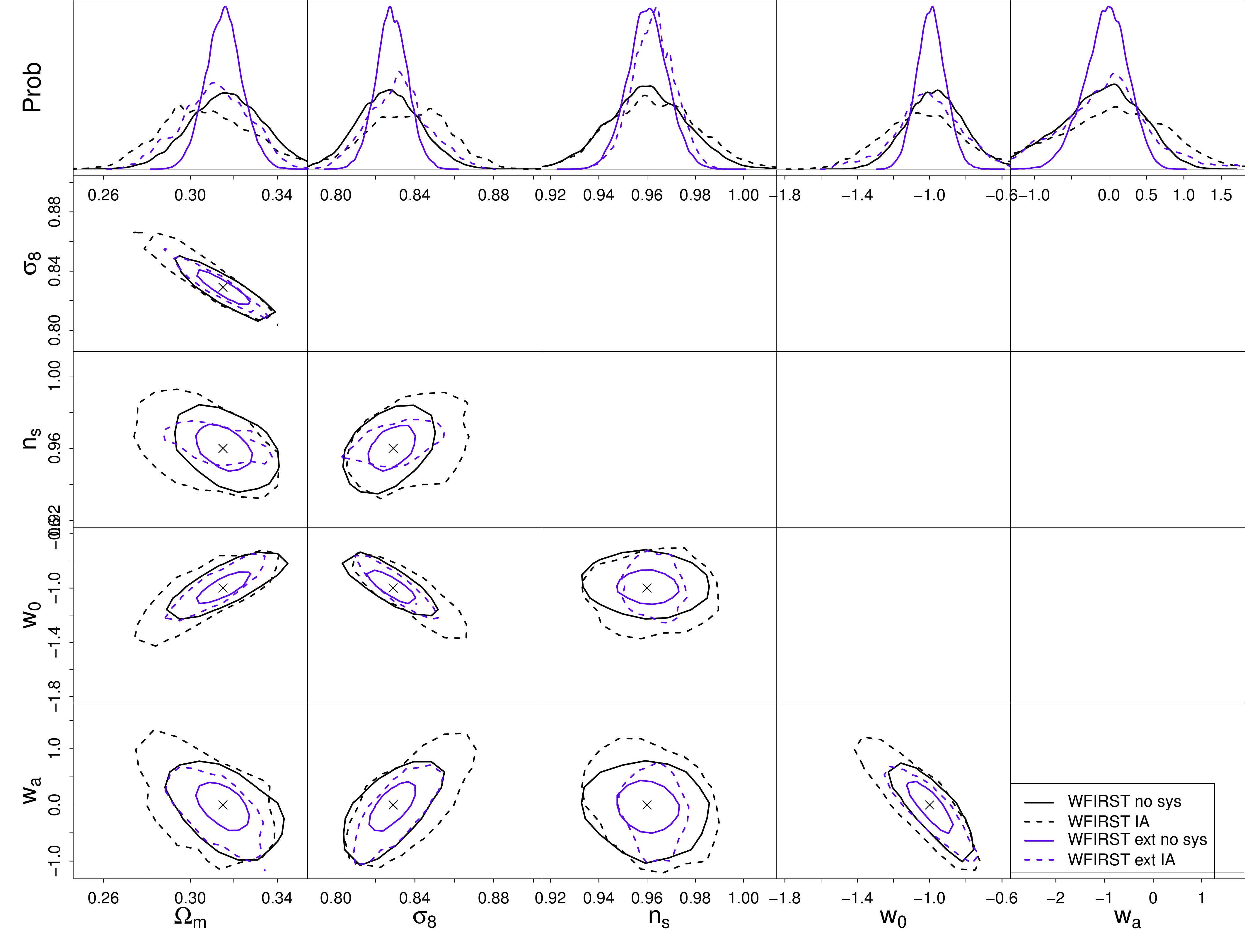}
\caption{Increase of WFIRST (nominal and extended mission) error bars when marginalizing over intrinsic alignment nuisance parameters \citep[see][for comparison]{Krause2016}.} \label{fi:IA}
\end{figure}

\begin{figure}
\includegraphics[width=14cm]{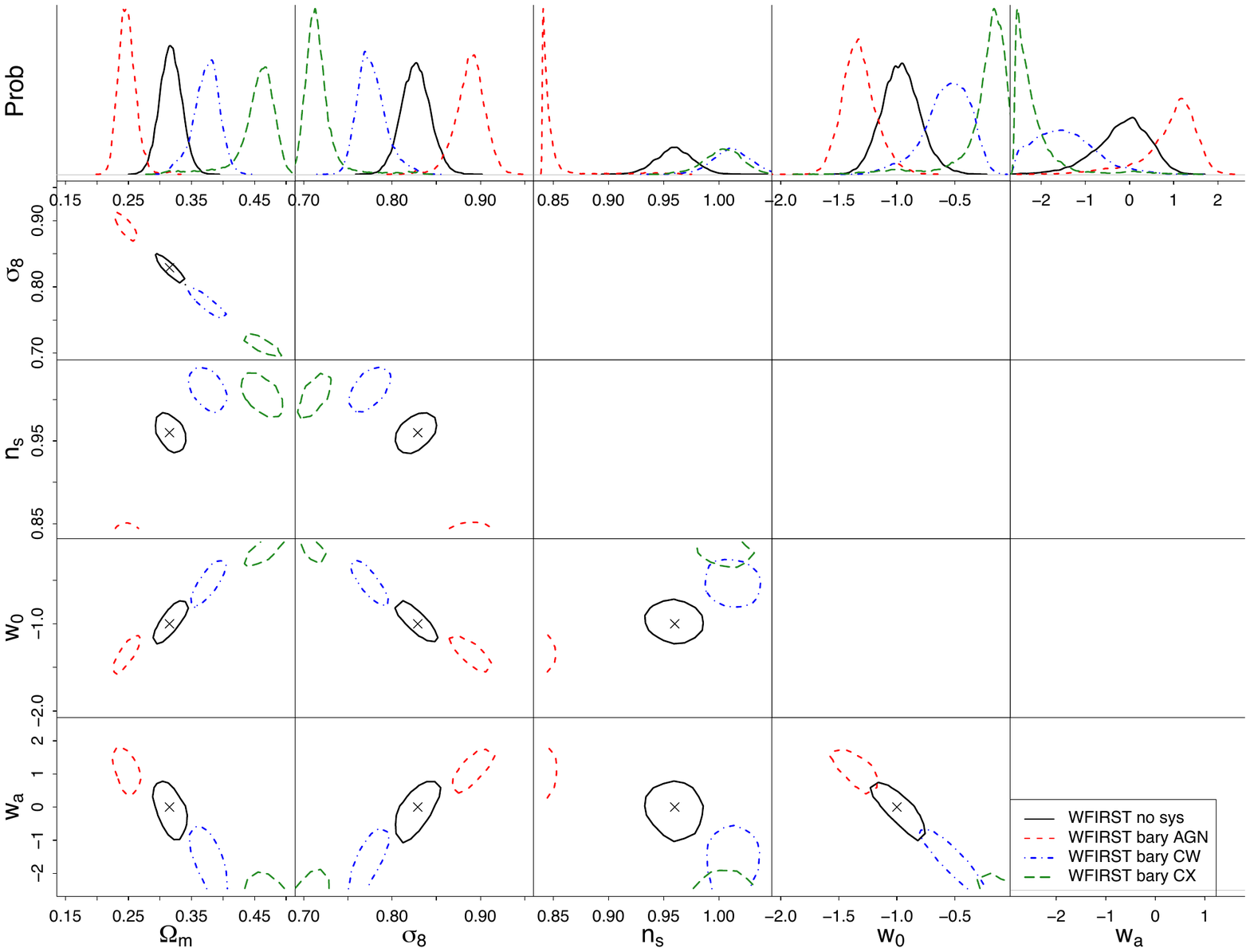}
\includegraphics[width=14cm]{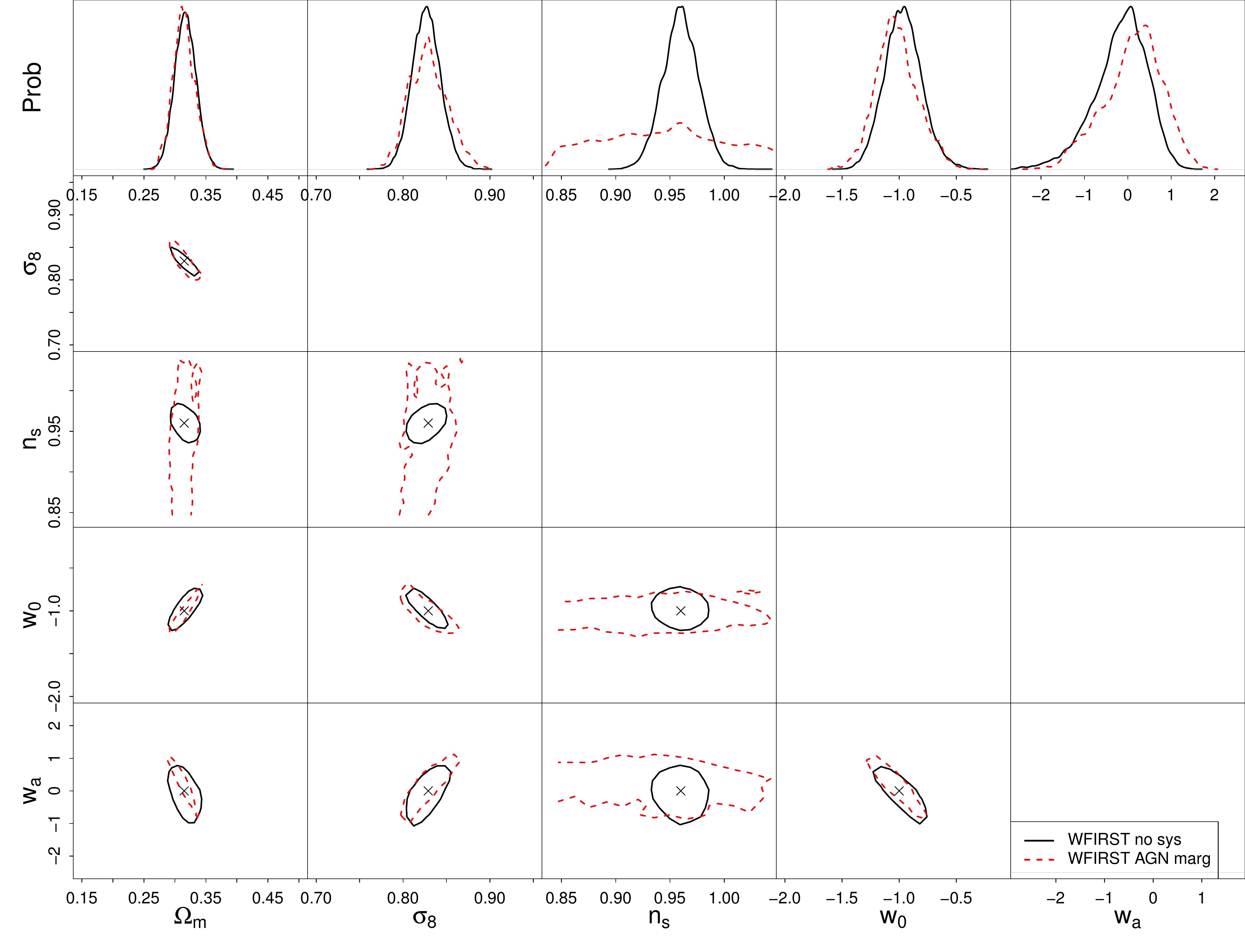}
\caption{Increase of WFIRST (nominal and extended mission) error bars when marginalizing over nuisance parameters modeling baryonic physics \citep[see][for details of the method]{Eifler2015}.}
         \label{fi:bary}
\end{figure}

\subsubsection{Modeling Weak lensing observables and covariances}
\label{sec:lensingbasics}

We begin this WFIRST HLIS section by forecasting the weak lensing science component only. We closely follow the weak lensing part of the forecasting machinery described in \cite{Krause2017}.

\paragraph{Shear Tomography Power Spectra} \CoLi obtains matter power spectra through calls to the Boltzmann \textsc{CLASS} code and it uses the \citet{tsn12} calibration of the \textsc{halofit} fitting function for the non-linear matter power spectrum \citep{smp03}. Time-dependent dark energy models ($w=w_0+(1-a)\,w_a$) are incorporated following the recipe of {\sc icosmo} \citep{rak11}, which in the non-linear regime interpolates Halofit between flat and open cosmological models \citep[also see][for more details]{shj10}.

Having obtained the density power spectra we calculate the shear power spectra as
\begin{equation}
\label{eq:pdeltatopkappa}
C ^{ij} (l) = \frac{9H_0^4 \om^2}{4c^4} \int_0^{\chi_\mr h}
\mr d \chi \, \frac{g^{i}(\chi) g^{j}(\chi)}{a^2(\chi)} \pd \left(\frac{l}{f_K(\chi)},\chi \right) \,,
\end{equation}
with $l$ being the 2D wave vector perpendicular to the line of sight, $\chi$ is the comoving coordinate, $\chi_\mr h$ is the comoving coordinate of the horizon, $a(\chi)$ is the scale factor, and $f_K(\chi)$ the comoving angular diameter distance (throughout set to $\chi$ since we assume a flat Universe). The lens efficiency $g^{i}$ is defined as an integral over the redshift distribution of source galaxies $n(\chi(z))$ in the $i^\mr{th}$ tomographic interval
\begin{equation}
\label{eq:redshift_distri}
g^{i}(\chi) = \int_\chi^{\chi_{\mr h}} \mr d \chi' n^{i} (\chi') \frac{f_K (\chi'-\chi)}{f_K (\chi')} \,.
\end{equation}
Since we chose five tomographic bins, the resulting data vector which enters the likelihood analysis consists of 15 tomographic shear power spectra, each with 12 logarithmically spaced bins ($l \in [100;5000]$), hence 180 data points overall. The limits of the tomographic $z$-bins are chosen such that each bin contains a similar number of galaxies.

\paragraph{Shear Covariances} Under the assumption that the 4pt-function of the shear field can be expressed in terms of 2pt-functions (so-called Gaussian shear field) the covariance of projected shear power spectra can be calculated as in \citep{huj04}
\be
\label{eq:covhujain}
\mr{Cov_G} \left( C^{ij} (l_1) C^{kl} (l_2) \right) = \langle \Delta C^{ij} (l_1) \, \Delta C^{kl} (l_2) \rangle  =  \frac{\delta_{l_1 l_2}}{ 2 f_\mr{sky} l_1 \Delta l_1}  \left[\bar C^{ik}(l_1) \bar C^{jl}(l_1) + \bar C^{il}(l_1) \bar C^{jk} (l_1) \right]\,,
\ee

with
\be
\label{details}
\bar C^{ij}(l_1)= C^{ij}(l_1)+ \delta_{ij} \frac{\sigma_\eps^2}{n^{i}} \,,
\ee
where the superscripts indicate the redshift bin; $n^{i}$ is the density of source galaxies in the $i$-th redshift bin; and $\sigma_\eps$ is the RMS of the shape noise.

Since non-linear structure growth at late time induces significant non-Gaussianities in the shear field, using the covariance of Eq.~(\ref{eq:covhujain}) in a likelihood analysis results in an underestimate of the errors on cosmological parameters. Therefore, the covariance must be amended by an additional term, i.e. $\mr{Cov}=\mr{Cov_G}+\mr{Cov_{NG}}$.  The non-Gaussian covariance is calculated from the convergence trispectrum $T_{\kappa}$ \citep{CH01,taj09}, and we include a sample variance term $T_{\kappa,\rm{HSV}}$ that describes scatter in power spectrum measurements due to large scale density modes \citep{tb07, sht09},
\be
 \mr{Cov_{NG}}(C^{ij}(l_1),C^{kl}(l_2)) =  \int_{|\mathbf l|\in l_1}\frac{d^2\mathbf l}{A(l_1)}\int_{|\mathbf l'|\in l_2}\frac{d^2\mathbf l'}{A(l_2)} \left[\frac{1}{\Omega_{\mr s}}T_{\kappa,0}^{ijkl}(\mathbf l,-\mathbf l,\mathbf l',-\mathbf l') + T_{\kappa,\rm{HSV}}^{ijkl}(\mathbf l,-\mathbf l,\mathbf l',-\mathbf l') \right] \,,
\ee
with $A(l_i) = \int_{|\mathbf l|\in l_i}d^2\mathbf l \approx 2 \pi l_i\Delta l_i$ the integration area associated with a power spectrum bin centered at $l_i$ and width $\Delta l_i$.

The convergence trispectrum $T_{\kappa,0}^{ijkl}$ (in the absence of finite volume effects) is defined as
\be
\label{eq:tri2}
T_{\kappa,0}^{ijkl} (\mathbf l_1,\mathbf l_2,\mathbf l_3,\mathbf l_4) = \left( \frac{3}{2} \frac{H_0^2}{c^2} \om \right)^{4} \int_0^{\chi_h} \d \chi \, \left( \frac{\chi}{a(\chi)}\right)^4  g^i g^j g^k g^l \times \chi^{-6} \, T_{\delta,0}  \left( \frac{\mathbf l_1}{\chi}, \frac{\mathbf l_2}{\chi}, \frac{\mathbf l_3}{\chi}, \frac{\mathbf l_4}{\chi}, z(\chi) \right) \,,
\ee
with $T_{\delta,0}$ the matter trispectrum (again, not including finite volume effects), and where we abbreviated $g^i=g^i(\chi)$.

We model the matter trispectrum using the halo model \citep{Seljak00, CS02}, which assumes that all matter is bound in virialized structures that are modeled as biased tracers of the density field. Within this model the statistics of the density field can be described by the dark matter distribution within halos on small scales, and is dominated by the clustering properties of halos and their abundance on large scales. In this model, the trispectrum splits into five terms describing the 4-point correlation within one halo (the \emph{one-halo} term $T^{\mr{1h}}$), between 2 to 4 halos (\emph{two-, three-, four-halo} term), and a so-called halo sample variance term $T_{\mr{HSV}}$, caused by fluctuations in the number of massive halos within the survey area,
\be
\label{eq:t}
T = T_0 + T_{\mr{HSV}} = \left[T_{\mr{1h}}+T_{\mr{2h}}+T_{\mr{3h}}+T_{\mr{4h}}\right]+T_{\mr{HSV}}\;.
\ee
The \emph{two-halo} term is split into two parts, representing correlations between two or three points in the first halo and two or one point in the second halo. As halos are the building blocks of the density field in the halo model approach, we need to choose models for their internal structure, abundance, and clustering, in order to build a model for the trispectrum.

Our implementation of the one-, two- and four-halo term contributions to the matter trispectrum follows \citet{CH01}, and we neglect the three-halo term as it is subdominant compared to the other terms at the scales of interest for this analysis. Specifically, we assume NFW halo profiles \citep{NFW} with the \citet{Bhattacharya11} fitting formula for the halo mass--concentration relation $c(M,z)$, and the \citet{Tinker2008} fitting functions for the halo mass function $\frac{ dn}{dM}$ and linear halo bias $b(M)$ (all evaluated at $\Delta = 200$), neglecting terms involving higher order halo biasing.

Within the halo model framework, the halo sample variance term is described by the change of the number of massive halos within the survey area due to survey-scale density modes. Following \citet{sht09}, it is calculated as

\begin{eqnarray}
T_{\kappa,\rm{HSV}}^{ijkl}(\mathbf l_1,-\mathbf l_1,\mathbf l_2,-\mathbf l_2)= \left(\frac{3}{2}\frac{H_0^2}{c^2}\Omega_{\mr m}\right)^4 &\times&  \int_0^{\chi_\mr h}d\chi \left(\frac{d^2 V}{d\chi d\Omega}\right)^2 \left(\frac{\chi}{a(\chi)}\right)^4 g^i g^j g^k g^l \nn \\
&\times&  \int d M \frac{d n}{d M} b(M)\left(\frac{M}{\bar{\rho}}\right)^2 |\tilde{u}(l_1/\chi, c(M,z(\chi))|^2 \nn \\
 &\times& \int d M' \frac{d n}{d M'} b(M')\left(\frac{M'}{\bar{\rho}}\right)^2 |\tilde{u}(l_2/\chi, c(M',z(\chi))|^2 \nn \\
 &\times&  \int_0^\infty \frac{k dk}{2\pi}P_\delta^{\mr{lin}}(k,z(\chi))|\tilde W(k\chi \Theta_{\mr s})|^2 \,.
\end{eqnarray}

\paragraph{Cosmological Parameter Studies} Our ability to model shear power spectra and non-Gaussian covariances means that we can forcast cosmological constraints for different survey scenarios. For all forecasts, we sample the joint parameter space of cosmological $\pco$ and nuisance parameters $\pnu$ and parameterize the joint likelihood as a multivariate Gaussian
\be
\label{eq:like}
\like (\D| \pco, \pnu) = N \, \times \, \exp \biggl( -\frac{1}{2} \underbrace{\left[ (\D -\M)^t \, \matC^{-1} \, (\D-\M) \right]}_{\chi^2(\pco, \pnu)}  \biggr) \,.
\ee
The model vector $\M$ is a function of cosmology and nuisance parameters, i.e. $\M=\M(\pco, \pnu)$ and the normalization constant $N=(2 \pi)^{-\frac{n}{2}} |C|^{-\frac{1}{2}}$ can be ignored under the assumption that the covariance is constant in parameter space. The assumption of a constant, known covariance matrix $\matC$ is an approximation to the correct approach of a cosmology dependent or estimated covariance \citep[see][for further details]{esh09, seh16}.
Given the likelihood function we can compute the posterior probability in parameter space from Bayes' theorem
\be
\label{eq:bayes}
\prob(\pco, \pnu|\D) \propto \probr (\pco, \pnu) \,\like (\D| \pco, \pnu),
\ee
where $\probr (\pco, \pnu)$ denotes the prior probability.

Equation \ref{eq:bayes} allows us to simulate realistic likelihood analyses for various survey scenarios. In particular we consider the following scenarios:
\begin{itemize}
\item Science gain when extending the survey to 10,000 deg$^2$, instead of 2,200 deg$^2$ (see Fig \ref{fi:extended} top)
\item Impact of combined uncertainties in shape measurements and photo-z measurements (see Fig \ref{fi:extended} bottom)
\item Impact of uncertainties in shape measurements only (see Fig \ref{fi:sys_obs} top)
\item Impact of uncertainties in photo-z estimation only (see Fig \ref{fi:sys_obs} bottom)
\item Impact of uncertainties in baryonic physics modeling (see Fig \ref{fi:bary} bottom)
\item Impact of uncertainties in intrinsic alignment modeling (see Fig \ref{fi:IA} bottom)
\end{itemize}

Shear calibration uncertainties are modeled as a Gaussian with $\sigma=0.005$ in each of the 10 tomographic bins, which altogether leads to 11 nuisance parameters. Redshift errors are modeled as Gaussian errors with a bias around mean zero and $\sigma=0.05$. In each tomographic bin these parameters (bias and $sigma$) again have Gaussian priors, i.e. $\Delta \mr{bias}=0.005$ and $\Delta \sigma=0.006$.

For intrinsic alignment we assume the non-linear alignment model, closely following the parameterization and implementation of \cite{Krause2016}, i.e. we marginalize over 11 nuisance parameters for amplitude, redshift dependence, and luminosity function uncertainties.

For the impact of baryonic physics modeling, we follow the work of \cite{Eifler2015}, where the authors
examined the impact of different baryonic scenarios and quantify the bias in
cosmological constraints for LSST and DES when the impact of baryons are neglected. We repeat this analysis for WFIRST (Figure~\ref{fi:bary}, upper) and
also account for the uncertainties due to baryonic scenarios in a subsequent
analysis (Figure~\ref{fi:bary}, lower). There, we project out modes that are
sensitive to baryonic physics in our analysis, which comes at the cost of
constraining power, in particular the spectral index $n_s$. We note, however, that
$n_s$ is strongly constrained by CMB measurements and that a corresponding prior
will recover the lost information.

\subsection{Expanding the Science Case - Multi-Probe Forecasts}
\label{sec:multi-probe}

\begin{summaryii}
Using our updates to the \CoLi package described in the previous sections, we have now implemented a full multi-probe analysis, an extremely promising area of active research. We are able to combine \emph{all} cosmological probes enabled by WFIRST HLS. This is one of the major goal of our unified HLS SIT. The likelihood analyses presented here are the start of a demanding study to optimally combine WFIRST internal observables and also to study synergies between WFIRST and other, external data sets, in particular LSST\@. Our tools will enable us to do so in the coming years. We also study new cosmological signals these multi-probe analysis will allow us to measure, such as the effect of primordial non-Gaussianity on galaxy shapes and synergies with CMB lensing.
\end{summaryii}

\begin{figure}
\includegraphics[width=15cm]{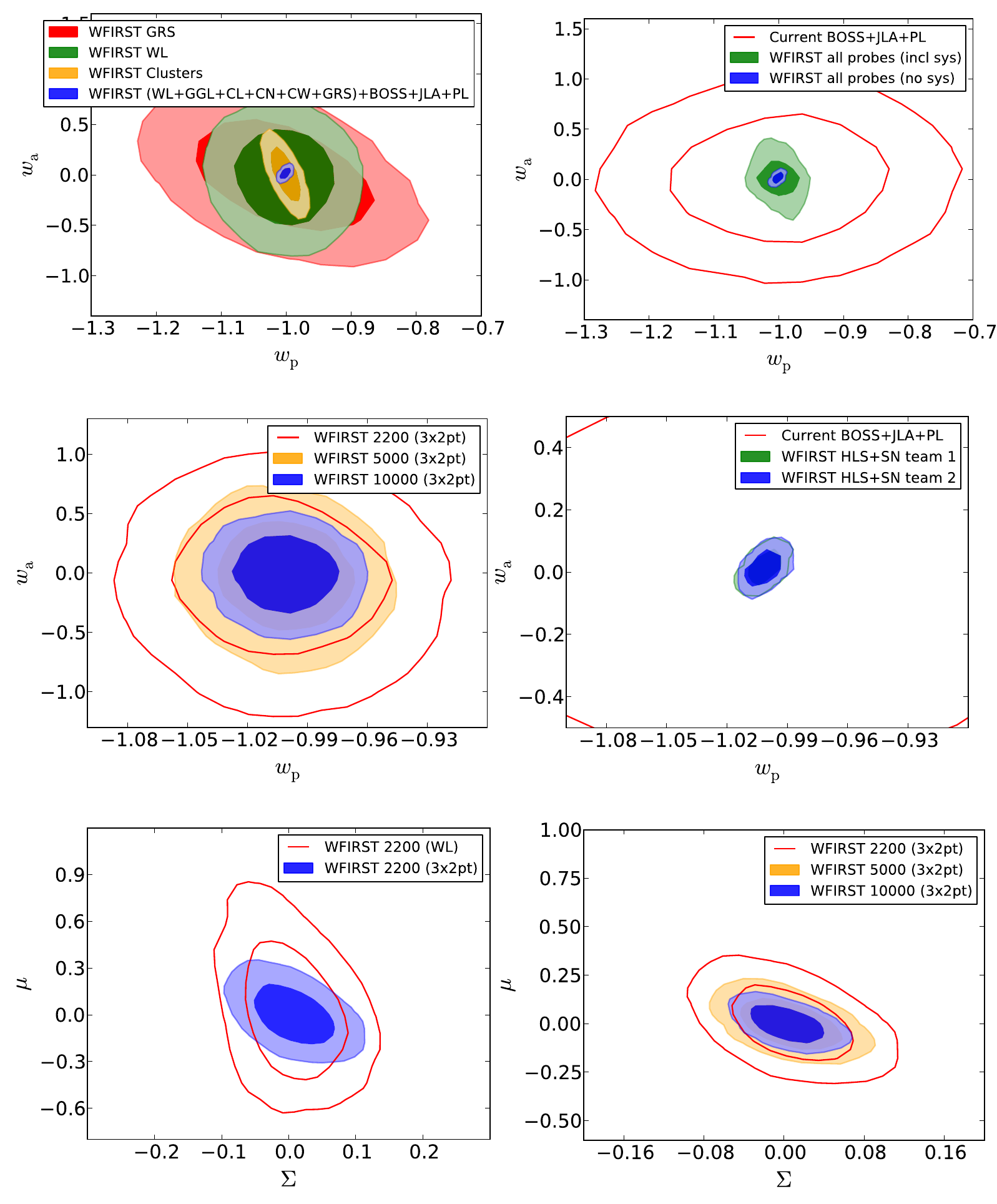}
\caption{WFIRST multi-probe studies, see text for details.}
\label{fi:multi}
\end{figure}

\subsubsection{\CoLi Multi-probe Analysis} Developing a multi-probe cosmology analysis framework is challenging given that the analysis pipeline for individual probes are each under continuous
development in order to meet the requirements of improving data quality. These
efforts have historically been largely independent from each other; as a
consequence, individual probe analyses utilize information from other probes to
offset/calibrate systematics. For example, one of the most important
astrophysical uncertainties for galaxy clustering is the relation of dark matter
halos and luminous galaxies. The joint analysis of clustering and galaxy-galaxy
lensing measurements removes uncertainties in this halo-galaxy connection when
constraining cosmology. However, cosmic shear analyses use the same
galaxy-galaxy lensing measurements to offset other systematic uncertainties
(e.g., intrinsic galaxy alignment). This is only one example of many systematics
mitigation conflicts that can occur when considering probes in isolation. A
multi-probe analysis framework must account for these correlated systematic
effects consistently across all probes; it cannot simply combine the optimal
versions of individual analyses and it must include a global, well-designed
systematics modeling and mitigation concept.

In this section we present a variety of simulated multi-probe WFIRST analyses.
Our results are summarized in Figure~\ref{fi:multi}. The upper left panel shows
the constraining power of the individual survey elements (galaxy-redshift
survey, weak lensing, galaxy clusters) in comparison to their combined
constraining power plus galaxy-galaxy lensing, photometric galaxy clustering,
and prior information from Planck, SN1a, BOSS\@. The likelihood analysis was
carried out in a 7-dimensional cosmological parameter space, excluding any
systematics. The right panel compares the aforementioned joint analysis (blue)
with the current state of the art (red) and we show how the contours increase
when including a realistic set of systematics in the analysis. These systematics
include uncertainties arising from shear and photo-$z$ calibration, cluster
mass-observable relation, galaxy intrinsic alignment, and galaxy bias. Modeling
and marginalizing over these systematics requires us to run a likelihood
analysis in 54 dimensional parameter space; the details of our analysis are very
similar to the LSST forecasts in~\citet{Krause2017}, but for WFIRST survey
parameters (survey area, number density, and redshift distribution).

The second row, left panel shows results from a simulated multi-probe analysis
that can be extracted from the imaging survey only. The so-called 3$\times$2pt analysis
consists of second-order statistics from weak lensing, photometric galaxy
clustering, and galaxy-galaxy lensing. We show the scaling of the error bars
(statistics only, no systematics) when going from the nominal survey area to a
5,000 and 10,000 deg$^2$ survey.

The second row, right panel shows again our full multi-probe analysis including
galaxy clusters and the spectroscopic survey component, i.e., BAO and RSD
measurements, and we also include information from the 2 WFIRST supernovae
teams. These team independently forecasted the WFIRST SN1a constraining power
and are in excellent agreement.

The lower row, left panel shows the constraining power of WFIRST weak lensing
and 3$\times$2pt for the nominal survey area on modified gravity parameters $\mu$ and
$\Sigma$ \citep[see e.g.,][for details]{jjk15, baa15}. The right panel shows the
constraining power when increasing the survey area again to 5,000 and 10,000
deg$^2$. The work on modified gravity extensions of \CoLi is led by JPL Postdocs Hironao Miyatake and Phil Bull in close collaboration with Tim Eifler and Elisabeth Krause and we plan to extend the results shown in Figure~\ref{fi:multi} to include further, recent developments in the field of modified gravity parameterization and classification. In particular, we plan to implement the ability to model modified gravity scenarios that belong to the class of Horndeski actions, which is the most general action for a single classical scalar field in the presence of gravity which does not result in any derivatives higher than second order in the equations of motion.

We emphasize that multi-probe studies are an area of active research and that
the likelihood analyses presented here represents a first step towards our goal of optimally combining WFIRST internal observables and to study synergies between WFIRST and other, external data sets, in particular LSST\@.

\subsubsection{Other Signals in Multi-Probe Analysis} Spergel and his group have been exploring  the use of galaxy shapes to constrain primordial non-Gaussianity \citep{Chisari:2016xki}. Working with former Princeton student, Elisa Chisari, and former Princeton postdocs Cora Dvorkin and Fabian Schmidt, they showed that
multi-tracer weak lensing observations could create anisotropic non-Gaussianity. WFIRST
will be a powerful instrument for this approach.  Correlations between intrinsic
galaxy shapes on large-scales arise due to the effect of the tidal field of
large-scale structure. Anisotropic primordial non-Gaussianity induces a distinct
scale-dependent imprint in these tidal alignments on large scales. Motivated by
the observational finding that the alignment strength of luminous red galaxies
depends on how galaxy shapes are measured, we study the use of two different
shape estimators as a multi-tracer probe of intrinsic alignments. We show, by
means of a Fisher analysis, that this technique promises a significant
improvement on anisotropic non-Gaussianity constraints over a single-tracer
method. For future weak lensing surveys, the uncertainty in the anisotropic
non-Gaussianity parameter, $A_2$, is forecast to be $\sigma(A_2)\simeq 50$, $\sim$ 40\% smaller than
currently available constraints from the bispectrum of the Cosmic Microwave
Background. This corresponds to an improvement of a factor of 4−5 over the
uncertainty from a single-tracer analysis.

\begin{figure}
\includegraphics[width=15cm]{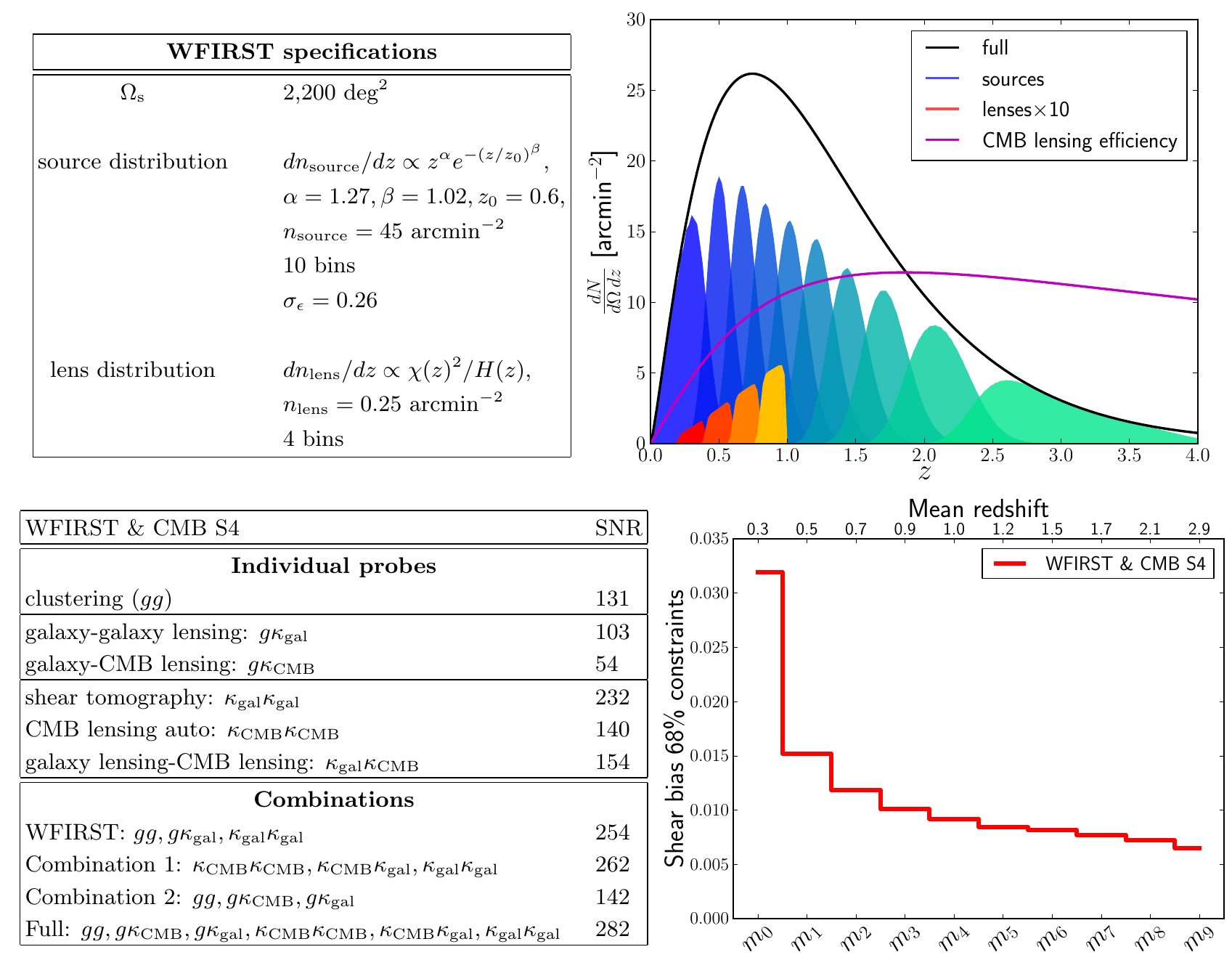}
\caption{Summary of the forecast for WFIRST and CMB S4 lensing. The top left panel summarizes our assumptions for
the WFIRST survey. The top right panel displays the photo-z bins assumed for the lens and source samples. The bottom left
panel summarizes the signal to noise ratios for the various probes and combinations. The bottom right panel shows the level
of shear calibration in WFIRST's ten source bins via CMB-S4 lensing. The calibration of the shear biases down 0.6-3.2\% in particular at high redshift bins, constitutes a promising alternative and an important cross-check to traditional simulation based approaches.}
\label{fi:CMB}
\end{figure}

Emmanuel Schaan has been working with Spergel and with the JPL group to
explore the use of CMB  lensing measurements as a tool to calibrate
multiplicative bias \citep{Schaan:2016ois}.  WFIRST  will require exquisite control over systematic
effects. In their paper, they address shear calibration and present the most
realistic forecast to date for WFIRST and CMB lensing from a stage 4 CMB experiment
(CMB S4). We use the CosmoLike code to simulate a joint analysis of all the
two-point functions of galaxy density, galaxy shear and CMB lensing convergence.
We include the full Gaussian and non-Gaussian covariances and explore the
resulting joint likelihood with Monte Carlo Markov Chains. We constrain shear
calibration biases while simultaneously varying cosmological parameters, galaxy
biases and photometric redshift uncertainties. We find that CMB lensing from CMB
S4 enables the calibration of the shear biases down 0.6-3.2\% in 10 bins for
WFIRST (see Fig. \ref{fi:CMB}). For a given lensing survey, the method works best at high redshift where
shear calibration is otherwise most challenging. This self-calibration is robust
to Gaussian photometric redshift uncertainties and to a reasonable level of
intrinsic alignment. It is also robust to changes in the beam and the
effectiveness of the component separation of the CMB experiment, and slowly
dependent on its depth, making it possible with third generation CMB experiments
such as AdvACT and SPT-3G, as well as the Simons Observatory.

\begin{summary}
Relying on our updates to the \CoLi package, we revised the cosmological
forecasts for the HLS spectroscopic and imaging survey. Using the unique \CoLi
capabilities, we conducted for the first time a full HLS multi-probe analysis in which we considered joint constraints between the HLS spectroscopic galaxy
sample, the HLS photometric galaxy samples, the HLS weak gravitational shear
signal, the HLS galaxy-galaxy lensing, the HLS galaxy clusters samples and the HLS SNe sample. Our
approach takes into account correlations between these probes as well as multiple systematic effects associated with each probe. We also combined WFIRST with other external surveys such as Planck and LSST. This is a milestone for our unified HLS SIT and we will investigate further the power and robustness of this multi-probe analysis in the coming years. We used this new capability to conduct trade studies to guide the development of the SRD. We also investigated new cosmological signals enabled by multi-probe studies, such as tests of general relativity.
\end{summary}

\section{Operations Model for the HLS and Evaluation of Trades (D7, D10, D11)}
\label{sec:operation}


\begin{summary}
Co-I Hirata is leading the development of the HLS observing plan, extending his
previous tools used for the SDT. These tools incorporate observing constraints
in the chosen orbit, an exposure-by-exposure observing sequence optimized with
detailed model of overheads, and tiling/coverage maps including field
distortions and curved sky effects. These tools treat both imaging and
spectroscopy with unified functions and scripts, and are well suited to joint
survey optimization when both hardware parameters (e.g., reaction wheel
orientations) and the observing program (e.g., depth vs.\ area) are considered.
This effort is coordinated with the scheduling and Design Reference Mission (DRM)
Working Groups. We are both providing an example detailed plan for the HLS to
the DRM working group, and cross-checking the Project's spreadsheet-level survey
calculators against our simulations. The HLS observing plan is also being
transferred to the Calibration Working Group, since the HLS observing strategy
feeds directly into the issue of self-calibration.
\end{summary}

\subsection{Survey Optimization Principles}
\label{sec:sur_opt}

The key constraint in survey optimization is the limited amount of observational time
available, since WFIRST is life-time limited and has multiple science focus areas.
For fixed instrumental capabilities and observing time, the primary
decisions on survey strategy are the trade between depth and total area, and
the balance between imaging and spectroscopy.
These decisions are driven by two major considerations:
\begin{itemize}
\item \emph{Pecision} -- to maximize the DE science return of WFIRST, taking advantage of synergies with other surveys; and
\item \emph{Accuracy} -- tight control of systematic uncertainties, to ensure correct DE measurements.
\end{itemize}
The combined expertise of our team in WL and GRS enables us to rapidly evaluate
these trades.

\paragraph{HLIS} The statistical power of a WL survey scales with the
total number of galaxy shape measurements, the product of the survey area and the effective surface density.
The SDT2015 report adopts an HLS imaging exposure time that yields roughly equal
contributions from read noise and sky noise in the most sensitive
filters, which approximately maximizes the total number of shape
measurements. This is a compromise between minimizing overheads and
read noise (which favors a ``deep'' mode), versus the shallow number
counts of resolved WL sources (shallower than $N\propto F^{-2}$, which
favors a ``wide'' mode). We re-examined the depth vs.\ area trade using
higher-fidelity tools (e.g., incorporation of shape measurement noise from realistic simulations, and updated
detector properties) and propagating the trades all the way to cosmological parameters
(using \CoLi).

\paragraph{HLSS} The depth vs.\ area trade for the GRS is driven by two
competing factors: for deep surveys, where the number density of galaxies $n$ is
large ($nP\gg 1$, where $P$ is the power spectrum at a given scale), the
information per unit area saturates; whereas for wide surveys, overheads reduce
inefficiency and galaxy shot noise inflates the statistical errors. In the SDT15
survey design, the GRS covers the same area as the HLS imaging survey ($\approx
2,200\deg^2$), to a $7\sigma$ limiting line flux of $\sim 10^{-16}
\erg\cm^{-2}\,{\rm s}^{-1}$ over most of the grism bandpass. This yields
approximately the largest number of galaxy redshifts for a fixed total observing
time. SDT15 found that doubling the survey area at fixed observing time (even
without additional imaging) {\it reduces} the precision of BAO measurements
because of the rapid increase in galaxy shot noise with decreasing spectroscopic
depth. However, this conclusion is sensitive to the luminosity function of
H$\alpha$ emitters at $z=1-2$, which remains uncertain
\cite{Mehta:2015,Pozzetti:2016}; Co-Is Teplitz and Wang lead our efforts to
reduce this uncertainty and feed the results into optimization of the GRS
(\S~\ref{sec:LF}).

\subsection{Snapshot of the HLS observing plan}
\label{ss:snapshot}

\begin{summaryii}
Our team provided a ``snapshot'' of the HLS observing sequence to the full FSWG
on April 19, 2017. This is by no means a final or even optimized version of the
HLS, but is a work in progress as a result of trades in Phase A, as well as the
recent decision to reduce the primary mission to 5 years.
\end{summaryii}

Major updates relative to the SDT plan have included:
\begin{itemize}
\item A Lissajous orbit around L2. This is presently a place-holder, as the exact orbit has not yet been selected (and would depend on the launch date), but it gives a possible sampling of Sun, Earth, and Moon constraints.
\item A rotated WFI (by 90$^\circ$ relative to the Cycle 6 design).
\item Recommended slew and settle times provided by the Project.
\item Faster detector readout (200 kHz instead of 100 kHz).
\item Changes to the exposure time and dithering strategy to accommodate a
5-year baseline mission (as is to be presented to the WIETR). Specifically, we
reduced exposure time to 140.2 s (imaging) and 297.0 s (spectroscopy); and
changed the dither pattern in J band. (We are working on checking this strategy
with image simulations, if it causes a problem we may have to revert.)
\item Implemented bright star avoidance (observations are skipped if there is an $H_{\rm AB}\le 3$ star within 6 arcmin of any SCA).
\end{itemize}
Known current issues with the snapshot plan include:
\begin{itemize}
\item
The SN and coronagraph programs in the code haven't been updated since the SDT
(except to cut the mission time by a factor of 5/6), even though they will
likely change significantly. As in the SDT report, the coronagraph has blocks of
time reserved. This will evolve in order to align the HLS plan with the other
groups, as well as any changes to the scheduling architecture that we are
directed to implement (e.g.\ block scheduling).
\item
We have begun putting the deep fields into this document, but right now they are
(i) not fully specified, (ii) the tiling is not optimized, and (iii) some roll
angles don't align with WFIRST constraints (hence didn't schedule). These issues
will be solved in the next snapshot.
\end{itemize}

We note that \emph{no policy decisions should be inferred from this sequence},
as these will come from a higher level.

The survey bounding box is 2,097 deg$^2$. The area covered with $\ge 3$ exposures
in every filter and the grism, including edge effects and holes around the
bright stars, is 1,947 deg$^2$. The time required for this version of the HLS is
394 days (imaging) + 215 days (spectroscopy).

\begin{figure}
\includegraphics[height=8in]{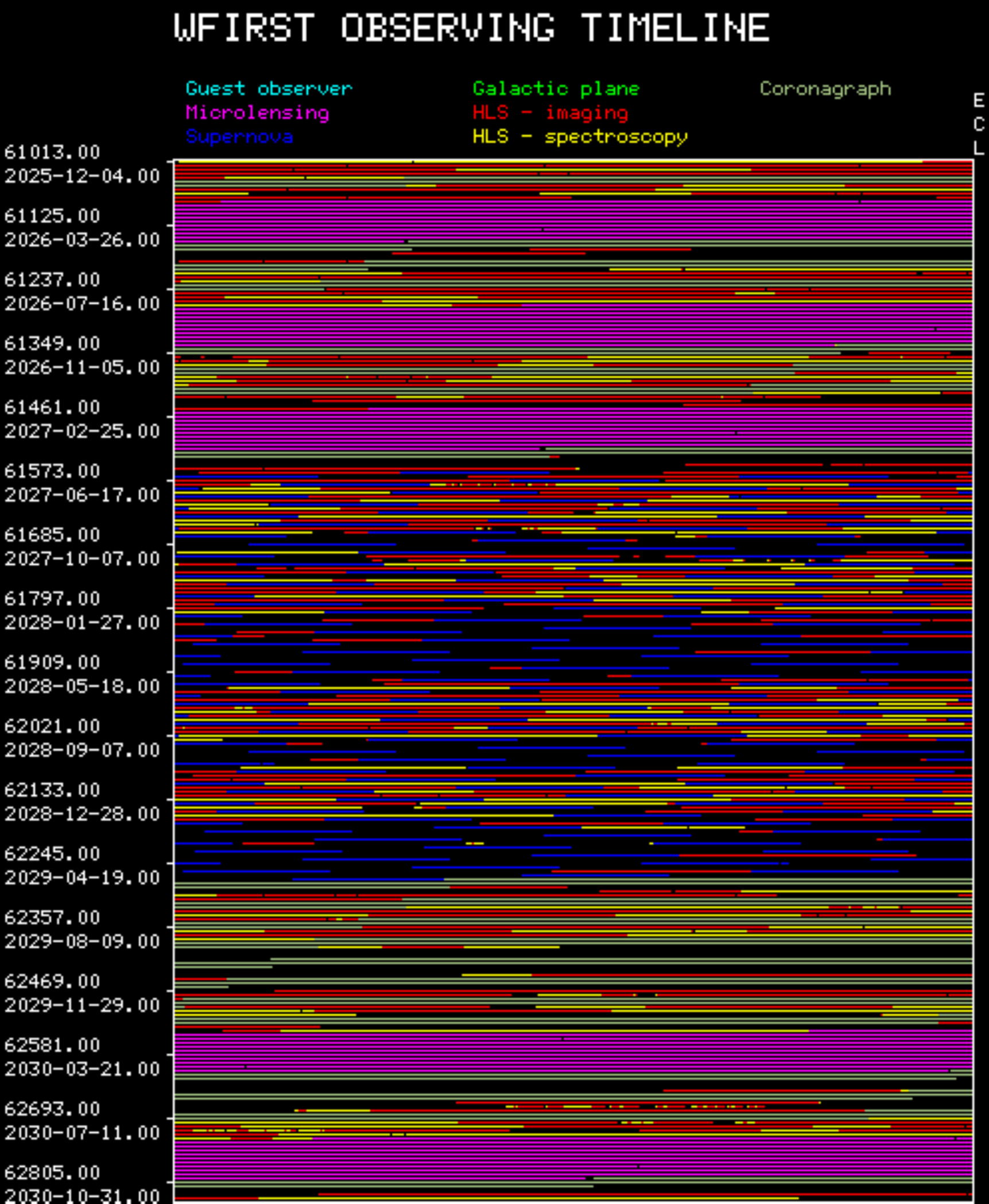}
\caption{\label{fig:observing_chart}Observing timeline. Each row represents 7 days of observations, and is color-coded according to the observing program. Note the microlensing seasons (magenta), supernova survey (blue: $\sim$5-day cadence), and HLS (red+yellow). Blank areas are not allocated. Labels on the left-hand side are shown every 16 weeks.}
\end{figure}

\begin{figure}
\includegraphics[width=5in]{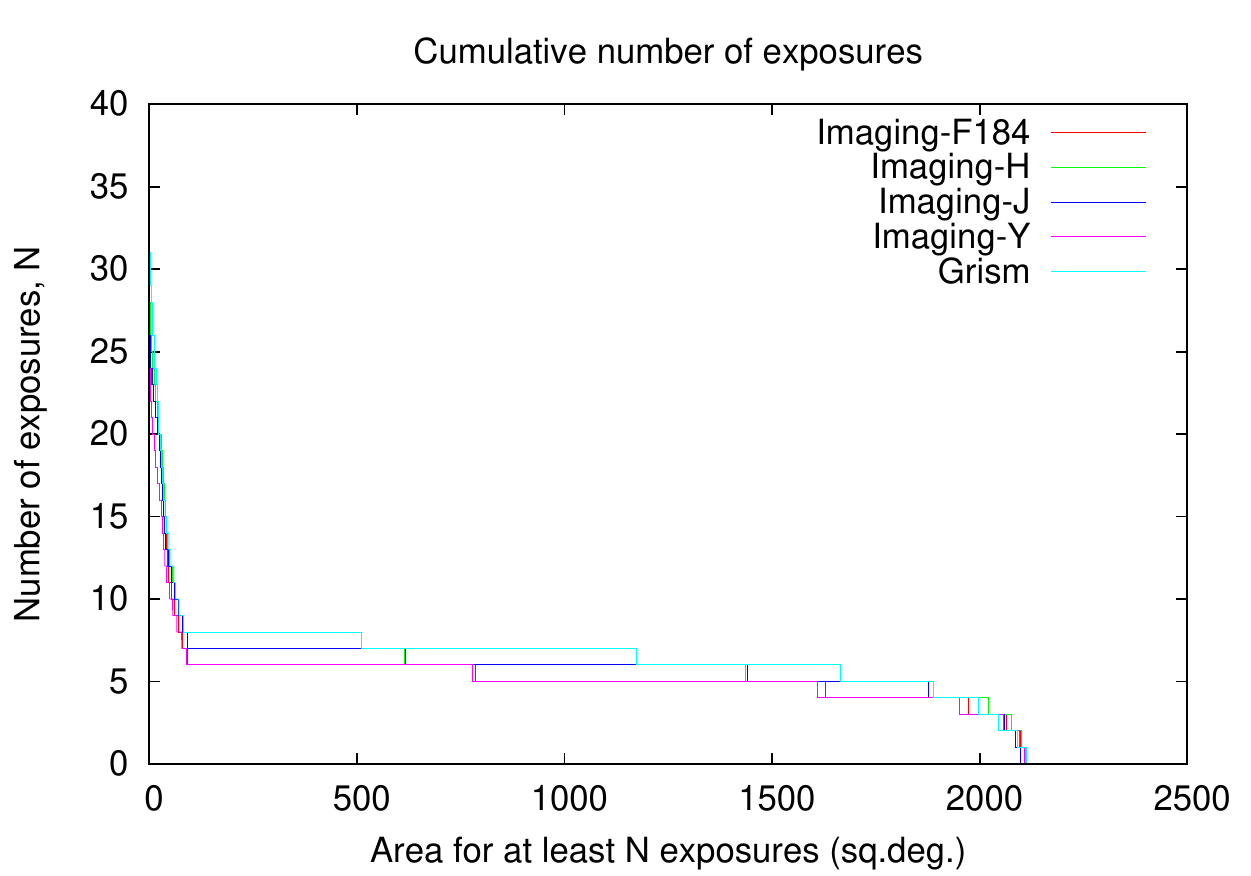}
\caption{\label{fig:hls_depth}The cumulative distribution of HLS exposure depths above a certain area. The pile-up with many exposures at small area is the result of the deep fields.}
\end{figure}

\begin{figure}
\includegraphics[width=5in]{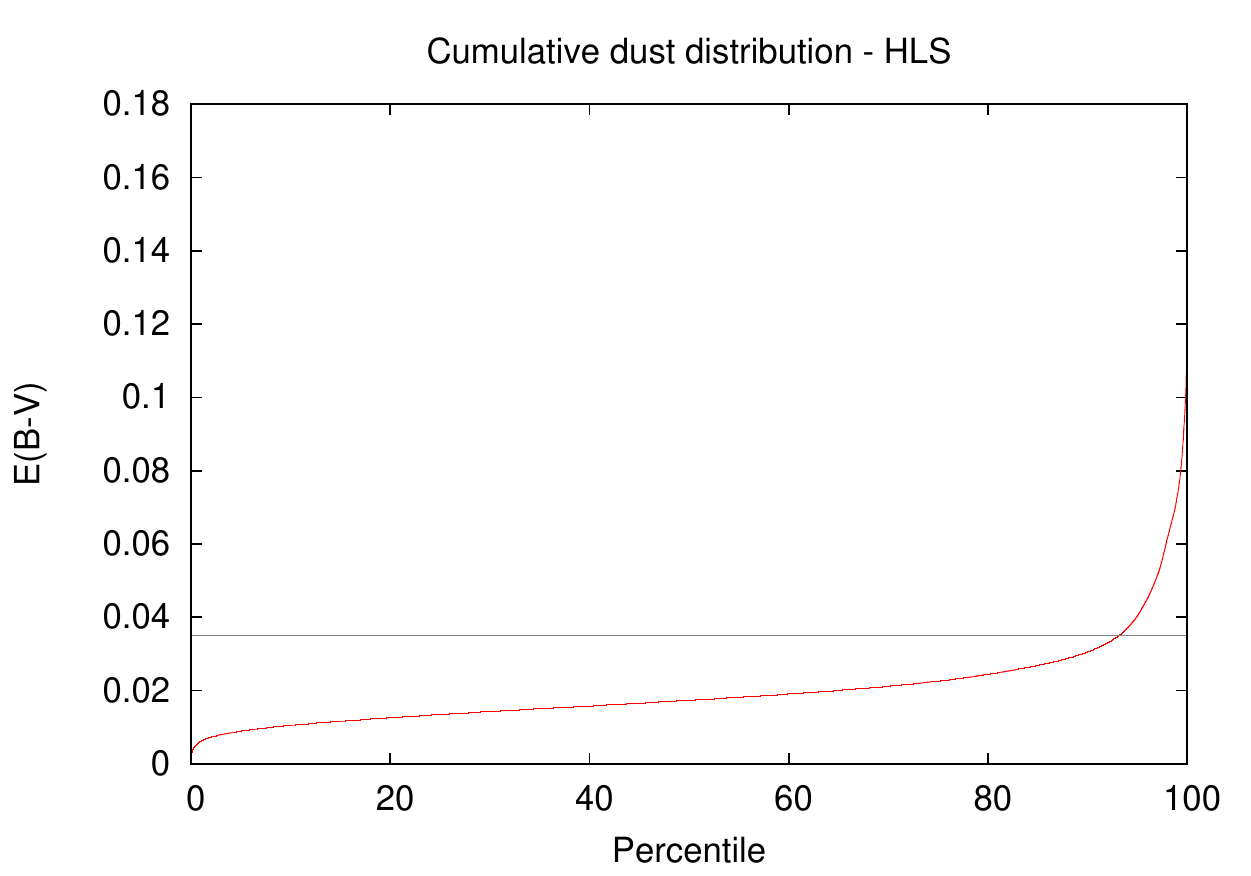}
\caption{\label{fig:hls_dust}The cumulative distribution of Galactic dust in the HLS.}
\end{figure}

\begin{figure}
\includegraphics[width=5in]{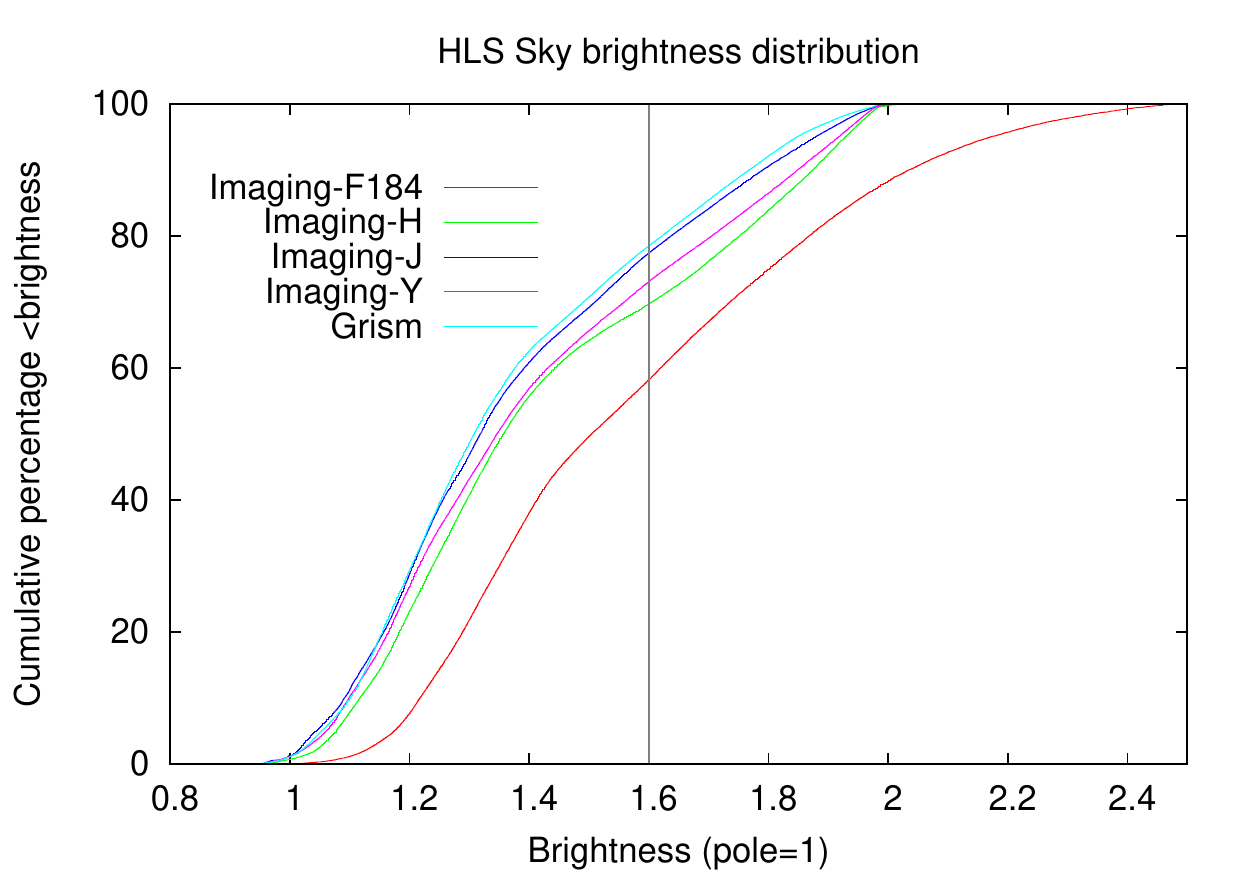}
\caption{\label{fig:hls_bright}The cumulative distribution of zodiacal light in the HLS.}
\end{figure}

\begin{figure}
\includegraphics[width=6in]{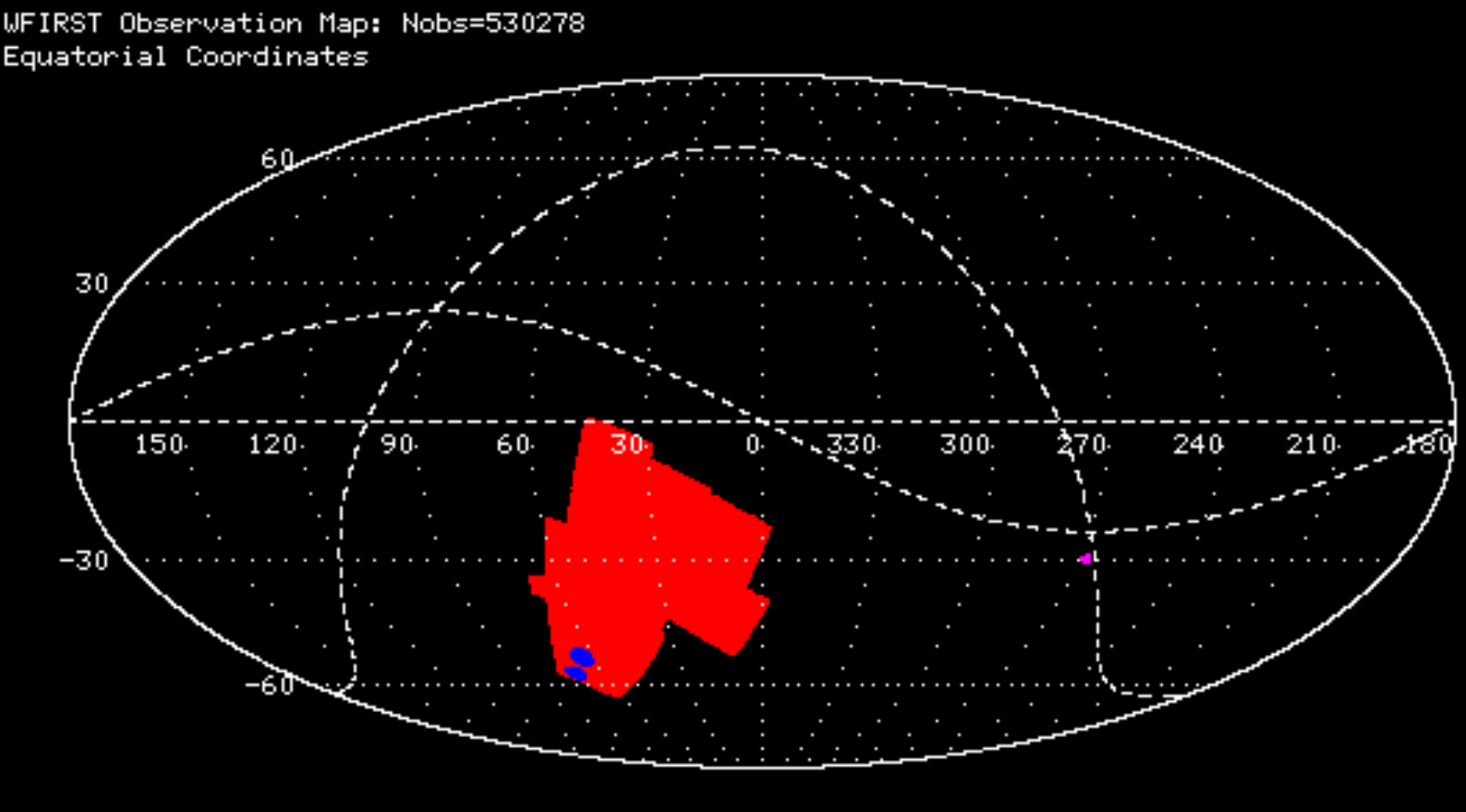}
\includegraphics[width=6in]{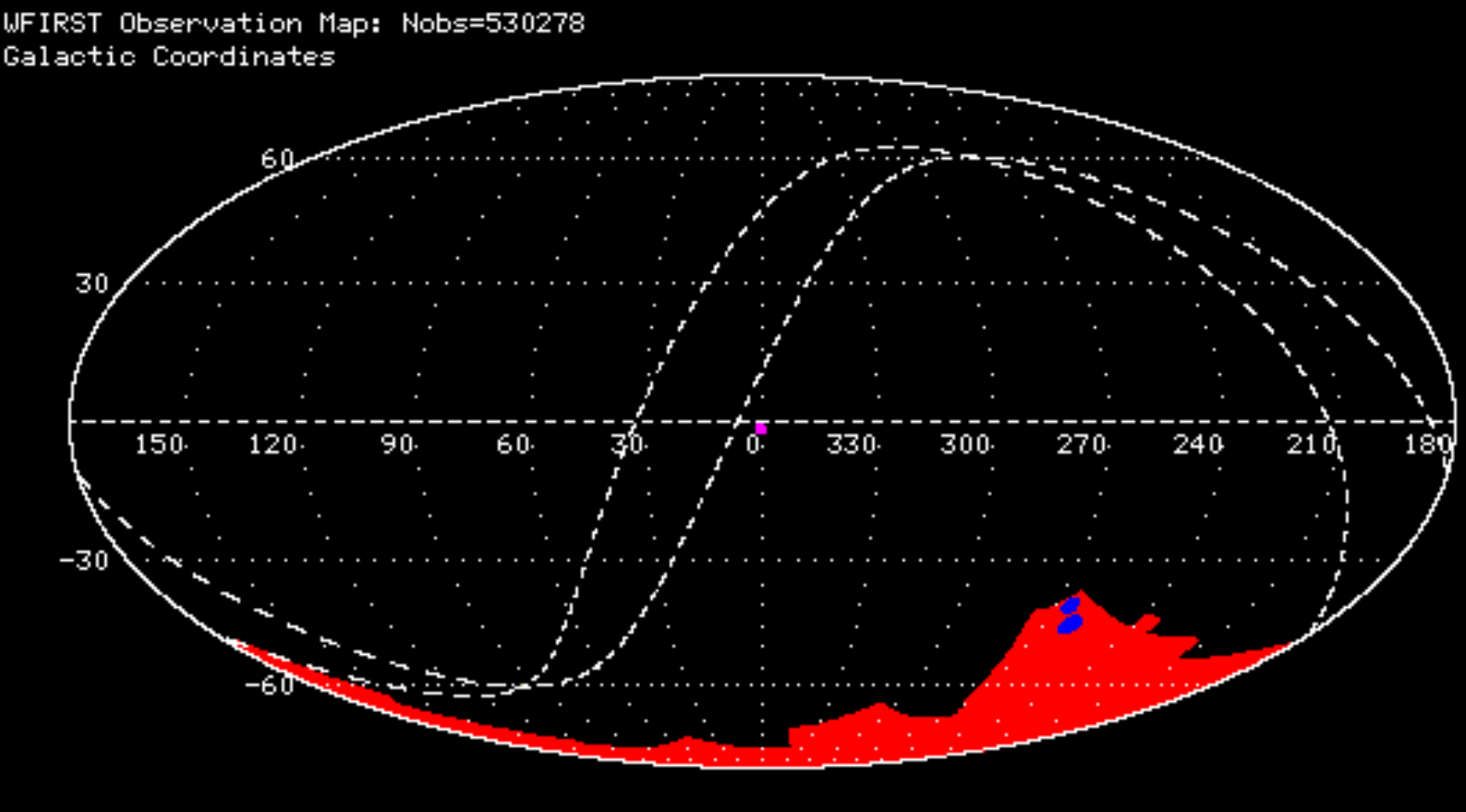}
\caption{\label{fig:footprint}The footprint of the HLS (red) in Equatorial (top) and Galactic (bottom) coordinates.}
\end{figure}

The scheduling tools output a set of charts, included in this package:
\begin{itemize}
\item
Fig.~\ref{fig:observing_chart}: Graphical display of the 5-year observing sequence.
\item
Fig.~\ref{fig:hls_depth}: HLS distribution of number of exposures in each filter.
\item
Fig.~\ref{fig:hls_dust}: HLS distribution of dust column [$E(B-V)$ in magnitudes]. Cosmological forecasts are based on a dust column of $E(B-V)=0.035$ mag.
\item
Fig.~\ref{fig:hls_bright}: HLS distribution of zodiacal light (normalized to 1 at the ecliptic poles averaged over the year). Cosmological forecasts are based on a zodiacal brightness of 1.60 (except for 1.75 in the F184 filter, which is the least sensitive to zodiacal light and therefore was scheduled at inferior times of year).
\item
Fig.~\ref{fig:footprint}: The footprint of the HLS on the sky. This is an area of ongoing optimization, as we consider the needs of the deep fields, overlap with LSST, and the fraction of the survey footprint accessible from Northern observatories such as Subaru.
\end{itemize}

\subsection{Further optimizations}

\begin{summaryii}
Our team plans to study further optimizations to the HLS -- including more
drastic changes such as multi-tiered surveys, or a significant re-balancing of
area vs.\ depth -- in Phase B. However, in preparation for SRR/MDR, our main
focus has been on demonstrating at least one survey configuration that meets
requirements, and the construction of tools that link the observing strategy to
calibration studies (\S\ref{sec:wl_calibration}) and image simulations (\S\ref{sec:hlis_image_sim}).
\end{summaryii}

The optimization of the WFIRST HLS will be tightly linked to operations simulations,
which inform the possible range of footprint area and location, depth in each filter (or grism), redundancy,
and temporal distribution of exposures. We propose a highly integrated approach, with the
operations simulations at the core, but with links to pixel-level simulations to assess required
redundancy, cosmological forecasting tools (\CoLi) to assess science reach, and comparison to the
observing regions of other surveys and telescopes to maximize synergies and meet requirements for
deep fields and photo-$z$ calibration.

\section{Community Engagement and External Data-sets (D10, D11, D12)}
\label{sec:engagement}
%
%

\begin{summary}
The coming decade will be an exciting time for cosmology. Before WFIRST launch,
major cosmological imaging surveys (KiDS, HSC, DES) and the DESI and PFS
spectroscopic surveys will significantly advance our current understanding.
WFIRST, Euclid and LSST  will then go further and survey the sky at optical and
infrared wavelengths, the  James Webb Space Telescope (JWST) and the Extremely
Large Telescopes (ELTs) will make very deep maps of the sky; eROSITA will survey
the X-ray sky; CMB-S4 will make a deep map of the millimeter sky; and the
Canadian Hydrogen Intensity Mapping Experiment (CHIME) and other radio surveys
will map the large-scale distribution of H$\,${\sc i}. One goal of our SIT is to
determine the analysis infrastructure and observations needed to achieve the
full potential of WFIRST in combination with these  surveys. This is best done
through a broad community effort that brings together scientists from these
complementary projects. For this purpose, we organized a first community
workshop focused on enabling the cosmological scientific synergies between
WFIRST and LSST.
\end{summary}

Our SIT naturally brings close connections to most of the
major current or planned cosmological experiments that will provide
the context for the WFIRST dark energy program. This includes the WMAP and Planck CMB missions,
the Sloan Digital Sky Survey (SDSS), the Baryon Oscillation Spectroscopic Survey (BOSS), the Dark Energy Survey (DES),
the Subaru Hyper Suprime-Cam (HSC) and Prime Focus Spectrograph (PFS)
projects, the Dark Energy Spectroscopic Instrument (DESI), the
Euclid mission and the Large Synoptic Survey Telescope (LSST) Dark
Energy Science Collaboration (DESC).




We also started to actively engage the community to identify and pursue the key areas
where WFIRST and the concurrent projects will provide new opportunities to
mitigate systematics and enhance the combined cosmological science return. We
started a series of open community workshops to incorporate the interplay
between major planned surveys and WFIRST into the WFIRST strategy, to identify:
(i) pre-launch observations, (ii) how these external data sets affect the WFIRST
observing strategy (e.g., deep fields) and the instrument, and (iii) the
software needed (to be built post-CDR) for combining these data sets.

\begin{figure}
\includegraphics[width=\textwidth,angle=0]{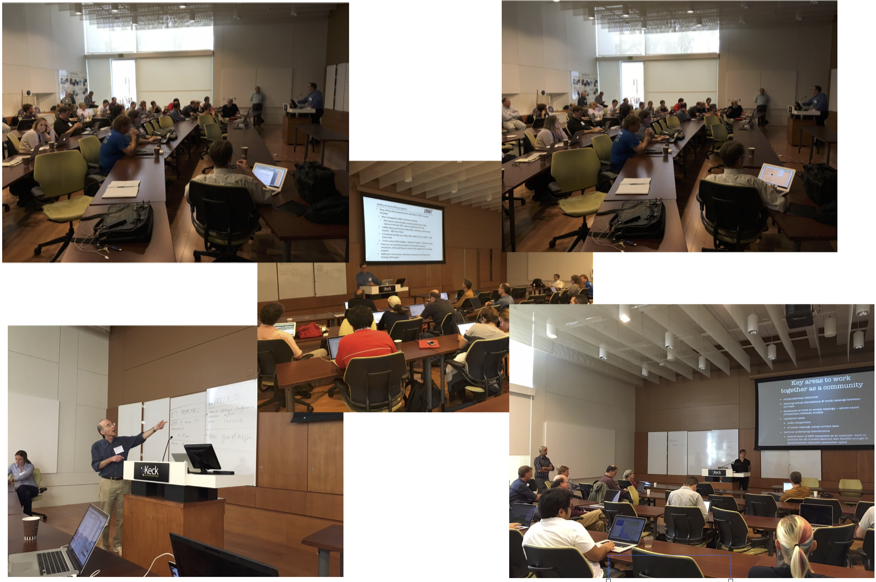}
\caption{\label{fig:pic_workshop}Pictures taken during the Pasadena WFIRST SIT - LSST DESC workshop on September 2016.}
\end{figure}

\paragraph*{First Community Workshop} The first of this meeting happened in
Pasadena in September 2016. It was focused on the synergies between the WFIRST
HLS Cosmology SIT and LSST Dark Energy Science Collaboration (DESC), both at the
science level but also at the implementation level. The meeting was co-organized
by Rachel Bean, Olivier Dor\'e, Steve Kahn and Jason Rhodes and was attended by
about 60 scientists from the two collaborations. The slides are available
\href{https://conference.ipac.caltech.edu/wfirst_lsst/}{here} and pictures can
be seen in Figure~\ref{fig:pic_workshop}. The conclusion of the lively and
productive workshop will be summarized in a white paper to be written shortly.

\paragraph*{Premise} The workshop was built on the premise that when considering
joint science return from LSST and WFIRST, \emph{the whole is greater than the
sum of the parts}, as already articulated in \citet{Jain:2015cpa}. In the
conclusion of the Jain et al. community white paper, it was articulated that the
scientific opportunity offered by the combination of data from LSST and WFIRST
(and Euclid) goes well beyond the science enabled by any one of the data sets
alone. The range in wavelength, angular resolution and redshift coverage that
these missions jointly span is remarkable. With major investments in LSST and
WFIRST (and partnership with ESA in Euclid), the US has an outstanding
scientific opportunity to carry out a combined analysis of these data sets. It
is imperative for us to seize it and, together with our European colleagues,
prepare for the defining cosmological pursuit of the 21st century.

As illustrated already in \S~\ref{sec:forecast}, the coming decade will be the
era of multi-probes/multi-survey science. The richer insights will come from
combining multiple probes (WL, GRS, GC) \emph{reliably}. If done properly,
multiple survey joint analysis will enable new calibration schemes (photo-$z$,
intrinsic alignment models, etc.) and new control of systematics (PSF effects,
contaminant control such as stars or interlopers).

\paragraph*{Joint Analysis} The main argument for conducting a single,
high-quality reference co-analysis exercise and carefully documenting the
results is the complexity and subtlety of systematics that define this
co-analysis. Falling back on many small efforts by different teams in selected
fields and for narrow goals will be inefficient, leading to significant
duplication of effort.  For much of the science, we will need to combine the
photometry across multiple wavelengths with varying spectral and spatial
resolution – a technical challenge. The joint analysis can be carried out in
ways that have different computational demands. The most technically demanding
joint analysis is to work with pixel level data of the entire area of overlap
between the surveys. Many of the goals of a joint analysis require such a
pixel-level analysis. If pixel-level joint analysis is not feasible,
catalog-level analysis can still be beneficial, say to obtain calibrations of
the lensing shear or the redshift distribution of galaxies. Hybrid efforts are
also potentially useful, for example using catalog level information from space
for deblending LSST galaxies, or using only a mutually agreed subset of the data
for calibration purposes. However the full benefits of jointly analyzing any two
of the surveys can be reaped only through pixel-level analysis \citep{Jain:2015cpa}.

\paragraph*{Possible Implementation} In the workshop, possible algorithmic
implementation of a joint analysis pipeline were discussed independently by
Peter Melchior and Michael Schneider. \citet{Melchior:2016asy} motivated the
need for complex galaxy morphology models and developed the relevant statistical
framework. For them, complex models become the norm. They are more flexible and properly
behaved. WFIRST benefits from LSST through color-morphology priors. LSST
benefits from sharp likelihood peaks in WFIRST bands. LSST benefits from WFIRST
through superior resolution. \citet{Schneider:2014rha} have developed a probabilistic
image reduction pipeline (forward modeling) motivated by challenges in
multi-epoch/multi-telescope combinations. They demonstrated improved shear
measurements for LSST using the HST Frontier field as a proxy for WFIRST.

\paragraph*{Enabling Synergies} It was recognized in the context of photo-$z$ for example, that
cooperation already exist to maximize utility and minimize duplication in
spectroscopic calibration samples. Capak and collaborators are building a public
calibration catalog for WFIRST and LSST \citep{Masters2017}. Also, relevant information for deep
surveys synchronization, is also discussed in this context and others (SNe,
shape calibration, grism calibration, guest observations). The detailed requirements for LSST + WFIRST photo-$z$ performance are however still to be articulated.

\paragraph*{Computing Needs} Large computing efforts are under way in LSST
(DESC) with multiple planned data challenges. Resources available for the scale
of simulations required (with possible exception of hydrodynamical simulations)
are secured. However, ressources needed for sharing simulations (e.g. Millennium
DB) and mock catalogs are required. They have historically greatly  expanded the
use of simulations in a broad range of applications. A 100 TB to 1 PB scale
infrastructure simply does not exist and data transfer will be a challenge (10
days for 100 TB at 1 Gb). There is great opportunity for future collaborations
(e.g., joint mock catalogs) and investments. Opportunity for collaborations and
investments. As we will discuss in \S\ref{sec:tacs}, members of our team are leading and participating in a Tri-Agency Cosmological Simulations (TACS) dedicated to this issue.

\paragraph*{Ressources} As already discussed in \citet{Jain:2015cpa}, the resources required to achieve this additional
science are outside of what is currently budgeted for LSST by NSF and DOE, and
for WFIRST (or Euclid) by NASA. Funding for this science would most naturally
emerge from coordination among all agencies involved, and would be closely
orchestrated scientifically and programmatically to optimize science returns. A
possible model would be to identify members of the science teams of each project
who would work together on the joint analysis. The analysis team would ideally
be coupled with an experienced science center acting as a focal point for the
implementation, and simultaneously preparing the public release and documentation for broadest access by the community.

\paragraph*{Future Workshop} The second in our series of SIT community workshop
will happen in the fall 2017 in Pasadena, synchronized with a FSWG, and will
identify (and discuss the enabling of) scientific synergies between the HLS and
other major surveys across all wavelengths.


\section{Other Science Investigation Team Contributions to WFIRST Mission (D11)}
\label{sec:other_contributions}

\begin{summary}
Our SIT engaged in multiple activities supporting the WFIRST mission and the Project Office not discussed above. We summarize them here:
\begin{enumerate}
\item Our SIT actively participated to two workshops dedicated to the WFI;
\item Our SIT contributed to evaluate the WFIRST Supernova Program;
\item Our SIT partially supported 15 post-doctoral researchers and one graduate student;
\item Our SIT published 13 scientific papers supporting our studies or extending the scientific case of WFIRST;
\item SIT team members are co-leading and participating in the Tri-agency Cosmological Simulation Task Force (TACS);
\item SIT team members are co-leading large observational efforts supporting WFIRST science goals;
\item Our team delivered to the community multiple high-value products and softwares.
\end{enumerate}
\end{summary}

\subsection{\emph{Princeton Meetings} Participation}

In addition to the active participation by the SIT leads in the 4 FSWG meetings that happened this year, team members participated in two focused so-called \emph{Princeton Meetings} organized by Prof. Spergel at Princeton University and  the newly established Center for Computing Astrophysics (CCA) at the Flatiron Institute in  New York, NY. These two meetings were focused respectively on planning the data processing need of the WFI given our collective experience with large astronomical dataset data processing, and ancillary science enabled by the HLS. Team members Lupton, Mandelbaum, Samushia, and von der Linden represented our SIT at these meetings and presented their thoughts on these topics.

\subsection{Contributions to Evaluation of the WFIRST Supernova Program}

In addition to our work on the HLS cosmology surveys,
team members Olivier Dor\'e, Chris Hirata, David Spergel,
and David Weinberg all contributed to evaluating strategies and
requirements for the WFIRST supernova program, providing a
sounding board for the two supernova SITs and synthesizing
information for project management.  Some of these contributions
took place in sessions of the WFIRST FSWG meetings and some
 in telecons and email exchanges with members of the
supernova teams.  Weinberg wrote an extensive referee report
for the Hounsell et al.\ paper (from the supernova SIT led by
Ryan Foley) on WFIRST supernova strategies.  Most importantly,
all four investigators participated in the ``supernova summit''
held at KIPAC in March 2017, reading background material from
the teams, participating in a day of presentations and discussions,
and writing a report for project management and the supernova SITs.

\subsection{Support of Postdoctoral Researchers and Graduate Students}

Our team has been very proactive at leveraging our collective involvement in other on-going large scale observational efforts such as the Dark Energy Survey (DES), and the SUBARU Hyper-Suprime Cam (HSC) survey, in addition to the upcoming ESA/NASA Euclid mission and the Large Synoptic Survey Telescope Dark Energy Science Collaboration (LSST DESC) where we have been able to create joint appointments with our WFIRST SIT and have assembled a team of very strong postdoctoral researchers. In particular, the following researchers joined our team and are partially supported by out SIT:
\begin{itemize}
\item Ivano Baronchelli (Caltech/IPAC), working with Harry Teplitz on updating measurements of the H$\alpha$ luminosity function using HST data;
\item Ami Choi (OSU), working with Chris Hirata and David Weinberg on image simulations for the WL analysis;
\item Shoubaneh Hemmati (IPAC/Caltech), working with Peter Capak on defining the photometric redshift requirements of the WFIRST WL investigation;
\item Albert Izard (JPL), working with Alina Kiessling on cosmological simulations and the requirements driven by the need to accurately compute covariance matrices;
\item Niall MacCrann (OSU), working with Chris Hirata and David Weinberg on image simulations for the WL analysis;
\item Elena Massara (LBL), working with Shirley Ho on generating light cone simulations for the GRS survey;
\item Alex Merson (Caltech/IPAC), working with Yun Wang and Andrew Benson on generating light cone simulations for the GRS survey;
\item Hironao Miyatake (JPL/Caltech), working with Jason Rhodes and Tim Eifler on including modified gravity and observational systematic effects in the cosmological parameter likelihood;
\item Andres Plazas Malagon (JPL/Caltech), working with Jason Rhodes and Charles Shapiro on computing requirements on detector imperfections driven by the WL survey;
\item Melanie Simet (UCR/JPL), working with Alina Kiessling and Jason Rhodes on the WL analysis;
\item Michael Troxel (OSU), working with Chris Hirata and David Weinberg on image simulations for the WL analysis;
\item Ying Zu (OSU), working with Chris Hirata and David Weinberg on simulations for the galaxy cluster investigation;
\item Chen He Heinrich (JPL), working with Olivier Dor\'e and Tim Eifler (starting in fall 2017);
\item Alice Pisani (Princeton), working with David Spergel on void statistics for the GRS survey (starting in fall 2017);
\item Hao-Yi "Heidi" Wu (OSU), working with Chris Hirata and David Weinberg on simulations for the galaxy cluster investigation (starting in fall 2017).
\end{itemize}

In addition, Arun Kannawadi, a graduate student at CMU supervised by Rachel Mandelbaum has been working on image simulation and shape measurement analysis. He is partially supported by our SIT.

\subsection{Relevant Scientific Publications by Team Members}

In addition to our work supporting the Project Office, we published our results in scientific journals and made them available on the arXiv. The following 13 scientific papers were published by our team members and were motivated by our studies:

\begin{enumerate}
\item \citet{2017MNRAS.467..928G}, ``Information Content of the Angular Multipoles of Redshift-Space Galaxy Bispectrum";
\item \citet{2016MNRAS.463.2708P}, ``Optimal weights for measuring redshift space distortions in multitracer galaxy catalogues";
\item \citet{2016MNRAS.463..467K}, ``Unbiased contaminant removal for 3D galaxy power spectrum measurements";
\item \citet{2016MNRAS.457..993P}, ``Estimating the power spectrum covariance matrix with fewer mock samples";
\item \citet{2016PASP..128j4001P}, ``The Effect of Detector Nonlinearity on WFIRST PSF Profiles for Weak Gravitational Lensing Measurements";
\item \citet{2017JInst..12C4009P}, "Nonlinearity and pixel shifting effects in HXRG infrared detectors";
\item \citet{2015MNRAS.449.2128K}, ``Can we use weak lensing to measure total mass profiles of galaxies on 20 kpc scales?";
\item \citet{Masters2017}, ``The Complete Calibration of the Color-Redshift Relation (C3R2) Survey: Survey Overview and Data Release 1";
\item \citet{Schaan:2016ois}, ``Looking through the same lens: shear calibration for LSST, Euclid \& WFIRST with stage 4 CMB lensing";
\item \citet{Chisari:2016xki}, ``Multitracing Anisotropic Non-Gaussianity with Galaxy Shapes";
\item \citet{Merson2018}, ``Predicting H$\alpha$ emission line galaxy counts for future galaxy redshift surveys", currently under review by MNRAS.
\item \citet{2017JInst..12C4009P}, ``Nonlinearity and pixel shifting effects in HXRG infrared detectors'', Journal of Instrumentation;
\item \citet{2017arXiv171206642P}, ``Laboratory measurement of the brighter-fatter effect in an H2RG infrared detector'', Accepted for publication in Publications of the Astronomical Society of the Pacific.

\end{enumerate}

\subsection{Participation to the Tri-agency  Cosmological Simulation Task Force (TACS)}
\label{sec:tacs}
The computational resources required to produce all of the cosmological simulations required for WFIRST are not yet well defined but they are known to be significant. WFIRST shares common goals with other upcoming cosmological surveys including LSST, Euclid, and DESI, so it makes sense to try to coordinate efforts between these projects. Team members have taken leadership positions in articulating and leading this effort.

Team members Kiessling and Heitmann have been asked to co-chair a Tri-Agency Cosmological Simulations (TACS) task force at the request of the Project leads from WFIRST, LSST, and Euclid. The co-chairs have formulated a charge for TACS that has been approved by the Tri-Agency (NASA, DOE, NSF), Tri-Project (WFIRST, LSST, Euclid) group (TAG). TAG has representation from each of the Agencies and Projects and is responsible for the coordination of joint data processing and cosmological simulations for the three Projects. JPL WFIRST Project Scientist Rhodes represents the WFIRST and Euclid Projects on the TAG, Program Scientist Benford represents NASA for WFIRST and WFI Adjutant (and SIT team member) Spergel represents the WFIRST FSWG. TACS consists of the two chairs, 9 task force members, and an advisory board of 7 people, with representation from WFIRST on the task force, advisory board, and from both of the co-chairs. Team member Mandelbaum is a member of the advisory board and team members Benson, Eifler, and Ho are task force members. EXPO SIT lead Robertson is also a member of the Advisory Board.

The primary goal of TACS is to investigate areas for coordination between WFIRST, LSST, Euclid, and DESI of supercomputing resources, supercomputing infrastructure, cosmological simulations, synthetic sky generation, systematics investigation, and workforce personnel. TACS will report their recommendations directly to TAG and the reports will be made public for transparency. The responsibility for implementing any recommendations or negotiating MOUs lies with the TAG. Work is currently underway in TACS and is anticipated to extend into FY18.

\subsection{Participation and Leadership Large Observational Program Supporting WFIRST Science Goals}

The WFIRST mission will provide an unprecedented survey (in depth and area) of the extragalactic sky at optical and near infrared wavelengths (0.5-2 $\mu$m).  However, data in the 3-5 $\mu$m wavelength range is an essential complement to measure stellar masses at $z>$3, probe the earliest sites of reionization, find the earliest quasars and structures, and see heavily obscured regions of galaxies and AGN.  Yet, besides Spitzer, no current or planned future mission is able to conduct deep yet wide area surveys at these wavelengths since JWST will be limited to a few square degrees of imaging over its lifetime. Co-I Capak is leading the Spitzer Legacy Survey (SLS) that will provide 20 square degrees of deep imaging. The Spitzer Legacy Survey enables new science and will improve the cosmological constraints provided by WFIRST mission.  The fields were chosen to be some of the most observable for missions at L2 (WFIRST, JWST, and Euclid), the North Ecliptic Pole (NEP) and the Chandra Deep Field South (CDFS).  The CDFS in particular is also the target of deep LSST data which will complement WFIRST for photometric redshifts.

Co-I Capak is also co-leading the Complete Calibration of the Color-Redshift Relation (C3R2) survey. As discussed in \S \ref{sec:wl_photoz}, a key goal of WFIRST is to measure the growth of structure with cosmic time from weak lensing analysis over large regions of the sky. Weak lensing cosmology will be challenging: in addition to highly accurate galaxy shape measurements, statistically robust and accurate photometric redshift (photo-$z$) estimates for billions of faint galaxies will be needed in order to reconstruct the three-dimensional matter distribution. The C3R2 survey is designed specifically to calibrate the empirical galaxy color-redshift relation to the Euclid, LSST and WFIRST. The C3R2 survey is obtaining multiplexed observations with Keck (DEIMOS, LRIS, and MOSFIRE), the Gran Telescopio Canarias (GTC; OSIRIS), and the Very Large Telescope (VLT; FORS2 and KMOS) of a targeted sample of galaxies most important for the redshift calibration. C3R2 focuses spectroscopic efforts on under-sampled regions of galaxy color space identified in previous work in order to minimize the number of spectroscopic redshifts needed to map the color-redshift relation to the required accuracy. Initial results include the 1283 high confidence redshifts obtained in the 2016A semester and released as Data Release 1 \citep{Masters2017}.

\subsection{Community Deliverables}

Throughout our studies, our SIT aims to release codes, products, and simulated data sets, with the goals of building awareness of and broad support for the WFIRST dark energy program and inspiring the community to develop methods and carry out investigations that will maximize the cosmological return from WFIRST. This year, we released the following products available mostly from our SIT team webpage \href{http://www.wfirst-hls-cosmology.org/products/}{[Link]}.

\subsubsection{\CoLi\ Monte-Carlo Markov Chains}

Cosmological parameter MCMC chains corresponding to forecasts for the current survey of the WFIRST High Latitude Survey, combining weak-gravitational lensing (WL), cluster counts (CC), and redshift space distortions (GRS). These chains were computed using the CosmoLike software \citep{Krause2016}.

\paragraph{Multi-probe cosmology forecasts (including SN from both SIT SN teams) with realistic systematic budgets} The tightest constraints on cosmological models including cosmic acceleration, modified laws of gravity, and neutrino physics will come from a joint analysis of multiple cosmological probes. We forecasted constraints using “traditional” single probe analyses for clusters, BAO+RSD, and weak lensing as well as a multi-probe analysis that utilizes these and several other observables that can be extracted from the data: galaxy-galaxy lensing, photometric galaxy clustering, cluster weak lensing, spectroscopic galaxy clustering, SN from WFIRST (forecasts from David Rubin and Dan Scolnic from the two SIT SNe teams), SN from the existing Joint Lightcurve Analysis, existing Baryon Acoustic Oscillation information from BOSS, and CMB information from Planck. Since multi-probe analyses are highly constraining they impose tight requirements on systematics control. These systematics include uncertainties in the estimation of galaxy shapes and redshifts (photo-$z$, spec-$z$), cluster mass calibration, galaxy bias, and intrinsic alignment. Uncertainties due to baryonic effects (SN and AGN feedback, cooling) are not included. It is critical over the coming years to study these uncertainties and to develop the capability to control these systematics at the level of WFIRST multi-probe analyses.

\paragraph{WFIRST modified gravity studies} While cosmic acceleration models have a relatively established parameterization, this is not true for modified gravity theories. We released and showed our first forecasts of modified gravity scenarios, where deviations from Einstein GR are parameterized through $\mu$ and $\Sigma$ (please see \citet{Joyce:2016vqv} for an introduction and \citet{Simpson:2012ra} for details about this particular parametrization). Constraints on these modified gravity parameters combining weak gravitational lensing, galaxy-galaxy lensing, photometric galaxy clustering joint analysis for the nominal WFIRST survey (2,200 deg.$^2$) and for 2 extended survey scenarios (5,000 deg.$^2$ and 10,000 deg.$^2$, respectively) were obtained. We also studied the difference of this multi-probe case with a Weak Lensing only analysis.

\subsubsection{A WFIRST module has been added to the GalSim package}

GalSim is an open-source software for simulating images of astronomical objects (stars, galaxies) in a variety of ways. The bulk of the calculations are carried out in C++, and the user interface is in python. In addition, the code can operate directly on ``config" files, for those users who prefer not to work in python. The impetus for the software package was a weak lensing community data challenge, called \href{http://great3challenge.info/}{[GREAT3]}.

However, the code has numerous additional capabilities beyond those needed for the challenge, and has been useful for a number of projects that needed to simulate high-fidelity galaxy images with accurate sizes and shears. For details of the GalSim algorithms and code validation, please see \citet{Rowe:2015}.

We have now added a specific module to accurately simulate WFIRST images. The GalSim software package including the WFIRST module is available \href{https://github.com/GalSim-developers/GalSim}{[here]}.The development of this WFIRST module is included in version 1.4 and precedes our SIT. It already include inter-pixel capacitance, persistence, reciprocity and spider pattern. It will be periodically updated by SIT members as the WFIRST hardware and survey parameters are adjusted. We expect the next update including the latest layout of the FPA to be released as part of GalSim v1.5 before the end of August 2017.

\subsubsection{CANDELS Based Mock WFIRST and LSST Catalogs}

In order to accurately simulate WFIRST photometric redshifts critical for the WL investigation, we transformed the CANDELS Catalog to the LSST and WFIRST color system with an LSST cut applied (Peter Capak, Shoubaneh Hemmati). The catalog includes photometry estimates in LSST (u,g,r,i,z), WFIRST(Y,J,H,F184W) and k band (AB magnitudes) as well as photometric redshifts, FWHM of F160W (pixel, 1 pixel = 0.06 arcsec), and F160W AB magnitude from the original CANDELS catalogs. All five CANDELS fields (GOODS-S, GOODS-N, EGS, UDS and COSMOS) are used here which cover $\sim$ 0.2 deg.$^2$ CANDELS photometric catalogs are published in GOODS-S, COSMOS, UDS and EGS and will be published in GOODS-N (therefore not available to the public yet). In cases where the photometry in a WFIRST or LSST filter could not be measured using the neighboring filters in CANDELS due to non detections, the magnitude is set to 99.0 and the limiting magnitude in the closest band is recorded as the error.

This catalog has been released internally to our SIT for testing and validation first. It has been shared with other SITs and is publicly released on our SIT website.

\subsubsection{Interloper Fraction Calculator}

We have also released a code that calculates the fraction of emission line galaxies that have emission lines (other than H$\alpha$ or [OIII]) that appear at the same wavelengths as the H$\alpha$ or [OIII] lines as observed by WFIRST. This could lead to an incorrect redshift estimate for those galaxies, whose population is referred to as the interloper fraction (Wong, Pullen, \& Ho 2016). Our Python-based software applies secondary line identification and photometric cuts to mock galaxy surveys in order to estimate the interloper fraction. We have also provided a module that is specifically designed to predict WFIRST interloper fractions, which is available \href{https://github.com/kazewong/Intercut}{here} on GitHub \citep{Wong:2016eku}.

\vspace{-0.25cm}
\section*{Acknowlegments}
\label{sec:acknowledgments}
\addcontentsline{toc}{section}{Acknowledgments}

We warmly thank our colleagues from the WFIRST Project Office and all the Science Investigation Teams for continuous constructive and stimulating interactions. We acknowledge use of the Annual Review of Astronomy and Astrophysics LaTeX template as a basis for our report. Part of this research was carried out at the Jet Propulsion Laboratory, California Institute of Technology, under a contract with the National Aeronautics and Space Administration. The decision to implement the WFIRST mission will not be finalized until NASA's completion of the National Environmental Policy Act (NEPA) process. This document is being made available for information purposes only.

\newpage

\section*{List of Acronyms and Abbreviations and References}
\addcontentsline{toc}{section}{References and List of Acronyms and Abbreviations}
\label{sec:acronyms}
\vspace{2 cm}

\begin{table*}[ht!]
  \small
  \begin{tabular}{@{}>{\raggedright}p{0.085\textwidth}>{\raggedright}p{0.39\textwidth}>{\raggedright}p{0.085\textwidth}>{\raggedright}p{0.39\textwidth}}

    $\sigma_m$ & -- rms amplitude of matter fluctuations
    & LSS & -- large scale structure \tabularnewline
    $\Omega_m$ & -- dimensionless density of the Universe
    & LSST & -- Large Synoptic Survey Telescope \tabularnewline
    $a$ \\ ACT & -- scale-factor of the Universe \\ -- Atacama Cosmology
    Telescope
    & NICMOS & -- HST Near Infrared Camera and Multi-Object Spectrometer
    \tabularnewline
    AdvACT & -- Advanced ACT
    & NIR & -- near-infrared \tabularnewline
    AFTA & -- Astrophysics Focused Telescope Asset
    & NRA & -- NASA Research Announcement \tabularnewline
    BAO & -- baryon acoustic oscillations
    & NRC & -- National Research Council \tabularnewline
    BOSS & -- Baryon Oscillation Spectroscopic Survey
    & NWNH \\ $P_m$ & -- New Worlds, New Horizons \\ -- matter power spectrum
    \tabularnewline
    CDR & -- critical design review
    & PFS & -- Subaru Prime Focus Spectrograph \tabularnewline
    CFHT & -- Canada-France-Hawaii Telescope
    & photo-$z$ & -- photometric redshift \tabularnewline
    CGL & -- cluster-galaxy lensing
    & PSF & -- point spread function \tabularnewline
    CHIME & -- Canadian Hydrogen Intensity Mapping Experiment
    & RSD \\ SAM & -- redshift-space distortions \\
    -- semi-analytic galaxy formation models \tabularnewline
    CL & -- galaxy clusters / cluster growth
    & SDSS & -- Sloan Digital Sky Survey \tabularnewline
    CMB & -- cosmic microwave background
    & SDT & -- Science Definition Team \tabularnewline
    CMB-S4 & -- CMB stage 4 experiment
    & SDT13 & -- 2013 WFIRST SDT report \tabularnewline
    $D(z)$ & -- distance-redshift relation
    & SDT15 & -- 2015 WFIRST SDT report \tabularnewline
    $D_A(z)$ & -- angular-diameter distance
    & SIT & -- Science Investigation Team \tabularnewline
    DE & -- dark energy
    & SMEX & -- NASA Small Explorer \tabularnewline
    DES & -- Dark Energy Survey
    & SN & -- supernovae \tabularnewline
    DESC & -- Dark Energy Science Collaboration
    & S/N & -- signal-to-noise \tabularnewline
    DESI \\ ELG \\ ELTs & -- Dark Energy Spectroscopic Instrument \\
    -- emission line galaxies \\ -- Extremely Large Telescopes
    & SPHEREx & -- Spectrophotometer for the History of the Universe, Epoch of
    Reionization, and Ices Explorer\tabularnewline
    eROSITA & -- extended Roentgen Survey with an Imaging Telescope Array
    & SPT-3G & -- South Pole Telescope Third-Generation Camera Survey
    \tabularnewline
    ETC & -- Exposure Time Calculator
    & STEP & -- Shear Testing Program \tabularnewline
    $f_g$ \\ FSWG & -- fluctuation growth rate \\ -- Formulation Science
    Working Group
    & STIS & HST Space Telescope Imaging Spectrograph \tabularnewline
    GEO & -- geostationary earth orbit
    & SZ & -- Sunyaev-Zeldovich \tabularnewline
    GGL & -- galaxy-galaxy lensing
    & $w(z)$ & -- dark energy equation-of-state \tabularnewline
    GREAT & -- Gravitational Lensing Accuracy Test
    & WFC3 & HST Wide-Field Camera 3 \tabularnewline
    GREAT3 & -- The third GREAT challenge
    & WFIRST & -- Wide-Field Infrared Survey Telescope \tabularnewline
    GRS \\ $H(z)$ & -- Galaxy Redshift Survey \\ -- Hubble parameter
    & WISPs & -- HST WFC3 IR Spectroscopic Parallel survey \tabularnewline
    HLS & -- High Latitude Survey
    & WL & -- weak lensing  \tabularnewline
    HOD & -- halo occupation distribution
    & WMAP & -- Wilkinson Microwave Anisotropy Probe \tabularnewline
    HSC & -- Subaru Hyper Suprime-Cam
    & WPS & -- WFIRST Preparatory Science \tabularnewline
    HST & -- Hubble Space Telescope
    & WSC & -- WFIRST Science Centers \tabularnewline
    IA & -- intrinsic galaxy alignments
    & $z$ & -- redshift \tabularnewline
    IPAC & -- Infrared Processing and Analysis Center
    & $z_p$ & -- pivot redshift \tabularnewline
    IPC & -- inter-pixel capacitance
    & & \tabularnewline
    IR & -- infrared
    & & \tabularnewline
    JDEM & -- Joint Dark Energy Mission
    & & \tabularnewline
    JWST & -- James Webb Space Telescope
    & & \tabularnewline
    KiDS & -- Kilo Degree Survey
    & & \tabularnewline
    L2 & -- Lagrange point 2 orbit
    & & \tabularnewline
    LF & -- luminosity function
    & & \tabularnewline

  \end{tabular}
\end{table*}

\clearpage
\newpage


\bibliographystyle{plainnat}
\bibliography{refs}

\begin{thebibliography}{100}
\providecommand{\natexlab}[1]{#1}
\providecommand{\url}[1]{\texttt{#1}}
\expandafter\ifx\csname urlstyle\endcsname\relax
  \providecommand{\doi}[1]{doi: #1}\else
  \providecommand{\doi}{doi: \begingroup \urlstyle{rm}\Url}\fi

\bibitem[Fai()]{Faisst2018}


\bibitem[{Atek} et~al.(2010){Atek}, {Malkan}, {McCarthy}, {Teplitz},
  {Scarlata}, {Siana}, {Henry}, {Colbert}, {Ross}, {Bridge}, {Bunker},
  {Dressler}, {Fosbury}, {Martin}, and {Shim}]{Atek:2010}
H.~{Atek}, M.~{Malkan}, P.~{McCarthy}, H.~I. {Teplitz}, C.~{Scarlata},
  B.~{Siana}, A.~{Henry}, J.~W. {Colbert}, N.~R. {Ross}, C.~{Bridge}, A.~J.
  {Bunker}, A.~{Dressler}, R.~A.~E. {Fosbury}, C.~{Martin}, and H.~{Shim}.
\newblock {The WFC3 Infrared Spectroscopic Parallel (WISP) Survey}.
\newblock \emph{\apj}, 723:\penalty0 104--115, November 2010.
\newblock \doi{10.1088/0004-637X/723/1/104}.

\bibitem[{Aubourg} et~al.(2015){Aubourg}, {Bailey}, {Bautista}, {Beutler},
  {Bhardwaj}, {Bizyaev}, {Blanton}, {Blomqvist}, {Bolton}, {Bovy},
  {Brewington}, {Brinkmann}, {Brownstein}, {Burden}, {Busca}, {Carithers},
  {Chuang}, {Comparat}, {Croft}, {Cuesta}, {Dawson}, {Delubac}, {Eisenstein},
  {Font-Ribera}, {Ge}, {Le Goff}, {Gontcho}, {Gott}, {Gunn}, {Guo}, {Guy},
  {Hamilton}, {Ho}, {Honscheid}, {Howlett}, {Kirkby}, {Kitaura}, {Kneib},
  {Lee}, {Long}, {Lupton}, {Maga{\~n}a}, {Malanushenko}, {Malanushenko},
  {Manera}, {Maraston}, {Margala}, {McBride}, {Miralda-Escud{\'e}}, {Myers},
  {Nichol}, {Noterdaeme}, {Nuza}, {Olmstead}, {Oravetz}, {P{\^a}ris},
  {Padmanabhan}, {Palanque-Delabrouille}, {Pan}, {Pellejero-Ibanez},
  {Percival}, {Petitjean}, {Pieri}, {Prada}, {Reid}, {Rich}, {Roe}, {Ross},
  {Ross}, {Rossi}, {Rubi{\~n}o-Mart{\'{\i}}n}, {S{\'a}nchez}, {Samushia},
  {G{\'e}nova-Santos}, {Sc{\'o}ccola}, {Schlegel}, {Schneider}, {Seo},
  {Sheldon}, {Simmons}, {Skibba}, {Slosar}, {Strauss}, {Thomas}, {Tinker},
  {Tojeiro}, {Vazquez}, {Viel}, {Wake}, {Weaver}, {Weinberg}, {Wood-Vasey},
  {Y{\`e}che}, {Zehavi}, {Zhao}, and {BOSS Collaboration}]{Aubourg:2015}
{\'E}.~{Aubourg}, S.~{Bailey}, J.~E. {Bautista}, F.~{Beutler}, V.~{Bhardwaj},
  D.~{Bizyaev}, M.~{Blanton}, M.~{Blomqvist}, A.~S. {Bolton}, J.~{Bovy},
  H.~{Brewington}, J.~{Brinkmann}, J.~R. {Brownstein}, A.~{Burden}, N.~G.
  {Busca}, W.~{Carithers}, C.-H. {Chuang}, J.~{Comparat}, R.~A.~C. {Croft},
  A.~J. {Cuesta}, K.~S. {Dawson}, T.~{Delubac}, D.~J. {Eisenstein},
  A.~{Font-Ribera}, J.~{Ge}, J.-M. {Le Goff}, S.~G.~A. {Gontcho}, J.~R. {Gott},
  J.~E. {Gunn}, H.~{Guo}, J.~{Guy}, J.-C. {Hamilton}, S.~{Ho}, K.~{Honscheid},
  C.~{Howlett}, D.~{Kirkby}, F.~S. {Kitaura}, J.-P. {Kneib}, K.-G. {Lee},
  D.~{Long}, R.~H. {Lupton}, M.~V. {Maga{\~n}a}, V.~{Malanushenko},
  E.~{Malanushenko}, M.~{Manera}, C.~{Maraston}, D.~{Margala}, C.~K. {McBride},
  J.~{Miralda-Escud{\'e}}, A.~D. {Myers}, R.~C. {Nichol}, P.~{Noterdaeme},
  S.~E. {Nuza}, M.~D. {Olmstead}, D.~{Oravetz}, I.~{P{\^a}ris},
  N.~{Padmanabhan}, N.~{Palanque-Delabrouille}, K.~{Pan},
  M.~{Pellejero-Ibanez}, W.~J. {Percival}, P.~{Petitjean}, M.~M. {Pieri},
  F.~{Prada}, B.~{Reid}, J.~{Rich}, N.~A. {Roe}, A.~J. {Ross}, N.~P. {Ross},
  G.~{Rossi}, J.~A. {Rubi{\~n}o-Mart{\'{\i}}n}, A.~G. {S{\'a}nchez},
  L.~{Samushia}, R.~T. {G{\'e}nova-Santos}, C.~G. {Sc{\'o}ccola}, D.~J.
  {Schlegel}, D.~P. {Schneider}, H.-J. {Seo}, E.~{Sheldon}, A.~{Simmons}, R.~A.
  {Skibba}, A.~{Slosar}, M.~A. {Strauss}, D.~{Thomas}, J.~L. {Tinker},
  R.~{Tojeiro}, J.~A. {Vazquez}, M.~{Viel}, D.~A. {Wake}, B.~A. {Weaver}, D.~H.
  {Weinberg}, W.~M. {Wood-Vasey}, C.~{Y{\`e}che}, I.~{Zehavi}, G.-B. {Zhao},
  and {BOSS Collaboration}.
\newblock {Cosmological implications of baryon acoustic oscillation
  measurements}.
\newblock \emph{\prd}, 92\penalty0 (12):\penalty0 123516, December 2015.
\newblock \doi{10.1103/PhysRevD.92.123516}.

\bibitem[{Becker} et~al.(2015){Becker}, {Troxel}, {MacCrann}, {Krause},
  {Eifler}, {Friedrich}, {Nicola}, {Refregier}, {Amara}, {Bacon}, {Bernstein},
  {Bonnett}, {Bridle}, {Busha}, {Chang}, {Dodelson}, {Erickson}, {Evrard},
  {Frieman}, {Gaztanaga}, {Gruen}, {Hartley}, {Jain}, {Jarvis}, {Kacprzak},
  {Kirk}, {Kravtsov}, {Leistedt}, {Rykoff}, {Sabiu}, {Sanchez}, {Seo},
  {Sheldon}, {Wechsler}, {Zuntz}, {Abbott}, {Abdalla}, {Allam}, {Armstrong},
  {Banerji}, {Bauer}, {Benoit-Levy}, {Bertin}, {Brooks}, {Buckley-Geer},
  {Burke}, {Capozzi}, {Carnero Rosell}, {Carrasco Kind}, {Carretero},
  {Castander}, {Crocce}, {Cunha}, {D'Andrea}, {da Costa}, {DePoy}, {Desai},
  {Diehl}, {Dietrich}, {Doel}, {Fausti Neto}, {Fernandez}, {Finley},
  {Flaugher}, {Fosalba}, {Gerdes}, {Gruendl}, {Gutierrez}, {Honscheid},
  {James}, {Kuehn}, {Kuropatkin}, {Lahav}, {Li}, {Lima}, {Maia}, {March},
  {Martini}, {Melchior}, {Miller}, {Miquel}, {Mohr}, {Nichol}, {Nord},
  {Ogando}, {Plazas}, {Reil}, {Romer}, {Roodman}, {Sako}, {Sanchez},
  {Scarpine}, {Schubnell}, {Sevilla-Noarbe}, {Smith}, {Soares-Santos},
  {Sobreira}, {Suchyta}, {Swanson}, {Tarle}, {Thaler}, {Thomas}, {Vikram},
  {Walker}, and {The DES Collaboration}]{Becker2015}
M.~R. {Becker}, M.~A. {Troxel}, N.~{MacCrann}, E.~{Krause}, T.~F. {Eifler},
  O.~{Friedrich}, A.~{Nicola}, A.~{Refregier}, A.~{Amara}, D.~{Bacon}, G.~M.
  {Bernstein}, C.~{Bonnett}, S.~L. {Bridle}, M.~T. {Busha}, C.~{Chang},
  S.~{Dodelson}, B.~{Erickson}, A.~E. {Evrard}, J.~{Frieman}, E.~{Gaztanaga},
  D.~{Gruen}, W.~{Hartley}, B.~{Jain}, M.~{Jarvis}, T.~{Kacprzak}, D.~{Kirk},
  A.~{Kravtsov}, B.~{Leistedt}, E.~S. {Rykoff}, C.~{Sabiu}, C.~{Sanchez},
  H.~{Seo}, E.~{Sheldon}, R.~H. {Wechsler}, J.~{Zuntz}, T.~{Abbott}, F.~B.
  {Abdalla}, S.~{Allam}, R.~{Armstrong}, M.~{Banerji}, A.~H. {Bauer},
  A.~{Benoit-Levy}, E.~{Bertin}, D.~{Brooks}, E.~{Buckley-Geer}, D.~L. {Burke},
  D.~{Capozzi}, A.~{Carnero Rosell}, M.~{Carrasco Kind}, J.~{Carretero}, F.~J.
  {Castander}, M.~{Crocce}, C.~E. {Cunha}, C.~B. {D'Andrea}, L.~N. {da Costa},
  D.~L. {DePoy}, S.~{Desai}, H.~T. {Diehl}, J.~P. {Dietrich}, P.~{Doel},
  A.~{Fausti Neto}, E.~{Fernandez}, D.~A. {Finley}, B.~{Flaugher},
  P.~{Fosalba}, D.~W. {Gerdes}, R.~A. {Gruendl}, G.~{Gutierrez},
  K.~{Honscheid}, D.~J. {James}, K.~{Kuehn}, N.~{Kuropatkin}, O.~{Lahav}, T.~S.
  {Li}, M.~{Lima}, M.~A.~G. {Maia}, M.~{March}, P.~{Martini}, P.~{Melchior},
  C.~J. {Miller}, R.~{Miquel}, J.~J. {Mohr}, R.~C. {Nichol}, B.~{Nord},
  R.~{Ogando}, A.~A. {Plazas}, K.~{Reil}, A.~K. {Romer}, A.~{Roodman},
  M.~{Sako}, E.~{Sanchez}, V.~{Scarpine}, M.~{Schubnell}, I.~{Sevilla-Noarbe},
  R.~C. {Smith}, M.~{Soares-Santos}, F.~{Sobreira}, E.~{Suchyta}, M.~E.~C.
  {Swanson}, G.~{Tarle}, J.~{Thaler}, D.~{Thomas}, V.~{Vikram}, A.~R. {Walker},
  and {The DES Collaboration}.
\newblock {Cosmic Shear Measurements with DES Science Verification Data}.
\newblock \emph{ArXiv e-prints}, July 2015.

\bibitem[{Benson}(2012)]{Benson2012}
A.~J. {Benson}.
\newblock {GALACTICUS: A semi-analytic model of galaxy formation}.
\newblock \emph{\na}, 17:\penalty0 175--197, February 2012.
\newblock \doi{10.1016/j.newast.2011.07.004}.

\bibitem[{Bernstein} and {Jarvis}(2002)]{bej02}
G.~M. {Bernstein} and M.~{Jarvis}.
\newblock {Shapes and Shears, Stars and Smears: Optimal Measurements for Weak
  Lensing}.
\newblock \emph{\aj}, 123:\penalty0 583--618, February 2002.
\newblock \doi{10.1086/338085}.

\bibitem[{Bhattacharya} et~al.(2011){Bhattacharya}, {Heitmann}, {White},
  {Luki{\'c}}, {Wagner}, and {Habib}]{Bhattacharya11}
S.~{Bhattacharya}, K.~{Heitmann}, M.~{White}, Z.~{Luki{\'c}}, C.~{Wagner}, and
  S.~{Habib}.
\newblock {Mass Function Predictions Beyond {$\Lambda$}CDM}.
\newblock \emph{\apj}, 732:\penalty0 122, May 2011.
\newblock \doi{10.1088/0004-637X/732/2/122}.

\bibitem[{Blake} and {Glazebrook}(2003)]{Blake03}
C.~{Blake} and K.~{Glazebrook}.
\newblock {Probing Dark Energy Using Baryonic Oscillations in the Galaxy Power
  Spectrum as a Cosmological Ruler}.
\newblock \emph{\apj}, 594:\penalty0 665--673, September 2003.
\newblock \doi{10.1086/376983}.

\bibitem[{Brammer} et~al.(2012){Brammer}, {van Dokkum}, {Franx}, {Fumagalli},
  {Patel}, {Rix}, {Skelton}, {Kriek}, {Nelson}, {Schmidt}, {Bezanson}, {da
  Cunha}, {Erb}, {Fan}, {F{\"o}rster Schreiber}, {Illingworth}, {Labb{\'e}},
  {Leja}, {Lundgren}, {Magee}, {Marchesini}, {McCarthy}, {Momcheva}, {Muzzin},
  {Quadri}, {Steidel}, {Tal}, {Wake}, {Whitaker}, and {Williams}]{Brammer:2012}
G.~B. {Brammer}, P.~G. {van Dokkum}, M.~{Franx}, M.~{Fumagalli}, S.~{Patel},
  H.-W. {Rix}, R.~E. {Skelton}, M.~{Kriek}, E.~{Nelson}, K.~B. {Schmidt},
  R.~{Bezanson}, E.~{da Cunha}, D.~K. {Erb}, X.~{Fan}, N.~{F{\"o}rster
  Schreiber}, G.~D. {Illingworth}, I.~{Labb{\'e}}, J.~{Leja}, B.~{Lundgren},
  D.~{Magee}, D.~{Marchesini}, P.~{McCarthy}, I.~{Momcheva}, A.~{Muzzin},
  R.~{Quadri}, C.~C. {Steidel}, T.~{Tal}, D.~{Wake}, K.~E. {Whitaker}, and
  A.~{Williams}.
\newblock {3D-HST: A Wide-field Grism Spectroscopic Survey with the Hubble
  Space Telescope}.
\newblock \emph{\apjs}, 200:\penalty0 13, June 2012.
\newblock \doi{10.1088/0067-0049/200/2/13}.

\bibitem[{Bridle} et~al.(2010){Bridle}, {Balan}, {Bethge}, {Gentile},
  {Harmeling}, {Heymans}, {Hirsch}, {Hosseini}, {Jarvis}, {Kirk}, {Kitching},
  {Kuijken}, {Lewis}, {Paulin-Henriksson}, {Sch{\"o}lkopf}, {Velander},
  {Voigt}, {Witherick}, {Amara}, {Bernstein}, {Courbin}, {Gill}, {Heavens},
  {Mandelbaum}, {Massey}, {Moghaddam}, {Rassat}, {R{\'e}fr{\'e}gier}, {Rhodes},
  {Schrabback}, {Shawe-Taylor}, {Shmakova}, {van Waerbeke}, and
  {Wittman}]{Bridle2010}
S.~{Bridle}, S.~T. {Balan}, M.~{Bethge}, M.~{Gentile}, S.~{Harmeling},
  C.~{Heymans}, M.~{Hirsch}, R.~{Hosseini}, M.~{Jarvis}, D.~{Kirk},
  T.~{Kitching}, K.~{Kuijken}, A.~{Lewis}, S.~{Paulin-Henriksson},
  B.~{Sch{\"o}lkopf}, M.~{Velander}, L.~{Voigt}, D.~{Witherick}, A.~{Amara},
  G.~{Bernstein}, F.~{Courbin}, M.~{Gill}, A.~{Heavens}, R.~{Mandelbaum},
  R.~{Massey}, B.~{Moghaddam}, A.~{Rassat}, A.~{R{\'e}fr{\'e}gier},
  J.~{Rhodes}, T.~{Schrabback}, J.~{Shawe-Taylor}, M.~{Shmakova}, L.~{van
  Waerbeke}, and D.~{Wittman}.
\newblock {Results of the GREAT08 Challenge: an image analysis competition for
  cosmological lensing}.
\newblock \emph{\mnras}, 405:\penalty0 2044--2061, July 2010.
\newblock \doi{10.1111/j.1365-2966.2010.16598.x}.

\bibitem[{Bull} et~al.(2015){Bull}, {Akrami}, {Adamek}, {Baker}, {Bellini},
  {Beltr{\'a}n Jim{\'e}nez}, {Bentivegna}, {Camera}, {Clesse}, {Davis}, {Di
  Dio}, {Enander}, {Finelli}, {Heavens}, {Heisenberg}, {Hu}, {Llinares},
  {Maartens}, {M{\"o}rtsell}, {Nadathur}, {Noller}, {Pasechnik}, {Pawlowski},
  {Pereira}, {Quartin}, {Ricciardone}, {Riemer-S{\o}rensen}, {Rinaldi},
  {Sakstein}, {Saltas}, {Salzano}, {Sawicki}, {Solomon}, {Spolyar}, {Starkman},
  {Steer}, {Tereno}, {Verde}, {Villaescusa-Navarro}, {von Strauss}, and
  {Winther}]{baa15}
P.~{Bull}, Y.~{Akrami}, J.~{Adamek}, T.~{Baker}, E.~{Bellini}, J.~{Beltr{\'a}n
  Jim{\'e}nez}, E.~{Bentivegna}, S.~{Camera}, S.~{Clesse}, J.~H. {Davis},
  E.~{Di Dio}, J.~{Enander}, F.~{Finelli}, A.~{Heavens}, L.~{Heisenberg},
  B.~{Hu}, C.~{Llinares}, R.~{Maartens}, E.~{M{\"o}rtsell}, S.~{Nadathur},
  J.~{Noller}, R.~{Pasechnik}, M.~S. {Pawlowski}, T.~S. {Pereira},
  M.~{Quartin}, A.~{Ricciardone}, S.~{Riemer-S{\o}rensen}, M.~{Rinaldi},
  J.~{Sakstein}, I.~D. {Saltas}, V.~{Salzano}, I.~{Sawicki}, A.~R. {Solomon},
  D.~{Spolyar}, G.~D. {Starkman}, D.~{Steer}, I.~{Tereno}, L.~{Verde},
  F.~{Villaescusa-Navarro}, M.~{von Strauss}, and H.~A. {Winther}.
\newblock {Beyond $\Lambda$CDM: Problems, solutions, and the road ahead}.
\newblock \emph{ArXiv e-prints}, December 2015.

\bibitem[{Calzetti} et~al.(2000){Calzetti}, {Armus}, {Bohlin}, {Kinney},
  {Koornneef}, and {Storchi-Bergmann}]{Calzetti00}
D.~{Calzetti}, L.~{Armus}, R.~C. {Bohlin}, A.~L. {Kinney}, J.~{Koornneef}, and
  T.~{Storchi-Bergmann}.
\newblock {The Dust Content and Opacity of Actively Star-forming Galaxies}.
\newblock \emph{\apj}, 533:\penalty0 682--695, April 2000.
\newblock \doi{10.1086/308692}.

\bibitem[{Charlot} and {Fall}(2000)]{Charlot00}
S.~{Charlot} and S.~M. {Fall}.
\newblock {A Simple Model for the Absorption of Starlight by Dust in Galaxies}.
\newblock \emph{\apj}, 539:\penalty0 718--731, August 2000.
\newblock \doi{10.1086/309250}.

\bibitem[Chisari et~al.(2016)Chisari, Dvorkin, Schmidt, and
  Spergel]{Chisari:2016xki}
Nora~Elisa Chisari, Cora Dvorkin, Fabian Schmidt, and David Spergel.
\newblock {Multitracing Anisotropic Non-Gaussianity with Galaxy Shapes}.
\newblock \emph{Phys. Rev.}, D94\penalty0 (12):\penalty0 123507, 2016.
\newblock \doi{10.1103/PhysRevD.94.123507}.

\bibitem[{Chuang} and {Wang}(2012)]{CW12}
C.-H. {Chuang} and Y.~{Wang}.
\newblock {Measurements of H(z) and D$_{A}$(z) from the two-dimensional
  two-point correlation function of Sloan Digital Sky Survey luminous red
  galaxies}.
\newblock \emph{\mnras}, 426:\penalty0 226--236, October 2012.
\newblock \doi{10.1111/j.1365-2966.2012.21565.x}.

\bibitem[{Colbert} et~al.(2013){Colbert}, {Teplitz}, {Atek}, {Bunker},
  {Rafelski}, {Ross}, {Scarlata}, {Bedregal}, {Dominguez}, {Dressler}, {Henry},
  {Malkan}, {Martin}, {Masters}, {McCarthy}, and {Siana}]{Colbert13}
J.~W. {Colbert}, H.~{Teplitz}, H.~{Atek}, A.~{Bunker}, M.~{Rafelski},
  N.~{Ross}, C.~{Scarlata}, A.~G. {Bedregal}, A.~{Dominguez}, A.~{Dressler},
  A.~{Henry}, M.~{Malkan}, C.~L. {Martin}, D.~{Masters}, P.~{McCarthy}, and
  B.~{Siana}.
\newblock {Predicting Future Space Near-IR Grism Surveys Using the WFC3
  Infrared Spectroscopic Parallels Survey}.
\newblock \emph{\apj}, 779:\penalty0 34, December 2013.
\newblock \doi{10.1088/0004-637X/779/1/34}.

\bibitem[Colbert et~al.(2013)]{Colbert:2013ita}
James~W. Colbert et~al.
\newblock {Predicting Future Space Near-IR Grism Surveys using the WFC3
  Infrared Spectroscopic Parallels Survey}.
\newblock \emph{Astrophys. J.}, 779:\penalty0 34, 2013.
\newblock \doi{10.1088/0004-637X/779/1/34}.

\bibitem[{Cooray} and {Sheth}(2002)]{CS02}
A.~{Cooray} and R.~{Sheth}.
\newblock {Halo models of large scale structure}.
\newblock \emph{\physrep}, 372:\penalty0 1--129, December 2002.
\newblock \doi{10.1016/S0370-1573(02)00276-4}.

\bibitem[Cooray and Hu(2001)]{CH01}
Asantha Cooray and Wayne Hu.
\newblock {Power spectrum covariance of weak gravitational lensing}.
\newblock \emph{Astrophys. J.}, 554:\penalty0 56--66, 2001.
\newblock \doi{10.1086/321376}.

\bibitem[Council(2010)]{NWNH2010}
National~Research Council.
\newblock \emph{New Worlds, New Horizons in Astronomy and Astrophysics}.
\newblock The National Academies Press, Washington, DC, 2010.
\newblock ISBN 978-0-309-15799-5.
\newblock URL
  \url{http://www.nap.edu/catalog/12951/new-worlds-new-horizons-in-astronomy-and-astrophysics}.

\bibitem[de~Putter et~al.(2013)de~Putter, Dor{\'e}, and
  Takada]{dePutter:2013xda}
Roland de~Putter, Olivier Dor{\'e}, and Masahiro Takada.
\newblock {The Synergy between Weak Lensing and Galaxy Redshift Surveys}.
\newblock 2013.

\bibitem[de~Putter et~al.(2014)de~Putter, Dor{\'e}, and Das]{dePutter:2013nha}
Roland de~Putter, Olivier Dor{\'e}, and Sudeep Das.
\newblock {Using Cross-Correlations to Calibrate Lensing Source Redshift
  Distributions: Improving Cosmological Constraints from Upcoming Weak Lensing
  Surveys}.
\newblock \emph{Astrophys. J.}, 780:\penalty0 185, 2014.
\newblock \doi{10.1088/0004-637X/780/2/185}.

\bibitem[{Dom{\'{\i}}nguez} et~al.(2013){Dom{\'{\i}}nguez}, {Siana}, {Henry},
  {Scarlata}, {Bedregal}, {Malkan}, {Atek}, {Ross}, {Colbert}, {Teplitz},
  {Rafelski}, {McCarthy}, {Bunker}, {Hathi}, {Dressler}, {Martin}, and
  {Masters}]{Dominguez13}
A.~{Dom{\'{\i}}nguez}, B.~{Siana}, A.~L. {Henry}, C.~{Scarlata}, A.~G.
  {Bedregal}, M.~{Malkan}, H.~{Atek}, N.~R. {Ross}, J.~W. {Colbert}, H.~I.
  {Teplitz}, \~M. {Rafelski}, P.~{McCarthy}, A.~{Bunker}, N.~P. {Hathi},
  A.~{Dressler}, C.~L. {Martin}, and D.~{Masters}.
\newblock {Dust Extinction from Balmer Decrements of Star-forming Galaxies at
  $0.75 \leqslant z \leqslant 1.5$ with Hubble Space
  Telescope/Wide-Field-Camera 3 Spectroscopy from the WFC3 Infrared
  Spectroscopic Parallel Survey\ }.
\newblock \emph{\apj}, 763:\penalty0 145, February 2013.
\newblock \doi{10.1088/0004-637X/763/2/145}.

\bibitem[{Eifler} et~al.(2009){Eifler}, {Schneider}, and {Hartlap}]{esh09}
T.~{Eifler}, P.~{Schneider}, and J.~{Hartlap}.
\newblock {Dependence of cosmic shear covariances on cosmology. Impact on
  parameter estimation}.
\newblock \emph{\aap}, 502:\penalty0 721--731, August 2009.
\newblock \doi{10.1051/0004-6361/200811276}.

\bibitem[{Eifler} et~al.(2014){Eifler}, {Krause}, {Schneider}, and
  {Honscheid}]{Eifler:2014}
T.~{Eifler}, E.~{Krause}, P.~{Schneider}, and K.~{Honscheid}.
\newblock {Combining probes of large-scale structure with COSMOLIKE}.
\newblock \emph{\mnras}, 440:\penalty0 1379--1390, May 2014.
\newblock \doi{10.1093/mnras/stu251}.

\bibitem[{Eifler} et~al.(2015){Eifler}, {Krause}, {Dodelson}, {Zentner},
  {Hearin}, and {Gnedin}]{Eifler2015}
T.~{Eifler}, E.~{Krause}, S.~{Dodelson}, A.~R. {Zentner}, A.~P. {Hearin}, and
  N.~Y. {Gnedin}.
\newblock {Accounting for baryonic effects in cosmic shear tomography:
  determining a minimal set of nuisance parameters using PCA}.
\newblock \emph{\mnras}, 454:\penalty0 2451--2471, December 2015.
\newblock \doi{10.1093/mnras/stv2000}.

\bibitem[{Ferland} et~al.(2013){Ferland}, {Porter}, {van Hoof}, {Williams},
  {Abel}, {Lykins}, {Shaw}, {Henney}, and {Stancil}]{Ferland13}
G.~J. {Ferland}, R.~L. {Porter}, P.~A.~M. {van Hoof}, R.~J.~R. {Williams},
  N.~P. {Abel}, M.~L. {Lykins}, G.~{Shaw}, W.~J. {Henney}, and P.~C. {Stancil}.
\newblock {The 2013 Release of Cloudy}.
\newblock \emph{\rmxaa}, 49:\penalty0 137--163, April 2013.

\bibitem[{Ferrara} et~al.(1999){Ferrara}, {Bianchi}, {Cimatti}, and
  {Giovanardi}]{Ferrara99}
A.~{Ferrara}, S.~{Bianchi}, A.~{Cimatti}, and C.~{Giovanardi}.
\newblock {An Atlas of Monte Carlo Models of Dust Extinction in Galaxies for
  Cosmological Applications}.
\newblock \emph{\apjs}, 123:\penalty0 437--445, August 1999.
\newblock \doi{10.1086/313244}.

\bibitem[{Foreman-Mackey} et~al.(2013){Foreman-Mackey}, {Hogg}, {Lang}, and
  {Goodman}]{Foreman-Mackey2013}
D.~{Foreman-Mackey}, D.~W. {Hogg}, D.~{Lang}, and J.~{Goodman}.
\newblock {emcee: The MCMC Hammer}.
\newblock \emph{\pasp}, 125:\penalty0 306--312, March 2013.
\newblock \doi{10.1086/670067}.

\bibitem[Friedrich et~al.(2016)Friedrich, Seitz, Eifler, and
  Gruen]{Friedrich:2015nga}
O.~Friedrich, S.~Seitz, T.~F. Eifler, and D.~Gruen.
\newblock {Performance of internal Covariance Estimators for Cosmic Shear
  Correlation Functions}.
\newblock \emph{Mon. Not. Roy. Astron. Soc.}, 456\penalty0 (3):\penalty0
  2662--2680, 2016.
\newblock \doi{10.1093/mnras/stv2833}.

\bibitem[{Gagrani} and {Samushia}(2017)]{2017MNRAS.467..928G}
P.~{Gagrani} and L.~{Samushia}.
\newblock {Information Content of the Angular Multipoles of Redshift-Space
  Galaxy Bispectrum}.
\newblock \emph{\mnras}, 467:\penalty0 928--935, May 2017.
\newblock \doi{10.1093/mnras/stx135}.

\bibitem[{Galametz} et~al.(2013){Galametz}, {Grazian}, {Fontana}, {Ferguson},
  {Ashby}, {Barro}, {Castellano}, {Dahlen}, {Donley}, {Faber}, {Grogin}, {Guo},
  {Huang}, {Kocevski}, {Koekemoer}, {Lee}, {McGrath}, {Peth}, {Willner},
  {Almaini}, {Cooper}, {Cooray}, {Conselice}, {Dickinson}, {Dunlop}, {Fazio},
  {Foucaud}, {Gardner}, {Giavalisco}, {Hathi}, {Hartley}, {Koo}, {Lai}, {de
  Mello}, {McLure}, {Lucas}, {Paris}, {Pentericci}, {Santini}, {Simpson},
  {Sommariva}, {Targett}, {Weiner}, {Wuyts}, and {the CANDELS
  Team}]{Galametz2013}
A.~{Galametz}, A.~{Grazian}, A.~{Fontana}, H.~C. {Ferguson}, M.~L.~N. {Ashby},
  G.~{Barro}, M.~{Castellano}, T.~{Dahlen}, J.~L. {Donley}, S.~M. {Faber},
  N.~{Grogin}, Y.~{Guo}, K.-H. {Huang}, D.~D. {Kocevski}, A.~M. {Koekemoer},
  K.-S. {Lee}, E.~J. {McGrath}, M.~{Peth}, S.~P. {Willner}, O.~{Almaini},
  M.~{Cooper}, A.~{Cooray}, C.~J. {Conselice}, M.~{Dickinson}, J.~S. {Dunlop},
  G.~G. {Fazio}, S.~{Foucaud}, J.~P. {Gardner}, M.~{Giavalisco}, N.~P. {Hathi},
  W.~G. {Hartley}, D.~C. {Koo}, K.~{Lai}, D.~F. {de Mello}, R.~J. {McLure},
  R.~A. {Lucas}, D.~{Paris}, L.~{Pentericci}, P.~{Santini}, C.~{Simpson},
  V.~{Sommariva}, T.~{Targett}, B.~J. {Weiner}, S.~{Wuyts}, and {the CANDELS
  Team}.
\newblock {CANDELS Multiwavelength Catalogs: Source Identification and
  Photometry in the CANDELS UKIDSS Ultra-deep Survey Field}.
\newblock \emph{\apjs}, 206:\penalty0 10, June 2013.
\newblock \doi{10.1088/0067-0049/206/2/10}.

\bibitem[{Geach} et~al.(2012){Geach}, {Sobral}, {Hickox}, {Wake}, {Smail},
  {Best}, {Baugh}, and {Stott}]{Geach2012}
J.~E. {Geach}, D.~{Sobral}, R.~C. {Hickox}, D.~A. {Wake}, I.~{Smail}, P.~N.
  {Best}, C.~M. {Baugh}, and J.~P. {Stott}.
\newblock {The clustering of H{$\alpha$} emitters at z=2.23 from HiZELS}.
\newblock \emph{\mnras}, 426:\penalty0 679--689, October 2012.
\newblock \doi{10.1111/j.1365-2966.2012.21725.x}.

\bibitem[{Giocoli} et~al.(2008){Giocoli}, {Pieri}, and {Tormen}]{Giocoli2008}
C.~{Giocoli}, L.~{Pieri}, and G.~{Tormen}.
\newblock {Analytical approach to subhalo population in dark matter haloes}.
\newblock \emph{\mnras}, 387:\penalty0 689--697, June 2008.
\newblock \doi{10.1111/j.1365-2966.2008.13283.x}.

\bibitem[Goodman and Weare(2010)]{Goodman2010}
Jonathan Goodman and Jonathan Weare.
\newblock Ensemble samplers with affine invariance.
\newblock \emph{Communications in Applied Mathematics and Computational
  Science}, 5\penalty0 (1):\penalty0 65--80, January 2010.
\newblock ISSN 2157-5452.
\newblock \doi{10.2140/camcos.2010.5.65}.
\newblock URL \url{http://dx.doi.org/10.2140/camcos.2010.5.65}.

\bibitem[{Guo} et~al.(2013){Guo}, {Ferguson}, {Giavalisco}, {Barro}, {Willner},
  {Ashby}, {Dahlen}, {Donley}, {Faber}, {Fontana}, {Galametz}, {Grazian},
  {Huang}, {Kocevski}, {Koekemoer}, {Koo}, {McGrath}, {Peth}, {Salvato},
  {Wuyts}, {Castellano}, {Cooray}, {Dickinson}, {Dunlop}, {Fazio}, {Gardner},
  {Gawiser}, {Grogin}, {Hathi}, {Hsu}, {Lee}, {Lucas}, {Mobasher}, {Nandra},
  {Newman}, and {van der Wel}]{Guo2013}
Y.~{Guo}, H.~C. {Ferguson}, M.~{Giavalisco}, G.~{Barro}, S.~P. {Willner},
  M.~L.~N. {Ashby}, T.~{Dahlen}, J.~L. {Donley}, S.~M. {Faber}, A.~{Fontana},
  A.~{Galametz}, A.~{Grazian}, K.-H. {Huang}, D.~D. {Kocevski}, A.~M.
  {Koekemoer}, D.~C. {Koo}, E.~J. {McGrath}, M.~{Peth}, M.~{Salvato},
  S.~{Wuyts}, M.~{Castellano}, A.~R. {Cooray}, M.~E. {Dickinson}, J.~S.
  {Dunlop}, G.~G. {Fazio}, J.~P. {Gardner}, E.~{Gawiser}, N.~A. {Grogin}, N.~P.
  {Hathi}, L.-T. {Hsu}, K.-S. {Lee}, R.~A. {Lucas}, B.~{Mobasher}, K.~{Nandra},
  J.~A. {Newman}, and A.~{van der Wel}.
\newblock {CANDELS Multi-wavelength Catalogs: Source Detection and Photometry
  in the GOODS-South Field}.
\newblock \emph{\apjs}, 207:\penalty0 24, August 2013.
\newblock \doi{10.1088/0067-0049/207/2/24}.

\bibitem[{Guzzo} et~al.(2008){Guzzo}, {Pierleoni}, {Meneux}, {Branchini}, {Le
  F{\`e}vre}, {Marinoni}, {Garilli}, {Blaizot}, {De Lucia}, {Pollo},
  {McCracken}, {Bottini}, {Le Brun}, {Maccagni}, {Picat}, {Scaramella},
  {Scodeggio}, {Tresse}, {Vettolani}, {Zanichelli}, {Adami}, {Arnouts},
  {Bardelli}, {Bolzonella}, {Bongiorno}, {Cappi}, {Charlot}, {Ciliegi},
  {Contini}, {Cucciati}, {de la Torre}, {Dolag}, {Foucaud}, {Franzetti},
  {Gavignaud}, {Ilbert}, {Iovino}, {Lamareille}, {Marano}, {Mazure}, {Memeo},
  {Merighi}, {Moscardini}, {Paltani}, {Pell{\`o}}, {Perez-Montero}, {Pozzetti},
  {Radovich}, {Vergani}, {Zamorani}, and {Zucca}]{Guzzo08}
L.~{Guzzo}, M.~{Pierleoni}, B.~{Meneux}, E.~{Branchini}, O.~{Le F{\`e}vre},
  C.~{Marinoni}, B.~{Garilli}, J.~{Blaizot}, G.~{De Lucia}, A.~{Pollo}, H.~J.
  {McCracken}, D.~{Bottini}, V.~{Le Brun}, D.~{Maccagni}, J.~P. {Picat},
  R.~{Scaramella}, M.~{Scodeggio}, L.~{Tresse}, G.~{Vettolani},
  A.~{Zanichelli}, C.~{Adami}, S.~{Arnouts}, S.~{Bardelli}, M.~{Bolzonella},
  A.~{Bongiorno}, A.~{Cappi}, S.~{Charlot}, P.~{Ciliegi}, T.~{Contini},
  O.~{Cucciati}, S.~{de la Torre}, K.~{Dolag}, S.~{Foucaud}, P.~{Franzetti},
  I.~{Gavignaud}, O.~{Ilbert}, A.~{Iovino}, F.~{Lamareille}, B.~{Marano},
  A.~{Mazure}, P.~{Memeo}, R.~{Merighi}, L.~{Moscardini}, S.~{Paltani},
  R.~{Pell{\`o}}, E.~{Perez-Montero}, L.~{Pozzetti}, M.~{Radovich},
  D.~{Vergani}, G.~{Zamorani}, and E.~{Zucca}.
\newblock {A test of the nature of cosmic acceleration using galaxy redshift
  distortions}.
\newblock \emph{\nat}, 451:\penalty0 541--544, January 2008.
\newblock \doi{10.1038/nature06555}.

\bibitem[{Heitmann} et~al.(2014){Heitmann}, {Lawrence}, {Kwan}, {Habib}, and
  {Higdon}]{Heitmann2014}
K.~{Heitmann}, E.~{Lawrence}, J.~{Kwan}, S.~{Habib}, and D.~{Higdon}.
\newblock {The Coyote Universe Extended: Precision Emulation of the Matter
  Power Spectrum}.
\newblock \emph{\apj}, 780:\penalty0 111, January 2014.
\newblock \doi{10.1088/0004-637X/780/1/111}.

\bibitem[{Heymans} et~al.(2006){Heymans}, {Van Waerbeke}, {Bacon}, {Berge},
  {Bernstein}, {Bertin}, {Bridle}, {Brown}, {Clowe}, {Dahle}, {Erben}, {Gray},
  {Hetterscheidt}, {Hoekstra}, {Hudelot}, {Jarvis}, {Kuijken}, {Margoniner},
  {Massey}, {Mellier}, {Nakajima}, {Refregier}, {Rhodes}, {Schrabback}, and
  {Wittman}]{Heymans2006}
C.~{Heymans}, L.~{Van Waerbeke}, D.~{Bacon}, J.~{Berge}, G.~{Bernstein},
  E.~{Bertin}, S.~{Bridle}, M.~L. {Brown}, D.~{Clowe}, H.~{Dahle}, T.~{Erben},
  M.~{Gray}, M.~{Hetterscheidt}, H.~{Hoekstra}, P.~{Hudelot}, M.~{Jarvis},
  K.~{Kuijken}, V.~{Margoniner}, R.~{Massey}, Y.~{Mellier}, R.~{Nakajima},
  A.~{Refregier}, J.~{Rhodes}, T.~{Schrabback}, and D.~{Wittman}.
\newblock {The Shear Testing Programme - I. Weak lensing analysis of simulated
  ground-based observations}.
\newblock \emph{\mnras}, 368:\penalty0 1323--1339, May 2006.
\newblock \doi{10.1111/j.1365-2966.2006.10198.x}.

\bibitem[{Heymans} et~al.(2012){Heymans}, {Van Waerbeke}, {Miller}, {Erben},
  {Hildebrandt}, {Hoekstra}, {Kitching}, {Mellier}, {Simon}, {Bonnett},
  {Coupon}, {Fu}, {Harnois D{\'e}raps}, {Hudson}, {Kilbinger}, {Kuijken},
  {Rowe}, {Schrabback}, {Semboloni}, {van Uitert}, {Vafaei}, and
  {Velander}]{Heymans2012}
C.~{Heymans}, L.~{Van Waerbeke}, L.~{Miller}, T.~{Erben}, H.~{Hildebrandt},
  H.~{Hoekstra}, T.~D. {Kitching}, Y.~{Mellier}, P.~{Simon}, C.~{Bonnett},
  J.~{Coupon}, L.~{Fu}, J.~{Harnois D{\'e}raps}, M.~J. {Hudson},
  M.~{Kilbinger}, K.~{Kuijken}, B.~{Rowe}, T.~{Schrabback}, E.~{Semboloni},
  E.~{van Uitert}, S.~{Vafaei}, and M.~{Velander}.
\newblock {CFHTLenS: the Canada-France-Hawaii Telescope Lensing Survey}.
\newblock \emph{\mnras}, 427:\penalty0 146--166, November 2012.
\newblock \doi{10.1111/j.1365-2966.2012.21952.x}.

\bibitem[{Hu} and {Jain}(2004)]{huj04}
W.~{Hu} and B.~{Jain}.
\newblock {Joint galaxy-lensing observables and the dark energy}.
\newblock \emph{\prd}, 70\penalty0 (4):\penalty0 043009, August 2004.
\newblock \doi{10.1103/PhysRevD.70.043009}.

\bibitem[{Huff} et~al.(2014){Huff}, {Eifler}, {Hirata}, {Mandelbaum},
  {Schlegel}, and {Seljak}]{Huff2014}
E.~M. {Huff}, T.~{Eifler}, C.~M. {Hirata}, R.~{Mandelbaum}, D.~{Schlegel}, and
  U.~{Seljak}.
\newblock {Seeing in the dark - II. Cosmic shear in the Sloan Digital Sky
  Survey}.
\newblock \emph{\mnras}, 440:\penalty0 1322--1344, May 2014.
\newblock \doi{10.1093/mnras/stu145}.

\bibitem[Izard et~al.(2016)Izard, Crocce, and Fosalba]{Izard:2015dja}
Albert Izard, Martin Crocce, and Pablo Fosalba.
\newblock {ICE-COLA: Towards fast and accurate synthetic galaxy catalogues
  optimizing a quasi $N$-body method}.
\newblock \emph{Mon. Not. Roy. Astron. Soc.}, 459\penalty0 (3):\penalty0
  2327--2341, 2016.
\newblock \doi{10.1093/mnras/stw797}.

\bibitem[Jain et~al.(2015)]{Jain:2015cpa}
B.~Jain et~al.
\newblock {The Whole is Greater than the Sum of the Parts: Optimizing the Joint
  Science Return from LSST, Euclid and WFIRST}.
\newblock 2015.

\bibitem[{Joyce} et~al.(2015){Joyce}, {Jain}, {Khoury}, and {Trodden}]{jjk15}
A.~{Joyce}, B.~{Jain}, J.~{Khoury}, and M.~{Trodden}.
\newblock {Beyond the cosmological standard model}.
\newblock \emph{\physrep}, 568:\penalty0 1--98, March 2015.
\newblock \doi{10.1016/j.physrep.2014.12.002}.

\bibitem[Joyce et~al.(2016)Joyce, Lombriser, and Schmidt]{Joyce:2016vqv}
Austin Joyce, Lucas Lombriser, and Fabian Schmidt.
\newblock {Dark Energy Versus Modified Gravity}.
\newblock \emph{Ann. Rev. Nucl. Part. Sci.}, 66:\penalty0 95--122, 2016.
\newblock \doi{10.1146/annurev-nucl-102115-044553}.

\bibitem[{Kalus} et~al.(2016){Kalus}, {Percival}, {Bacon}, and
  {Samushia}]{2016MNRAS.463..467K}
B.~{Kalus}, W.~J. {Percival}, D.~J. {Bacon}, and L.~{Samushia}.
\newblock {Unbiased contaminant removal for 3D galaxy power spectrum
  measurements}.
\newblock \emph{\mnras}, 463:\penalty0 467--476, November 2016.
\newblock \doi{10.1093/mnras/stw2008}.

\bibitem[{Kannawadi} et~al.(2016){Kannawadi}, {Shapiro}, {Mandelbaum},
  {Hirata}, {Kruk}, and {Rhodes}]{Kannawadi2016}
A.~{Kannawadi}, C.~A. {Shapiro}, R.~{Mandelbaum}, C.~M. {Hirata}, J.~W. {Kruk},
  and J.~D. {Rhodes}.
\newblock {The Impact of Interpixel Capacitance in CMOS Detectors on PSF Shapes
  and Implications for WFIRST}.
\newblock \emph{\pasp}, 128\penalty0 (9):\penalty0 095001, September 2016.
\newblock \doi{10.1088/1538-3873/128/967/095001}.

\bibitem[{Kitching} et~al.(2012){Kitching}, {Balan}, {Bridle}, {Cantale},
  {Courbin}, {Eifler}, {Gentile}, {Gill}, {Harmeling}, {Heymans}, {Hirsch},
  {Honscheid}, {Kacprzak}, {Kirkby}, {Margala}, {Massey}, {Melchior},
  {Nurbaeva}, {Patton}, {Rhodes}, {Rowe}, {Taylor}, {Tewes}, {Viola},
  {Witherick}, {Voigt}, {Young}, and {Zuntz}]{Kitching2012}
T.~D. {Kitching}, S.~T. {Balan}, S.~{Bridle}, N.~{Cantale}, F.~{Courbin},
  T.~{Eifler}, M.~{Gentile}, M.~S.~S. {Gill}, S.~{Harmeling}, C.~{Heymans},
  M.~{Hirsch}, K.~{Honscheid}, T.~{Kacprzak}, D.~{Kirkby}, D.~{Margala}, R.~J.
  {Massey}, P.~{Melchior}, G.~{Nurbaeva}, K.~{Patton}, J.~{Rhodes}, B.~T.~P.
  {Rowe}, A.~N. {Taylor}, M.~{Tewes}, M.~{Viola}, D.~{Witherick}, L.~{Voigt},
  J.~{Young}, and J.~{Zuntz}.
\newblock {Image analysis for cosmology: results from the GREAT10 Galaxy
  Challenge}.
\newblock \emph{\mnras}, 423:\penalty0 3163--3208, July 2012.
\newblock \doi{10.1111/j.1365-2966.2012.21095.x}.

\bibitem[{Kobayashi} et~al.(2015){Kobayashi}, {Leauthaud}, {More}, {Okabe},
  {Laigle}, {Rhodes}, and {Takeuchi}]{2015MNRAS.449.2128K}
M.~I.~N. {Kobayashi}, A.~{Leauthaud}, S.~{More}, N.~{Okabe}, C.~{Laigle},
  J.~{Rhodes}, and T.~T. {Takeuchi}.
\newblock {Can we use weak lensing to measure total mass profiles of galaxies
  on 20 kpc scales?}
\newblock \emph{\mnras}, 449:\penalty0 2128--2143, May 2015.
\newblock \doi{10.1093/mnras/stv424}.

\bibitem[{Krause} and {Eifler}(2017)]{Krause2017}
E.~{Krause} and T.~{Eifler}.
\newblock {CosmoLike - Cosmological Likelihood Analyses for Photometric Galaxy
  Surveys}.
\newblock \emph{MNRAS accepted}, January 2017.

\bibitem[{Krause} et~al.(2013){Krause}, {Hirata}, {Martin}, {Neill}, and
  {Wyder}]{Krause2013}
E.~{Krause}, C.~M. {Hirata}, C.~{Martin}, J.~D. {Neill}, and T.~K. {Wyder}.
\newblock {Halo occupation distribution modelling of green valley galaxies}.
\newblock \emph{\mnras}, 428:\penalty0 2548--2564, January 2013.
\newblock \doi{10.1093/mnras/sts221}.

\bibitem[{Krause} et~al.(2016){Krause}, {Eifler}, and {Blazek}]{Krause2016}
E.~{Krause}, T.~{Eifler}, and J.~{Blazek}.
\newblock {The impact of intrinsic alignment on current and future cosmic shear
  surveys}.
\newblock \emph{\mnras}, 456:\penalty0 207--222, February 2016.
\newblock \doi{10.1093/mnras/stv2615}.

\bibitem[{Mandelbaum} et~al.(2015){Mandelbaum}, {Rowe}, {Armstrong}, {Bard},
  {Bertin}, {Bosch}, {Boutigny}, {Courbin}, {Dawson}, {Donnarumma}, {Fenech
  Conti}, {Gavazzi}, {Gentile}, {Gill}, {Hogg}, {Huff}, {Jee}, {Kacprzak},
  {Kilbinger}, {Kuntzer}, {Lang}, {Luo}, {March}, {Marshall}, {Meyers},
  {Miller}, {Miyatake}, {Nakajima}, {Ngol{\'e} Mboula}, {Nurbaeva}, {Okura},
  {Paulin-Henriksson}, {Rhodes}, {Schneider}, {Shan}, {Sheldon}, {Simet},
  {Starck}, {Sureau}, {Tewes}, {Zarb Adami}, {Zhang}, and
  {Zuntz}]{Mandelbaum2015}
R.~{Mandelbaum}, B.~{Rowe}, R.~{Armstrong}, D.~{Bard}, E.~{Bertin}, J.~{Bosch},
  D.~{Boutigny}, F.~{Courbin}, W.~A. {Dawson}, A.~{Donnarumma}, I.~{Fenech
  Conti}, R.~{Gavazzi}, M.~{Gentile}, M.~S.~S. {Gill}, D.~W. {Hogg}, E.~M.
  {Huff}, M.~J. {Jee}, T.~{Kacprzak}, M.~{Kilbinger}, T.~{Kuntzer}, D.~{Lang},
  W.~{Luo}, M.~C. {March}, P.~J. {Marshall}, J.~E. {Meyers}, L.~{Miller},
  H.~{Miyatake}, R.~{Nakajima}, F.~M. {Ngol{\'e} Mboula}, G.~{Nurbaeva},
  Y.~{Okura}, S.~{Paulin-Henriksson}, J.~{Rhodes}, M.~D. {Schneider},
  H.~{Shan}, E.~S. {Sheldon}, M.~{Simet}, J.-L. {Starck}, F.~{Sureau},
  M.~{Tewes}, K.~{Zarb Adami}, J.~{Zhang}, and J.~{Zuntz}.
\newblock {GREAT3 results - I. Systematic errors in shear estimation and the
  impact of real galaxy morphology}.
\newblock \emph{\mnras}, 450:\penalty0 2963--3007, July 2015.
\newblock \doi{10.1093/mnras/stv781}.

\bibitem[{Massey} et~al.(2007){Massey}, {Heymans}, {Berg{\'e}}, {Bernstein},
  {Bridle}, {Clowe}, {Dahle}, {Ellis}, {Erben}, {Hetterscheidt}, {High},
  {Hirata}, {Hoekstra}, {Hudelot}, {Jarvis}, {Johnston}, {Kuijken},
  {Margoniner}, {Mandelbaum}, {Mellier}, {Nakajima}, {Paulin-Henriksson},
  {Peeples}, {Roat}, {Refregier}, {Rhodes}, {Schrabback}, {Schirmer}, {Seljak},
  {Semboloni}, and {van Waerbeke}]{Massey2007}
R.~{Massey}, C.~{Heymans}, J.~{Berg{\'e}}, G.~{Bernstein}, S.~{Bridle},
  D.~{Clowe}, H.~{Dahle}, R.~{Ellis}, T.~{Erben}, M.~{Hetterscheidt}, F.~W.
  {High}, C.~{Hirata}, H.~{Hoekstra}, P.~{Hudelot}, M.~{Jarvis}, D.~{Johnston},
  K.~{Kuijken}, V.~{Margoniner}, R.~{Mandelbaum}, Y.~{Mellier}, R.~{Nakajima},
  S.~{Paulin-Henriksson}, M.~{Peeples}, C.~{Roat}, A.~{Refregier}, J.~{Rhodes},
  T.~{Schrabback}, M.~{Schirmer}, U.~{Seljak}, E.~{Semboloni}, and L.~{van
  Waerbeke}.
\newblock {The Shear Testing Programme 2: Factors affecting high-precision
  weak-lensing analyses}.
\newblock \emph{\mnras}, 376:\penalty0 13--38, March 2007.
\newblock \doi{10.1111/j.1365-2966.2006.11315.x}.

\bibitem[{Masters} et~al.(2015){Masters}, {Capak}, {Stern}, {Ilbert},
  {Salvato}, {Schmidt}, {Longo}, {Rhodes}, {Paltani}, {Mobasher}, {Hoekstra},
  {Hildebrandt}, {Coupon}, {Steinhardt}, {Speagle}, {Faisst}, {Kalinich},
  {Brodwin}, {Brescia}, and {Cavuoti}]{Masters2015}
D.~{Masters}, P.~{Capak}, D.~{Stern}, O.~{Ilbert}, M.~{Salvato}, S.~{Schmidt},
  G.~{Longo}, J.~{Rhodes}, S.~{Paltani}, B.~{Mobasher}, H.~{Hoekstra},
  H.~{Hildebrandt}, J.~{Coupon}, C.~{Steinhardt}, J.~{Speagle}, A.~{Faisst},
  A.~{Kalinich}, M.~{Brodwin}, M.~{Brescia}, and S.~{Cavuoti}.
\newblock {Mapping the Galaxy Color-Redshift Relation: Optimal Photometric
  Redshift Calibration Strategies for Cosmology Surveys}.
\newblock \emph{\apj}, 813:\penalty0 53, November 2015.
\newblock \doi{10.1088/0004-637X/813/1/53}.

\bibitem[{Masters} et~al.(2016){Masters}, {Faisst}, and {Capak}]{Masters2016}
D.~{Masters}, A.~{Faisst}, and P.~{Capak}.
\newblock {A Tight Relation between N/O Ratio and Galaxy Stellar Mass Can
  Explain the Evolution of Strong Emission Line Ratios with Redshift}.
\newblock \emph{\apj}, 828:\penalty0 18, September 2016.
\newblock \doi{10.3847/0004-637X/828/1/18}.

\bibitem[{Masters} et~al.(2017){Masters}, {Stern}, {Cohen}, {Capak}, {Rhodes},
  {Castander}, and {Paltani}]{Masters2017}
D.~{Masters}, D.~{Stern}, J.~{Cohen}, P.~{Capak}, J.~{Rhodes}, F.~{Castander},
  and S.~{Paltani}.
\newblock {The Complete Calibration of the Color-Redshift Relation (C3R2)
  Survey: Survey Overview and Data Release 1}.
\newblock \emph{ArXiv e-prints}, April 2017.

\bibitem[{Mehta} et~al.(2015){Mehta}, {Scarlata}, {Colbert}, {Dai}, {Dressler},
  {Henry}, {Malkan}, {Rafelski}, {Siana}, {Teplitz}, {Bagley}, {Beck}, {Ross},
  {Rutkowski}, and {Wang}]{Mehta:2015}
V.~{Mehta}, C.~{Scarlata}, J.~W. {Colbert}, Y.~S. {Dai}, A.~{Dressler},
  A.~{Henry}, M.~{Malkan}, M.~{Rafelski}, B.~{Siana}, H.~I. {Teplitz},
  M.~{Bagley}, M.~{Beck}, N.~R. {Ross}, M.~{Rutkowski}, and Y.~{Wang}.
\newblock {Predicting the Redshift 2 H{$\alpha$} Luminosity Function Using
  [OIII] Emission Line Galaxies}.
\newblock \emph{\apj}, 811:\penalty0 141, October 2015.
\newblock \doi{10.1088/0004-637X/811/2/141}.

\bibitem[Melchior and Goulding(2016)]{Melchior:2016asy}
Peter Melchior and Andy~D. Goulding.
\newblock {Filling the gaps: Gaussian mixture models from noisy, truncated or
  incomplete samples}.
\newblock 2016.

\bibitem[{Merson} et~al.(2018){Merson}, {Wang}, {Benson}, {Faisst}, {Masters},
  {Kiessling}, and {Rhodes}]{Merson2018}
A.~{Merson}, Y.~{Wang}, A.~{Benson}, A.~{Faisst}, D.~{Masters}, A.~{Kiessling},
  and J.~{Rhodes}.
\newblock {Predicting H{$\alpha$} emission-line galaxy counts for future galaxy
  redshift surveys}.
\newblock \emph{\mnras}, 474:\penalty0 177--196, February 2018.
\newblock \doi{10.1093/mnras/stx2649}.

\bibitem[{Navarro} et~al.(1997){Navarro}, {Frenk}, and {White}]{NFW}
J.~F. {Navarro}, C.~S. {Frenk}, and S.~D.~M. {White}.
\newblock {A Universal Density Profile from Hierarchical Clustering}.
\newblock \emph{\apj}, 490:\penalty0 493--+, December 1997.
\newblock \doi{10.1086/304888}.

\bibitem[{Nayyeri} et~al.(2017){Nayyeri}, {Hemmati}, {Mobasher}, {Ferguson},
  {Cooray}, {Barro}, {Faber}, {Dickinson}, {Koekemoer}, {Peth}, {Salvato},
  {Ashby}, {Darvish}, {Donley}, {Durbin}, {Finkelstein}, {Fontana}, {Grogin},
  {Gruetzbauch}, {Huang}, {Khostovan}, {Kocevski}, {Kodra}, {Lee}, {Newman},
  {Pacifici}, {Pforr}, {Stefanon}, {Wiklind}, {Willner}, {Wuyts}, {Castellano},
  {Conselice}, {Dolch}, {Dunlop}, {Galametz}, {Hathi}, {Lucas}, and
  {Yan}]{Nayyeri2017}
H.~{Nayyeri}, S.~{Hemmati}, B.~{Mobasher}, H.~C. {Ferguson}, A.~{Cooray},
  G.~{Barro}, S.~M. {Faber}, M.~{Dickinson}, A.~M. {Koekemoer}, M.~{Peth},
  M.~{Salvato}, M.~L.~N. {Ashby}, B.~{Darvish}, J.~{Donley}, M.~{Durbin},
  S.~{Finkelstein}, A.~{Fontana}, N.~A. {Grogin}, R.~{Gruetzbauch}, K.~{Huang},
  A.~A. {Khostovan}, D.~{Kocevski}, D.~{Kodra}, B.~{Lee}, J.~{Newman},
  C.~{Pacifici}, J.~{Pforr}, M.~{Stefanon}, T.~{Wiklind}, S.~P. {Willner},
  S.~{Wuyts}, M.~{Castellano}, C.~{Conselice}, T.~{Dolch}, J.~S. {Dunlop},
  A.~{Galametz}, N.~P. {Hathi}, R.~A. {Lucas}, and H.~{Yan}.
\newblock {CANDELS Multi-wavelength Catalogs: Source Identification and
  Photometry in the CANDELS COSMOS Survey Field}.
\newblock \emph{\apjs}, 228:\penalty0 7, January 2017.
\newblock \doi{10.3847/1538-4365/228/1/7}.

\bibitem[O'Connell et~al.(2016)O'Connell, Eisenstein, Vargas, Ho, and
  Padmanabhan]{OConnell:2015src}
Ross O'Connell, Daniel Eisenstein, Mariana Vargas, Shirley Ho, and Nikhil
  Padmanabhan.
\newblock {Large covariance matrices: smooth models from the two-point
  correlation function}.
\newblock \emph{Mon. Not. Roy. Astron. Soc.}, 462\penalty0 (3):\penalty0
  2681--2694, 2016.
\newblock \doi{10.1093/mnras/stw1821}.

\bibitem[{Padmanabhan} et~al.(2016){Padmanabhan}, {White}, {Zhou}, and
  {O'Connell}]{Padmanabhan2016}
N.~{Padmanabhan}, M.~{White}, H.~H. {Zhou}, and R.~{O'Connell}.
\newblock {Estimating sparse precision matrices}.
\newblock \emph{\mnras}, 460:\penalty0 1567--1576, August 2016.
\newblock \doi{10.1093/mnras/stw1042}.

\bibitem[{Pearson} and {Samushia}(2016)]{2016MNRAS.457..993P}
D.~W. {Pearson} and L.~{Samushia}.
\newblock {Estimating the power spectrum covariance matrix with fewer mock
  samples}.
\newblock \emph{\mnras}, 457:\penalty0 993--999, March 2016.
\newblock \doi{10.1093/mnras/stw062}.

\bibitem[{Pearson} et~al.(2016){Pearson}, {Samushia}, and
  {Gagrani}]{2016MNRAS.463.2708P}
D.~W. {Pearson}, L.~{Samushia}, and P.~{Gagrani}.
\newblock {Optimal weights for measuring redshift space distortions in
  multitracer galaxy catalogues}.
\newblock \emph{\mnras}, 463:\penalty0 2708--2715, December 2016.
\newblock \doi{10.1093/mnras/stw2177}.

\bibitem[{Plazas} et~al.(2016){Plazas}, {Shapiro}, {Kannawadi}, {Mandelbaum},
  {Rhodes}, and {Smith}]{2016PASP..128j4001P}
A.~A. {Plazas}, C.~{Shapiro}, A.~{Kannawadi}, R.~{Mandelbaum}, J.~{Rhodes}, and
  R.~{Smith}.
\newblock {The Effect of Detector Nonlinearity on WFIRST PSF Profiles for Weak
  Gravitational Lensing Measurements}.
\newblock \emph{\pasp}, 128\penalty0 (10):\penalty0 104001, October 2016.
\newblock \doi{10.1088/1538-3873/128/968/104001}.

\bibitem[{Plazas} et~al.(2017{\natexlab{a}}){Plazas}, {Shapiro}, {Smith},
  {Rhodes}, and {Huff}]{2017JInst..12C4009P}
A.~A. {Plazas}, C.~{Shapiro}, R.~{Smith}, J.~{Rhodes}, and E.~{Huff}.
\newblock {Nonlinearity and pixel shifting effects in HXRG infrared detectors}.
\newblock \emph{Journal of Instrumentation}, 12:\penalty0 C04009, April
  2017{\natexlab{a}}.
\newblock \doi{10.1088/1748-0221/12/04/C04009}.

\bibitem[{Plazas} et~al.(2017{\natexlab{b}}){Plazas}, {Shapiro}, {Smith},
  {Huff}, and {Rhodes}]{2017arXiv171206642P}
A.~A. {Plazas}, C.~A. {Shapiro}, R.~{Smith}, E.~{Huff}, and J.~{Rhodes}.
\newblock {Laboratory measurement of the brighter-fatter effect in an H2RG
  infrared detector}.
\newblock \emph{ArXiv e-prints}, December 2017{\natexlab{b}}.

\bibitem[{Pozzetti} et~al.(2016){Pozzetti}, {Hirata}, {Geach}, {Cimatti},
  {Baugh}, {Cucciati}, {Merson}, {Norberg}, and {Shi}]{Pozzetti:2016}
L.~{Pozzetti}, C.~M. {Hirata}, J.~E. {Geach}, A.~{Cimatti}, C.~{Baugh},
  O.~{Cucciati}, A.~{Merson}, P.~{Norberg}, and D.~{Shi}.
\newblock {Modelling the number density of H{$\alpha$} emitters for future
  spectroscopic near-IR space missions}.
\newblock \emph{\aap}, 590:\penalty0 A3, May 2016.
\newblock \doi{10.1051/0004-6361/201527081}.

\bibitem[{Rauscher}(2015)]{Rauscher2015}
B.~J. {Rauscher}.
\newblock {Teledyne H1RG, H2RG, and H4RG Noise Generator}.
\newblock \emph{\pasp}, 127:\penalty0 1144, November 2015.
\newblock \doi{10.1086/684082}.

\bibitem[{Refregier} et~al.(2011){Refregier}, {Amara}, {Kitching}, and
  {Rassat}]{rak11}
A.~{Refregier}, A.~{Amara}, T.~D. {Kitching}, and A.~{Rassat}.
\newblock {iCosmo: an interactive cosmology package}.
\newblock \emph{\aap}, 528:\penalty0 A33+, April 2011.
\newblock \doi{10.1051/0004-6361/200811112}.

\bibitem[{Rowe} et~al.(2015){Rowe}, {Jarvis}, {Mandelbaum}, {Bernstein},
  {Bosch}, {Simet}, {Meyers}, {Kacprzak}, {Nakajima}, {Zuntz}, {Miyatake},
  {Dietrich}, {Armstrong}, {Melchior}, and {Gill}]{Rowe:2015}
B.~T.~P. {Rowe}, M.~{Jarvis}, R.~{Mandelbaum}, G.~M. {Bernstein}, J.~{Bosch},
  M.~{Simet}, J.~E. {Meyers}, T.~{Kacprzak}, R.~{Nakajima}, J.~{Zuntz},
  H.~{Miyatake}, J.~P. {Dietrich}, R.~{Armstrong}, P.~{Melchior}, and M.~S.~S.
  {Gill}.
\newblock {GALSIM: The modular galaxy image simulation toolkit}.
\newblock \emph{Astronomy and Computing}, 10:\penalty0 121--150, April 2015.
\newblock \doi{10.1016/j.ascom.2015.02.002}.

\bibitem[{Sato} et~al.(2009){Sato}, {Hamana}, {Takahashi}, {Takada}, {Yoshida},
  {Matsubara}, and {Sugiyama}]{sht09}
M.~{Sato}, T.~{Hamana}, R.~{Takahashi}, M.~{Takada}, N.~{Yoshida},
  T.~{Matsubara}, and N.~{Sugiyama}.
\newblock {Simulations of Wide-Field Weak Lensing Surveys. I. Basic Statistics
  and Non-Gaussian Effects}.
\newblock \emph{\apj}, 701:\penalty0 945--954, August 2009.
\newblock \doi{10.1088/0004-637X/701/2/945}.

\bibitem[Schaan et~al.(2016)Schaan, Krause, Eifler, Dor\'é, Miyatake, Rhodes,
  and Spergel]{Schaan:2016ois}
Emmanuel Schaan, Elisabeth Krause, Tim Eifler, Olivier Dor\'é, Hironao
  Miyatake, Jason Rhodes, and David~N. Spergel.
\newblock {Looking through the same lens: shear calibration for LSST, Euclid \&
  WFIRST with stage 4 CMB lensing}.
\newblock 2016.

\bibitem[Schneider et~al.(2015)Schneider, Hogg, Marshall, Dawson, Meyers, Bard,
  and Lang]{Schneider:2014rha}
Michael~D. Schneider, David~W. Hogg, Philip~J. Marshall, William~A. Dawson,
  Joshua Meyers, Deborah~J. Bard, and Dustin Lang.
\newblock {Hierarchical probabilistic inference of cosmic shear}.
\newblock \emph{Astrophys. J.}, 807\penalty0 (1):\penalty0 87, 2015.
\newblock \doi{10.1088/0004-637X/807/1/87}.

\bibitem[{Schrabback} et~al.(2010){Schrabback}, {Hartlap}, {Joachimi},
  {Kilbinger}, {Simon}, {Benabed}, {Brada{\v c}}, {Eifler}, {Erben},
  {Fassnacht}, {High}, {Hilbert}, {Hildebrandt}, {Hoekstra}, {Kuijken},
  {Marshall}, {Mellier}, {Morganson}, {Schneider}, {Semboloni}, {van Waerbeke},
  and {Velander}]{shj10}
T.~{Schrabback}, J.~{Hartlap}, B.~{Joachimi}, M.~{Kilbinger}, P.~{Simon},
  K.~{Benabed}, M.~{Brada{\v c}}, T.~{Eifler}, T.~{Erben}, C.~D. {Fassnacht},
  F.~W. {High}, S.~{Hilbert}, H.~{Hildebrandt}, H.~{Hoekstra}, K.~{Kuijken},
  P.~J. {Marshall}, Y.~{Mellier}, E.~{Morganson}, P.~{Schneider},
  E.~{Semboloni}, L.~{van Waerbeke}, and M.~{Velander}.
\newblock {Evidence of the accelerated expansion of the Universe from weak
  lensing tomography with COSMOS}.
\newblock \emph{\aap}, 516:\penalty0 A63, June 2010.
\newblock \doi{10.1051/0004-6361/200913577}.

\bibitem[{Seljak}(2000)]{Seljak00}
U.~{Seljak}.
\newblock {Analytic model for galaxy and dark matter clustering}.
\newblock \emph{\mnras}, 318:\penalty0 203--213, October 2000.
\newblock \doi{10.1046/j.1365-8711.2000.03715.x}.

\bibitem[{Sellentin} and {Heavens}(2016)]{seh16}
E.~{Sellentin} and A.~F. {Heavens}.
\newblock {Parameter inference with estimated covariance matrices}.
\newblock \emph{\mnras}, 456:\penalty0 L132--L136, February 2016.
\newblock \doi{10.1093/mnrasl/slv190}.

\bibitem[{Seo} and {Eisenstein}(2003)]{Seo03}
H.-J. {Seo} and D.~J. {Eisenstein}.
\newblock {Probing Dark Energy with Baryonic Acoustic Oscillations from Future
  Large Galaxy Redshift Surveys}.
\newblock \emph{\apj}, 598:\penalty0 720--740, December 2003.
\newblock \doi{10.1086/379122}.

\bibitem[{Simet} and {Mandelbaum}(2015)]{2015MNRAS.449.1259S}
M.~{Simet} and R.~{Mandelbaum}.
\newblock {Background sky obscuration by cluster galaxies as a source of
  systematic error for weak lensing}.
\newblock \emph{\mnras}, 449:\penalty0 1259--1269, May 2015.
\newblock \doi{10.1093/mnras/stv313}.

\bibitem[Simpson et~al.(2013)]{Simpson:2012ra}
Fergus Simpson et~al.
\newblock {CFHTLenS: Testing the Laws of Gravity with Tomographic Weak Lensing
  and Redshift Space Distortions}.
\newblock \emph{Mon. Not. Roy. Astron. Soc.}, 429:\penalty0 2249, 2013.
\newblock \doi{10.1093/mnras/sts493}.

\bibitem[{Smith} et~al.(2003){Smith}, {Peacock}, {Jenkins}, {White}, {Frenk},
  {Pearce}, {Thomas}, {Efstathiou}, and {Couchman}]{smp03}
R.~E. {Smith}, J.~A. {Peacock}, A.~{Jenkins}, S.~D.~M. {White}, C.~S. {Frenk},
  F.~R. {Pearce}, P.~A. {Thomas}, G.~{Efstathiou}, and H.~M.~P. {Couchman}.
\newblock {Stable clustering, the halo model and non-linear cosmological power
  spectra}.
\newblock \emph{\mnras}, 341:\penalty0 1311--1332, June 2003.
\newblock \doi{10.1046/j.1365-8711.2003.06503.x}.

\bibitem[{Sobral} et~al.(2013){Sobral}, {Smail}, {Best}, {Geach}, {Matsuda},
  {Stott}, {Cirasuolo}, and {Kurk}]{Sobral13}
D.~{Sobral}, I.~{Smail}, P.~N. {Best}, J.~E. {Geach}, Y.~{Matsuda}, J.~P.
  {Stott}, M.~{Cirasuolo}, and J.~{Kurk}.
\newblock {A large H{$\alpha$} survey at z = 2.23, 1.47, 0.84 and 0.40: the 11
  Gyr evolution of star-forming galaxies from HiZELS}.
\newblock \emph{\mnras}, 428:\penalty0 1128--1146, January 2013.
\newblock \doi{10.1093/mnras/sts096}.

\bibitem[{Spergel} et~al.(2013){Spergel}, {Gehrels}, {Breckinridge}, {Donahue},
  {Dressler}, {Gaudi}, {Greene}, {Guyon}, {Hirata}, {Kalirai}, {Kasdin},
  {Moos}, {Perlmutter}, {Postman}, {Rauscher}, {Rhodes}, {Wang}, {Weinberg},
  {Centrella}, {Traub}, {Baltay}, {Colbert}, {Bennett}, {Kiessling},
  {Macintosh}, {Merten}, {Mortonson}, {Penny}, {Rozo}, {Savransky},
  {Stapelfeldt}, {Zu}, {Baker}, {Cheng}, {Content}, {Dooley}, {Foote},
  {Goullioud}, {Grady}, {Jackson}, {Kruk}, {Levine}, {Melton}, {Peddie},
  {Ruffa}, and {Shaklan}]{Spergel2013}
D.~{Spergel}, N.~{Gehrels}, J.~{Breckinridge}, M.~{Donahue}, A.~{Dressler},
  B.~S. {Gaudi}, T.~{Greene}, O.~{Guyon}, C.~{Hirata}, J.~{Kalirai}, N.~J.
  {Kasdin}, W.~{Moos}, S.~{Perlmutter}, M.~{Postman}, B.~{Rauscher},
  J.~{Rhodes}, Y.~{Wang}, D.~{Weinberg}, J.~{Centrella}, W.~{Traub},
  C.~{Baltay}, J.~{Colbert}, D.~{Bennett}, A.~{Kiessling}, B.~{Macintosh},
  J.~{Merten}, M.~{Mortonson}, M.~{Penny}, E.~{Rozo}, D.~{Savransky},
  K.~{Stapelfeldt}, Y.~{Zu}, C.~{Baker}, E.~{Cheng}, D.~{Content}, J.~{Dooley},
  M.~{Foote}, R.~{Goullioud}, K.~{Grady}, C.~{Jackson}, J.~{Kruk}, M.~{Levine},
  M.~{Melton}, C.~{Peddie}, J.~{Ruffa}, and S.~{Shaklan}.
\newblock {Wide-Field InfraRed Survey Telescope-Astrophysics Focused Telescope
  Assets WFIRST-AFTA Final Report}.
\newblock \emph{ArXiv e-prints}, May 2013.

\bibitem[{Spergel} et~al.(2015){Spergel}, {Gehrels}, {Baltay}, {Bennett},
  {Breckinridge}, {Donahue}, {Dressler}, {Gaudi}, {Greene}, {Guyon}, {Hirata},
  {Kalirai}, {Kasdin}, {Macintosh}, {Moos}, {Perlmutter}, {Postman},
  {Rauscher}, {Rhodes}, {Wang}, {Weinberg}, {Benford}, {Hudson}, {Jeong},
  {Mellier}, {Traub}, {Yamada}, {Capak}, {Colbert}, {Masters}, {Penny},
  {Savransky}, {Stern}, {Zimmerman}, {Barry}, {Bartusek}, {Carpenter}, {Cheng},
  {Content}, {Dekens}, {Demers}, {Grady}, {Jackson}, {Kuan}, {Kruk}, {Melton},
  {Nemati}, {Parvin}, {Poberezhskiy}, {Peddie}, {Ruffa}, {Wallace}, {Whipple},
  {Wollack}, and {Zhao}]{Spergel2015}
D.~{Spergel}, N.~{Gehrels}, C.~{Baltay}, D.~{Bennett}, J.~{Breckinridge},
  M.~{Donahue}, A.~{Dressler}, B.~S. {Gaudi}, T.~{Greene}, O.~{Guyon},
  C.~{Hirata}, J.~{Kalirai}, N.~J. {Kasdin}, B.~{Macintosh}, W.~{Moos},
  S.~{Perlmutter}, M.~{Postman}, B.~{Rauscher}, J.~{Rhodes}, Y.~{Wang},
  D.~{Weinberg}, D.~{Benford}, M.~{Hudson}, W.-S. {Jeong}, Y.~{Mellier},
  W.~{Traub}, T.~{Yamada}, P.~{Capak}, J.~{Colbert}, D.~{Masters}, M.~{Penny},
  D.~{Savransky}, D.~{Stern}, N.~{Zimmerman}, R.~{Barry}, L.~{Bartusek},
  K.~{Carpenter}, E.~{Cheng}, D.~{Content}, F.~{Dekens}, R.~{Demers},
  K.~{Grady}, C.~{Jackson}, G.~{Kuan}, J.~{Kruk}, M.~{Melton}, B.~{Nemati},
  B.~{Parvin}, I.~{Poberezhskiy}, C.~{Peddie}, J.~{Ruffa}, J.~K. {Wallace},
  A.~{Whipple}, E.~{Wollack}, and F.~{Zhao}.
\newblock {Wide-Field InfrarRed Survey Telescope-Astrophysics Focused Telescope
  Assets WFIRST-AFTA 2015 Report}.
\newblock \emph{ArXiv e-prints}, March 2015.

\bibitem[{Springel} et~al.(2005){Springel}, {White}, {Jenkins}, {Frenk},
  {Yoshida}, {Gao}, {Navarro}, {Thacker}, {Croton}, {Helly}, {Peacock},
  {Cole}\, {Thomas}, {Couchman}, {Evrard}, {Colberg}, and {Pearce}]{Springel05}
V.~{Springel}, S.~D.~M. {White}, A.~{Jenkins}, C.~S. {Frenk}, N.~{Yoshida},
  L.~{Gao}, J.~{Navarro}, R.~{Thacker}, D.~{Croton}, J.~{Helly}, J.~A.
  {Peacock}, S.~{Cole}\, P.~{Thomas}, H.~{Couchman}, A.~{Evrard}, J.~{Colberg},
  and F.~{Pearce}.
\newblock {Simulations of the formation, evolution and clustering of galaxies
  and quasars}.
\newblock \emph{\nat}, 435:\penalty0 629--636, June 2005.
\newblock \doi{10.1038/nature03597}.

\bibitem[{Stefanon} et~al.(2017){Stefanon}, {Yan}, {Mobasher}, {Barro},
  {Donley}, {Fontana}, {Hemmati}, {Koekemoer}, {Lee}, {Lee}, {Nayyeri}, {Peth},
  {Pforr}, {Salvato}, {Wiklind}, {Wuyts}, {Ashby}, {Castellano}, {Conselice},
  {Cooper}, {Cooray}, {Dolch}, {Ferguson}, {Galametz}, {Giavalisco}, {Guo},
  {Willner}, {Dickinson}, {Faber}, {Fazio}, {Gardner}, {Gawiser}, {Grazian},
  {Grogin}, {Kocevski}, {Koo}, {Lee}, {Lucas}, {McGrath}, {Nandra}, {Newman},
  and {van der Wel}]{Stefanon2017}
M.~{Stefanon}, H.~{Yan}, B.~{Mobasher}, G.~{Barro}, J.~L. {Donley},
  A.~{Fontana}, S.~{Hemmati}, A.~M. {Koekemoer}, B.~{Lee}, S.-K. {Lee},
  H.~{Nayyeri}, M.~{Peth}, J.~{Pforr}, M.~{Salvato}, T.~{Wiklind}, S.~{Wuyts},
  M.~L.~N. {Ashby}, M.~{Castellano}, C.~J. {Conselice}, M.~C. {Cooper}, A.~R.
  {Cooray}, T.~{Dolch}, H.~{Ferguson}, A.~{Galametz}, M.~{Giavalisco},
  Y.~{Guo}, S.~P. {Willner}, M.~E. {Dickinson}, S.~M. {Faber}, G.~G. {Fazio},
  J.~P. {Gardner}, E.~{Gawiser}, A.~{Grazian}, N.~A. {Grogin}, D.~{Kocevski},
  D.~C. {Koo}, K.-S. {Lee}, R.~A. {Lucas}, E.~J. {McGrath}, K.~{Nandra}, J.~A.
  {Newman}, and A.~{van der Wel}.
\newblock {CANDELS Multi-wavelength Catalogs: Source Identification and
  Photometry in the CANDELS Extended Groth Strip}.
\newblock \emph{\apjs}, 229:\penalty0 32, April 2017.
\newblock \doi{10.3847/1538-4365/aa66cb}.

\bibitem[{Stickley} et~al.(2016){Stickley}, {Capak}, {Masters}, {de Putter},
  {Dor{\'e}}, and {Bock}]{stickley2016}
N.~R. {Stickley}, P.~{Capak}, D.~{Masters}, R.~{de Putter}, O.~{Dor{\'e}}, and
  J.~{Bock}.
\newblock {An Empirical Approach to Cosmological Galaxy Survey Simulation:
  Application to SPHEREx Low-Resolution Spectroscopy}.
\newblock \emph{ArXiv e-prints}, June 2016.

\bibitem[{Takada} and {Bridle}(2007)]{tb07}
M.~{Takada} and S.~{Bridle}.
\newblock {Probing dark energy with cluster counts and cosmic shear power
  spectra: including the full covariance}.
\newblock \emph{New Journal of Physics}, 9:\penalty0 446, December 2007.
\newblock \doi{10.1088/1367-2630/9/12/446}.

\bibitem[{Takada} and {Jain}(2009)]{taj09}
M.~{Takada} and B.~{Jain}.
\newblock {The impact of non-Gaussian errors on weak lensing surveys}.
\newblock \emph{\mnras}, 395:\penalty0 2065--2086, June 2009.
\newblock \doi{10.1111/j.1365-2966.2009.14504.x}.

\bibitem[{Takahashi} et~al.(2012){Takahashi}, {Sato}, {Nishimichi}, {Taruya},
  and {Oguri}]{tsn12}
R.~{Takahashi}, M.~{Sato}, T.~{Nishimichi}, A.~{Taruya}, and M.~{Oguri}.
\newblock {Revising the Halofit Model for the Nonlinear Matter Power Spectrum}.
\newblock \emph{\apj}, 761:\penalty0 152, December 2012.
\newblock \doi{10.1088/0004-637X/761/2/152}.

\bibitem[Tassev et~al.(2013)Tassev, Zaldarriaga, and Eisenstein]{Tassev:2013pn}
Svetlin Tassev, Matias Zaldarriaga, and Daniel Eisenstein.
\newblock {Solving Large Scale Structure in Ten Easy Steps with COLA}.
\newblock \emph{JCAP}, 1306:\penalty0 036, 2013.
\newblock \doi{10.1088/1475-7516/2013/06/036}.

\bibitem[{The Dark Energy Survey Collaboration} et~al.(2015){The Dark Energy
  Survey Collaboration}, {Abbott}, {Abdalla}, {Allam}, {Amara}, {Annis},
  {Armstrong}, {Bacon}, {Banerji}, {Bauer}, {Baxter}, {Becker},
  {Benoit-L{\'e}vy}, {Bernstein}, {Bernstein}, {Bertin}, {Blazek}, {Bonnett},
  {Bridle}, {Brooks}, {Bruderer}, {Buckley-Geer}, {Burke}, {Busha}, {Capozzi},
  {Carnero Rosell}, {Carrasco Kind}, {Carretero}, {Castander}, {Chang},
  {Clampitt}, {Crocce}, {Cunha}, {D'Andrea}, {da Costa}, {Das}, {DePoy},
  {Desai}, {Diehl}, {Dietrich}, {Dodelson}, {Doel}, {Drlica-Wagner},
  {Efstathiou}, {Eifler}, {Erickson}, {Estrada}, {Evrard}, {Fausti Neto},
  {Fernandez}, {Finley}, {Flaugher}, {Fosalba}, {Friedrich}, {Frieman},
  {Gangkofner}, {Garcia-Bellido}, {Gaztanaga}, {Gerdes}, {Gruen}, {Gruendl},
  {Gutierrez}, {Hartley}, {Hirsch}, {Honscheid}, {Huff}, {Jain}, {James},
  {Jarvis}, {Kacprzak}, {Kent}, {Kirk}, {Krause}, {Kravtsov}, {Kuehn},
  {Kuropatkin}, {Kwan}, {Lahav}, {Leistedt}, {Li}, {Lima}, {Lin}, {MacCrann},
  {March}, {Marshall}, {Martini}, {McMahon}, {Melchior}, {Miller}, {Miquel},
  {Mohr}, {Neilsen}, {Nichol}, {Nicola}, {Nord}, {Ogando}, {Palmese}, {Peiris},
  {Plazas}, {Refregier}, {Roe}, {Romer}, {Roodman}, {Rowe}, {Rykoff}, {Sabiu},
  {Sadeh}, {Sako}, {Samuroff}, {S{\'a}nchez}, {Sanchez}, {Seo},
  {Sevilla-Noarbe}, {Sheldon}, {Smith}, {Soares-Santos}, {Sobreira}, {Suchyta},
  {Swanson}, {Tarle}, {Thaler}, {Thomas}, {Troxel}, {Vikram}, {Walker},
  {Wechsler}, {Weller}, {Zhang}, and {Zuntz}]{DES2015}
{The Dark Energy Survey Collaboration}, T.~{Abbott}, F.~B. {Abdalla},
  S.~{Allam}, A.~{Amara}, J.~{Annis}, R.~{Armstrong}, D.~{Bacon}, M.~{Banerji},
  A.~H. {Bauer}, E.~{Baxter}, M.~R. {Becker}, A.~{Benoit-L{\'e}vy}, R.~A.
  {Bernstein}, G.~M. {Bernstein}, E.~{Bertin}, J.~{Blazek}, C.~{Bonnett}, S.~L.
  {Bridle}, D.~{Brooks}, C.~{Bruderer}, E.~{Buckley-Geer}, D.~L. {Burke}, M.~T.
  {Busha}, D.~{Capozzi}, A.~{Carnero Rosell}, M.~{Carrasco Kind},
  J.~{Carretero}, F.~J. {Castander}, C.~{Chang}, J.~{Clampitt}, M.~{Crocce},
  C.~E. {Cunha}, C.~B. {D'Andrea}, L.~N. {da Costa}, R.~{Das}, D.~L. {DePoy},
  S.~{Desai}, H.~T. {Diehl}, J.~P. {Dietrich}, S.~{Dodelson}, P.~{Doel},
  A.~{Drlica-Wagner}, G.~{Efstathiou}, T.~F. {Eifler}, B.~{Erickson},
  J.~{Estrada}, A.~E. {Evrard}, A.~{Fausti Neto}, E.~{Fernandez}, D.~A.
  {Finley}, B.~{Flaugher}, P.~{Fosalba}, O.~{Friedrich}, J.~{Frieman},
  C.~{Gangkofner}, J.~{Garcia-Bellido}, E.~{Gaztanaga}, D.~W. {Gerdes},
  D.~{Gruen}, R.~A. {Gruendl}, G.~{Gutierrez}, W.~{Hartley}, M.~{Hirsch},
  K.~{Honscheid}, E.~M. {Huff}, B.~{Jain}, D.~J. {James}, M.~{Jarvis},
  T.~{Kacprzak}, S.~{Kent}, D.~{Kirk}, E.~{Krause}, A.~{Kravtsov}, K.~{Kuehn},
  N.~{Kuropatkin}, J.~{Kwan}, O.~{Lahav}, B.~{Leistedt}, T.~S. {Li}, M.~{Lima},
  H.~{Lin}, N.~{MacCrann}, M.~{March}, J.~L. {Marshall}, P.~{Martini}, R.~G.
  {McMahon}, P.~{Melchior}, C.~J. {Miller}, R.~{Miquel}, J.~J. {Mohr},
  E.~{Neilsen}, R.~C. {Nichol}, A.~{Nicola}, B.~{Nord}, R.~{Ogando},
  A.~{Palmese}, H.~V. {Peiris}, A.~A. {Plazas}, A.~{Refregier}, N.~{Roe}, A.~K.
  {Romer}, A.~{Roodman}, B.~{Rowe}, E.~S. {Rykoff}, C.~{Sabiu}, I.~{Sadeh},
  M.~{Sako}, S.~{Samuroff}, C.~{S{\'a}nchez}, E.~{Sanchez}, H.~{Seo},
  I.~{Sevilla-Noarbe}, E.~{Sheldon}, R.~C. {Smith}, M.~{Soares-Santos},
  F.~{Sobreira}, E.~{Suchyta}, M.~E.~C. {Swanson}, G.~{Tarle}, J.~{Thaler},
  D.~{Thomas}, M.~A. {Troxel}, V.~{Vikram}, A.~R. {Walker}, R.~H. {Wechsler},
  J.~{Weller}, Y.~{Zhang}, and J.~{Zuntz}.
\newblock {Cosmology from Cosmic Shear with DES Science Verification Data}.
\newblock \emph{ArXiv e-prints}, July 2015.

\bibitem[{Tinker} et~al.(2008){Tinker}, {Kravtsov}, {Klypin}, {Abazajian},
  {Warren}, {Yepes}, {Gottl{\"o}ber}, and {Holz}]{Tinker2008}
J.~{Tinker}, A.~V. {Kravtsov}, A.~{Klypin}, K.~{Abazajian}, M.~{Warren},
  G.~{Yepes}, S.~{Gottl{\"o}ber}, and D.~E. {Holz}.
\newblock {Toward a Halo Mass Function for Precision Cosmology: The Limits of
  Universality}.
\newblock \emph{\apj}, 688:\penalty0 709-728, December 2008.
\newblock \doi{10.1086/591439}.

\bibitem[{van Dokkum} et~al.(2011){van Dokkum}, {Brammer}, {Fumagalli},
  {Nelson}, {Franx}, {Rix}, {Kriek}, {Skelton}, {Patel}, {Schmidt}, {Bezanson},
  {Bian}, {da Cunha}, {Erb}, {Fan}, {F{\"o}rster Schreiber}, {Illingworth},
  {Labb{\'e}}, {Lundgren}, {Magee}, {Marchesini}, {McCarthy}, {Muzzin},
  {Quadri}, {Steidel}, {Tal}, {Wake}, {Whitaker}, and
  {Williams}]{VanDokkum:2011}
P.~G. {van Dokkum}, G.~{Brammer}, M.~{Fumagalli}, E.~{Nelson}, M.~{Franx},
  H.-W. {Rix}, M.~{Kriek}, R.~E. {Skelton}, S.~{Patel}, K.~B. {Schmidt},
  R.~{Bezanson}, F.~{Bian}, E.~{da Cunha}, D.~K. {Erb}, X.~{Fan},
  N.~{F{\"o}rster Schreiber}, G.~D. {Illingworth}, I.~{Labb{\'e}},
  B.~{Lundgren}, D.~{Magee}, D.~{Marchesini}, P.~{McCarthy}, A.~{Muzzin},
  R.~{Quadri}, C.~C. {Steidel}, T.~{Tal}, D.~{Wake}, K.~E. {Whitaker}, and
  A.~{Williams}.
\newblock {First Results from the 3D-HST Survey: The Striking Diversity of
  Massive Galaxies at z $>$ 1}.
\newblock \emph{\apjl}, 743:\penalty0 L15, December 2011.
\newblock \doi{10.1088/2041-8205/743/1/L15}.

\bibitem[{Wang}(2008)]{Wang08}
Y.~{Wang}.
\newblock {Differentiating dark energy and modified gravity with galaxy
  redshift surveys}.
\newblock \emph{\jcap}, 5:\penalty0 021, May 2008.
\newblock \doi{10.1088/1475-7516/2008/05/021}.

\bibitem[{Wang} et~al.(2013){Wang}, {Chuang}, and {Hirata}]{Wang2013}
Y.~{Wang}, C.-H. {Chuang}, and C.~M. {Hirata}.
\newblock {Towards more realistic forecasting of dark energy constraints from
  galaxy redshift surveys}.
\newblock \emph{\mnras}, 430:\penalty0 2446--2453, April 2013.
\newblock \doi{10.1093/mnras/stt068}.

\bibitem[Wong et~al.(2016)Wong, Pullen, and Ho]{Wong:2016eku}
Kaze Wong, Anthony Pullen, and Shirley Ho.
\newblock {Filtering interlopers from galaxy surveys}.
\newblock 2016.

\end{thebibliography}

\clearpage
\newpage

\end{document}